\documentclass[referee, usenatbib]{mn2e}

\usepackage{epsfig}
\usepackage{color}
\usepackage{rotating}
\usepackage{amssymb}
\usepackage{graphicx}
\usepackage{subfigure}
\usepackage{journals}
\usepackage{amsmath}
\usepackage{lscape}
\usepackage{array}
\usepackage{comment}
\usepackage{hyperref}

\DeclareRobustCommand{\ion}[2]{%
\relax\ifmmode
\ifx\testbx\f@series
{\mathbf{#1\,\mathsc{#2}}}\else
{\mathrm{#1\,\mathsc{#2}}}\fi
\else\textup{#1\,{\mdseries\textsc{#2}}}%
\fi}

\def\Ha{H$\alpha$~}
\def\HI{\ion{H}{i}~} 
\def\HII{\ion{H}{ii}~} 
\def\HeII{\ion{He}{ii}~}
\def\CIV{\ion{C}{iv}~}

\def\deg{\hbox{$^\circ$}}
\def\kms{km~s$^{-1}$}

\title[\HI imaging of dwarf star-forming galaxies] {\HI imaging of dwarf star-forming galaxies:\\ Masses, morphologies and gas deficiencies}
\author[Jaiswal et al.]{S.~Jaiswal$^{1}$\thanks{E-mail: sumit@shao.ac.cn} and
A.~Omar$^{2}$\\ 
$^{1}$Shanghai Astronomical Observatory, Key Laboratory of Radio Astronomy, Chinese Academy of Sciences,\\ 
80 Nandan Road, Shanghai 200030, China\\
$^{2}$Aryabhatta Research Institute of Observational Sciences (ARIES), 
Manora Peak, Nainital, 263002, India}

\date{Accepted ---. Received ---; in original form \today}
\pagerange{\pageref{firstpage}--\pageref{lastpage}}
\pubyear{2011}

\begin{document}
\label{firstpage}
\maketitle

\begin{abstract}
The GMRT observations of the \HI 21~cm-line emission from 13~nearby dwarf star-forming galaxies are presented. These galaxies are selected from the catalogues of Wolf-Rayet galaxies having very young ($\le$10~Myr) star formation. The ranges of star formation rates and stellar masses of the sample galaxies are 0.03 -- 1.7~$M_\odot~{\rm yr}^{-1}$ and 0.04 -- 22.3~$\times 10^8~M_\odot$, respectively. The \HI line emission is detected from 12~galaxies with peak column density $>1\times10^{21}$~cm$^{-2}$. The 3$\sigma$~\HI column density sensitivities per channel width of 7~\kms ~for low ($60''\times60''$) resolution images are in the range 0.8--1.9~$\times~10^{19}$~cm$^{-2}$. The  \HI channel images, moment images, global profiles and mass surface density profiles are presented here. The average value of the peak \HI mass surface density is estimated to be $\sim$2.5~M$_{\odot}$~pc$^{-2}$, which is significantly less compared to that in massive spiral galaxies. The scaling relations of $(M_{stars} + M_{\rm H\,I} + M_{\rm He})$ vs $M_{dyn}$, gas fraction vs $M_B$, $M_{\rm H\,I}$ vs $M_{stars}$, \ion{H}{i}-to-stellar mass ratio vs $M_{stars}$, and $M_{\rm H\,I}$ vs $D_{\rm H\,I}$ for the sample galaxies are estimated. These scaling relations can be used to constraint the key parameters in the galaxy evolution models. These galaxies are residing in group environment with galaxy density up to 8~galaxy~Mpc$^{-3}$. A \HI mass deficiency (with $DEF_{\rm HI} > 0.3$) is noticed in majority of galaxies for their optical diameters as compared to galaxies in field environments. Clear signatures of tidal interactions in these galaxies could be inferred using the \HI images. Isolated \HI clouds without known optical counterparts are seen in the vicinity of several galaxies. \HI emission envelope is found to be having an offset from the optical envelope in several galaxies. Consistent with the previous studies on galaxy evolution in group environments, tidal interactions seem to play an important role in triggering recent star formation.

\end{abstract}


\section{Introduction} 
Dwarf galaxies in the nearby Universe are often characterized by high gas content, low oxygen abundance and compact or irregular morphology with absolute $B$-band magnitude $\gtrsim-18$ \citep{Sargent1970ApJ...162L.155S, Thuan1981ApJ...247..823T, Papaderos1996A&AS..120..207P, Telles1997MNRAS.286..183T, Kunth2000A&ARv..10....1K, Cairos2001ApJS..136..393C, Bergvall2002A&A...390..891B, Cairos2003ApJ...593..312C, Noeske2003A&A...410..481N, Gil2003ApJS..147...29G, Gil2005ApJS..156..345G, Amor2009A&A...501...75A, Bergvall2012ASSP...28..175B, Micheva2013MNRAS.431..102M}.  The dwarf irregular (dI) and blue compact dwarf galaxies (BCD), both having disk-like morphology, show ongoing star formation. The star formation in BCDs is normally dominated in a few compact regions with intense blue color while dI galaxies show nearly uniformly distributed star-forming activities across the disk \citep{Heller1999MNRAS.304....8H, van2001AJ....121.2003V}. The optical emission in these galaxies is dominated by nebular emission lines \citep[e.g.][]{Markarian1981Afz....17..619M, MacAlpine1981ApJS...45..113M, Wasilewski1983ApJ...272...68W, Pesch1995ApJS...98...41P, Gil2003ApJS..147...29G}. The oxygen abundance (12 + log(O/H)) in dwarf galaxies show a linear relation with the absolute optical magnitude in the sense that the low metallicity galaxies are often optically faint \citep{Lequeux1979A&A....80..155L, Skillman1989ApJ...347..875S}. The stellar population in these galaxies has a mix of young and old stars \citep{Tolstoy1999Ap&SS.265..199T, Zhao2011AJ....141...68Z, Mart2012MNRAS.420.1294M}, however, the radiation from the young stellar population can  dominate the total energy outputs from these galaxies \citep[e.g.][]{Cairos2001ApJS..136..393C, Amor2009A&A...501...75A}.

The hierarchical models of galaxy formation consider dwarf galaxies as the building blocks of larger galaxies, which can be formed through merger and accretion of dwarf galaxies \citep[e.g.][]{Shlosman2013seg..book..555S, Amorisco2014Natur.507..335A, Deason2014ApJ...794..115D}. Majority of dwarf galaxies show low oxygen abundance (7.0 $\leq$ 12 + log(O/H) $\leq$ 8.4) indicating a relatively pristine interstellar medium (ISM) compared to large galaxies. This could in turn indicate a relatively low star formation efficiency in dwarf galaxies over past several giga-years. Understanding star formation in dwarf galaxies is therefore valuable for studies related to evolution of galaxies. The star-forming dwarf galaxies contain a large reservoir of gas and can sustain star formation over several giga-years \citep{Hunter1985ApJS...58..533H, van2001AJ....121.2003V, Hunter2004AJ....128.2170H}. It is important to identify active mechanisms in dwarf galaxies responsible for star formation trigger and sustainment. This issue can be possibly better addressed by selecting galaxies which show evidence of very recent star formation. A subset of \HII galaxies known as Wolf-Rayet (WR) galaxies offer a unique possibility to study very early phases of star formation and its triggering and propagation mechanisms \citep{Schaerer1998ApJ...497..618S}. The WR galaxies contain a very young ($\le 5$ Myr) population of massive (M $\ge$ 8 M$_{\odot}$) WR stars, which are soon to end their lives in supernovae explosions within the next few Myr \citep{Meynet2005A&A...429..581M}. The presence of WR stars in these galaxies ensures that the most recent phase of star formation is only a few Myr old and hence star formation triggering mechanism may be traced easily.

Several star formation trigger mechanisms have been proposed and observationally confirmed in dwarf galaxies. The galaxy-galaxy tidal interactions appear to play a fundamental role in triggering star formation, both in massive spirals \citep{Koribalski1996ASPC..106..238K} and in low mass dwarf galaxies \citep{Mendez1999AJ....118.2723M}. Many star-forming galaxies show features of tidal interactions in their optical and near-infrared (NIR) images \citep[e.g.][]{Larson1978ApJ...219...46L, Nikolic2004MNRAS.355..874N, Sanders1996ARA&A..34..749S, Genzel1998ApJ...498..579G}. However, a small subset of star-forming galaxies does not show any ongoing tidal interaction in their optical images, which questions tidal interactions as the sole mechanism for the star-formation \citep[e.g.][]{Balkowski1981A&A....97..223B, Campos1993AJ....106.1784C, Telles1995MNRAS.275....1T, Noeske2001A&A...371..806N}. It is worth to point out here that some dwarfs known as tidal dwarf galaxies are considered to be formed out of the ejected material in large galaxy-galaxy collisions and mergers \citep{Schweizer1978IAUS...77..279S, Mirabel1991A&A...243..367M, Mirabel1992A&A...256L..19M, Hernquist1992Natur.360..105H, Bournaud2010ASPC..423..177B, Duc2012ASSP...28..305D}.  Such galaxies are usually characterized by high oxygen abundance for their optical magnitudes \citep[e.g.][]{Duc2012ASSP...28..305D}.

The star formation histories of several dwarf galaxies, decoded from optical photometric and spectroscopic observations, indicate that the star formation can be well characterized by episodic bursts with quiescent periods of several hundreds of Myr \citep[e.g.][]{Krueger1995A&A...303...41K, Mas1999A&A...349..765M}.  The star formation can also be propagated and sustained over long periods through self-propagating stochastic process \citep[e.g.][]{van2001AJ....121.2003V}. Gas compression by shocks due to mass loss by means of galactic winds and subsequent cooling of the medium can also be a viable mechanism for the initiation of star formation \citep{Thuan1991mss..book..183T, Hirashita2000PASJ...52..107H}. The possible mechanisms for the starburst in galaxies are discussed in \citet{Gallagher1993ASPC...35..463G}. It may be noted that dynamical features such as spiral density waves and bars can also sustain star formation in large spiral galaxies \citep{Clarke2006MNRAS.371..530C, Friedli1993A&A...268...65F}.  Alternative mechanisms other than tidal interaction do not seem to provide adequate explanations for the starburst in dwarf galaxies \citep[e.g.][]{Lopez2010A&A...521A..63L, Jaiswal2016MNRAS.462...92J}. The tidal interactions in galaxies are best traced in the \HI 21 cm-line imaging as the gaseous disks in the outer regions of galaxies can be easily perturbed in a tidal encounter. A large number of \HI emission studies have demonstrated that the \HI imaging often provides evidences of interactions which are undetectable at optical wavelengths \citep[e.g.][]{Simkin1987Sci...235.1367S, Ekta2008MNRAS.391..881E, English2010AJ....139..102E, Lopez2012MNRAS.419.1051L, Lelli2014MNRAS.445.1694L, McQuinn2015MNRAS.450.3886M, Hibbard2001ASPC..240..657H}. An optically faint or low-mass companion galaxy in the vicinity can also be traced via detection of interaction features in the \HI images.

We present here \HI imaging of 13 nearby dwarf WR galaxies using the Giant Meter-wave Radio Telescope (GMRT) to examine occurrence of tidal interaction features in dwarf galaxies. Previous similar studies on dwarf galaxies indicates a high rate of detection of morphological distortions indicating recent tidal interactions \citep[e.g.][]{Lequeux1980A&A....91..269L, Viallefond1983ApJ...269..444V, Brinks1988MNRAS.231P..63B, Taylor1993AJ....105..128T, Taylor1995ApJS...99..427T, van1998AJ....116.1186V, Meurer1998MNRAS.300..705M, van2001AJ....122..121V, Begum2006MNRAS.365.1220B, Ramya2011ApJ...728..124R, Roychowdhury2012MNRAS.426..665R, Patra2016MNRAS.456.2467P, McNichols2016ApJ...832...89M}. This \HI study is in continuation of the optical study of the same sample of dwarf WR galaxies, in which tidal interactions features were identified based on optical continuum and \Ha imaging \citep{Jaiswal2013JApA...34..247J, Jaiswal2016MNRAS.462...92J}. These galaxies are selected from the catalogues of WR galaxies of \citet{Schaerer1999A&AS..136...35S} and \citet{Brinchmann2008A&A...485..657B}. The \HI observations of  the star-forming dwarf galaxies also allow us to study kinematics of ISM and various correlations, viz., Tully-Fisher relations \citep{Tully1977A&A....54..661T}, gas mass - size, and gas fraction - optical luminosity involving \HI mass and optical properties. Such correlations are well known for the field spiral galaxies. Often, a comparison of these properties with those for the field galaxies allows to study role of environment in galaxy evolution \citep[e.g.][]{Haynes1984AJ.....89..758H, Denes2014MNRAS.444..667D, Bradford2015ApJ...809..146B, Bradford2016ApJ...832...11B}. The gas dynamics also plays a crucial role in the evolution of low-mass galaxies \citep{Larson1974MNRAS.166..585L, Dekel1986ApJ...303...39D, Tajiri2002A&A...389..367T}. The \HI velocity field can also be used to carry out detailed mass modeling of galaxies. A comparison of the known gas mass correlations with galaxy size and optical luminosity with those for the field galaxies can be used to infer galaxy evolution in different galaxy density environments such as groups and clusters \citep{Haynes1984AJ.....89..758H}. It is worth to note that while single dish \HI observations are plentiful, high resolution \HI imaging studies using radio interferometers for dwarf galaxies are relatively fewer. We present here the \HI images and various correlations involving multiple integrated properties and gas mass in dwarf galaxies. 


\section{Sample properties} 

\begin{figure*}
\centering
\begin{tabular}{cc}
\includegraphics[angle=0,width=0.5\linewidth]{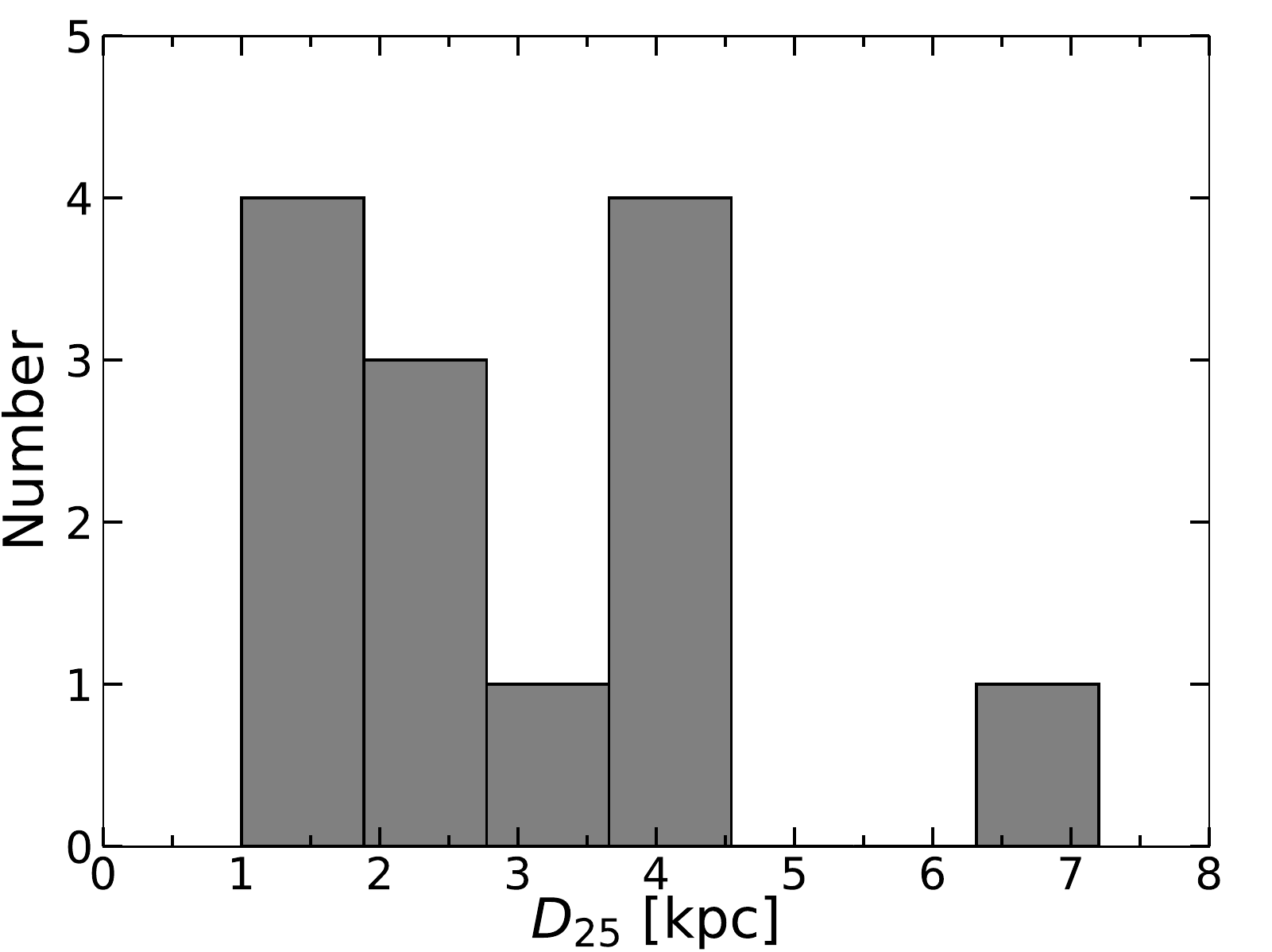}
\includegraphics[angle=0,width=0.5\linewidth]{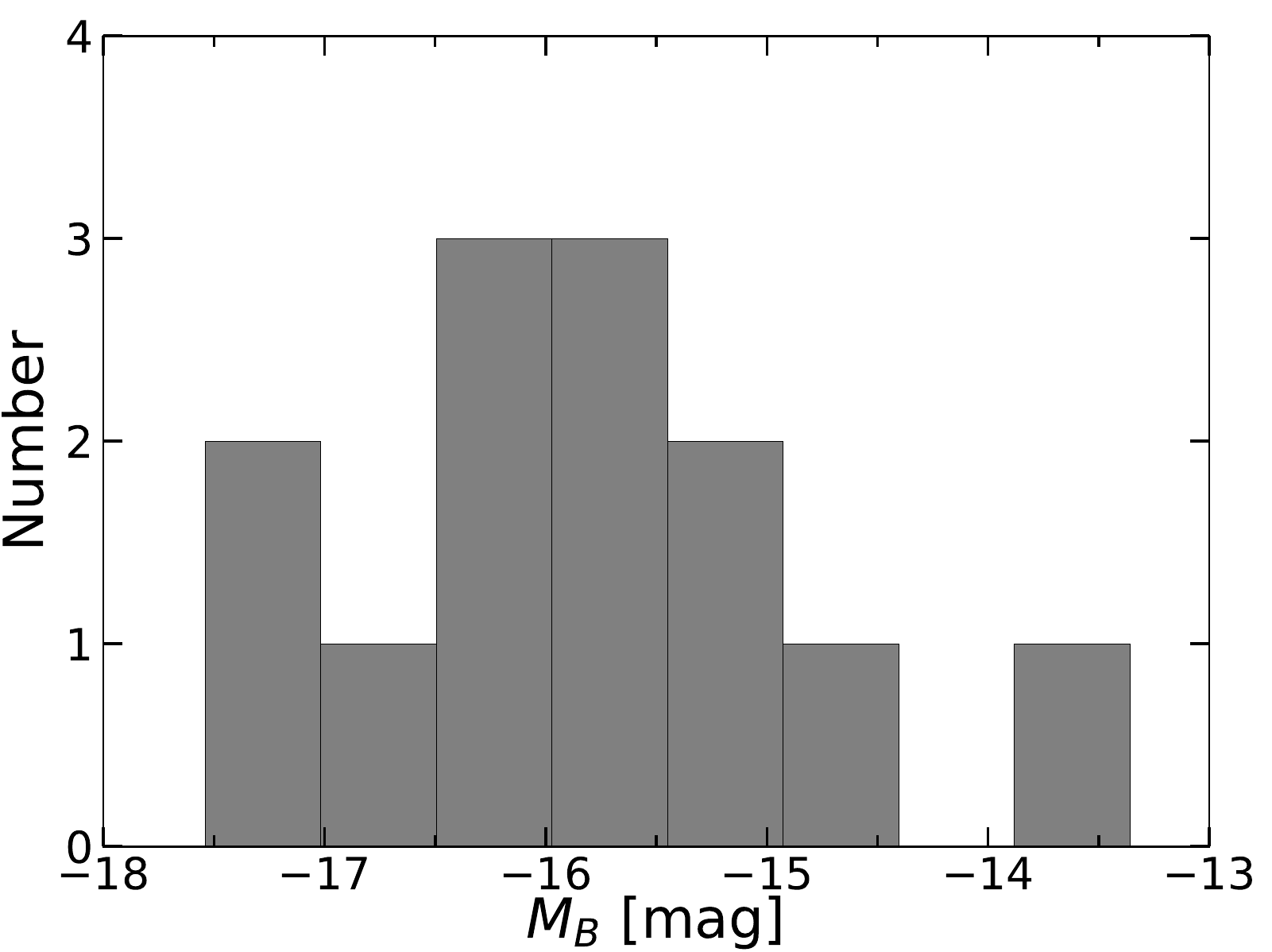}\\
\includegraphics[angle=0,width=0.5\linewidth]{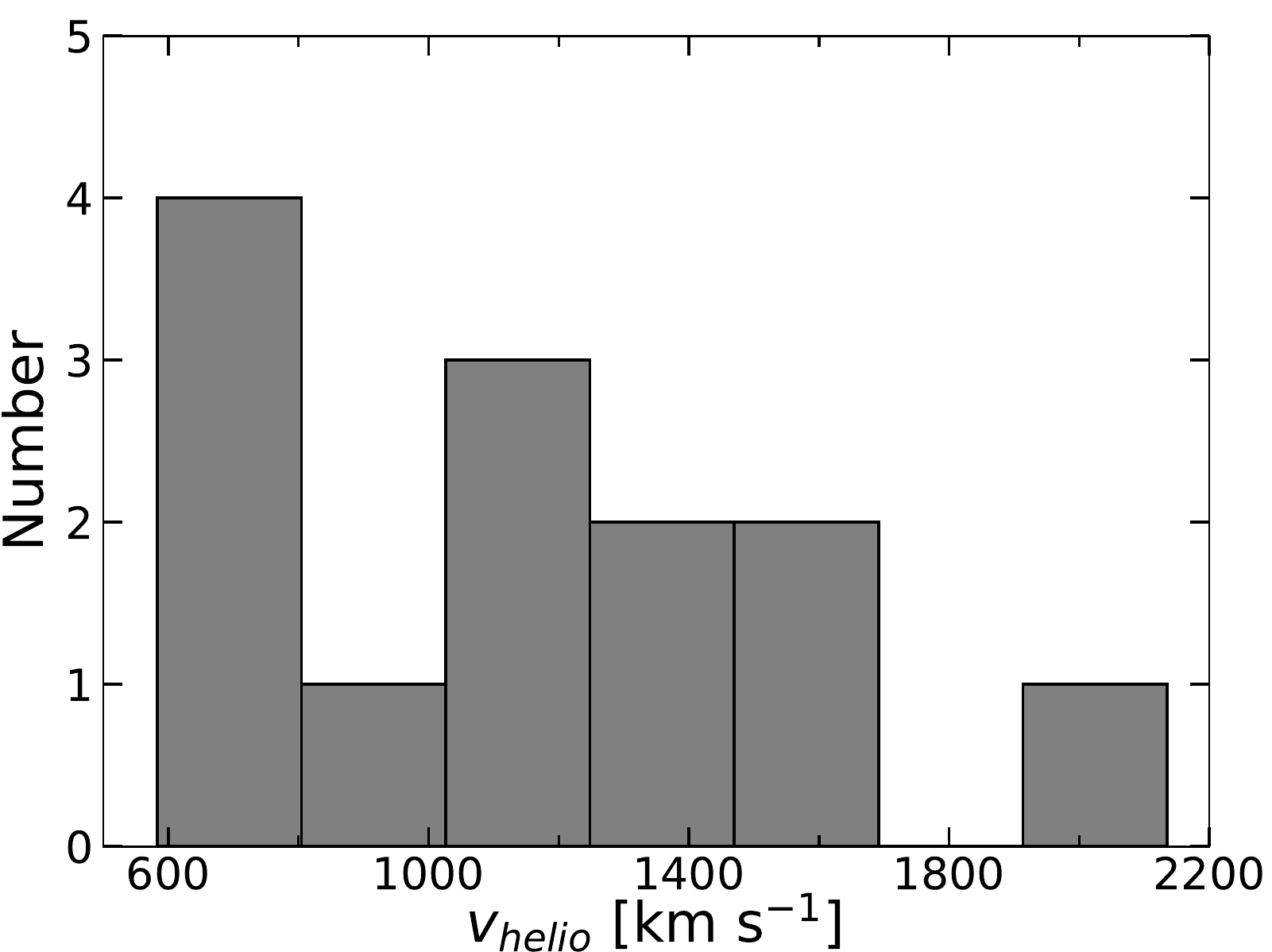}
\includegraphics[angle=0,width=0.5\linewidth]{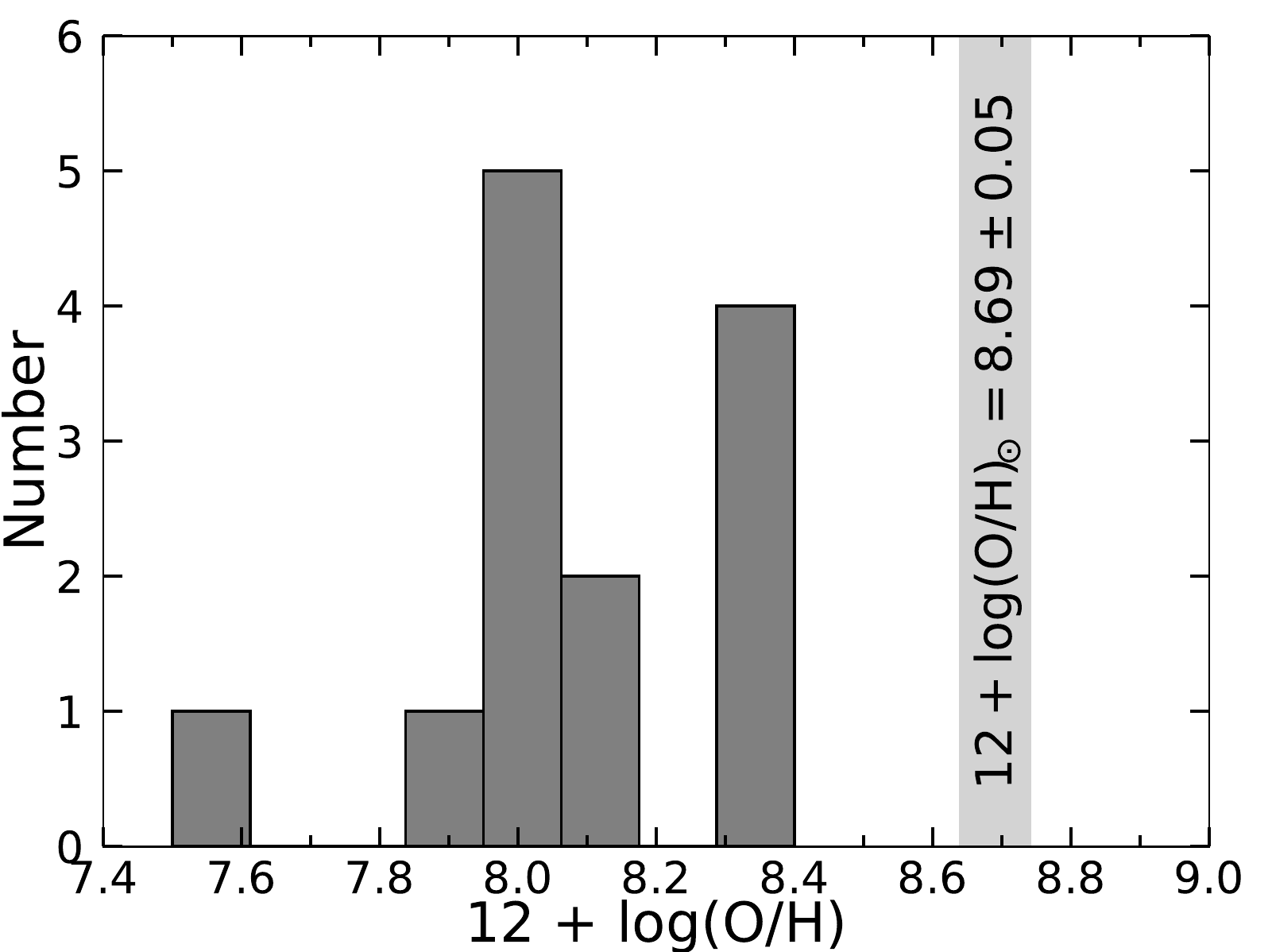}\\
\includegraphics[angle=0,width=0.5\linewidth]{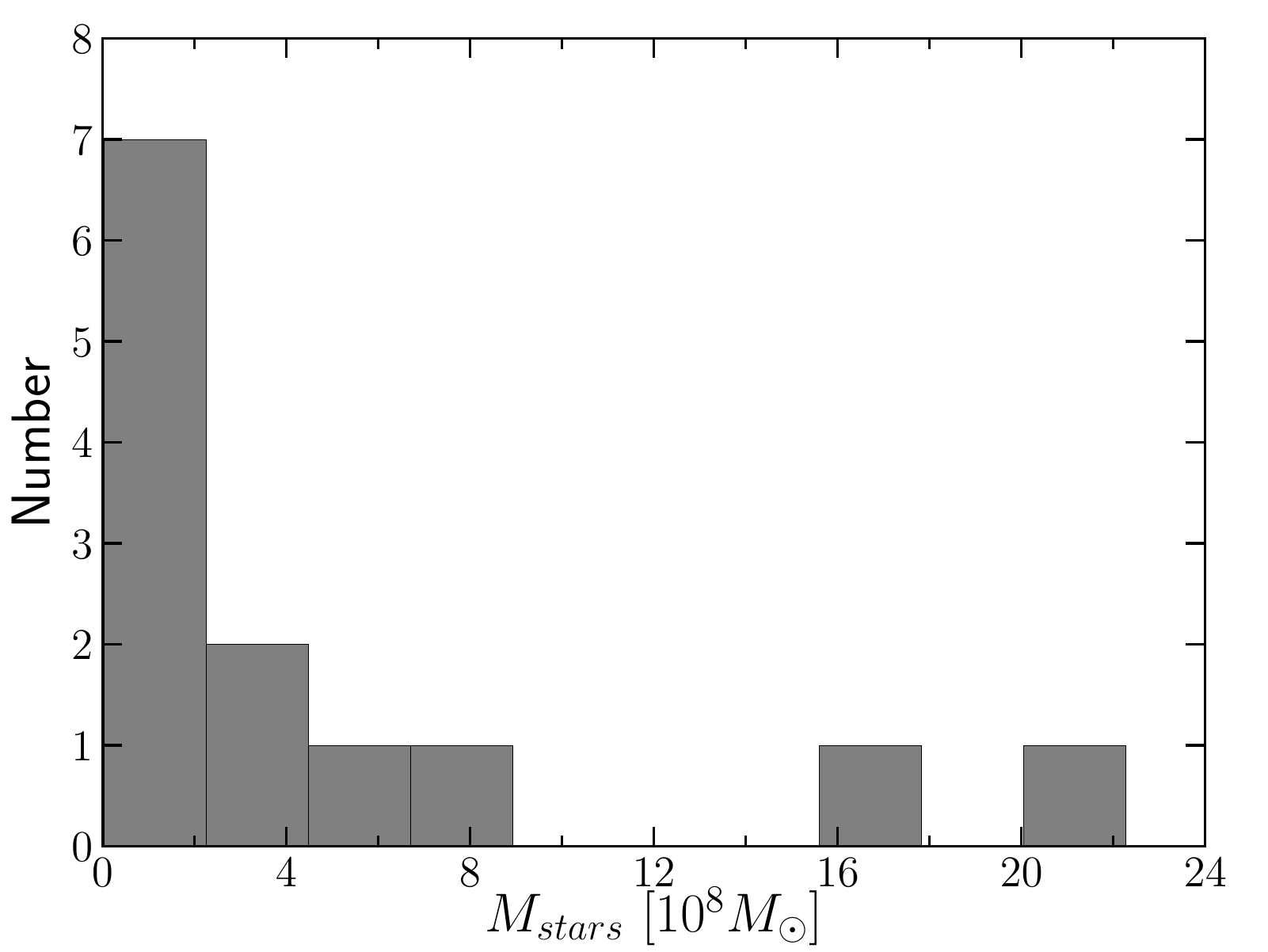}
\end{tabular}
\caption[ ]{The histograms showing linear sizes, absolute $B$-magnitudes, radial velocities, gas-phase metallicities and stellar masses of the galaxies in our sample.}
\label{histogram}
\end{figure*}

We used the catalogues of WR galaxies \citep[e.g.][]{Allen1976MNRAS.177...91A, Osterbrock1982ApJ...261...64O, Conti1991ApJ...377..115C} prepared by \citet{Schaerer1999A&AS..136...35S} and \citet{Brinchmann2008A&A...485..657B} to construct a sample. \citet{Schaerer1999A&AS..136...35S} catalogue includes 139 WR galaxies collected from the literature, while \citet{Brinchmann2008A&A...485..657B} catalogue includes 570 WR galaxies and extragalactic \HII regions categorized using the Sloan Digital Sky Survey (SDSS) Data Release-6 \citep{Adelman2008ApJS..175..297A}. Almost all the objects in the former catalogue that fall in the SDSS sky coverage are detected in the later also. The WR galaxies are identified with broad optical emission lines (\HeII 4686 \AA ~and \CIV 5808 \AA) in these catalogues. These broad emission lines attributed to the presence of substantial population ($10^2$--$10^5$) of WR stars in the galaxy \citep[e.g.][]{Kunth1986A&A...169...71K}. The WR stars are short-lived massive stars \citep{Meynet2005A&A...429..581M} and therefore their presence in a large number suggests a young ($<10$~Myr) on-going star formation in the galaxy. The WR galaxies are found in almost all morphological types, starting from low-mass blue compact dwarfs (BCDs) and irregular galaxies to massive spirals and luminous merging IRAS galaxies\citep[e.g.][]{Conti1991ApJ...377..115C, Schaerer1999A&AS..136...35S, Brinchmann2008A&A...485..657B}. We considered only dwarf WR galaxies up to $\sim$25 Mpc distance, apparent magnitude brighter than 17~mag in the SDSS $r$-band and declination $> -25\deg$ to carry out this study. Out of these dwarf WR galaxies, we randomly selected 13 objects to study under the present work. This sample is described in \citet{Jaiswal2016MNRAS.462...92J}. The basic properties of the galaxies are given in Table~\ref{sample}. The source coordinates and morphological type are taken from NASA/IPAC Extragalactic Database (NED). The $B$-band absolute magnitude is estimated using $B$-band apparent magnitude given in Table~\ref{phot}. The star formation rates (SFRs) are estimated using our narrowband \Ha line observations of these galaxies \citep{Jaiswal2013JApA...34..247J,Jaiswal2016MNRAS.462...92J}. The \Ha based SFR is sensitive to the recent star formation in these galaxies. The BCD galaxies and WR galaxies have been studied before. Most of such studies on BCDs were not targeted particularly for WR galaxies. The large sample of 20 WR galaxies studied by \citet{Lopez2010A&A...521A..63L} has majority of galaxies with SFR $>$~1~$M_\odot~{\rm yr}^{-1}$. In comparison, our sample is dominated by SFR $<$~1~$M_\odot~{\rm yr}^{-1}$. Several other similar studies in the past were also mainly focused on galaxies with high star formation rates. Therefore, our sample is unique in terms of low SFR and complimentary to previous studies on WR and BCD galaxies. Figure~\ref{histogram} shows the histograms of $D_{25}$ (see Table~\ref{phot}) linear size, $B$-band absolute magnitude, radial velocity, oxygen abundance and stellar mass of the selected galaxies. The stellar masses of the sample galaxies are estimated in Section~\ref{sec:mass}. The galaxies in this sample have sizes $<8$ kpc with majority having sizes $<5$ kpc. The optical magnitude $M_{B}$ is greater than $-18$ and oxygen abundances are sub-solar ($12 + {\rm log(O/H)}_\odot = 8.69 \pm 0.05$ \citep{Asplund2009ARA&A..47..481A}) in a range $7.5 - 8.4$. Majority of galaxies in the sample have stellar masses $<8.5\times 10^8 M_\odot$.

\begin{table*}
\centering
\caption{Basic properties of the galaxies in the sample}
\scalebox{0.9}{
\begin{tabular}{lcccccccc}
\hline
\hline
Name & RA (J2000) & DEC (J2000)           & Type & $M_B$   & $v_{\mathrm{helio}}$ & 12 + log(O/H) & $(\int S{\rm d}v)_{SD}$ & SFR\\ 
     & h m s      & $^{\circ}$ $'$ $''$   &      & [mag]   & [km/s]               &               & [Jy km s$^{-1}$] & [$M_\odot yr^{-1}$]\\
\hline 
\noalign{\smallskip}
MRK~996      & 01 27 35.5 & $-$06 19 36 & BCD             & $-17.32 \pm 0.12$ & $1622 \pm 10^a$ & $\sim 8.0^a$  & $1.04 \pm 0.16^n$ & $0.40 \pm 0.10$\\
UGCA~116     & 05 55 42.6 & $+$03 23 32 & compact Irr pec & $-16.31 \pm 0.17$ & $789 \pm 4^b$   & $8.09 \pm 0.02^h$ & $16.2-19.2$   & $1.73 \pm 0.25$\\
UGCA~130     & 06 42 15.5 & $+$75 37 33 & Irr             & $-15.63 \pm 0.06$ & $792 \pm 5^b$   & $8.04 \pm 0.04^i$ & $0.80-2.54$   & $0.05 \pm 0.01$\\
MRK~22       & 09 49 30.3 & $+$55 34 47 & BCD             & $-15.71 \pm 0.08$ & $1551 \pm 12^c$ & $8.04 \pm 0.01^j$ & $1.22 \pm 0.11^o$ & $0.04 \pm 0.01$\\
IC~2524      & 09 57 32.8 & $+$33 37 11 & S               & $-16.86 \pm 0.20$ & $1450 \pm 3$    & $8.39 \pm 0.03$   & $3.94 \pm 1.10^p$ & $0.06 \pm 0.02$\\
KUG~1013+381 & 10 16 24.5 & $+$37 54 46 & BCD             & $-15.37 \pm 0.04$ & $1173 \pm 3$    & $7.50 \pm 0.01$   & $1.51 \pm 0.39^q$ & $0.11 \pm 0.01$\\
CGCG~038-051 & 10 55 39.2 & $+$02 23 45 & dIrr            & $-14.48 \pm 0.06$ & $1021 \pm 2$    & $7.91 \pm 0.08$   & $3.5 \pm 0.6^r$  & $0.05 \pm 0.01$\\
IC~2828      & 11 27 10.9 & $+$08 43 52 & Im              & $-15.84 \pm 0.08$ & $1039 \pm 12^d$ & $8.33 \pm 0.01$   & $\sim 2.7^s$  & $0.12 \pm 0.02$\\
UM~439       & 11 36 36.8 & $+$00 48 58 & Irr             & $-16.27 \pm 0.06$ & $1099 \pm 4^e$  & $8.08 \pm 0.03^k$ & $5.50-7.23$   & $0.08 \pm 0.01$\\
I~SZ~59      & 11 57 28.0 & $-$19 37 27 & S0              & $-17.54 \pm 0.23$ & $2135 \pm 18^f$ & $\sim 8.4^l$  & ---  & $0.94 \pm 0.18$ \\
SBS~1222+614 & 12 25 05.4 & $+$61 09 11 & dIrr            & $-14.97 \pm 0.09$ & $706 \pm 2^c$   & $8.00 \pm 0.02^m$ & $\sim 6.2^t$ & $0.09 \pm 0.01$\\
UGC~9273     & 14 28 10.8 & $+$13 33 06 & Im              & $-16.45 \pm 0.19$ & $1289 \pm 5^g$  & $8.33 \pm 0.01$   & $1.57 \pm 0.22^u$ & $0.04 \pm 0.01$\\
MRK~475      & 14 39 05.4 & $+$36 48 22 & BCD             & $-13.36 \pm 0.04$ & $583 \pm 2^b$   & $7.97 \pm 0.02^j$ & $0.15 \pm 0.04^v$ & $0.025 \pm 0.003$\\
\noalign{\smallskip}    
\hline
\end{tabular}}
\begin{flushleft}
{\footnotesize References for $M_B$: see Section~\ref{sec:hipar}.\linebreak}\\

{\footnotesize {References for $v_{\mathrm{helio}}$: $^a$ \citet{Thuan1996ApJ...463..120T}, $^b$ Third Reference Catalogue (RC3: \citet{de1991rc3..book.....D}), $^c$ \citet{Thuan1999A&AS..139....1T}, $^d$ \citet{Smoker2000A&A...361...19S}, $^e$ \citet{Comte1999Ap.....42..149C}, $^f$ \citet{Firth2006MNRAS.372.1856F}, $^g$ \citet{Schneider1990ApJS...72..245S}. The $v_{\mathrm{helio}}$ values for other galaxies are taken from SDSS data release-7 \citep[DR7:][]{Abazajian2009ApJS..182..543A}.\linebreak}}\\

{\footnotesize {References for 12 + log(O/H): \quad  $^h$ \citet{Guseva2000ApJ...531..776G}, $^i$ \citet{Izotov1998ApJ...500..188I}, $^j$ \citet{Izotov1994ApJ...435..647I}, $^k$ \citet{Zhao2013ApJ...764...44Z}, $^l$ \citet{Kunth1985A&A...142..411K}, $^m$ \citet{Ekta2010MNRAS.406.1238E}. The oxygen abundances for other galaxies are taken from \citet{Brinchmann2008A&A...485..657B}.\linebreak}}\\

{\footnotesize {References for $(\int S{\rm d}v)_{SD}$: $^n$~\citet{Thuan1996ApJ...463..120T}, $^o$~\citet{Salzer2002AJ....124..191S}, $^p$~Third Reference Catalogue (RC3: \citet{de1991rc3..book.....D}), $^q$~\citet{Pustilnik2007A&A...464..859P}, $^r$~our measurement using HIPASS data \citep{Barnes2001MNRAS.322..486B}, $^s$~\citet{Stierwalt2009AJ....138..338S}, $^t$~This emission is from the nearby source MCG~+10-18-044 \citep{Huchtmeier2000A&AS..141..469H} (see Section~\ref{note_SBS}), $^u$~\citet{Schneider1990ApJS...72..245S}, $^v$~\citet{Huchtmeier2005A&A...434..887H}. The galaxies which have large variation in single dish \HI flux taken from different measurements, a range is provided.}}
\end{flushleft}
\label{sample}
\normalsize
\end{table*}

\section{Observations and data analysis}
\subsection{Optical data}

The \Ha and $R$-band observations were carried out using the 1.3-m Devasthal Fast Optical Telescope (DFOT) and 2-m IUCAA Girawali Observatory (IGO) telescope in the years 2011 - 2013. The details of the optical observations and data reduction are provided in \citet{Jaiswal2016MNRAS.462...92J}. The \Ha images have total exposure time in the range $30 - 235$ minutes while the R-band images have exposures in the range $15 - 50$ minutes. The point spread function has FWHM values in the range $1.2'' - 2.6''$. The CCD images were processed following the standard tasks of IRAF (Image Reduction and Analysis facility) software developed by {\it National Optical Astronomy Observatory}. The \Ha images have surface brightness sensitivities in the range $0.4 - 1.8 ~\times 10^{-16} ~\rm erg~s^{-1}~cm^{-2}~arcsec^{-2}$ 
for 3$\sigma$ detection. These optical images are used here for overlaying \HI contours.

\subsection{Radio data}

The \HI observations of 13 galaxies were carried out using the GMRT \citep{Swarup1991CuSc...60...95S}. The GMRT is a Y-shaped interferometric array of thirty fully steerable parabolic dishes of 45-m diameter each. The half power beam width of the GMRT antennas is $\sim24'$ at 1.4~GHz. The GMRT has a mix of both short and long baselines, making it sensitive to diffuse emission of extent as much as $7'$ while having a maximum possible resolution of
$\sim2''$ at 1.4~GHz. The observations of the galaxies were carried out interspersed with the observations of a suitable nearby phase calibrator every 45 minutes and observations of at least one of the primary flux calibrators 3C48, 3C147, and 3C286. The phase calibrators were selected from the Very Large Array list of calibrators within an angular distance of 10\deg ~to the target galaxy. The bandpass calibration was performed using the primary flux calibrators. A summary of the observational parameters are given in Table~\ref{obs_radio}. The data were taken in right and left circular polarization sensitive for total intensity. The digital correlator back-end was configured to use either 4.2 or 8 MHz bandwidth in 256 or 512 spectral channels. The centre frequency of the band-width was set corresponding to the redshifted \HI emission line frequency from the galaxy.

The data were reduced following the standard calibration and imaging methods using the Astronomical Image Processing System \citep[AIPS:][]{Greisen2003ASSL..285..109G} developed by the National Radio Astronomical Observatory. The data were calibrated for amplitude, phase and frequency response for all the antennas and separately for each of the two polarizations. The data were also self-calibrated in phase and amplitude using the continuum emission formed in a 100-channel averaged dataset. The resulting complex gain corrections for the antennas were applied to all the frequency channels. The continuum emission common to all the channels was subtracted from the line-data using the `UVLIN' task in AIPS. The self-calibrated and continuum subtracted line-data were used to make the image cubes at different resolutions by selecting appropriate spatial frequency range, taper and robustness parameter as implemented in the AIPS task `IMAGR'. The image cubes were generated and de-convolved using the CLEAN algorithm as implemented in the AIPS task IMAGR. The Doppler frequency shift corrections arising due to the orbital and spin motions of the Earth relative to the galaxy were applied on the image cubes to bring all the velocities in the Helio-centric frame of reference. The frequency was converted to velocity following the optical definition. The images were corrected for the primary beam of the GMRT antenna using the AIPS task PBCOR. The channel images were further analysed for estimating various parameters as described below. 

\begin{table*}
\centering
\caption{Summary of the GMRT observations}
\scalebox{0.8}{
\begin{tabular}{lcccccccc}
\hline
\hline
Galaxy name  & Date & Integration time & Central Frequency & Bandwidth & No. of channels & Channel width & Flux calibrator(s) & Phase calibrator(s)\\
             &      & [hrs]            & [MHz]             & [MHz]     & & [\kms] &  & \\
\hline 
MRK~996      & 2009 Jul 31 & 5.1 & 1412.8 & 8.0 & 256 & 6.7 & 3C48  & 0059+001 \\
UGCA~116     & 2011 May 29 & 3.5 & 1416.7 & 4.2 & 512 & 1.7 & 3C147 & 0532+075 \\
UGCA~130     & 2011 May 28 & 3.8 & 1416.9 & 4.2 & 256 & 3.5 & 3C147 & 0410+769 \\
MRK~22       & 2009 Jul 31 & 5.5 & 1413.1 & 8.0 & 256 & 6.7 & 3C48  & 0834+555 \\
IC~2524      & 2011 Dec 04 & 4.0 & 1413.6 & 4.2 & 512 & 1.7 & 3C147 & 0958+324 \\
KUG~1013+381 & 2011 Dec 05 & 6.3 & 1414.9 & 4.2 & 512 & 1.7 & 3C48, 3C286  & 0958+324 \\
CGCG~038-051 & 2011 Nov 28 & 3.2 & 1415.7 & 4.2 & 512 & 1.7 & 3C147, 3C286 & 1119-030 \\
IC~2828      & 2011 Dec 05 & 4.0 & 1415.5 & 4.2 & 512 & 1.7 & 3C147, 3C286 & 1120+143 \\
UM~439       & 2009 Aug 02 & 5.0 & 1415.2 & 8.0 & 256 & 6.7 & 3C147, 3C286 & 1130-148 \\
I~SZ~59      & 2009 Aug 01 & 4.8 & 1410.4 & 8.0 & 256 & 6.7 & 3C147, 3C286 & 1130-148 \\
SBS~1222+614 & 2011 Dec 02 & 3.6 & 1417.1 & 4.2 & 512 & 1.7 & 3C286 & 1313+675 \\
UGC~9273     & 2011 Nov 28 & 4.7 & 1413.3 & 4.2 & 512 & 1.7 & 3C286 & 1445+099 \\
MRK~475      & 2009 Jul 29 & 5.6 & 1417.7 & 8.0 & 256 & 6.7 & 3C286, 3C48 & 3C286, 1416+347 \\
\hline
\end{tabular}}
\label{obs_radio}
\end{table*}

\subsubsection{\HI channel maps}\label{channels}

The velocity-resolved \HI image cubes were made at three different angular resolutions for the galaxies. We convolved these image cubes appropriately to make the synthesized beam circular. These image cubes are designated as low ($60''\times 60''$), intermediate ($20''\times 20''$) and high ($8''\times 8''$) resolution cubes. The velocity resolution for the intermediate and high resolution cubes are equal to the channel-width given in the observing log (Table~\ref{obs_radio}). The low-resolution cubes are made with a smoothed velocity resolution of $\sim$7~\kms ~for all the galaxies.  The 3$\sigma$ column density sensitivities per channel for the low resolution image cubes are estimated in the range $0.8-1.9~\times 10^{19}$ cm$^{-2}$. The sensitivities at 3$\sigma$ level of the image cubes are given in Table~\ref{impar}. The peak \HI column density estimated from the high-resolution total \HI map are also given in Table~\ref{impar}. The \HI channel images are presented at low angular resolution since these images are maximally sensitive to tidal interaction features such as gaseous tails and bridges in galaxies. The mid and high resolution \HI images provide a visual correlation with the star-forming regions through the optical R-band and \Ha images of the galaxy. 

\begin{table*}
\centering
\caption{Sensitivity parameters of image cubes.}
\begin{tabular}{lccccc}
\hline
\hline
Galaxy name & Low resolution cube & Intermediate resolution cube & High resolution cube & Column density & Peak column\\
            & RMS noise & RMS noise & RMS noise & sensitivity & density detected\\
   & [mJy/beam] & [mJy/beam] & [mJy/beam] & [10$^{19}$ cm$^{-2}$] & [10$^{21}$ cm$^{-2}$]\\
\hline
MRK~996      & 1.3 & 1.0 & 0.8 & 0.8 & 3.3 \\
UGCA~116     & 2.3 & 3.0 & 2.8 & 1.5 & 11.4\\
UGCA~130     & 2.9 & 2.4 & 1.8 & 1.9 & 2.5 \\
MRK~22       & 1.5 & 1.0 & 1.0 & 1.0 & 1.8 \\
IC~2524      & 1.9 & 2.6 & 2.2 & 1.2 & 7.2 \\
KUG~1013+381 & 1.4 & 2.0 & 1.5 & 0.9 & 1.1 \\
CGCG~038-051 & 2.1 & 2.7 & 2.3 & 1.4 & 3.1 \\
IC~2828      & 1.8 & 2.3 & 2.3 & 1.2 & 2.0 \\
UM~439       & 1.6 & 1.4 & 1.2 & 1.0 & 5.9 \\
I~SZ~59      & 2.0 & 1.2 & 1.0 & 1.3 & 3.0 \\
SBS~1222+614 & 2.1 & 3.1 & 2.3 & 1.4 & --- \\
UGC~9273     & 2.0 & 3.0 & 2.3 & 1.3 & 1.6 \\
MRK~475      & 2.0 & 1.1 & 1.0 & 1.3 & 2.2 \\
\hline
\end{tabular}
\label{impar}
\end{table*}

\subsubsection{Global \HI profile, \HI width and systemic velocity}\label{sec_parm}
The integrated (global) \HI profiles were obtained using the GMRT primary beam corrected flux density estimated within the region showing \HI emission in the velocity channels of the low resolution image cube. The \HI flux integral ($\int S{\rm d}v$) was estimated by summing the primary beam corrected flux densities in the channel images and multiplying it with the channel width. The errors in the estimates are computed using the rms in the image and flux scale calibration error taken as 5\%. A Gaussian fit to the \HI global profile was made to estimate the \HI line-width using the Groningen Image Processing System \citep[GIPSY:][]{van1992ASPC...25..131V} task PROFGLOB. GIPSY is distributed by the Kapteyn Astronomical Institute, Groningen, Netherlands. The line-widths were estimated at the two edges of the \HI profile as $W_{20}$ and $W_{50}$ at the 20\% and 50\% levels of the peak flux density respectively. These widths were corrected for the broadening due to the finite instrumental velocity resolution using the method described in \citet[][see Appendix~\ref{app}]{Verheijen2001A&A...370..765V}. The corrected widths are listed in Table~\ref{radio_par}. We have not attempted to further correct the \HI line-widths for the turbulent (random) motions in the gas \citep[][]{Tully1985ApJS...58...67T}. The line-widths corrected for the galaxy inclination can be taken as twice of the maximum rotation velocity of the galaxy to obtain an estimate of total dynamical mass \citep[e.g.][see the Appendix~\ref{app}]{Courteau2014RvMP...86...47C}. As \HI and optical morphology of most of the galaxies in the sample were found to be very irregular and also the presence of tidal features can impact drastically the fitting ellipses to isophotes in measuring the disk inclination angle, no inclination correction was applied in estimating the galaxy dynamical mass. The dynamical masses estimated are therefore at their lower limits.

We also obtained central rotational velocities, namely $V_{20}$ and $V_{50}$, estimated using the \HI profiles at 20\% and 50\% levels respectively. The systemic velocity $V_{sys}$ of a galaxy was taken as the mean of $V_{20}$ and $V_{50}$ and the error as the absolute value of half of the difference between $V_{20}$ and $V_{50}$. The distance to the galaxy was estimated using $V_{sys}$. The \HI masses of the galaxies were estimated from the \HI flux integral using the relation given in Appendix~\ref{app}.

\subsubsection{\HI mass surface density profile}
The mean radial \HI mass surface density profiles for the galaxies were estimated by integrating the low-resolution total \HI maps in concentric tilted rings using the GIPSY task ELLINT. The initial geometrical parameters of these tilted rings were determined by isophotal fitting of the optical disks of the galaxies. The surface density profiles were scaled to obtain the \HI mass surface density profiles using the total \HI mass estimated in the Section~\ref{sec_parm}. The profiles were corrected for the projection effects to obtain the face-on mass surface densities. We have estimated the \HI disk radius, $R_{\rm HI}$, as the distance from the galaxy centre at which the \HI mass surface density falls to 1~$M_\odot~{\rm pc}^{-2}$. The error in $R_{\rm HI}$ was estimated as the half of the difference between its values for the approaching and receding sides of the galaxy, separately. Note that $R_{\rm HI}$ for UGCA~116 is measured along the axis of its tidal features ($PA = 115\deg$) in the optical image. The dynamical masses of the galaxies were estimated within their corresponding \HI disk radius.

\subsubsection{Total \HI maps and velocity fields}
The \HI moment maps were generated from the channel images containing \HI emission and two additional channels on both the sides. The noise or the emission-free pixels in the channel images were first blanked by applying a flux cutoff at $3-4$ times the rms noise in the images. The blanking was done to avoid inclusion of emission-free pixels to the moment maps. The blanking level was determined after smoothing the channel images in both spatial and velocity axes. A boxcar smoothing over three channels was applied in the velocity axis and a Gaussian kernel with $FWHM \sim 2$ times that of the synthesized beam was applied along the spatial axes. The smoothing improves signal to noise ratio and helps in better estimation of blanking level. It is worth to point out that the smoothing was applied only for blanking level determination and not for creating moment maps, which were created with native resolutions in the image cubes. The moment-0 (total \HI image) and moment-1 (\HI velocity fields) maps were generated for all the galaxies using the AIPS task MOMNT. The moment-1 maps were generated using intensity weighted average velocity estimates at each pixels.

\section{Results}
The GMRT \HI line images of the observed dwarf star-forming galaxies are presented in Figures~\ref{MRK996-1}-\ref{MRK475-2} in Appendix. Out of the 13 observed galaxies, \HI emissions were detected from 12 galaxies. In the galaxy SBS~1222+614, no \HI emission was detected within the sensitivity limit of the observations. However, the \HI emission is detected from a nearby dwarf irregular galaxy MCG~+10-18-044 at an angular distance of $\sim5'.5$, south of SBS~1222+614. We have provided here the \HI images of MCG~+10-18-044. The radio image panel for each galaxy contains the low-resolution \HI channel map overlying on the grey-scale optical $R$-band image of the galaxy, the low-resolution \HI column density map overlying on the grey-scale optical $R$-band image, the intermediate-resolution \HI column density map overlying on the grey-scale optical $R$-band image, the high resolution \HI column density map overlying on the grey-scale optical \Ha image of the galaxy, the intermediate-resolution moment-1 map showing the velocity field of the galaxy, the global \HI line profile obtained using the low-resolution image cube and the \HI mass surface density profile obtained using the low-resolution \HI intensity map. The weak \HI emission from the galaxy MRK~475 is not clearly detected in the low-resolution cube, however, detected in the intermediate-resolution channel map, which is presented in the panel. The channel map and the corresponding low-resolution column density map for each galaxy have the same scale. The radio image panels (b), (c) and (d) also have the same scale. The details of these images and the estimated \HI parameters are given below.

\subsection{\HI morphology}
The moment maps at different resolutions provide complementary information, such as the low-resolution images provide large-scale dynamics, global extent of \ion{H}{i}, effect of the environment and dark matter halo properties, while the high-resolution images trace the local phenomena like star formation and its feedback, and finer details of the \HI morphology near the central regions of the galaxies. The low-resolution \HI column density contour image plotted on the grey-scale optical $R$-band image shows the overall \HI morphology of the galaxy with respect to its optical morphology. These images are sensitive to detect tidal interaction features and nearby companion, if any. The intermediate resolution \HI column density contours plotted over the grey-scale optical $R$-band image provides visual correlation with the stellar disk and detection of lopsidedness features. The intermediate resolution moment-1 color maps show the velocity field of the galaxies. In most of the cases, the velocity field is found to be slightly irregular or disturbed. It was not possible to obtain spatially resolved reliable rotation curves of these galaxies due to lack of adequate SNR at high angular resolution. The high resolution \HI column density contours plotted on grey-scale \Ha image for the galaxies provides correlation of \HI with the star forming regions \citep[e.g.][]{van1998AJ....116.1186V}. The star-forming regions in these dwarf galaxies are mostly located in the central regions. In some cases, asymmetries in the \HI distribution with respect to the optical image of the galaxy were seen.

\subsection{\HI parameters}\label{sec:hipar}
\begin{figure}
\centering
\includegraphics[angle=0,width=0.7\linewidth]{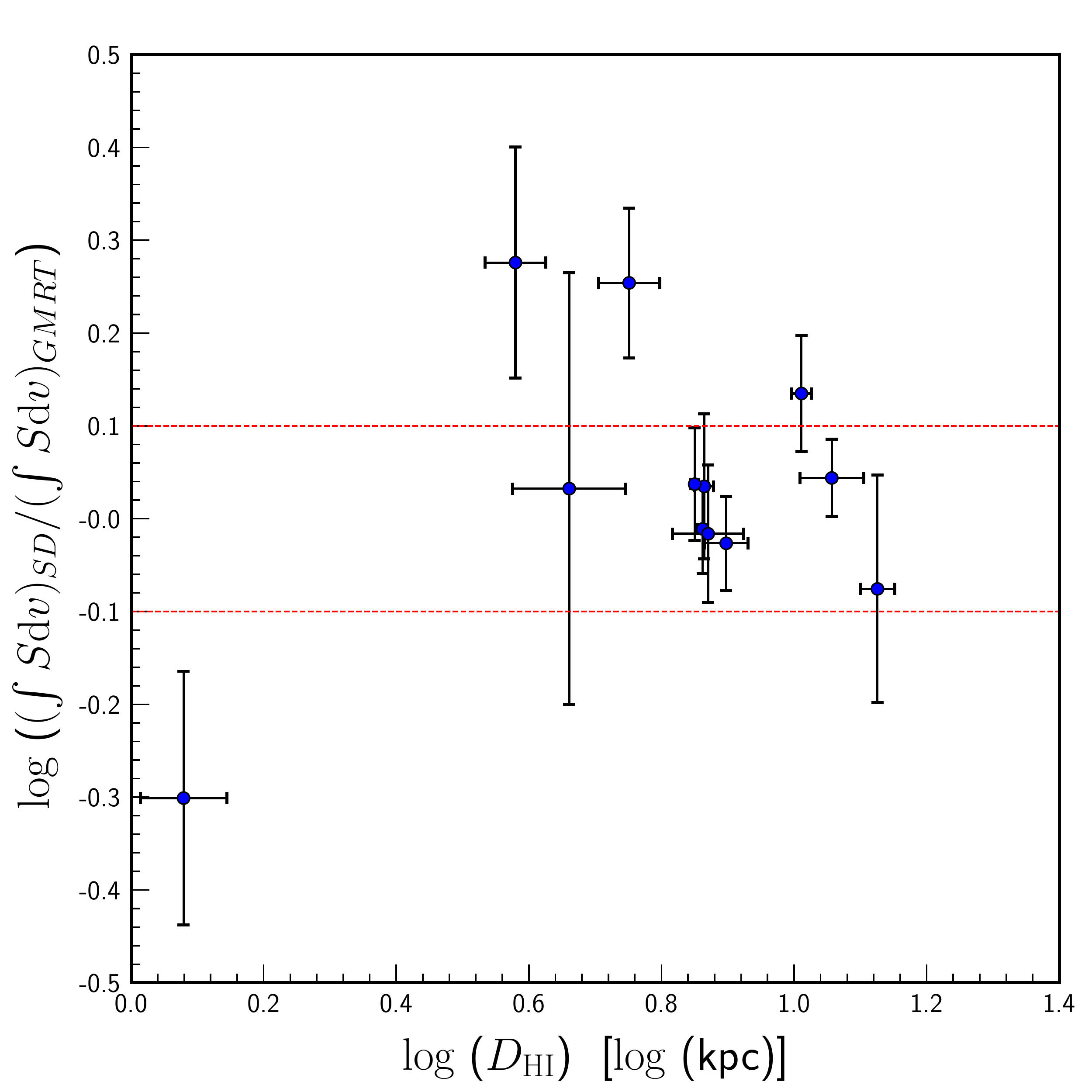}
\caption[ ]{Comparison of single-dish \HI flux integral with the GMRT flux integral for the  galaxies studied here. The dotted lines are showing deviations in the flux integral ratio at $\pm 0.1$ dex.}
\label{sd}
\end{figure}

\begin{figure}
\centering
\includegraphics[angle=0,width=0.7\linewidth]{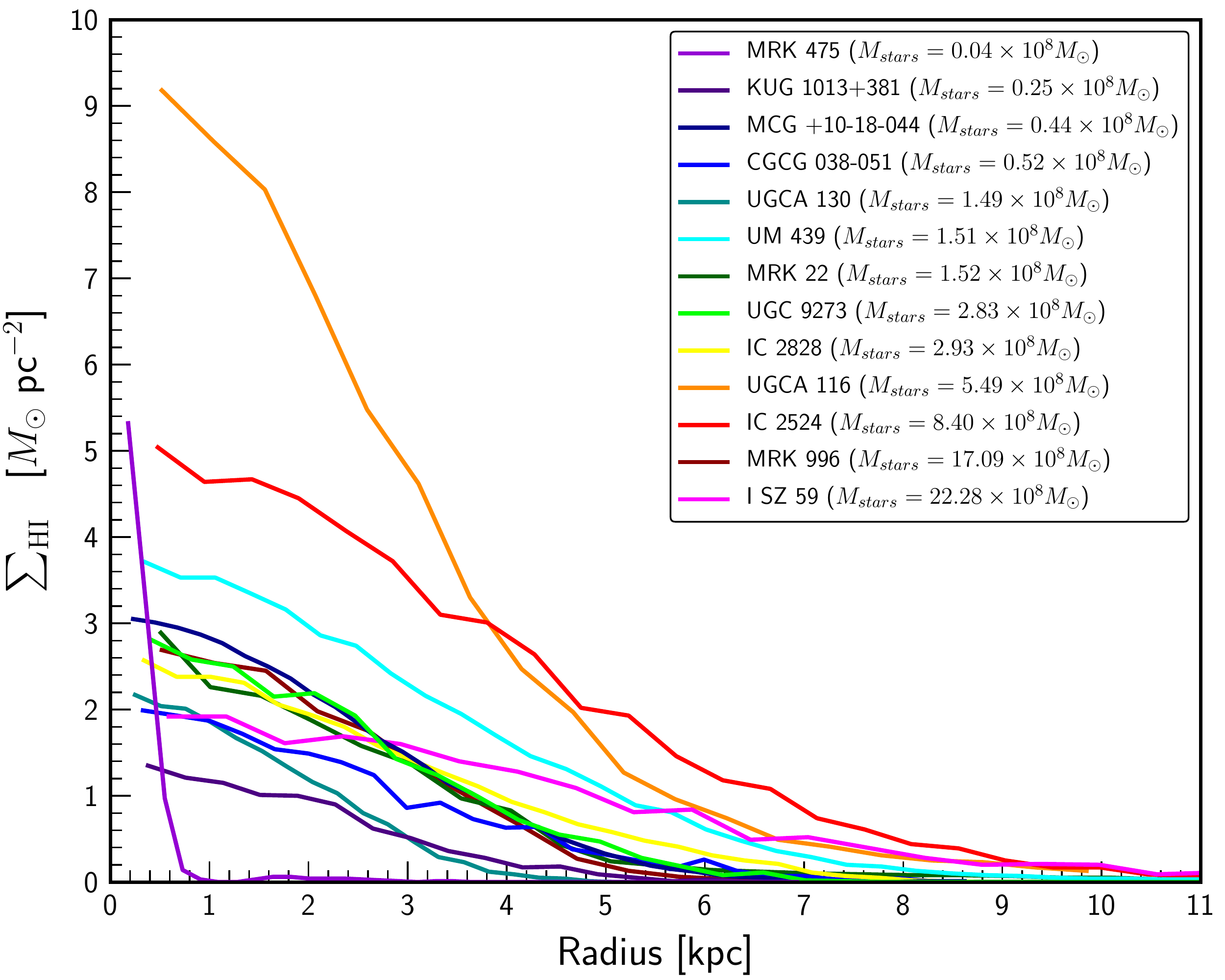}
\caption[ ]{The face-on \HI mass surface density profiles of the galaxies. The color coding is done in order of increasing stellar mass.}
\label{surface}
\end{figure}

The estimated \HI parameters are given in Table~\ref{radio_par}. The columns in this table stand for the galaxy name, \HI flux integral, corrected line-widths at 20\% and 50\% levels, optical inclination angle, galaxy rotation velocity, galaxy systemic velocity, \HI mass, \HI disk radius and \HI mass to blue light ratio, respectively. The \HI mass to blue light ratio is a scale-free \HI parameter as it is independent of the distance to the galaxy. The blue luminosities of the galaxies were estimated using the $B$-band magnitudes from literature (see, Table~\ref{phot}) and the extinction corrections from \citet{Jaiswal2016MNRAS.462...92J} in the relation $L_B/L_{B,\odot} = 10^{0.4(M_{B,\odot} - M_B)}$ with $M_{B,\odot} =5.515 \pm 0.020$ mag \citep{Ram2012ApJ...752....5R, Pecaut2013ApJS..208....9P}. The comparison of the \HI flux integral estimated using the GMRT observation and that from the single-dish measurements for the sample galaxies available in the literature is shown in Figure~\ref{sd}. In many cases, the two measurements are different even within uncertainties. In the plot of logarithm of \HI flux integral ratios, majority of galaxies show deviation $<$0.1 dex. A few galaxies show deviations up to 0.3 dex. Considering relatively large errors in the single-dish measurements, this comparison indicates that the GMRT measurements are fairly accurate and do not suffer from the interferometric flux-loss for the sizes of the galaxies studies here. The \HI mass surface-density profiles of the sample galaxies are shown in Figure~\ref{surface}. The average peak surface density of these galaxies is $\sim 2.5~M_\odot~{\rm pc}^{-2}$. The optical disk radius at 25 mag~arcsec$^{-2}$ surface brightness in the $B$-band image of the sample galaxies are given table~\ref{phot} from the HyperLEDA database \citep{Paturel1991A&A...243..319P, Paturel1997A&AS..124..109P}. The error-weighted average value of the \HI diameter to optical diameter ratio ($D_{\rm HI}/D_{25}$) is estimated to be $1.9 \pm 0.1$, which is higher than those estimated for spiral galaxies in field and group environment \citep{Broeils1997A&A...324..877B, Verheijen2001A&A...370..765V, Omar2005JApA...26....1O}. It is worth to point that due to strong tidal features in UGCA~116, the \HI diameter of UGCA~116 is poorly constrained. The \HI radius for UGCA~116 is estimated by assuming a face-on system. The \HI radius is taken along the tidal tail, hence the ratio of its \HI diameter to its optical diameter is significantly different from the average value.

\begin{table*}
\centering
\caption{\HI parameters.}
\scalebox{0.95}{
\begin{tabular}{lccccccc}
\hline
\hline
Galaxy name  & $\int S{\rm d}v$ & $W_{20}$      & $W_{50}$      & $V_{sys}$     & $M_{\rm HI}$         & $R_{\rm HI}$ & $\frac{M_{\rm HI}}{L_B}$ \\
             & [Jy km s$^{-1}$] & [km s$^{-1}$] & [km s$^{-1}$] & [km s$^{-1}$] & [10$^{8}$ $M_\odot$] & [kpc]        & $\left[\frac{M_\odot}{L_{B,\odot}}\right]$ \\
\hline
MRK~996        & $0.96 \pm 0.09$ & $152 \pm 2$ & $105 \pm 2$ & $1618 \pm 10$& $1.05 \pm 0.11$ & $3.66 \pm 0.39$ & $0.08 \pm 0.01$\\
UGCA~116       & $16.00 \pm 0.72$& $148 \pm 2$ & $109 \pm 3$ & $802 \pm 7$  & $4.32 \pm 0.27$ & $5.70 \pm 0.63$ & $0.81 \pm 0.14$\\
UGCA~130       & $1.55 \pm 0.19$ & $67 \pm 1$  & $57 \pm 1$  & $788 \pm 1$  & $0.40 \pm 0.05$ & $2.29 \pm 0.45$ & $0.14 \pm 0.02$\\
MRK~22         & $1.12 \pm 0.12$ & $125 \pm 3$ & $103 \pm 1$ & $1563 \pm 4$ & $1.15 \pm 0.13$ & $3.54 \pm 0.05$ & $0.37 \pm 0.05$\\
IC~2524        & $4.69 \pm 0.19$ & $182 \pm 1$ & $135 \pm 5$ & $1471 \pm 7$ & $4.26 \pm 0.18$ & $6.68 \pm 0.40$ & $0.48 \pm 0.09$\\
KUG~1013+381   & $0.72 \pm 0.09$ & $73 \pm 1$  & $48 \pm 2$  & $1171 \pm 4$ & $0.41 \pm 0.05$ & $1.90 \pm 0.20$ & $0.18 \pm 0.02$\\
CGCG~038-051   & $1.95 \pm 0.14$ & $96 \pm 1$  & $78 \pm 2$  & $1029 \pm 2$ & $0.87 \pm 0.07$ & $2.82 \pm 0.30$ & $0.87 \pm 0.09$\\
IC~2828        & $2.87 \pm 0.17$ & $66 \pm 2$  & $46 \pm 3$  & $1043 \pm 1$ & $1.31 \pm 0.08$ & $3.95 \pm 0.30$ & $0.38 \pm 0.04$\\
UM~439         & $4.67 \pm 0.21$ & $125 \pm 1$ & $101 \pm 4$ & $1094 \pm 4$ & $2.35 \pm 0.12$ & $5.13 \pm 0.18$ & $0.45 \pm 0.04$\\
I~SZ~59        & $1.66 \pm 0.18$ & $153 \pm 1$ & $126 \pm 2$ & $1816 \pm 5$ & $2.30 \pm 0.26$ & $4.75 \pm 0.15$ & $0.14 \pm 0.03$\\
MCG~+10-18-044 & $5.34 \pm 0.24$ & $84 \pm 1$  & $69 \pm 2$  & $702 \pm 1$  & $1.10 \pm 0.05$ & $3.64 \pm 0.06$ & $1.16 \pm 0.19$\\
UGC~9273       & $1.63 \pm 0.16$ & $102 \pm 1$ & $58 \pm 3$  & $1275 \pm 5$ & $1.11 \pm 0.11$ & $3.71 \pm 0.46$ & $0.18 \pm 0.04$\\
MRK~475        & $0.30 \pm 0.05$ & $55 \pm 1$  & $40 \pm 1$  & $564 \pm 1$  & $0.04 \pm 0.01$ & $0.60 \pm 0.09$ & $0.11 \pm 0.03$\\
\hline
\end{tabular}}
\label{radio_par}
\end{table*}

\begin{table*}
\centering
\caption{Photometric data.}
\scalebox{0.8}{
\begin{tabular}{lccccc}
\hline
\hline
Galaxy name & $B$   & $g-i$ & $i$   & $R_{25}$ & Galaxy density \\
            & [mag] & [mag] & [mag] & [kpc]    & [Mpc$^{-3}$]   \\
\hline
MRK~996        & $15.01 \pm 0.03^a$ & $0.82 \pm 0.15^d$ & $14.16 \pm 0.15^d$ & $1.93 \pm 0.40^e$ & 4.54 \\
UGCA~116       & $15.48 \pm 0.13^b$ & $0.93 \pm 0.05$   & $15.80 \pm 0.04$   & $0.96 \pm 0.10^f$ & 1.19 \\
UGCA~130       & $15.13 \pm 0.04^a$ & $0.43 \pm 0.15^d$ & $14.74 \pm 0.14^d$ & $0.92 \pm 0.18^g$ & 2.86 \\
MRK~22         & $15.99 \pm 0.04^a$ & $0.22 \pm 0.02$   & $15.69 \pm 0.01$   & $1.05 \pm 0.36^g$ & 4.54 \\
IC~2524        & $14.90 \pm 0.20^b$ & $0.52 \pm 0.01$   & $14.16 \pm 0.01$   & $2.11 \pm 0.33^g$ & 6.21 \\
KUG~1013+381   & $16.11 \pm 0.01^c$ & $-0.44 \pm 0.01$  & $16.36 \pm 0.01$   & $0.79 \pm 0.28^g$ & 3.34 \\
CGCG~038-051   & $16.53 \pm 0.03^c$ & $0.22 \pm 0.03$   & $16.02 \pm 0.02$   & $1.00 \pm 0.24^g$ & 4.77 \\
IC~2828        & $15.43 \pm 0.02^c$ & $0.47 \pm 0.02$   & $14.62 \pm 0.01$   & $2.02 \pm 0.24^h$ & 5.01 \\
UM~439         & $14.77 \pm 0.03^a$ & $0.07 \pm 0.01$   & $14.76 \pm 0.01$   & $1.72 \pm 0.24^g$ & 4.06 \\
I~SZ~59        & $15.02 \pm 0.04^a$ & $0.84 \pm 0.05^d$ & $14.14 \pm 0.03^d$ & $3.60 \pm 0.40^g$ & 6.92 \\
SBS~1222+614   & $14.96 \pm 0.07^c$ & $-0.14 \pm 0.07$  & $14.87 \pm 0.04$   & $0.88 \pm 0.22^i$ & 7.16 \\
MCG~+10-18-044 & $16.01 \pm 0.01^c$ & $0.22 \pm 0.02$   & $15.51 \pm 0.02$   & $1.53 \pm 0.16^j$ & 7.40 \\
UGC~9273       & $14.88 \pm 0.18^b$ & $0.47 \pm 0.02$   & $14.91 \pm 0.02$   & $2.06 \pm 0.28^g$ & 1.43 \\
MRK~475        & $16.20 \pm 0.03^a$ & $-0.46 \pm 0.03$  & $16.62 \pm 0.03$   & $0.50 \pm 0.20^i$ & 3.82 \\
\hline
\end{tabular}}
\begin{flushleft}
{\footnotesize {References for $B$-band magnitudes: $^a$ \citet{Gil2003ApJS..147...29G}, $^b$ Third Reference Catalogue \citep[RC3:][]{de1991rc3..book.....D}, $^c$ B-band magnitude estimated using SDSS magnitudes and Lupton transformation equation.\linebreak}}\\

{\footnotesize {The $(g-i)$ colors and $i$-band apparent magnitudes super-scripted by `$d$' are calculated using Johnson $B$ and Cousins $R$ band magnitudes from \citet{Gil2003ApJS..147...29G}, and Lupton transformation equations. The $(g-i)$ colors and $i$-band magnitudes for other galaxies are taken from the SDSS.\linebreak}}\\

{\footnotesize References for $R_{25}$ radius: $^e$ \citet{MacGillivray1988ESOC...28..389M}, $^f$ \citet{Salzer2002AJ....124..191S}, $^g$ \citet{Paturel2000A&AS..146...19P}, $^h$ \citet{Vorontsov1968TrSht..38....1V}, $^i$ \citet{Adelman2007ApJS..172..634A}, $^j$ \citet{Huchtmeier2000A&AS..141..469H}.}
\end{flushleft}
\label{phot}
\end{table*}

\section{Discussions}

\subsection{Interaction features in galaxies}\label{morpho}
The \HI morphologies indicate that the majority of these galaxies have features indicative of tidal interaction such as lopsidedness, tails and bridges, plumes and irregular velocity field. In many cases, the centre of \HI emission is offset from that of the \Ha line or optical continuum emission. Often additional \HI components with no optical counterpart are found in the vicinity of the galaxy. As the dynamical ages of such tidal features are typically 100 Myr, these features often trace very recent tidal encounters which may be responsible for the young and massive star-formation in the galaxies. The following notes summarize main features seen in the \HI images of the individual galaxies:\\

\subsubsection{MRK~996}
The \HI emission is seen associated with the galaxy and also from the separate regions in east-south and north of the galaxy. The \HI emission from the galaxy shows extension towards the north. These isolated \HI emission regions are very likely some \HI clouds or faint dwarf galaxies interacting with MRK~996. The global \HI profile shows lopsidedness. The intermediate resolution column density map shows mild warping of the \HI contours, which is also an indication  of recent interaction.\\

\subsubsection{UGCA~116}
It is an interacting pair with well-developed \HI tidal tail and bridge as evident from the channel images and the total \HI image. The velocity field has smooth gradient along the tidal structures. This galaxy pair is \HI rich and is most likely in a merger stage, as also reported previously by \citet{van1998AJ....116.1186V}.\\

\subsubsection{UGCA~130}
The galaxy is also designated as MRK~5. The low-resolution \HI image shows mild warping of the \HI contours in the outer region. The velocity field appears disturbed, indicative of ongoing interaction as also speculated in \citet{Lopez2010A&A...521A..63L}. The \HI envelope shows diffuse extension towards north with a possibility of isolated \HI clouds in the vicinity.\\

\subsubsection{MRK~22}
The \HI image shows an ongoing interaction with a diffuse tidal tail-like structure extending towards south. The velocity field is highly irregular. A bright \HI region is also seen towards the north. It is very likely that MRK~22 is interacting with another galaxy or \HI cloud in the vicinity. Alternatively, it could be a merger system as double nuclei is indicated  \citep{Mazzarella1991AJ....101.2034M}. \citet{Paswan2018MNRAS.473.4566P} recently found a significant metallicity difference between the two nuclei, which also suggests that MRK~22 could be a merging system of two galaxies. The global \HI profile shows multiple peaks.\\

\subsubsection{IC~2524}
The galaxy shows a large region of low column density \HI gas extending towards the south-east of the galaxy. The high resolution \HI image reveals lopsidedness with a cometary shape, which can also be interpreted as a tidal tail. The velocity field shows smooth gradient. The galaxy could be accreting gas or interacting with some other faint galaxy, not identified in the optical images. The global \HI profile shows lopsidedness. The peak of the \HI emission has an offset from the galaxy centre.\\

\subsubsection{KUG~1013+381} 
It also shows signs of interactions with an isolated \HI cloud or optically unidentified faint galaxy towards the east of the galaxy. The velocity field appears disturbed with no obvious sign of rotation in a disk. \HI emissions from the spiral galaxy UGC~5540 at $\sim 8'$ angular distance towards south of the galaxy KUG~1013+381 was also detected. The \HI images of UGC~5540 are not presented here.\\

\subsubsection{CGCG~038-051} 
The velocity field appears disturbed and lopsided. Two bright \HI regions are seen near the center. It is possible that the galaxy has recently encountered tidal interaction or alternatively it is a merger system, which appears more consistent with the optical morphology showing multiple nuclei and cometary shape.\\

\subsubsection{IC~2828}
The \HI morphology of this galaxy shows anomalous velocity \HI emissions and as a result the velocity field appears highly irregular and disturbed. The \HI morphology also indicates a warp. It is very likely a merging system as indicated by the velocity field.\\

\subsubsection{UM~439}
The \HI map indicates some low column density gas extending towards north-east. An extension is also seen in the opposite side (south-west). This morphology is indicative of tail-bridge structure. It is likely that this galaxy is undergoing a tidal interaction with some unidentified galaxy, as also predicted by \citet{Taylor1995ApJS...99..427T} and \citet{van1998AJ....116.1186V}.\\

\subsubsection{I~SZ~59}
The velocity field is highly irregular and disturbed. The \HI map also appear disturbed and lopsided with peak of the \HI emission having an offset from the the galaxy centre. Some \HI clouds with anomalous velocity are detected in channel maps. It is very likely that this galaxy has recently experienced some tidal interaction or is accreting gas.\\

\subsubsection{SBS~1222+614}\label{note_SBS}
The \HI emission from SBS~1222+614 is not detected within the sensitivity of the observations. The \HI mass upper limit to the galaxy is $3 \times 10^6 M_\odot$, assuming a minimum line-width of 20 \kms~ and 3$\sigma$ detection. However, \HI emission is detected from a nearby galaxy MCG~+10-18-044 at an angular distance of $\sim5'.5$ southward of SBS~1222+614. MCG~+10-18-044 is a dwarf irregular galaxy with low optical surface brightness \citep{Huchtmeier2000A&AS..141..469H}. A weak \Ha emission is detected from MCG~+10-18-044, giving $SFR = 0.007 \pm 0.003~ M_\odot {\rm yr}^{-1}$. \HI emission is also detected from a region at an angular distance of $\sim4'.5$ southward of MCG~+10-18-044. The single-dish \HI integral flux of MCG~+10-18-044 from the 100-m radio telescope at Effelsberg is $\sim$6.2~Jy~\kms ~\citep{Huchtmeier2000A&AS..141..469H}. Our estimate is in good agreement with this value after considering the flux contribution from all the \HI emission regions. The velocity field of MCG~+10-18-044 is irregular. The \HI images clearly indicate the presence of tidal tails in MCG~+10-18-044. It is very likely that the galaxies in this region are undergoing tidal interactions.\\

\subsubsection{UGC~9273}
Two \HI clouds having no obvious optical counterparts are detected towards the north and north-east of the galaxy. It is very likely that the galaxy is interacting with these \HI clouds, which may be associated with a faint dwarf galaxy. The \HI disk is asymmetric with respect to the stellar disk with an offset from the optical disk. Two \HI regions are seen near the center. The velocity field is very irregular and disturbed. The high resolution \HI emissions are found to be associated with the diffused \Ha emission regions. We have also detected \HI emission from the spiral galaxy UGC~9275 at an angular distance of $\sim 2'$ east and $\sim 14'$ north of the galaxy UGC~9273. The \HI images of UGC~9275 will be presented elsewhere.\\

\subsubsection{MRK~475}
Using the Nan\c{c}ay 300-m telescope, \citet{Huchtmeier2005A&A...434..887H} detected $0.15 \pm 0.04$ Jy~\kms ~of \HI line flux integral from MRK~475. Our detection is around 2 times of this value. Due to the low \HI content and compact size, this galaxy is not clearly detected in the low resolution image cube. However, the \HI emission is detected in intermediate and high resolution cubes. An elongated \HI feature is seen along the south of the galaxy. The peak of the \HI emission has an offset from the optical center. The velocity field appears disturbed.

\subsection{Tidal interaction in dwarf WR galaxies}
The \HI  morphologies of all the galaxies in the sample were visually checked for signatures of tidal interactions and presence of \HI clouds in the vicinity. A summary of the prominent optical and radio morphological features seen in the WR galaxies in our sample is provided in Table~\ref{morphology}. The \Ha and optical morphology column in this table is taken from \citet{Jaiswal2016MNRAS.462...92J}. The interaction probability in this table is inferred based on various morphological features. The presence of multiple nuclei, arcs and tidal tails are generally considered as a signature of recent tidal encounter \citep[e.g.][]{Beck1999AJ....117..190B, Lopez2004ApJS..153..243L, Matsui2012ApJ...746...26M, Adamo2012MNRAS.426.1185A}. Galaxies showing \HI or optical tidal tails, irregular velocity field, presence of \HI clouds in the vicinity or distinct multiple nuclei and a disturbed optical envelope are termed as highly probable. Galaxies showing lopsidedness or asymmetric light distribution along with any other feature such as probable multiple nuclei or arcs/plumes are termed as moderately probable. Galaxies without any significant disturbed optical or \HI morphology are labeled with low interaction probability. It should be noted that the galaxies termed as having low probability may show some interaction features such as bar, mild lopsidedness, or nuclear star formation. However, due to the absence of any strong feature of recent tidal interaction, we preferred to label such objects with low interaction probability.

In the present sample, we find that three galaxies show lopsided or disturbed radio continuum emission based on analyses presented in \citet{Jaiswal2016MNRAS.462...92J}. Based on the present study, six galaxies show lopsidedness in \HI and nine galaxies are found to have disturbed \HI morphologies and velocity fields. The lopsidedness in galaxies in general has been studied previously and is believed to be caused by tidal interactions \citep[e.g.][]{Jog1997ApJ...488..642J,Zaritsky1997ApJ...477..118Z,Angiras2006MNRAS.369.1849A}. Majority of dwarf galaxies in the present sample are showing intense nuclear star-burst consistent with the previous results \citep[e.g.][]{Strickland1999MNRAS.306...43S,Adamo2011MNRAS.415.2388A}. It has also been predicted via N-body simulations \citep[e.g.][]{Hernquist1995ApJ...448...41H} that the tidal interaction and mergers between galaxies can channel gas towards the centre of galaxy as a result of loss of angular momentum. This gas can give rise to nuclear starburst in galaxies \citep{Mihos1994ApJ...425L..13M}. The optical and radio morphological features seen in our study are also highly suggestive of prevalence of tidal interactions and mergers in the WR galaxies. A total of 12 galaxies are inferred here to be tidally interacting based on optical and \HI morphologies. The remaining one galaxies, namely SBS~1222+614, is not detected in \HI line imaging. However, the detected galaxy MCG~+10-18-044 in the vicinity of SBS~1222+614 is found to have some interaction features. These results are in agreement with the previous studies on the dwarf WR galaxies, where tidal interactions were inferred \citep{Mendez1999AJ....118.2723M,Lopez2008A&A...491..131L,Lopez2010A&A...521A..63L}. The main cause of starburst in several dwarf and blue compact galaxies appear to be tidal interactions with a nearby companion or low mass \HI cloud without any known optical companion. The optical companion may be an ultra-faint or low surface brightness dwarf galaxy not detected in present optical images. The \HI clouds may also be anomalous velocity \HI clouds falling onto galaxy as a result of past merger or tidal interaction with a galaxy. Alternatively, some of these clouds may be a result of accretion of gas from the inter-galactic medium. It is not possible to favour or reject a possibility based on the present observations. Combined with previous studies on the dwarf WR galaxies, our analysis suggests that overall more than three-fourths of WR galaxies show obvious signatures of recent tidal encounters and mergers.

Recently \citet{Teich2016ApJ...832...85T} and \citet{McNichols2016ApJ...832...89M} have performed a ``Survey of \HI in Extremely Low-mass Dwarfs (SHIELD)'' having 12 objects observed through the Karl G. Jansky Very Large Array (VLA) at a variety of spatial and spectral resolutions. They also found the offsets between \HI and star formation peaks in galaxies with the lowest total \HI masses. SHIELD galaxies' dynamics are found to be equally supported by rotational and pressure contributions. However, in the present sample, most of the galaxies are showing disturbed velocity field.

\begin{table*}
\caption{Optical and radio morphologies of galaxies}
\begin{tabular}{lcccc}
\hline 
\hline
Name & Morphological & {\Ha \& optical} & {H{\sc i}/radio morphology} & Interaction \\
     & type          & {morphology}     & {\& other features}         & {probability} \\
\hline 
\hline
MRK~996      & BCD      & misaligned \Ha and        & lopsided, interacting      & high     \\
             &          & old stellar disk          &                            &          \\
\hline
UGCA~116     & Irr pec  & \Ha arcs and plumes,  & tails and bridges, & high \\
             &          & cometary, tidal tails & merger             &      \\
\hline
UGCA~130     & Irr      & double nuclei, & disturbed,       & moderate \\
             &          & cometary       & past interaction &          \\
\hline
MRK~22       & BCD      & \Ha arcs,     & tails, irregular velocity field, & high \\
             &          & double nuclei & merger                           &      \\
\hline
IC~2524      & S        & nuclear SF & mild tail/bridge, lopsided & high \\
             &          &            &                            &      \\
\hline
KUG~1013+381 & BCD      & nuclear SF & tail, disturbed velocity field, & high \\
             &          &            & interacting pair                &      \\
\hline
CGCG~038-051 & dIrr     & multiple nuclei?, & disturbed, lopsided & high \\
             &          & asymmetric        &                     &      \\
\hline
IC~2828      & Im       & two nuclei?, \Ha arcs and plumes, & Irregular velocity field, & high \\
             &          & cometary, irregular               & anomalous \HI regions     &      \\
\hline
UM~439       & Irr      & multiple nuclei?, \Ha arcs and plumes, & mild tail/bridge, & high \\
             &          & irregular                              & \HI lopsided      &      \\
\hline
I~SZ~59      & S0       & nuclear SF,        & irregular velocity field, &  high \\
             &          & elongated envelope & anomalous \HI regions     &       \\
\hline
SBS~1222+614 & dIrr     & nuclear SF, \Ha arcs, & Not detected in \HI & low \\
             &          & irregular             &                     &     \\
\hline
MCG~+10-18-044 & dIrr   & \Ha arcs, lopsided & irregular velocity field,  & high \\
               &        &                    & Nearby companion/\HI cloud &      \\
\hline
UGC~9273     & Im       & \Ha ring?, & \HI cloud, interacting & high \\
             &          & lopsided   &                        &      \\
\hline
MRK~475      & BCD      & nuclear SF, & Irregular velocity field & high \\
             &          & lopsided    &                     &     \\
\hline
\hline
\end{tabular}
\label{morphology}
\end{table*}

\subsection{Mass analysis}\label{sec:mass}
The main mass components of a late-type galaxy are gas, dust, stars and dark matter. The total mass of gas ($M_{gas}$), dust ($M_{dust}$) and stars ($M_{stars}$) is jointly termed as the luminous mass or baryonic mass. The stellar mass contribution to the baryonic mass generally dominates over the gas mass in bright spiral galaxies. However, in dwarf star-forming galaxies the gas mass can be a large fraction of the stellar mass. The dust mass remains only a few percent of the gas mass \citep[e.g.][]{Draine2007ApJ...663..866D, Clemens2013MNRAS.433..695C} and therefore can be neglected while estimating the baryonic mass. An estimate of the dark matter mass ($M_{dark}$) can be done by subtracting the baryonic mass ($M_{bary}$) from the total dynamical mass ($M_{dyn}$), which is estimated here within the \HI disk radius of the galaxy. The masses of different components of a galaxy constraint the galaxy properties and therefore provide an indication of the galaxy formation and evolution \citep[e.g.][]{Lopez2010A&A...521A..63L}.

The estimations of the gas, dust and stellar masses require analyses of different radiations coming from these components. The gas mass can be calculated by adding the neutral atomic hydrogen (\ion{H}{i}) mass, molecular hydrogen (H$_2$) mass and helium (He) mass, neglecting very small contributions from the metals in the gaseous phase and the ionized gas \citep{Remy2014A&A...563A..31R}. The \HI mass in the galaxies in the present sample are estimated using the GMRT \HI flux integral and are given in Table~\ref{radio_par}. The H$_2$ gas mass is normally measured using the CO emission lines. The amount of H$_2$ gas in dwarf star-forming galaxies can be substantial. However, no CO observations are available for the galaxies in our sample and hence we could not include H$_2$ mass in the present analysis. The atomic gas masses (see, Table~\ref{mass}) are estimated as $M_{\rm H\,I} + M_{\rm He} = 1.4M_{\rm H\,I}$ following \citet{Bradford2015ApJ...809..146B}.

The stellar mass of a galaxy can be estimated by multiplying its optical or near-infrared (NIR) luminosity with a value of stellar mass-to-light (M/L) ratio known for different wavebands. The mass-to-light ratio of stellar population varies with age and metallicity \citep{Schombert2009AJ....137..528S, Schombert2014PASA...31...36S}. It is therefore an approximation to assume a single, constant mass-to-light ratio for estimating the stellar mass in a galaxy as the galaxy is likely to have stars of different ages and metallicity. A slightly sophisticated approach for estimating the stellar mass is to use the color-dependent mass-to-light ratio derived using a stellar population synthesis model. The variation of mass-to-light ratio with optical color is found to be minimum in the NIR bands in comparison to the optical wavebands \citep[e.g.][]{McGaugh2014AJ....148...77M}. \citet{Taylor2011MNRAS.418.1587T} have found that the relation between $(g-i)$ SDSS color and mass-to-light ratio in the $i$-band is very tight for a sample of galaxies. Therefore, we also used $(g-i)$ color and the $i$-band mass-to-light ratio relation for the galaxies in the present sample to estimate stellar mas following the \citet[][see Appendix~\ref{app}]{Bell2003ApJS..149..289B} relation. The scatter in this correlation is nearly 0.1 dex. Note that \citet{Bell2003ApJS..149..289B} relation to estimate the stellar mass uses \citet{Salpeter1955ApJ...121..161S} initial mass function (IMF) modified with \citet{Bell2001ApJ...550..212B} procedure, in order to provide the maximum possible stellar M/L ratio (at a given color) which can truly constrain the galaxy rotation velocity, while the normal \citet{Salpeter1955ApJ...121..161S} IMF overpredicts the galaxy rotation velocity through higher number of low-mass faint stars and thereby larger value of the stellar M/L ratio. The $(g-i)$ colors and $i$-band apparent magnitudes (see Table~\ref{phot}) for 10 galaxies are taken from the SDSS catalogue data release-7 \citep{Abazajian2009ApJS..182..543A}. The $g$ and $i$ band magnitudes for the remaining three galaxies (MRK~996, UGCA~130 and I~SZ~59), which do not have SDSS data, are extrapolated from the Johnson $B$ and Cousins $R$ band magnitudes \citep{Gil2003ApJS..147...29G} with the help of Lupton transformation relations (\url{http://www.sdss3.org/dr8/algorithms/sdssUBVRITransform.php#Lupton2005}). The $i$-band luminosities are estimated in a similar manner as the $B$-band luminosities, considering $M_{i,\odot} = 4.57 \pm 0.03$~mag (\url{http://www.sdss3.org/dr8/algorithms/ugrizVegaSun.php}). The masses of different mass components of the galaxies are given in Table~\ref{mass}. It can be seen from this table that the gas masses of some galaxies are greater than their stellar masses. The errors in the masses are estimated using error propagation of the quantities in the formula of corresponding mass given in the Appendix~\ref{app}.

\begin{table*}
\centering
\caption{Mass components of the galaxies}
\begin{tabular}{lccccc}
\hline
\hline
Galaxy name & $M_{\rm H\,I} + M_{\rm He}$ & $M_{stars}$ & gas fraction & $M_{dyn}$ & DEF$_{\rm HI}$\\
            & [$10^{8} M_\odot$] & [$10^{8} M_\odot$] &  & [$10^{8} M_\odot$] & \\
\hline
MRK~996        & $1.47 \pm 0.15$ & $17.09 \pm 3.37$ & $0.08 \pm 0.01$ & $23.30 \pm 2.64$  & $0.93 \pm 0.27$ \\
UGCA~116       & $6.05 \pm 0.38$ & $5.49 \pm 0.72$  & $0.52 \pm 0.04$ & $39.11 \pm 4.83$  & $-0.07 \pm 0.14$\\
UGCA~130       & $0.56 \pm 0.07$ & $1.49 \pm 0.27$  & $0.27 \pm 0.05$ & $4.30 \pm 0.86$   & $0.94 \pm 0.26$\\
MRK~22         & $1.61 \pm 0.18$ & $1.52 \pm 0.17$  & $0.51 \pm 0.07$ & $21.69 \pm 0.52$  & $0.56 \pm 0.44$\\
IC~2524        & $5.96 \pm 0.25$ & $8.40 \pm 0.89$  & $0.42 \pm 0.02$ & $70.31 \pm 6.70$  & $0.37 \pm 0.21$\\
KUG~1013+381   & $0.57 \pm 0.07$ & $0.25 \pm 0.03$  & $0.70 \pm 0.10$ & $2.53 \pm 0.34$   & $0.85 \pm 0.46$\\
CGCG~038-051   & $1.22 \pm 0.10$ & $0.52 \pm 0.06$  & $0.70 \pm 0.07$ & $9.91 \pm 1.17$   & $0.65 \pm 0.31$\\
IC~2828        & $1.83 \pm 0.11$ & $2.93 \pm 0.32$  & $0.38 \pm 0.03$ & $4.83 \pm 0.73$   & $0.86 \pm 0.16$\\
UM~439         & $3.29 \pm 0.17$ & $1.51 \pm 0.19$  & $0.69 \pm 0.05$ & $30.22 \pm 2.62$  & $0.52 \pm 0.18$\\
I~SZ~59        & $3.22 \pm 0.36$ & $22.28 \pm 3.32$ & $0.13 \pm 0.02$ & $43.55 \pm 1.95$  & $0.94 \pm 0.16$\\
SBS~1222+614   & --- & $0.41 \pm 0.05$ & --- & --- & --- \\
MCG~+10-18-044 & $1.54 \pm 0.07$ & $0.44 \pm 0.06$  & $0.78 \pm 0.05$ & $10.01 \pm 0.60$  & $0.79 \pm 0.14$\\
UGC~9273       & $1.55 \pm 0.15$ & $2.83 \pm 0.31$  & $0.35 \pm 0.04$ & $7.21 \pm 1.16$   & $0.95 \pm 0.18$\\
MRK~475        & $0.06 \pm 0.01$ & $0.04 \pm 0.01$  & $0.60 \pm 0.12$ & $0.55 \pm 0.09$   & $1.61 \pm 0.52$\\
\hline
\end{tabular}
\label{mass}
\end{table*}

\begin{figure}
\centering
\includegraphics[angle=0,width=0.7\linewidth]{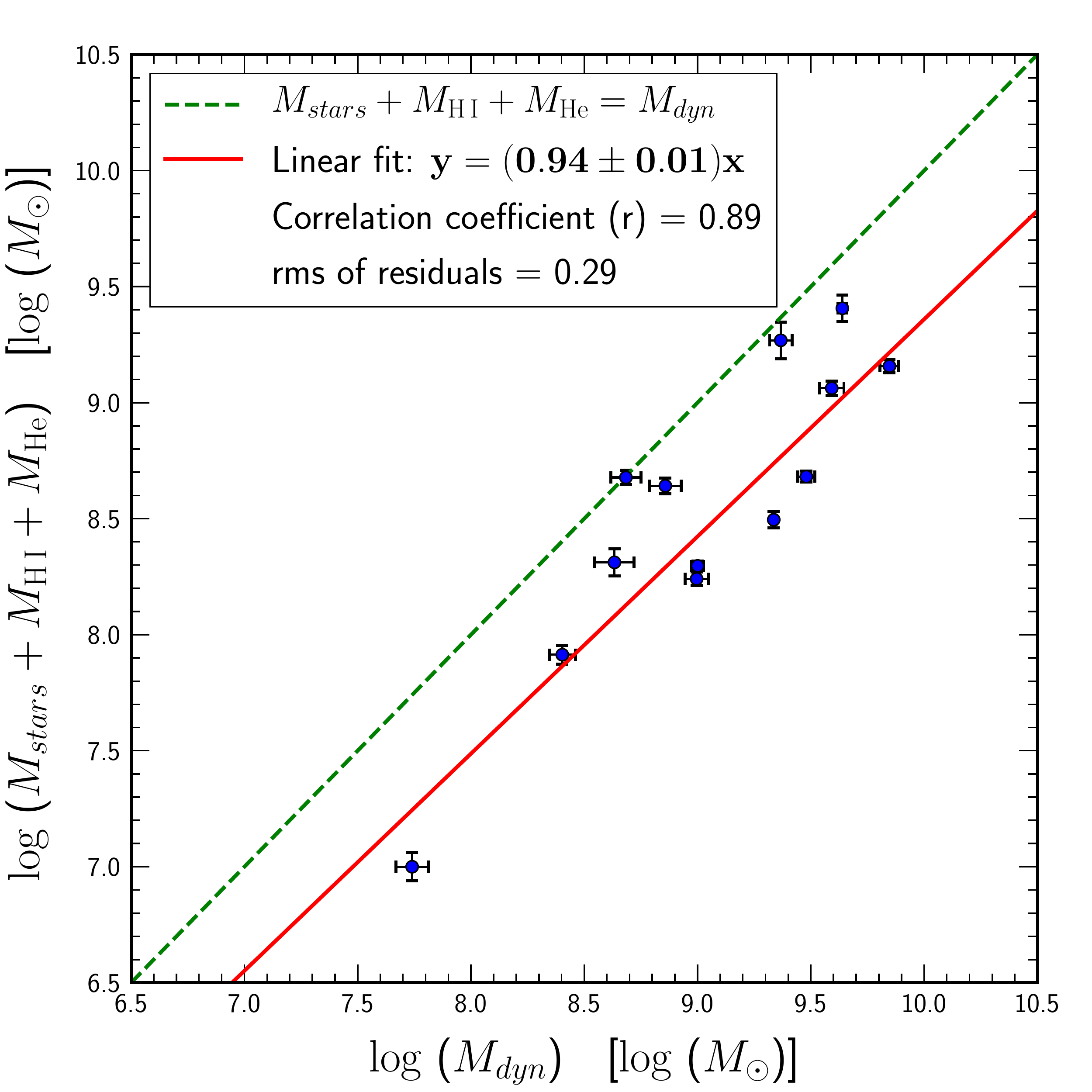}
\caption[ ]{The relation between baryonic mass and total dynamical mass within the \HI disk radius. The solid line is a fit to the data-points and the dashed line corresponds to equal baryonic and dynamical masses.}
\label{Dark}
\end{figure}

\begin{figure}
\centering
\includegraphics[angle=0,width=0.7\linewidth]{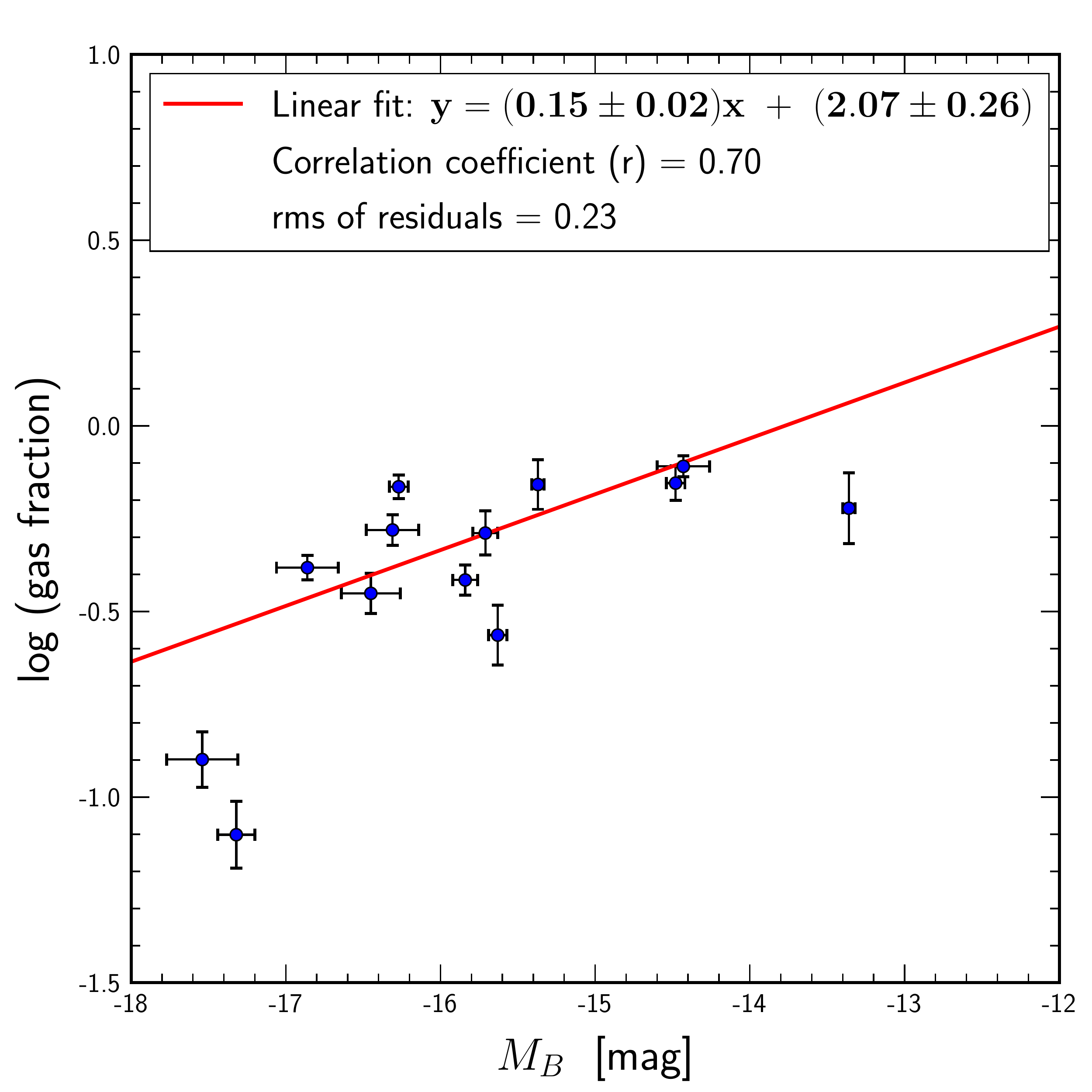}
\caption[ ]{The variation of gas fraction with the absolute $B$-band magnitude.}
\label{GasFrac}
\end{figure}

We plotted the baryonic mass $(M_{stars} + M_{\rm H\,I} + M_{\rm He})$ against the dynamical mass and carried out a linear fit in Figure~\ref{Dark}. We applied the condition that the baryonic mass tends to zero at the zero dynamical mass, i.e. the intercept of the linear fit on the $y$-axis is fixed to zero. This plot indicates that the baryonic mass and the dynamical mass are well correlated with the correlation coefficient $r=0.89$, which in turn shows a little or devoid of dark matter within the \HI disk radius of these systems. We also find that the \HI disk diameter up to 1~$M_\odot~{\rm pc}^{-2}$ \HI mass surface density is larger than the stellar disk diameter for the galaxies in the sample, consistent with previous studies. We also noticed that the gas fraction, given as the ratio of the gas mass $(M_{\rm H\,I} + M_{\rm He})$ and the baryonic mass are directly correlated with their $B$-band absolute magnitude (see Figure~\ref{GasFrac}), which is related to the young stellar population in the galaxies. The error-weighted average gas fraction is found to be $0.24 \pm 0.01$, which is in agreement with the lower limit of $0.15$ in \citet{Bradford2015ApJ...809..146B} study of non-isolated low-mass galaxies suggesting that the low gas fraction in these systems are caused by the environmental effects (see Section~\ref{content}). The error-weighted average baryon fraction ($(M_{stars} + M_{\rm H\,I} + M_{\rm He})/M_{dyn} = 0.18 \pm 0.01$) for our sample is also in agreement with \citet{Bradford2015ApJ...809..146B} results.

\begin{figure}
\centering
\includegraphics[angle=0,width=0.49\linewidth]{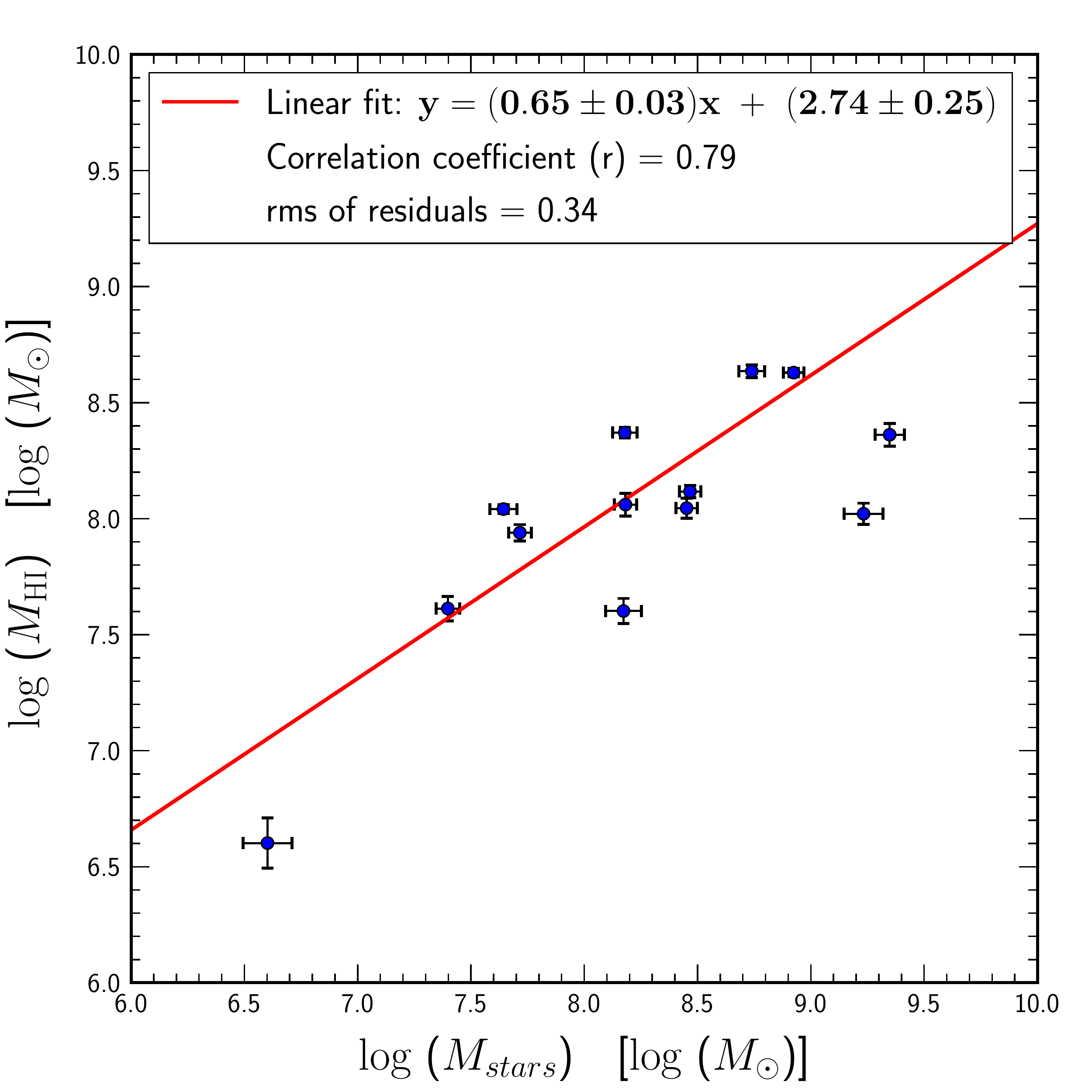}
\includegraphics[angle=0,width=0.49\linewidth]{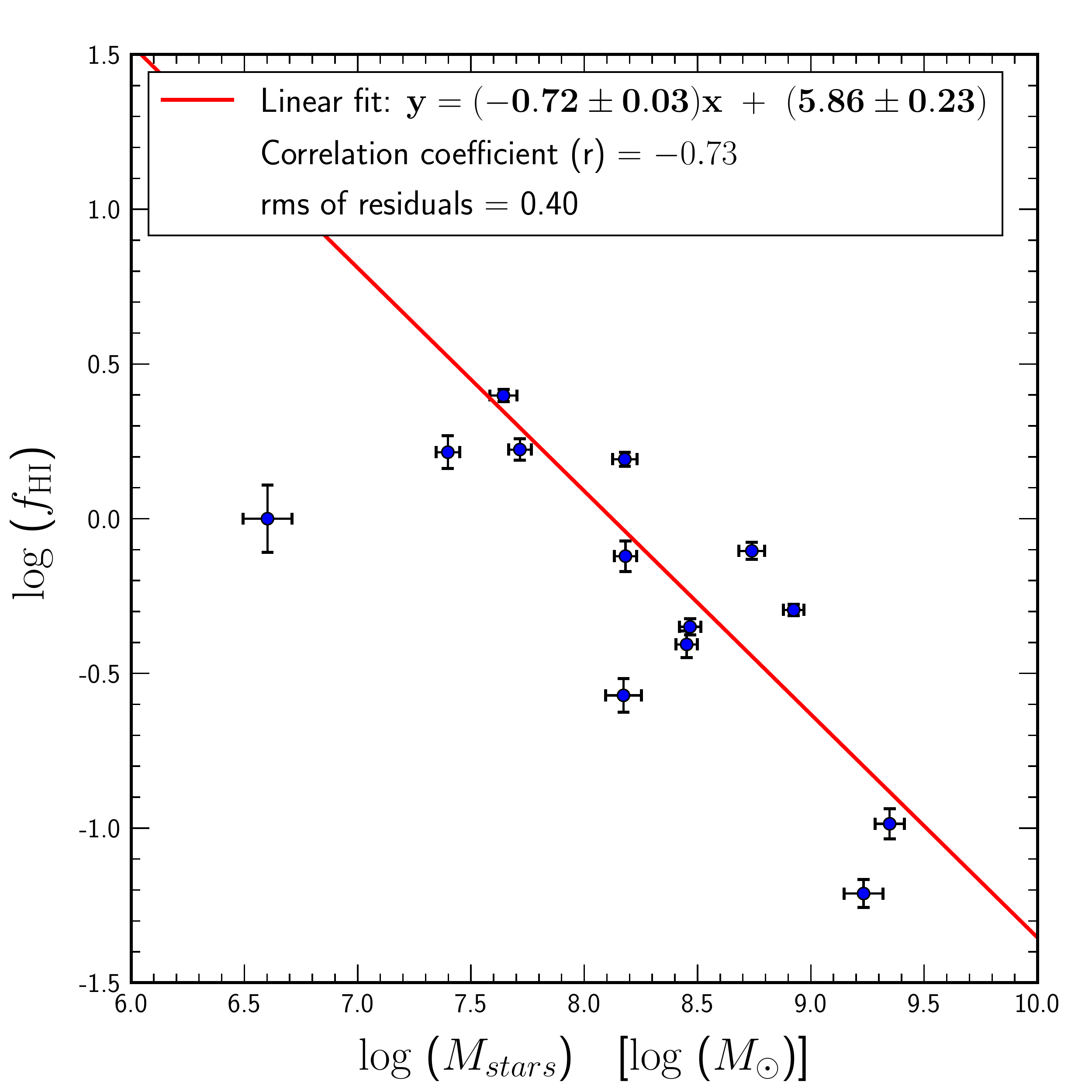}
\caption[ ]{The variation of \HI mass and \ion{H}{i}-to-stellar mass fraction with the stellar mass.}
\label{HIstars}
\end{figure}

The \ion{H}{i}-stellar mass relation is well known tool to provide an indication of the history of gas accretion in conjunction with star formation \citep[e.g.][]{Huang2012ApJ...756..113H, Maddox2015MNRAS.447.1610M, McQuinn2015ApJ...802...66M, Parkash2018ApJ...864...40P}. The \HI mass is found to be directly correlated with the stellar mass of galaxies in the sample and the \ion{H}{i}-to-stellar mass fraction ($f_{\rm H\,I} = \frac{M_{\rm H\,I}}{M_{stars}}$) is anti-correlated with the stellar mass (figure~\ref{HIstars}). These correlations are consistent with the previous studies \citep[e.g.][]{Wang2017MNRAS.472.3029W, Parkash2018ApJ...864...40P}. The average \ion{H}{i}-to-stellar mass fraction for the sample galaxies is found to be $0.14 \pm 0.01$.

\begin{figure}
\centering
\includegraphics[angle=0,width=0.7\linewidth]{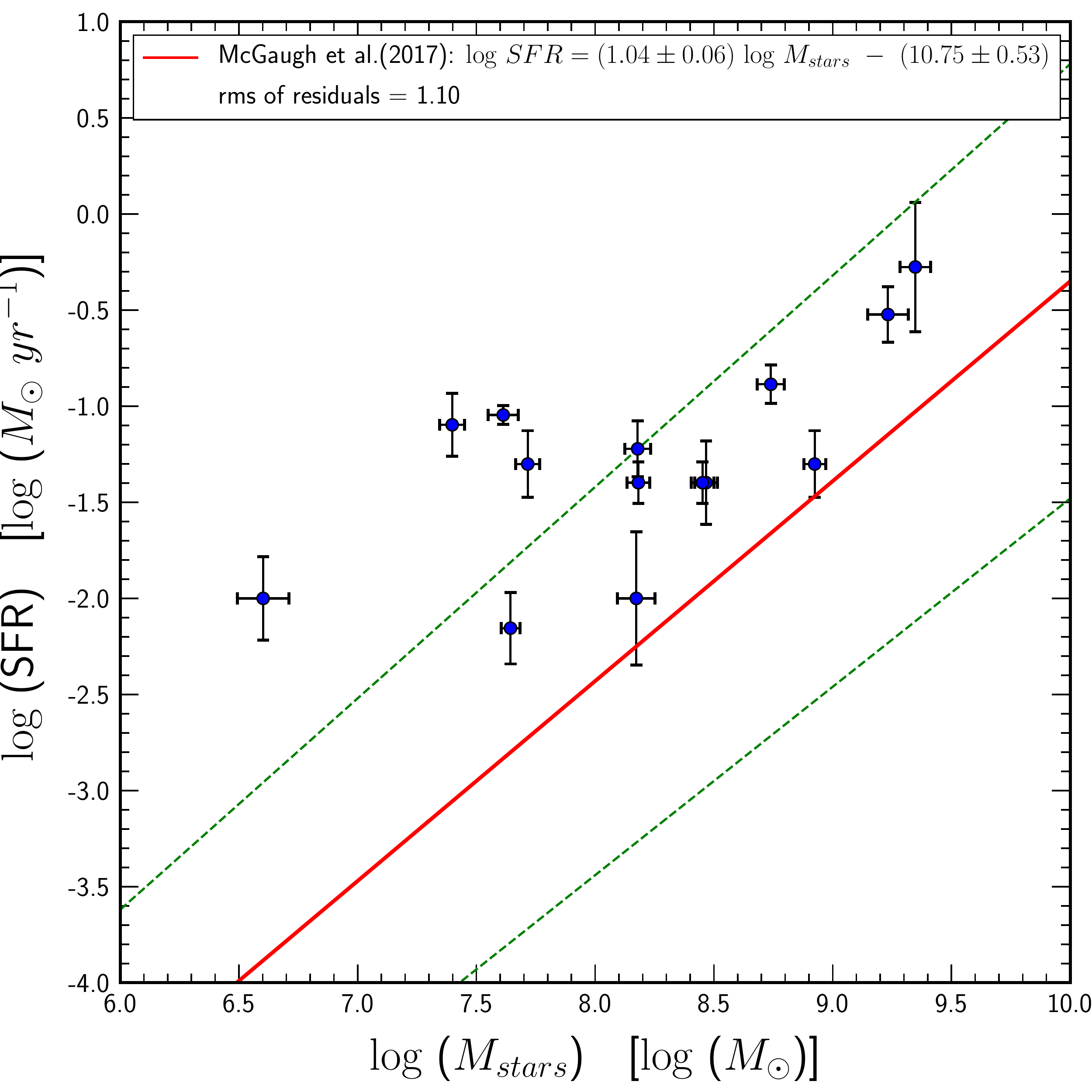}
\caption[ ]{The SFR and stellar mass relation for the sample galaxies.}
\label{SFRstars}
\end{figure}

The SFR-stellar mass relation for star-forming galaxies is referred as the main sequence of galaxies \citep{Noeske2007ApJ...660L..43N,Speagle2014ApJS..214...15S,Kurczynski2016ApJ...820L...1K}. This relation suggests possible range of SFRs at a given stellar mass for a particular redshift and can be used to study the star formation history of galaxies. We have compared our sample with a resembling sample, both in mass and redshift, of 56 LSB galaxies from \citet{McGaugh2017ApJ...851...22M} in terms of the SFR-stellar mass relation. Figure~\ref{SFRstars} shows the SFR vs stellar mass plot for our sample galaxies along with the SFR-stellar mass relation obtained from \citet{McGaugh2017ApJ...851...22M}. Our sample galaxies are found to have elevated SFRs compared to the reference sample. This result is consistent with the expectation as we have selected the galaxies from the catalogs of WR galaxies.

\subsection{\HI content}\label{content}
The environment of a galaxy plays an important role in deciding the gas content and the morphology of a galaxy \citep[e.g.][]{Fasano2000ApJ...542..673F, Goto2003MNRAS.346..601G}. It is found observationally that the galaxies located near the cluster centres have on average less \HI content in comparison to those at the outskirts of the clusters or field galaxies of the same morphological type and size \citep{Giovanelli1985ApJ...292..404G, Haynes1986ApJ...306..466H, Solanes2001ApJ...548...97S, Gavazzi2006A&A...449..929G}. The deficiency of \HI in the galaxies within the cluster environment can be understood using several gas-removal mechanisms, for example tidal interactions \citep{Mihos2005ApJ...631L..41M} and ram-pressure stripping \citep{Gunn1972ApJ...176....1G}. The process of gas removal via tidal interactions depends upon galaxy density and relative velocity of the encounters. The tidal interactions for the galaxies moving slow in a high galaxy density region (e.g., in a group environment) will be more effective in removing gas from the galaxies. The gas removal via ram-pressure depends on density of matter in the inter-galactic medium (IGM) and velocity with which the galaxy is moving in the medium in the sense that at higher density and velocity (e.g., in a cluster environment), the process will be more effective. These processes affect the \HI morphology in the form of \HI tails and bridges, displaced \HI from the disk, lopsided and truncated \HI disks \citep[e.g.][]{Chung2007ApJ...659L.115C, Hibbard2001ASPC..240..657H, Denes2016MNRAS.455.1294D}. The removed \HI content of a galaxy affected by the environment through the gas removal processes can be quantified as the \HI deficiency parameter \citep{Giovanelli1983AJ.....88..881G}: 
$DEF_{\rm HI} = {\rm log} (M_{\rm HI, exp}) - {\rm log} (M_{\rm HI, obs})$,
where $M_{\rm HI, obs}$ is the observed \HI mass of the galaxy and $M_{\rm HI, exp}$ is the expected \HI mass of the galaxy obtained using some \ion{H}{i}-optical scaling relation for a sample of isolated galaxies. We have used the convention that the galaxies with $DEF_{\rm HI} > 0.3$ are considered to be \ion{H}{i}-deficient. Here, $DEF_{\rm HI} = 0.3$ indicates two times less \HI than the isolated galaxy of the same morphological type and size in the field environment. There are various \ion{H}{i}-optical scaling relation available for the estimation of expected \HI mass for a galaxy of a particular morphological type \citep{Haynes1984AJ.....89..758H, Chamaraux1986A&A...165...15C, Solanes1996ApJ...461..609S, Catinella2010MNRAS.403..683C, Toribio2011ApJ...732...93T, Denes2014MNRAS.444..667D}. It is suggested \citep[e.g.][]{Haynes1984AJ.....89..758H} to use the distance-independent quantities like $M_{\rm HI}/L$ and $M_{\rm HI}/D^2$, where $L$ is luminosity and $D$ is diameter in some optical waveband, in order to obtain \HI deficiency in a galaxy. The diameters are less affected by the extinction and its uncertainty in comparison to the luminosities. It has been shown \citet{Haynes1984AJ.....89..758H}  that $M_{\rm HI}/D^2$ is a better parameter to represent the \HI deficiency in comparison to $M_{\rm HI}/L$ by measuring the uncertainty in these quantities for individual galaxies. They also noted that $M_{\rm HI}/D^2$, in contrast to $M_{\rm HI}/L$, is nearly independent of the morphological type of the galaxy. This is because $M_{\rm HI}$ and $D$ are the disk properties only, while $L$ includes the contribution due to the (gas-free) spheroidal bulge whose contribution to the optical luminosity varies with the morphological type of the galaxy. We therefore use \HI mass and optical diameter scaling relation to estimate the expected \HI mass of the galaxies and the \HI deficiency parameter. Based on the above arguments and the availability of the photometric data, we have therefore used the scaling relation between the \HI mass and the $B$-band isophotal diameter at 25~mag~arcsec$^{-2}$ surface brightness derived using an un-biased sample of galaxies given by \citet{Denes2014MNRAS.444..667D} for 844 galaxies having \HI masses between $10^8$--$10^{10.5}$~$M_\odot$.

\begin{figure*}
\centering
\includegraphics[angle=0,width=0.6\linewidth]{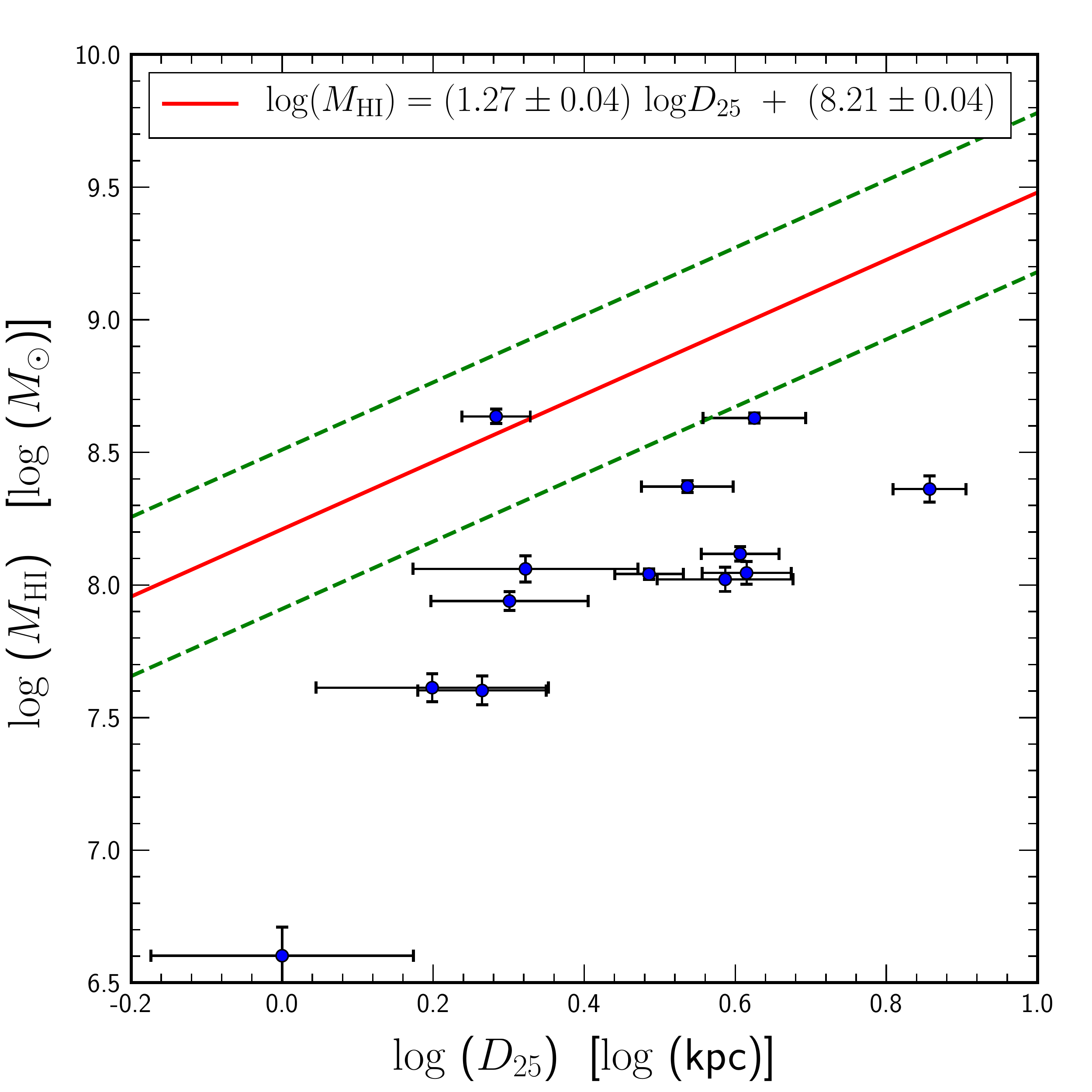}
\caption[ ]{\HI mass is plotted against $B$-band isophotal diameter showing the \HI deficiency of the sample galaxies. The solid line indicates the \HI mass and $B$-band diameter scaling relation from \citet{Denes2014MNRAS.444..667D} and the dashed lines indicate $DEF_{\rm HI}=\pm 0.3$.}
\label{DEF1}
\end{figure*}

\begin{figure*}
\centering
\includegraphics[angle=0,width=0.6\linewidth]{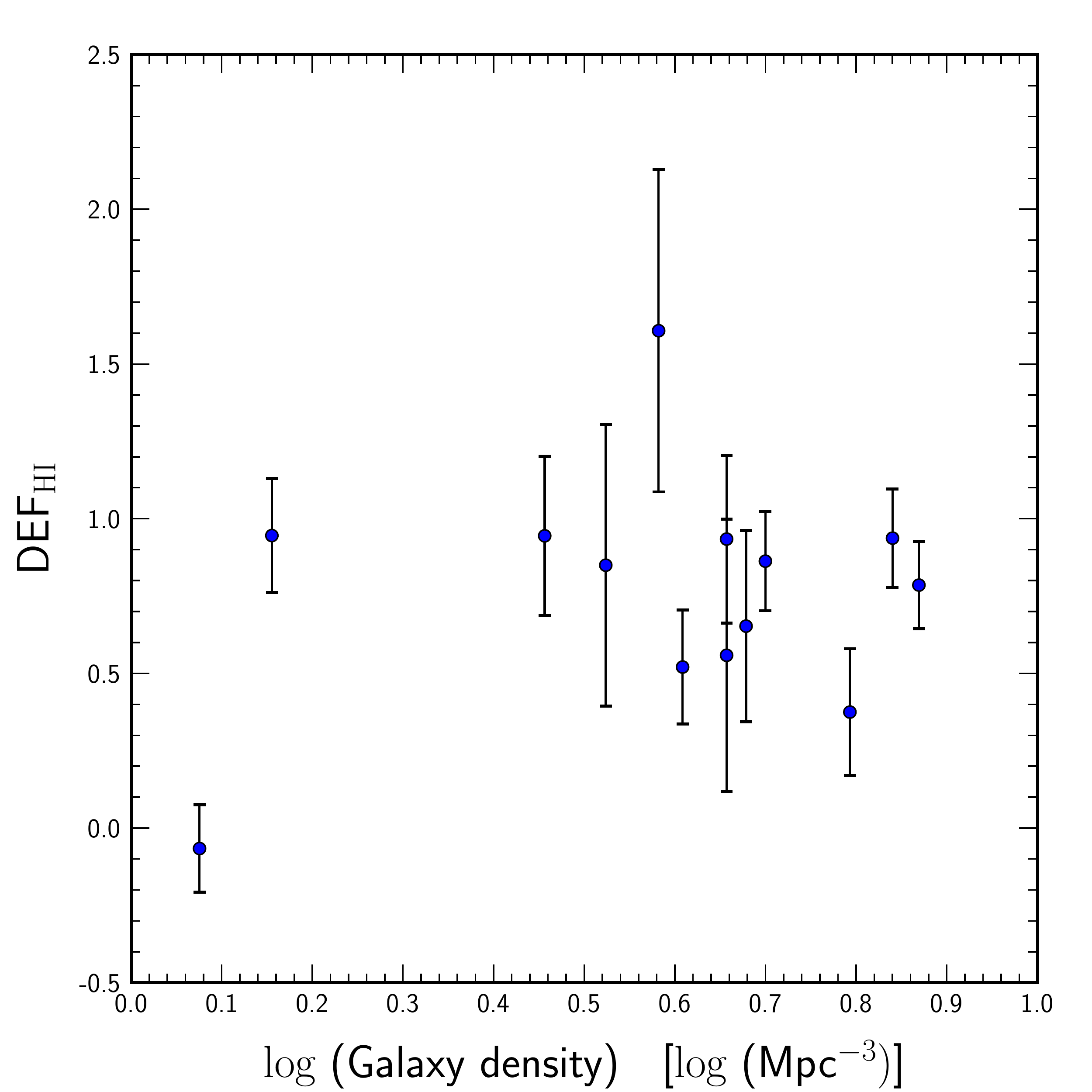}
\caption[ ]{The \HI deficiency plotted against the galaxy density in the vicinity.}
\label{DEF2}
\end{figure*}

Figure~\ref{DEF1} shows the plot of \HI mass versus $B$-band isophotal diameter. The solid line indicates the \HI mass and $B$-band diameter scaling relation from \citet{Denes2014MNRAS.444..667D} and the dashed lines indicate $1\sigma$ scatter of $\pm 0.3$. It can be seen that most of the galaxies in our sample are \HI deficient. Only UGCA~116 is found to have its normal \HI content. The effect of environment on the deficiency parameter is investigated by plotting the galaxy density around the sample galaxies versus the deficiency parameter (Figure~\ref{DEF2}). The galaxy density is defined here as the number of galaxies within the co-moving volume having 1~Mpc projected radius and $\pm 500$~km~s$^{-1}$ radial velocity range around the target galaxy. The environmental galaxy densities around each of the sample galaxies are taken directly from the NED tool(\url{https://ned.ipac.caltech.edu/forms/denv.html}) and are given in Table~\ref{phot}. The NED-measured environment uses the spectroscopic redshift information of the sources given in the literature and optical/NIR/radio line survey catalogues, for example, SDSS catalogues \citep{York2000AJ....120.1579Y}, Third Reference Catalogue of bright galaxies \citep[RC3:][]{de1991rc3..book.....D}, The Updated Zwicky Catalog \citep[UZC:][]{Falco1999PASP..111..438F}, NOAO Fundamental Plane Survey \citep{Smith2004AJ....128.1558S}, The DEEP2 Galaxy Redshift Survey \citep[DEEP2:][]{Davis2003SPIE.4834..161D}, The CFA Redshift Survey \citep{Huchra1999ApJS..121..287H}, The 6dF Galaxy Survey \citep{Jones2009MNRAS.399..683J}, The Southern Sky Redshift Survey \citep{da1998AJ....116....1D}, The Smithsonian Hectospec Lensing Survey \citep[SHELS:][]{Geller2016ApJS..224...11G}, 2MASS Redshift Survey \citep{Huchra2012ApJS..199...26H}, Neutral Hydrogen Surveys \citep{Schneider1992ApJS...81....5S, Theureau1998A&AS..130..333T, Barnes2001MNRAS.322..486B, Paturel2003A&A...412...57P, Tully2008ApJ...676..184T, Haynes2011AJ....142..170H}. As our sample galaxies are nearby systems, the redshifts of most of their environment sources are known accurately. Estimated galaxy densities from NED are only a lower limit due to a possibility of unidentified members surrounding the galaxies in the sample. Our aim here was to assess the environment, which appears to be similar to groups of galaxies for almost all WR galaxies studied here. The number of group members within $\sim$1~Mpc radius for the WR galaxies studied here is 5--31. The galaxy densities are similar to typical group environments such as those in the Eridanus groups of galaxies \citep{Omar2005JApA...26...71O} and \citet{Sengupta2006MNRAS.369..360S}. The primary cause of \HI deficiency in these groups are identified as tidal interactions. It is noticed that UGCA~116 having no deficiency is lying in the lowest galaxy density ($\sim$1.2~Mpc$^{-3}$) region and all the galaxies with significant deficiency are residing in a relatively higher galaxy density ($< 8$~Mpc$^{-3}$) regions. This trend is similar to that noticed in \citet{Omar2005JApA...26...71O}.

\begin{figure*}
\centering
\includegraphics[angle=0,width=0.6\linewidth]{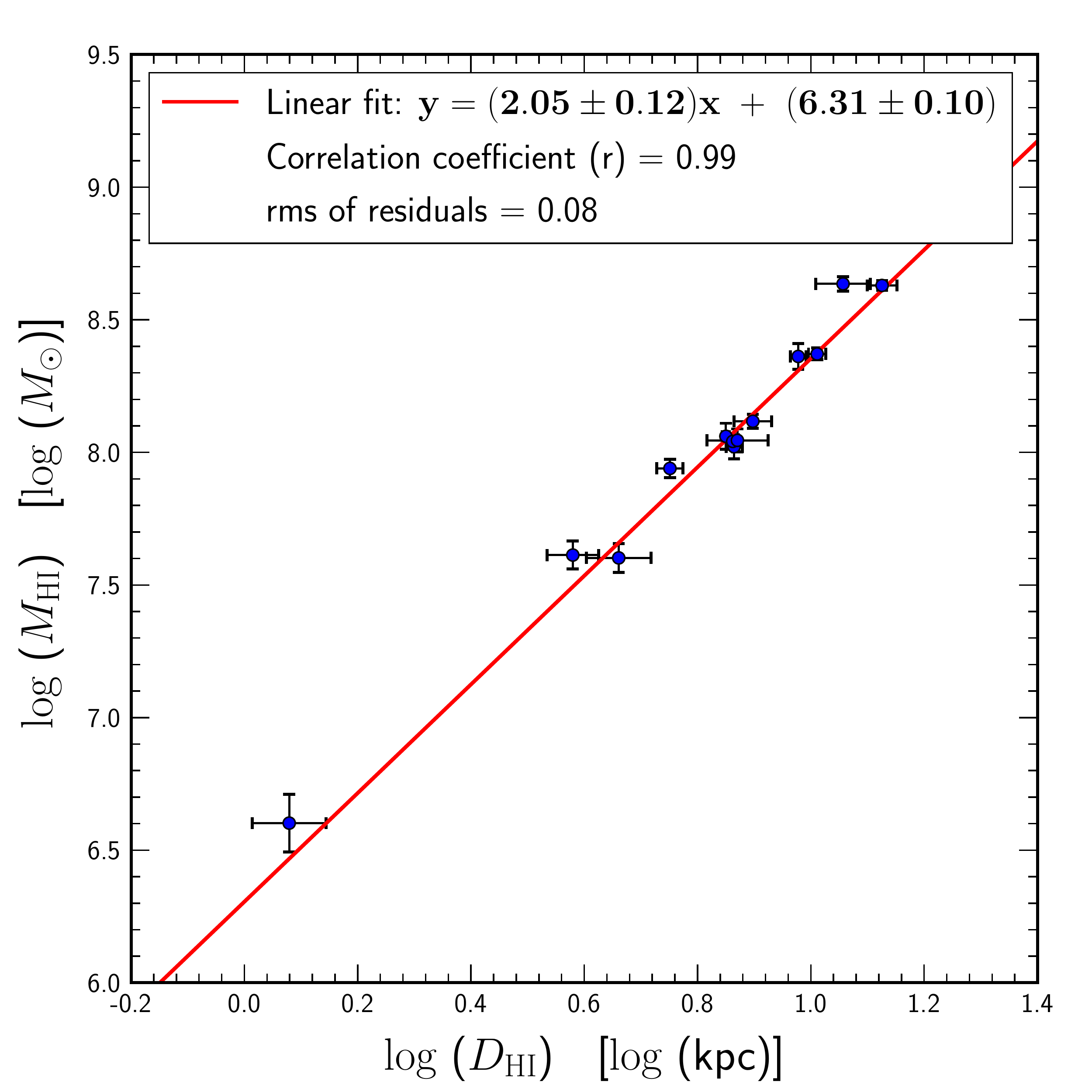}
\caption[ ]{The variation of \HI mass with \HI disk diameter showing a very tight relation.}
\label{HIdia}
\end{figure*}

The variation of the \HI mass with the \HI diameter is plotted in Figure~\ref{HIdia}. The \HI diameter is estimated at the limiting \HI mass surface density as $1~M_\odot~{\rm pc}^{-2}$. The \HI masses of the galaxies are tightly correlated with their \HI disk diameter with a correlation coefficient of $1$ and a scatter of $0.08$. The linear-fit to the galaxies in our sample has the form of ${\rm log} M_{\rm HI} = (2.05 \pm 0.12)~{\rm log} D_{\rm HI} + (6.31 \pm 0.10)$, in good agreement with the previous studies \citep{Haynes1984AJ.....89..758H, Broeils1997A&A...324..877B, Verheijen2001A&A...370..765V, Omar2005JApA...26....1O}. It is worth to point out that the late-type galaxies in the Eridanus group studied by \citet{Omar2005JApA...26...71O} show \HI deficiency, however, average as well as peak \HI mass surface densities of the majority of the Eridanus galaxies are much higher and comparable to the massive spirals in the field environments. The dwarf galaxies studied here, on the other hand, have low \HI mass surface density throughout the \HI disk although the \HI disk extends far beyond the optical disk. The \HI mass surface densities of the galaxies studied here are however comparable to those in dwarf irregular galaxies \citep{Begum2008MNRAS.386.1667B}. A tight relation between \HI mass and \HI diameter indicates that average \HI mass density is fairly constant in all the galaxy types.

The \HI content of a galaxy is an important parameter to study galaxy evolution \citep{Haynes1984AJ.....89..758H}. For example, the gas-rich late-type galaxies in high-density regions such as in galaxy clusters can be transformed into the gas-poor galaxies over time, under the effect of several gas removal mechanisms, such as mergers and tidal interactions \citep{Mihos2005ApJ...631L..41M}, ram pressure stripping  \citep{Gunn1972ApJ...176....1G}, turbulent or viscous stripping \citep{Nulsen1982MNRAS.198.1007N}, thermal evaporation \citep{Cowie1977Natur.266..501C}, starvation \citep{Larson1980ApJ...237..692L} and harassment \citep{Moore1996Natur.379..613M}. However, in low density environment such as loose groups, the gas removal is generally not very intense as compared to that in the galaxy clusters. The main mechanisms for the gas removal in groups is identified as tidal interaction \citep[e.g.][]{Kern2008MNRAS.384..305K, Rasmussen2008MNRAS.388.1245R} or a combination of ram-pressure stripping and tidal interaction \citep[e.g.][]{Davis1997AJ....114..613D, Mayer2006MNRAS.369.1021M, Rasmussen2012ApJ...757..122R}. As the galaxies in our sample are in a group-like environment, it is therefore expected that the tidal interaction mechanism is responsible for the \HI deficiency of the sample galaxies. It is also consistent with the \HI morphologies discussed in Section~\ref{morpho}.

\section{Summary}
The GMRT \HI 21-cm line images of a sample of 13 dwarf star-forming galaxies with WR emission line features were presented. The morphological features in the \HI line images of these galaxies were analysed along with the already published optical \Ha line and $R$-band morphological features. The analysis of masses of the different components of the galaxy was performed by including optical data from archive. The \HI deficiency was also estimated for the galaxies. Our main findings are summarized below:

\begin{enumerate}
\item[1.] The estimated \HI masses, \HI mass surface density, \HI mass to blue light ratios and widths of the \HI profiles of the galaxies are found to be similar to those for other dwarf star-forming galaxies. The \HI masses are estimated in the range 0.04--4.32~$\times~10^{8}$~M$_{\odot}$. The \HI mass surface density of the galaxies are fairly constant at a value of 2.5~M$_{\odot}$~pc$^{-2}$.

\item[2.] Almost all the galaxies in our sample show features of tidal interactions or presence of additional \HI clouds without obvious optical counterparts in the vicinity of the galaxies. These clouds could be remnant of a past merger or associated with a ultra-faint galaxy. Several galaxies show that the centers of the \HI emission differ from that of the \Ha line or optical continuum emission. The optical images of the galaxies also reveal tidal interactions. It is very likely that the tidal interaction has triggered  massive star formation in the WR galaxies. 

\item[3.] The baryonic mass $(M_{stars} + M_{\rm H\,I} + M_{\rm He})$ is strongly correlated with the dynamical mass within \HI disk radius. The gas fraction is well correlated with the absolute $B$-band magnitude. The gas masses are comparable to or larger than the stellar masses. A clear \ion{H}{i}-stellar mass relation exists for the galaxies in the sample. The sample galaxies show elevated SFRs when compared with the star-forming main-sequence of a sample of LSB galaxies with similar stellar mass-range and redshifts.

\item[4.] Majority of the galaxies studied here are found to be \HI deficient. The galaxies in our sample are residing in a group like environment with other galaxies in the vicinity. The \HI deficiency is also very likely caused by galaxy-galaxy tidal interactions, as inferred from the optical and \HI morphologies.\\

\end{enumerate}

The present study focuses on the dwarf galaxies having fewer star-forming regions and therefore easier to analyze the role of star formation in galaxy evolution in comparison to the large galaxies. This work is very important in verifying the star formation triggering through tidal interaction process, especially in case when optical image do not show any feature of interaction. The scaling relations of different mass components of the galaxy constraints the key parameters of galaxy evolution. The estimation of the velocity dispersion in interacting dwarf galaxies requires high sensitivity \HI observation with high spatial and velocity resolution using the future Square Kilometer Array (SKA). The SKA observation will also help in detail investigation of \HI gas removal/accretion mechanism that is responsible for \HI deficiency/access.

\section*{Acknowledgements}
We would like to thank the referee for his/her valuable comments and suggestions which have improved our manuscript. This work is supported by the National Key R\&D Programme of China (2018YFA0404603) and the Chinese Academy of Sciences (CAS, 114231KYSB20170003). SJ is supported by the CAS-PIFI (grant No. 2020PM0057) postdoctoral fellowship. We thank the staff of the GMRT that made these observations possible. GMRT is run by the National Centre for Radio Astrophysics of the Tata Institute of Fundamental Research. We thank the staff of DFOT for their support during our observations. DFOT is run by Aryabhatta Research Institute of Observational Sciences with support from the Department of Science and Technology, Govt. of India. We wish to acknowledge the IUCAA/IGO staff for their support during our observations. IRAF (Image Reduction and Analysis facility) is distributed by NOAO which is operated by AURA Inc., under cooperative agreement with NSF. This research has made use of the NASA/IPAC Extragalactic Database (NED) which is operated by the Jet Propulsion Laboratory, California Institute of Technology, under contract with the National Aeronautics and Space Administration. This research has made use of NASA's Astrophysics Data System. We acknowledge the usage of the HyperLEDA database (http://leda.univ-lyon1.fr). This publication makes use of data products from the Two Micron All Sky Survey, which is a joint project of the University of Massachusetts and the Infrared Processing and Analysis Center/California Institute of Technology, funded by the National Aeronautics and Space Administration and the National Science Foundation. We acknowledge the use of the SDSS. Funding for the SDSS and SDSS-II has been provided by the Alfred P. Sloan Foundation, the Participating Institutions, the National Science Foundation, the U.S. Department of Energy, the National Aeronautics and Space Administration, the Japanese Monbukagakusho, the Max Planck Society, and the Higher Education Funding Council for England. The SDSS Web Site is http://www.sdss.org/. The SDSS is managed by the Astrophysical Research Consortium for the Participating Institutions. The Participating Institutions are the American Museum of Natural History, Astrophysical Institute Potsdam, University of Basel, University of Cambridge, Case Western Reserve University, University of Chicago, Drexel University, Fermilab, the Institute for Advanced Study, the Japan Participation Group, Johns Hopkins University, the Joint Institute for Nuclear Astrophysics, the Kavli Institute for Particle Astrophysics and Cosmology, the Korean Scientist Group, the Chinese Academy of Sciences (LAMOST), Los Alamos National Laboratory, the Max-Planck-Institute for Astronomy (MPIA), the Max-Planck-Institute for Astrophysics (MPA), New Mexico State University, Ohio State University, University of Pittsburgh, University of Portsmouth, Princeton University, the United States Naval Observatory, and the University of Washington.

\section{Data availability}
The \HI 21-cm line data underlying this article are available in GMRT Online Archive, at \url{https://naps.ncra.tifr.res.in/goa/data/search} with proposal codes: 16\_281, 20\_060 and 21\_032.

\bibliography{references}

\appendix
\section{Necessary Formulas}\label{app}
\begin{itemize}
\item The flux density (mJy beam$^{-1}$) is converted to \HI column density (cm$^{-2}$) using the relation \citep{Spitzer1978ppim.book.....S}:
\begin{equation}
N_{\rm HI}(\alpha,\delta) = \frac{1.1 \times 10^{21}}{\theta_a \times \theta_b} \Delta v \sum_j S_j(\alpha,\delta)
\end{equation}
where $\theta_a$ and $\theta_b$ are the synthesized beam size measured in arcsec along the major and minor axes, respectively. $S_j$ is the \HI flux density (mJy beam$^{-1}$) in the $j^{th}$ channel and $\Delta v$ is the channel velocity resolution in \kms.

\item The dynamical mass within a radius $R_{\rm HI}$ of a galaxy with line-width $W_{50}$ is estimated using the relation:
\begin{equation}
\frac{M_{dyn}}{M_\odot} = 2.31 \times 10^5 \left(\frac{R_{\rm HI}}{\rm kpc}\right) \left(\frac{W_{50}/2}{\rm km~s^{-1}}\right)^{\!\!2}
\end{equation}
where the correction due to galaxy inclination is ignored for our highly disturbed galaxy sample.

\item The flux integral (Jy~\kms) is converted to \HI mass ($M_\odot$) using the relation \citep{Roberts1962AJ.....67..437R}:
\begin{equation}
\frac{M_{\rm HI}}{M_\odot} = 2.36 \times 10^5 \left(\frac{d}{\rm Mpc} \right)^{\!\!2} \left(\frac{\int S~{\rm d}v}{\rm Jy~km~s^{-1}}\right)
\end{equation}
where $d ~(= V_{sys}/H_0)$ is the distance to the galaxy in Mpc.

\item The line widths corrected for the instrumental velocity resolution $\Delta V$ in \kms~ are given by the relations \citep{Verheijen2001A&A...370..765V}:
\begin{equation}
\begin{aligned}
W_{20}^r &= W_{20} - 35.8\cdot\left[\sqrt{1 + \left(\frac{\Delta
V}{23.5}\right)^2} - 1 \right]\\
W_{50}^r &= W_{50} - 23.5\cdot\left[\sqrt{1 + \left(\frac{\Delta
V}{23.5}\right)^2} - 1 \right]
\end{aligned}
\end{equation}

\item \citet{Bell2003ApJS..149..289B} relation to estimate the stellar mass using $i$-band luminosity and ($g-i$) color is given as:
\begin{equation}
{\rm log} \left[\left(\frac{M_{star}}{L_i}\right)\left(\frac{L_{i,\odot}}{M_\odot}\right)\right] = -0.152 + 0.518(g - i)
\end{equation}
This relation has the uncertainty of 0.1 dex. The error in stellar mass is estimated using error propagation as:
\begin{equation}
\sigma_{M_{star}} = M_{star} \sqrt{\left(\frac{\sigma_{L_i}}{L_i}\right)^2 + \left(\frac{0.518\times \sigma_{(g-i)}}{(g-i)}\right)^2 + 0.1^2}
\end{equation}

\end{itemize}


\begin{figure*}
\begin{center}
\includegraphics[angle=0,width=1\linewidth]{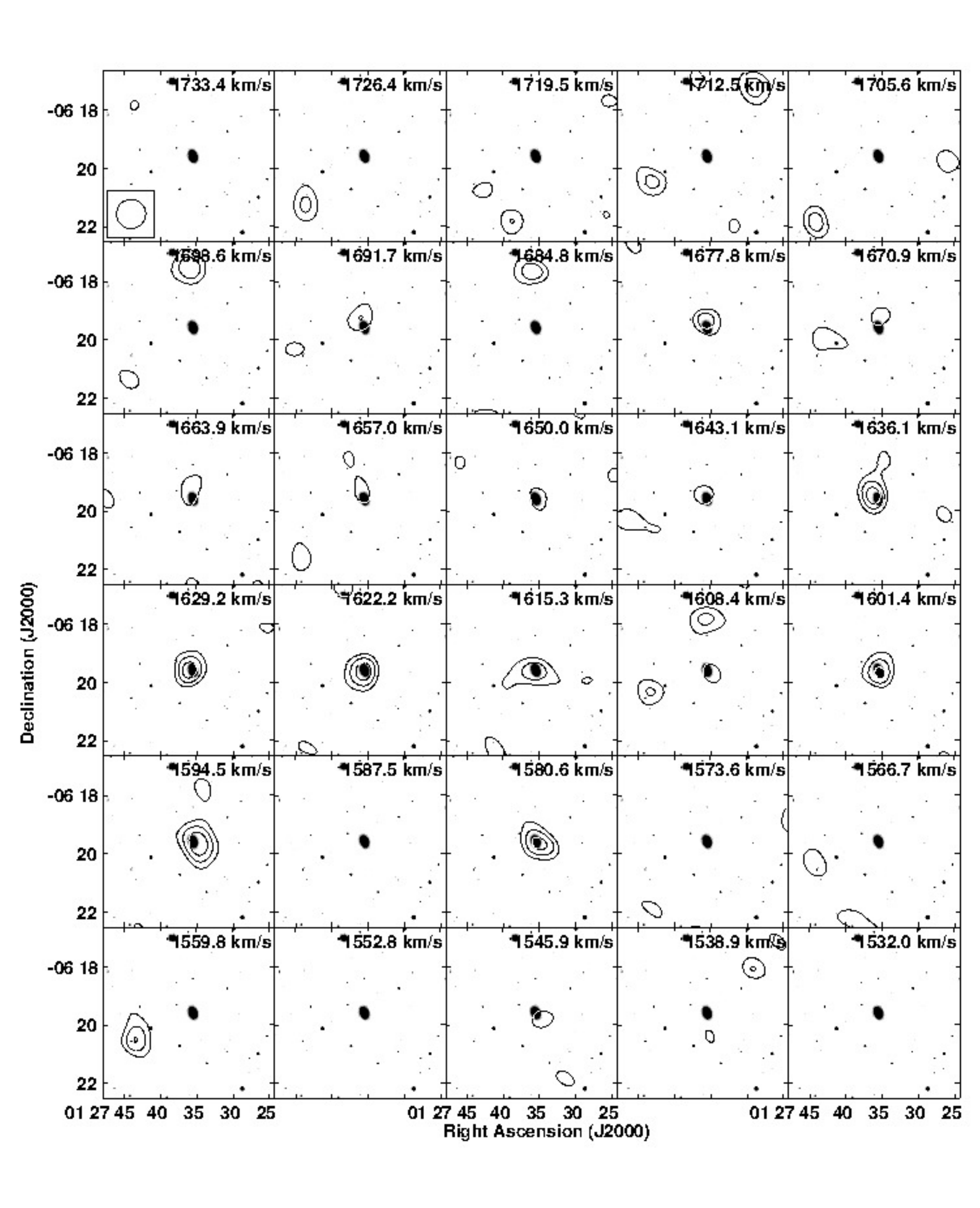}
\end{center}
\caption{The \HI contours from the low resolution channel images overlaid upon the grey scale optical $r$-band image of MRK~996. The contours representing \HI emission flux are drawn at $2.5\sigma\times n$~mJy/Beam; n=1,1.5,2,3,4,6.}
\label{MRK996-1}
\end{figure*}
\begin{figure}
\setlength{\unitlength}{1cm}
\begin{picture}(12,25) 
\put(0,17.9){\hbox{\includegraphics[scale=0.39]{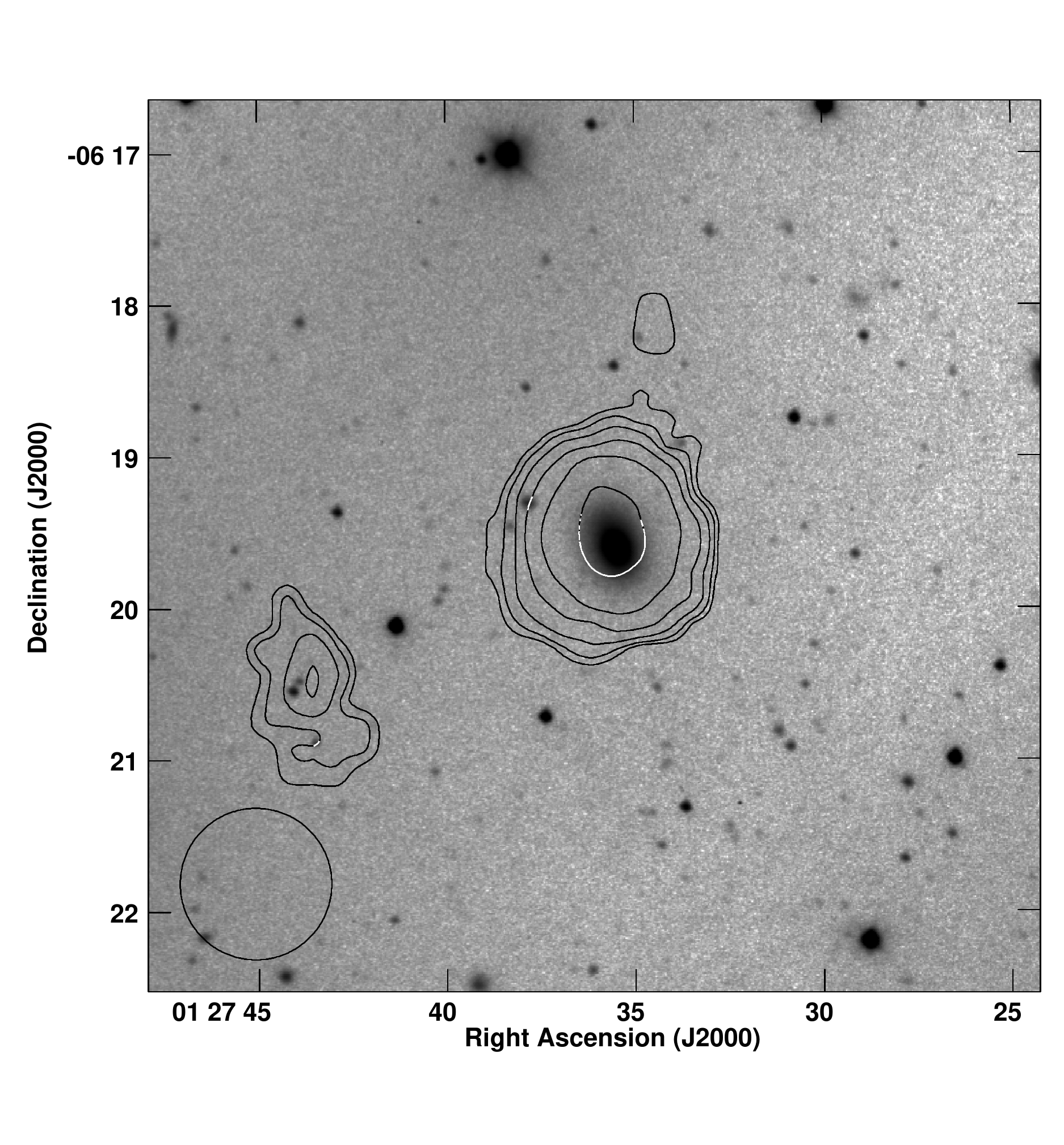}}} 
\put(1.3,24.8){(a)} 
\put(8.3,17.87){\hbox{\includegraphics[scale=0.407]{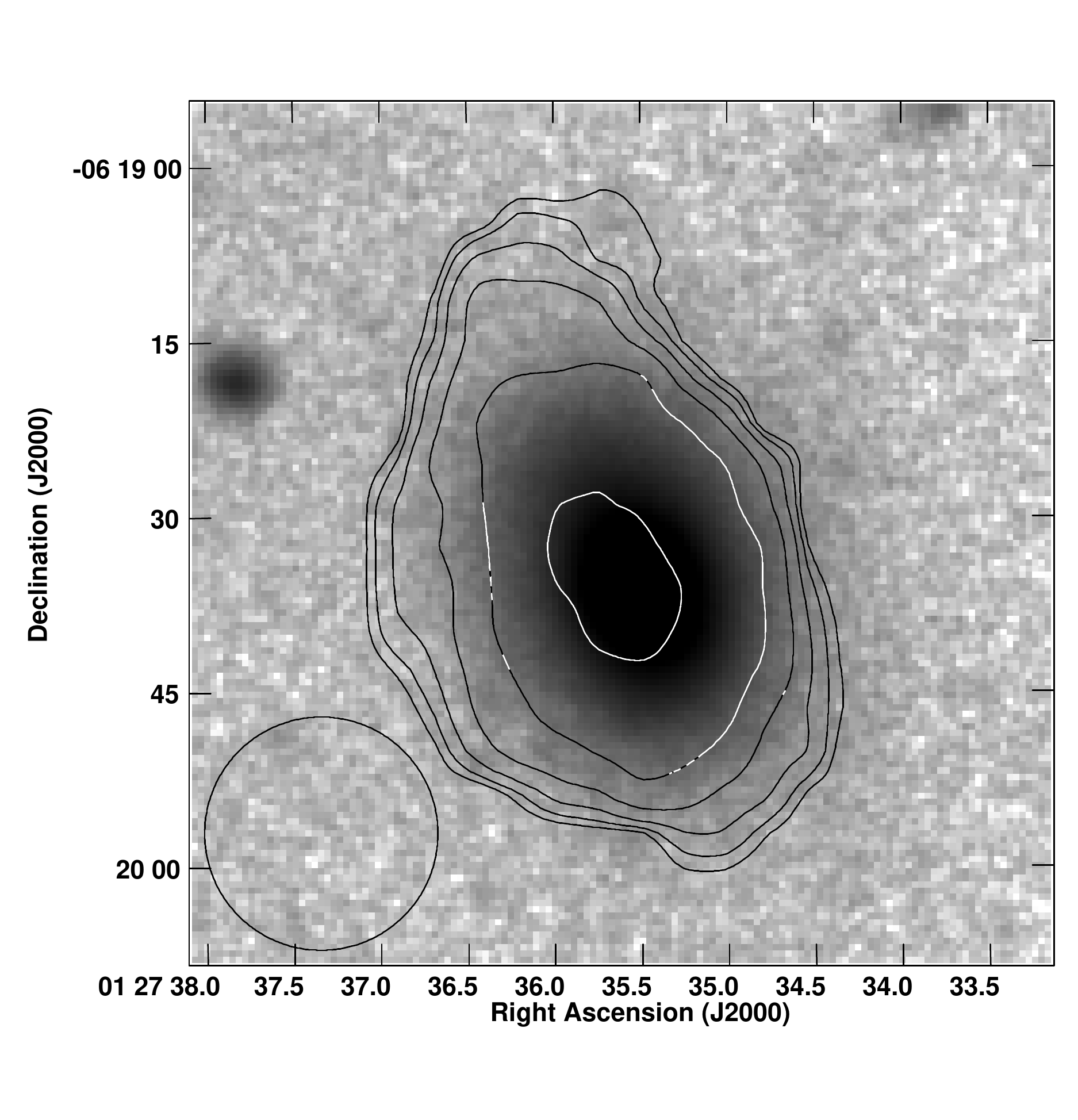}}}
\put(9.9,24.8){(b)} 
\put(-0.3,10.8){\hbox{\includegraphics[scale=0.407]{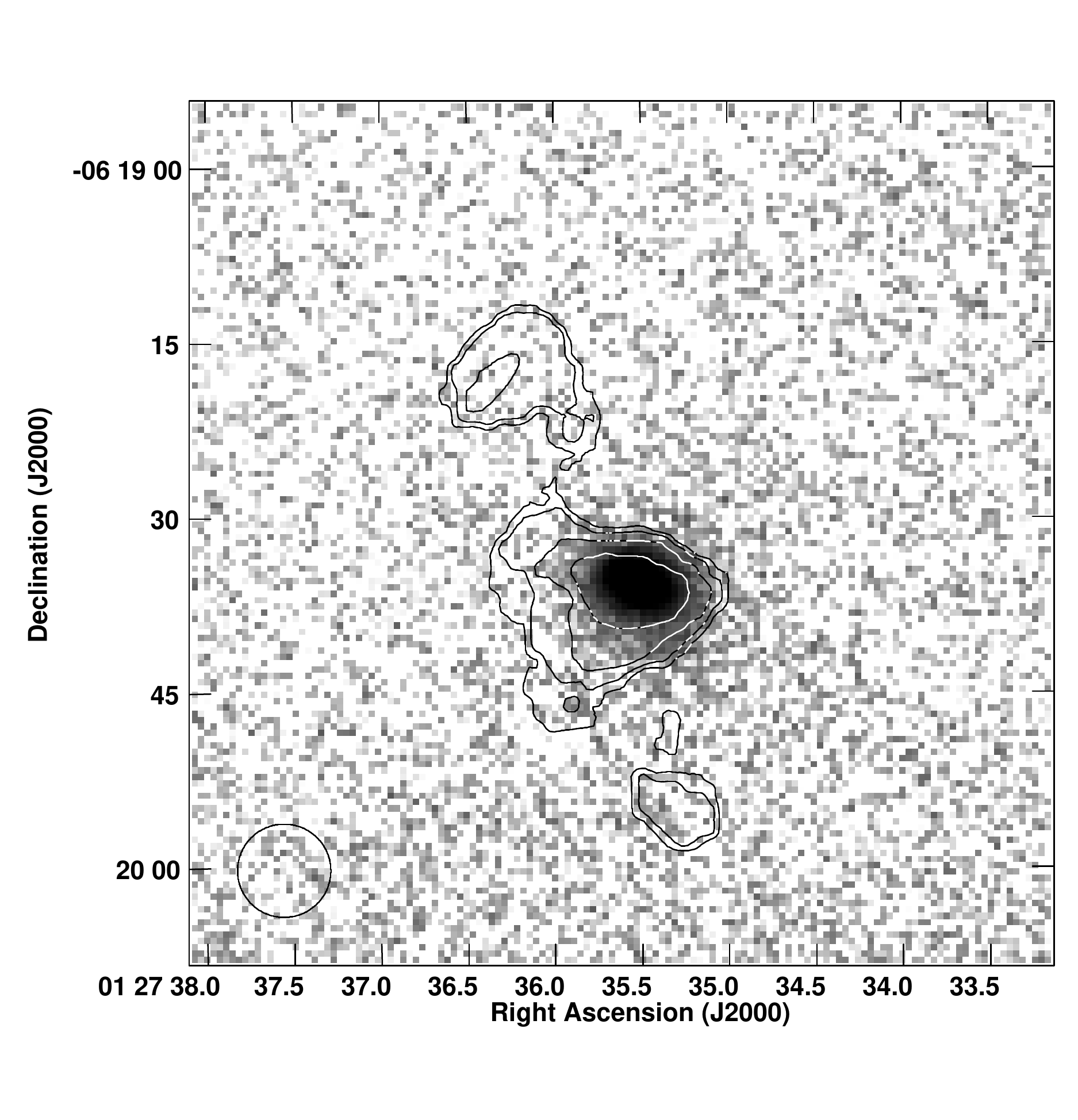}}}
\put(1.3,17.6){(c)} 
\put(8.19,11.05){\hbox{\includegraphics[scale=0.60]{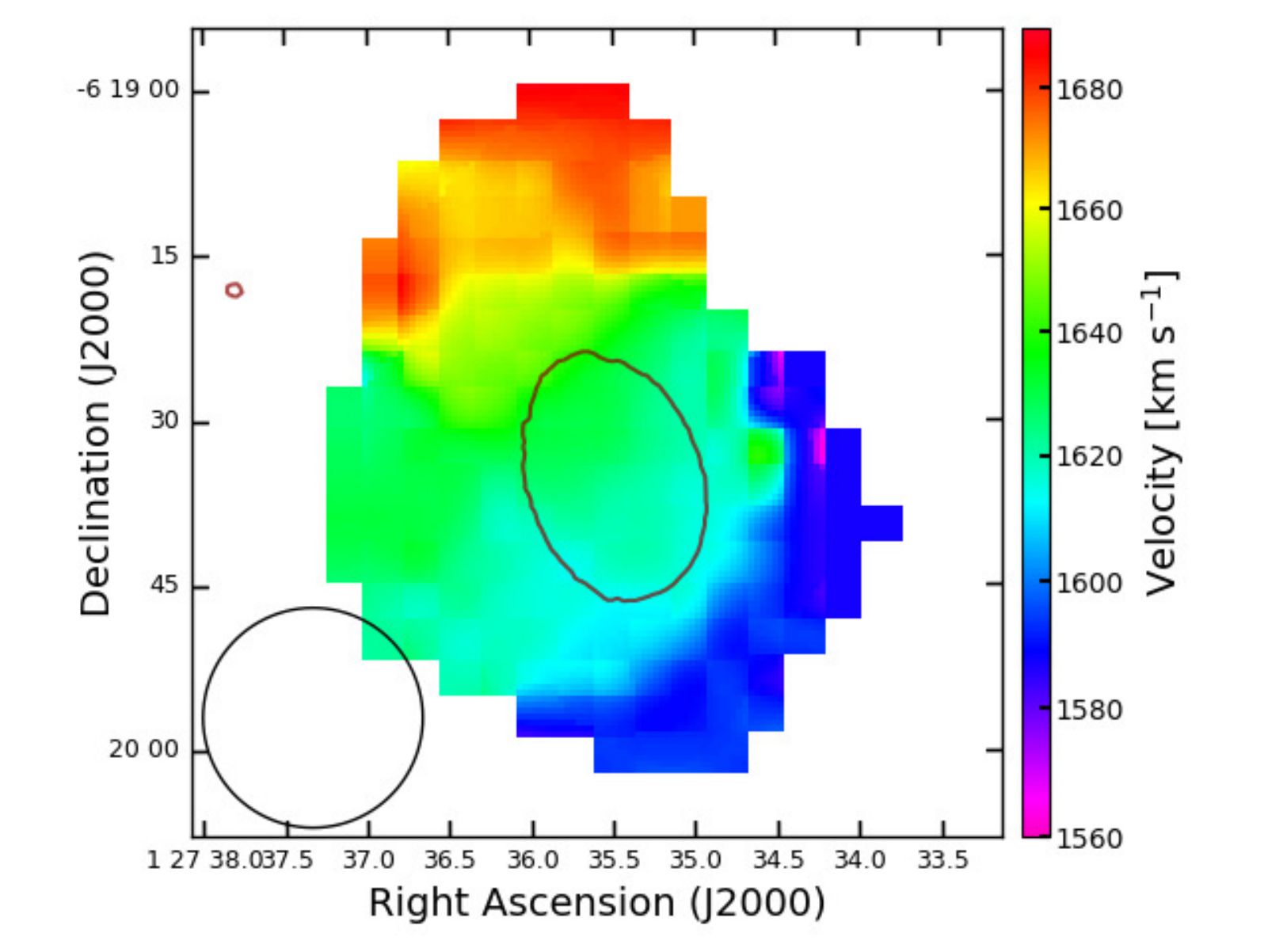}}}
\put(9.9,17.6){(d)} 
\put(0.32,3.9){\hbox{\includegraphics[scale=0.288]{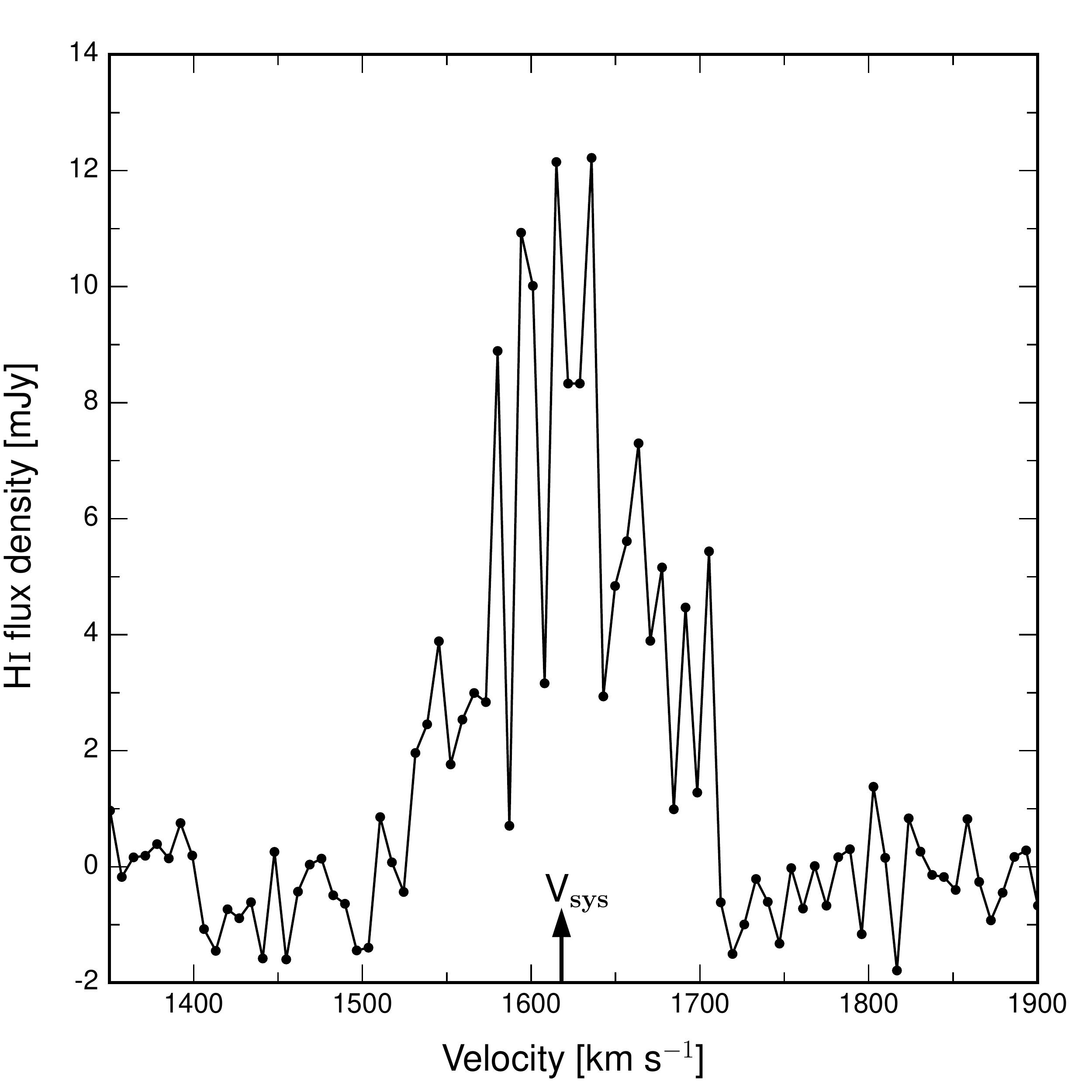}}} 
\put(1.3,10.4){(e)} 
\put(8.73,3.9){\hbox{\includegraphics[scale=0.288]{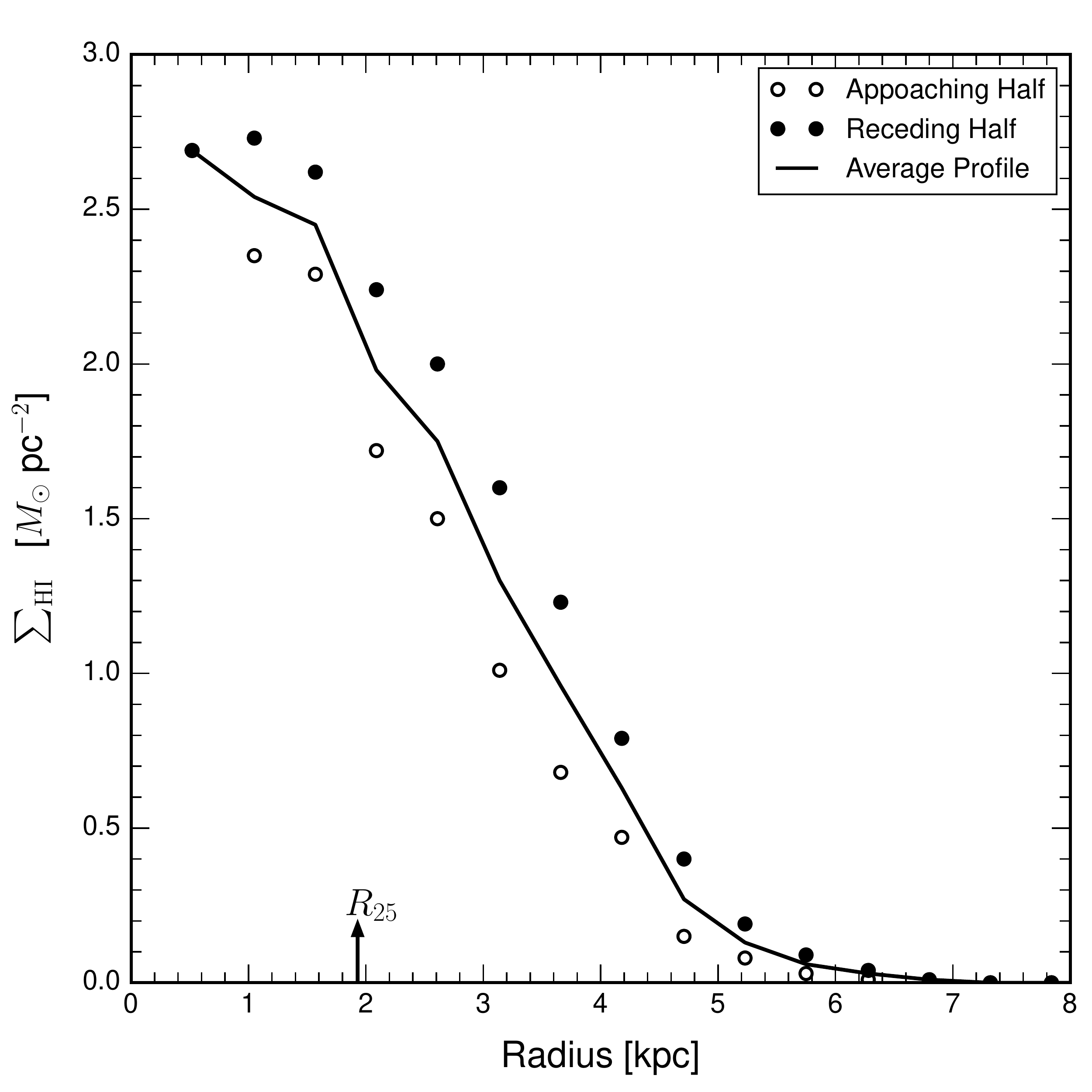}}} 
\put(9.9,10.4){(f)} 
\end{picture} 
\vspace{-4.3cm}
\caption{(a) The low resolution \HI column density contours of MRK~996 overlaid upon its grey scale optical $r$-band image. The contour levels are $0.3 \times n$, where $n=1,2,4,8,16,32$ in units of $10^{19}$~cm$^{-2}$. (b) The intermediate resolution \HI column density contours overlaid upon the grey scale optical $r$-band image. The contour levels are $2.8 \times n$ in units of $10^{19}$~cm$^{-2}$. (c) The high resolution \HI column density contours overlaid upon the grey scale \Ha line image. The contour levels are $20.9 \times n$ in units of $10^{19}$~cm$^{-2}$. (d) The intermediate resolution moment-1 map, showing the velocity field, with an overlying optical $r$-band outer contour. The circle at the bottom of each image is showing the synthesized beam. The average FWHM seeing during the optical observation was $\sim 2''.4$. (e) The global \HI profile obtained using the low resolution \HI images. The arrow at the abscissa denotes the systemic \HI velocity. (f) The \HI mass surface density profile obtained using the low resolution \HI map. The arrow at the abscissa denotes the $B$-band optical disk radius.}
\label{MRK996-2}
\end{figure}

\begin{figure*}
\begin{center}
\includegraphics[angle=0,width=1\linewidth]{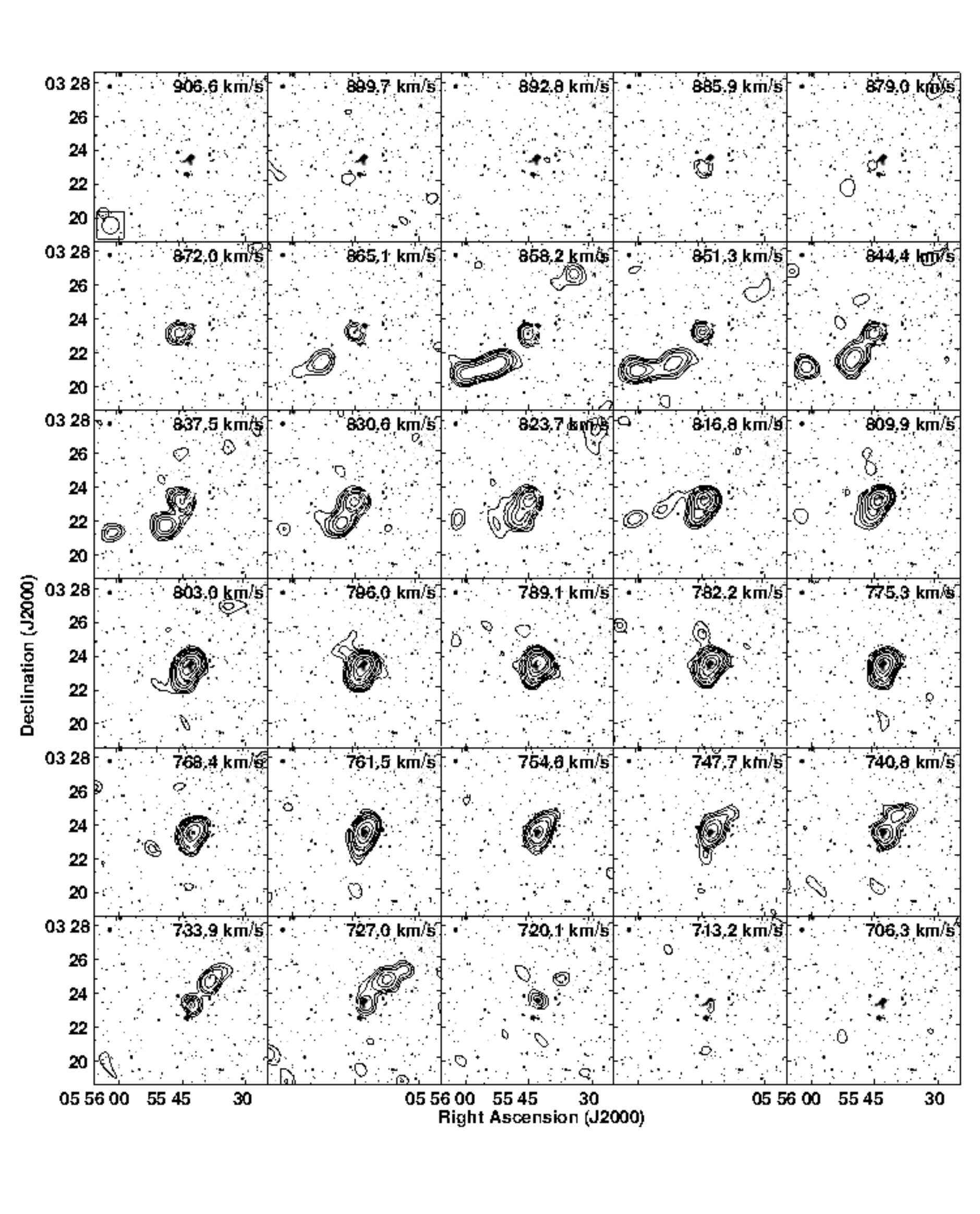}
\end{center}
\caption{The \HI contours from the low resolution channel images overlaid upon the grey scale optical $r$-band image of UGCA~116. The contours representing \HI emission flux are drawn at $2.5\sigma\times n$~mJy/Beam; n=1,1.5,2,3,4,6.}
\end{figure*}
\begin{figure}
\setlength{\unitlength}{1cm}
\begin{picture}(12,25) 
\put(0,18.2){\hbox{\includegraphics[scale=0.38]{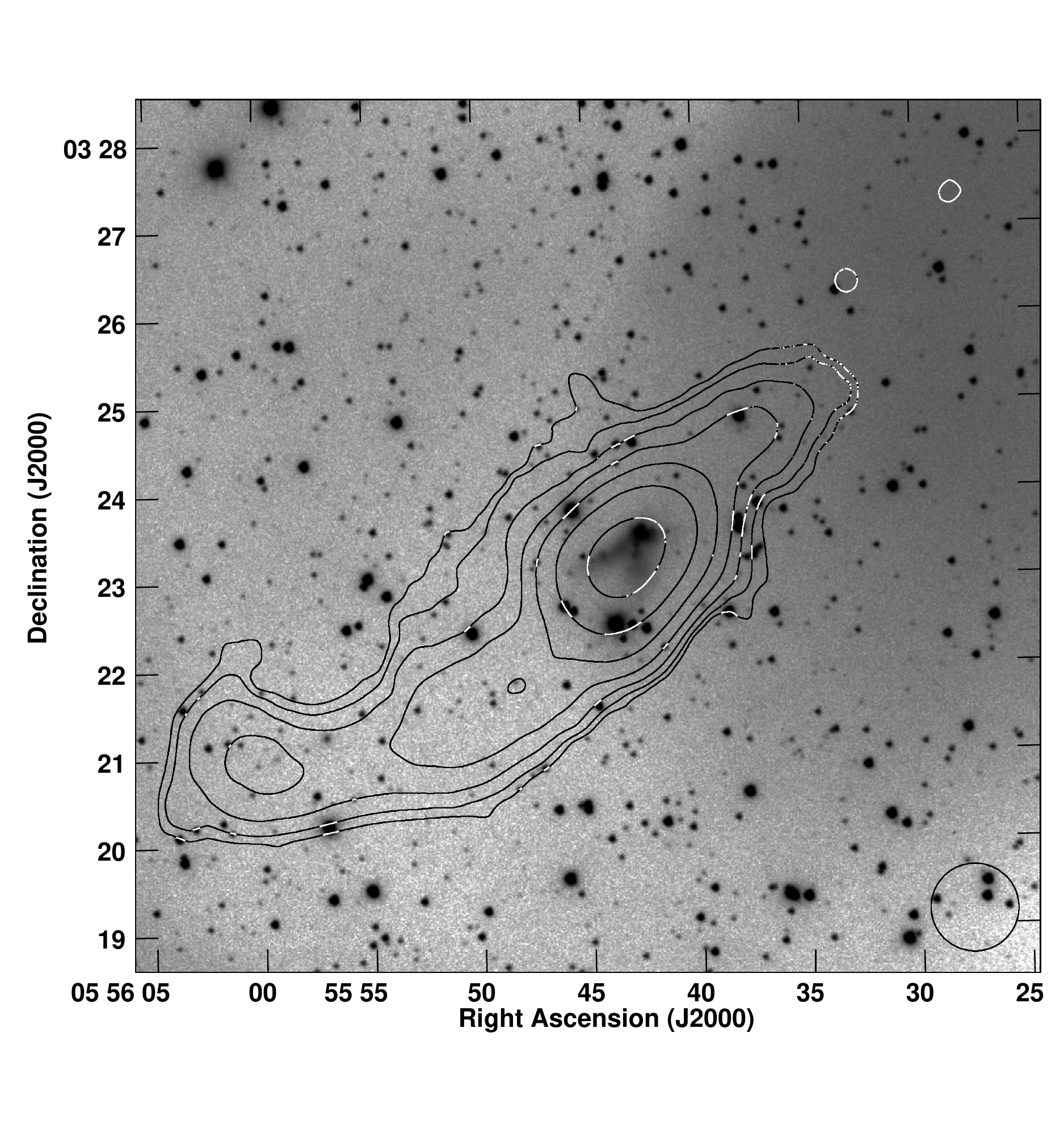}}} 
\put(1.2,24.8){(a)} 
\put(8.3,18.0){\hbox{\includegraphics[scale=0.44]{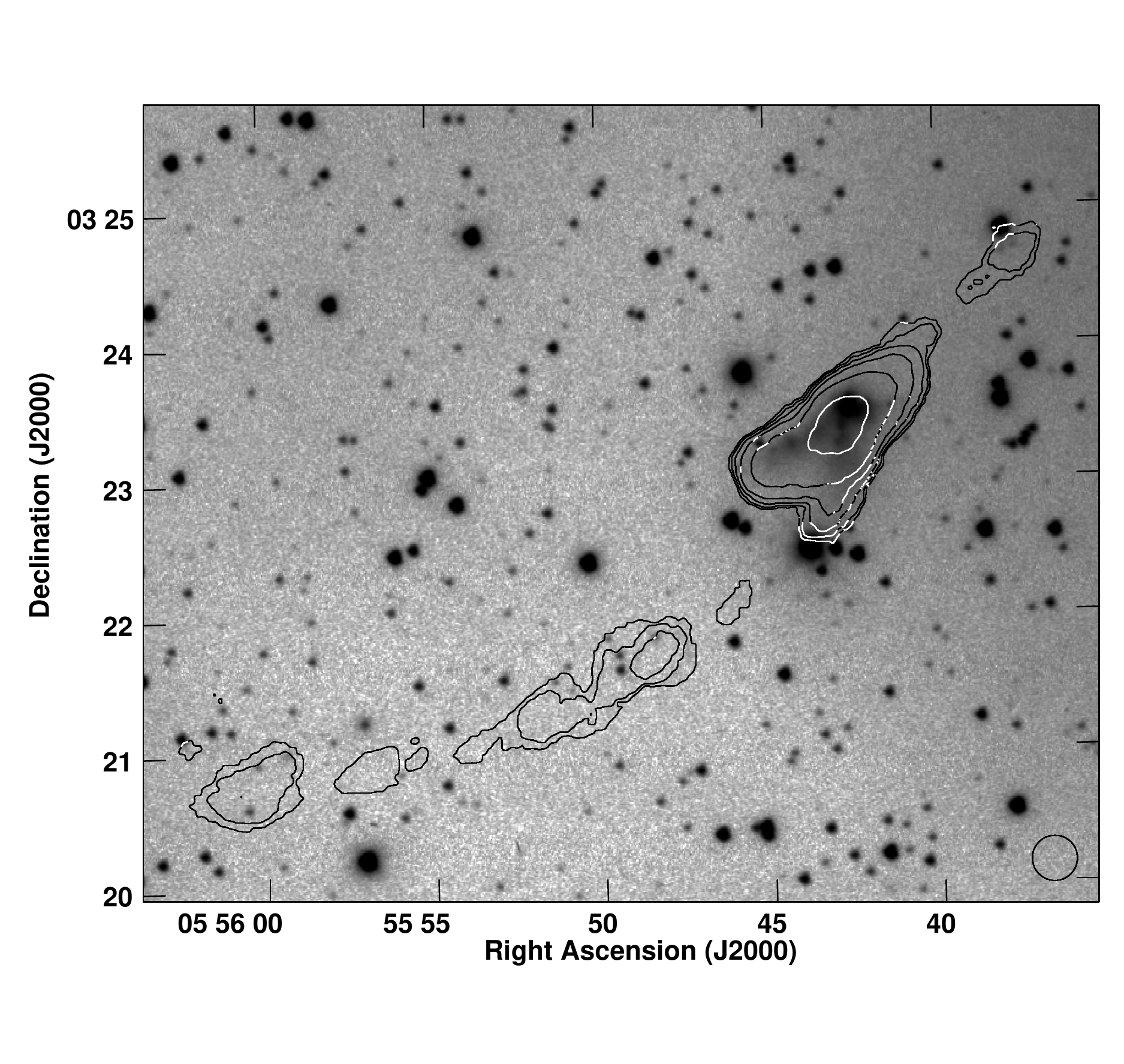}}}
\put(9.7,24.8){(b)} 
\put(-0.35,11.0){\hbox{\includegraphics[scale=0.44]{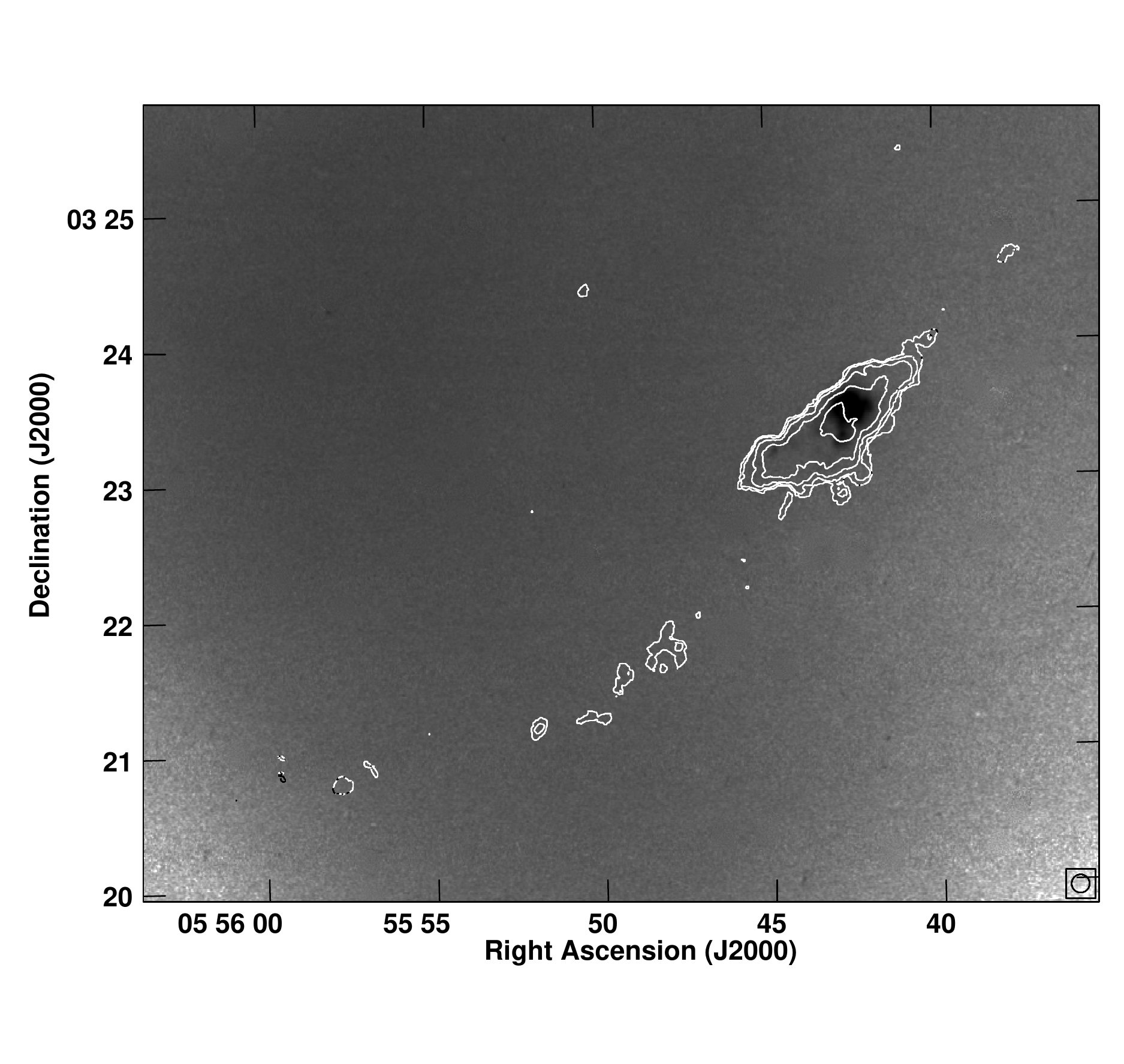}}}
\put(1.0,17.7){(c)} 
\put(8.43,11.27){\hbox{\includegraphics[scale=0.59]{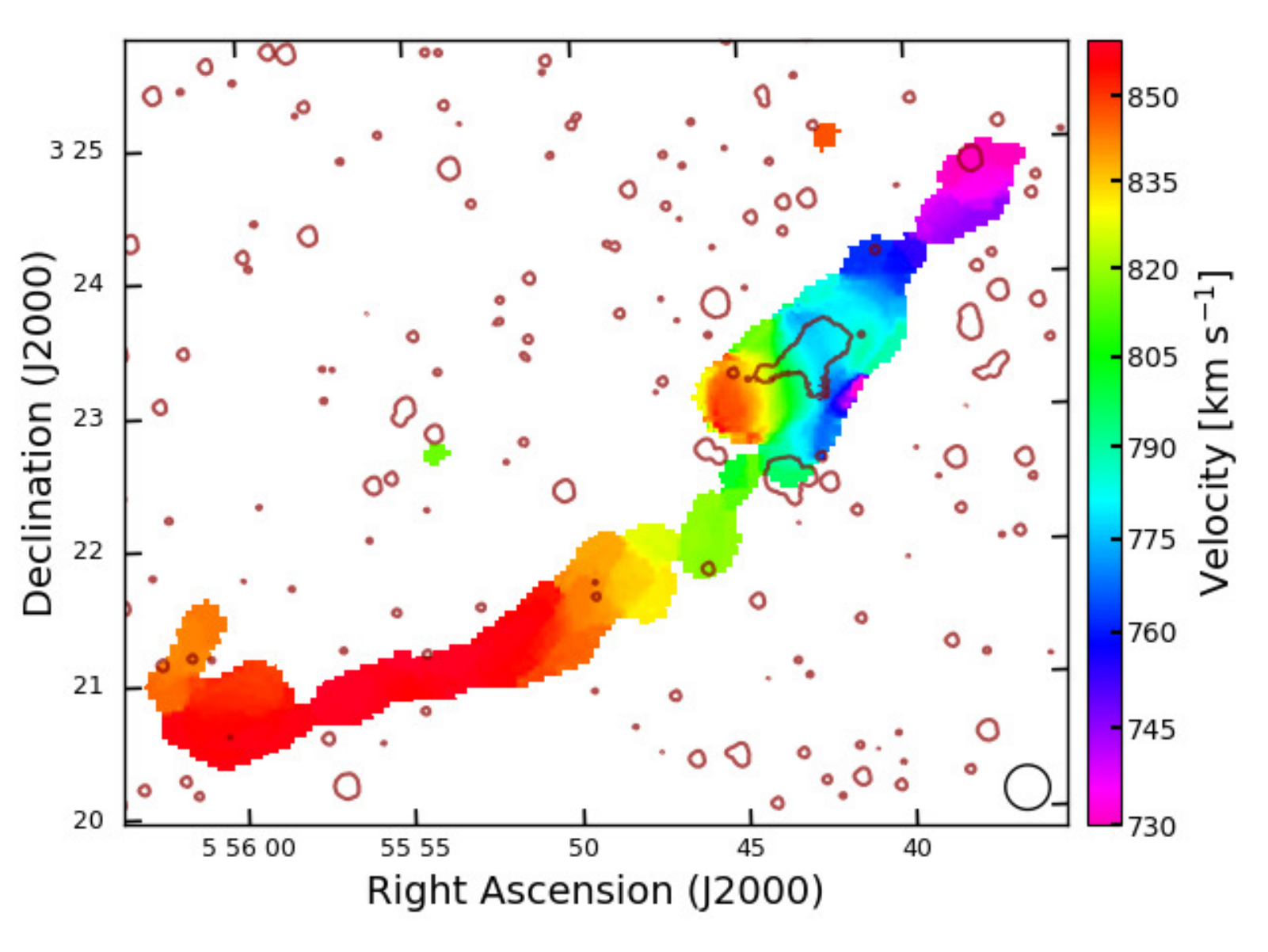}}}
\put(9.6,17.7){(d)} 
\put(0.19,3.9){\hbox{\includegraphics[scale=0.295]{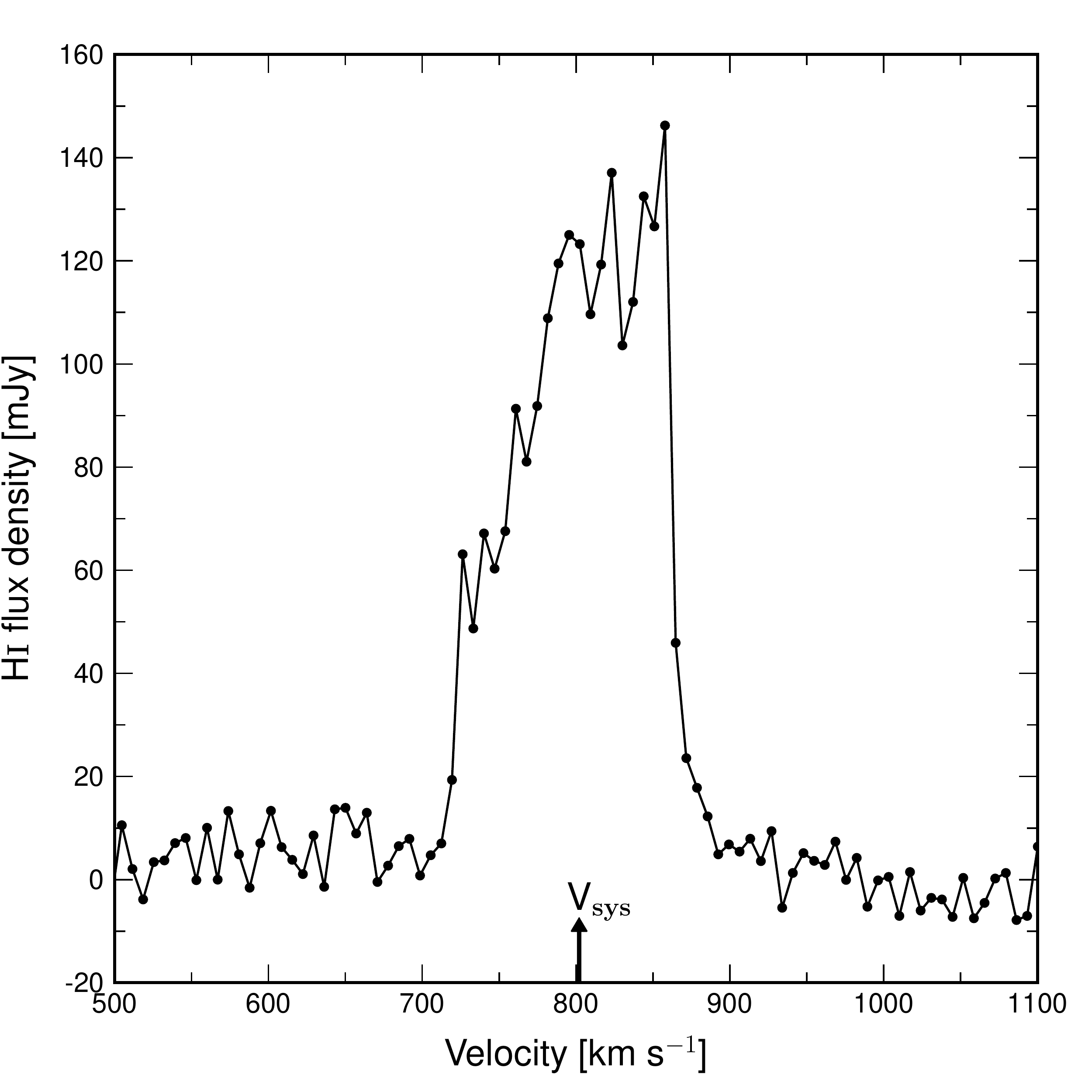}}} 
\put(1.3,10.6){(e)} 
\put(8.5,3.9){\hbox{\includegraphics[scale=0.295]{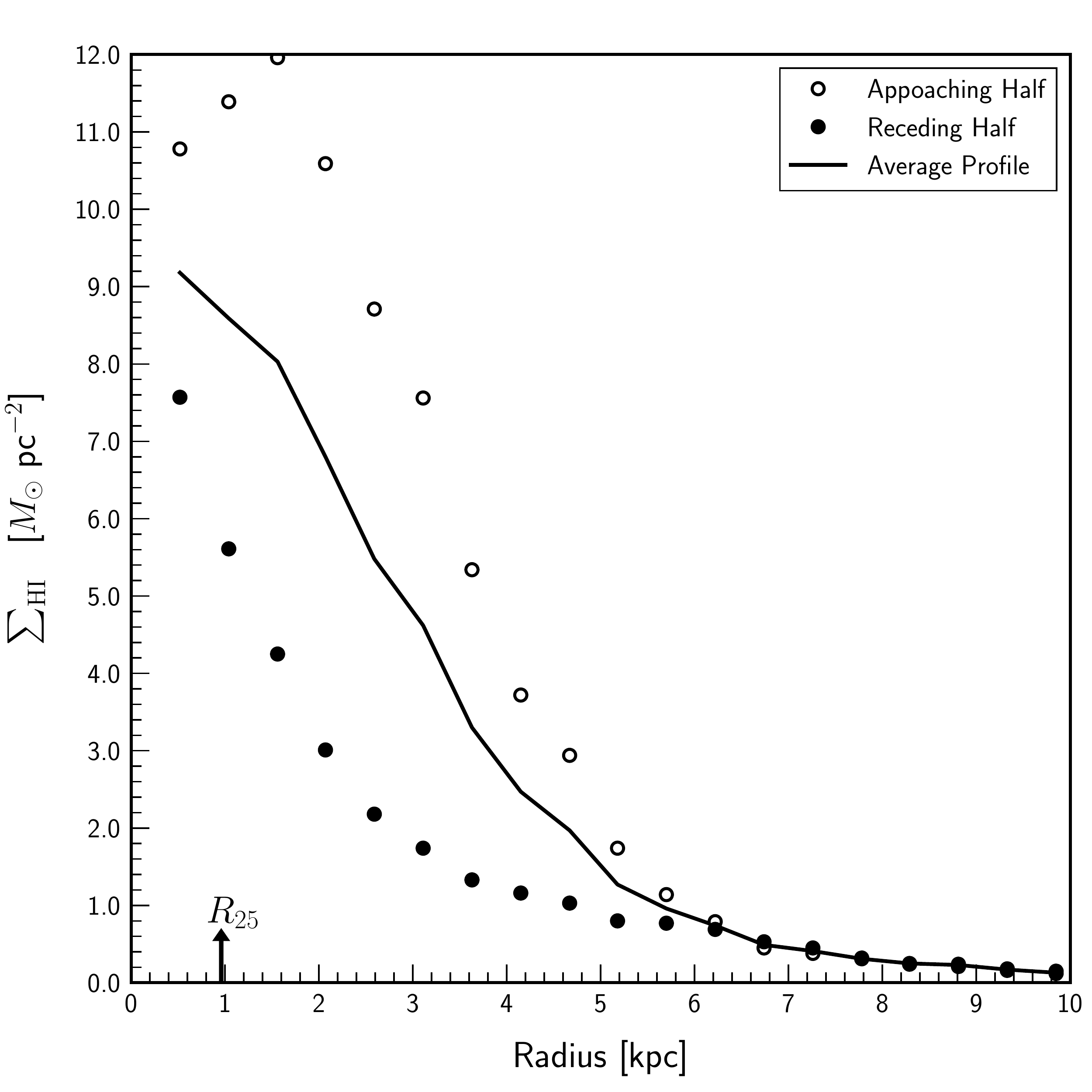}}} 
\put(9.7,10.6){(f)} 
\end{picture} 
\vspace{-4.2cm}
\caption{(a) The low resolution \HI column density contours of UGCA~116 overlaid upon its grey scale optical $r$-band image. The contour levels are $1.9 \times n$, where $n=1,2,4,8,16,32$ in units of $10^{19}$~cm$^{-2}$. (b) The intermediate resolution \HI column density contours overlaid upon the grey scale optical $r$-band image. The contour levels are $11.2 \times n$ in units of $10^{19}$~cm$^{-2}$. (c) The high resolution \HI column density contours overlaid upon the grey scale \Ha line image. The contour levels are $52.3 \times n$ in units of $10^{19}$~cm$^{-2}$. (d) The intermediate resolution moment-1 map, showing the velocity field, with an overlying optical $r$-band outer contour. The circle at the bottom of each image is showing the synthesized beam. The average FWHM seeing during the optical observation was $\sim 2''.2$. (e) The global \HI profile obtained using the low resolution \HI images. The arrow at the abscissa shows the systemic \HI velocity. (f) The \HI mass surface density profile obtained using the low resolution \HI map. The arrow at the abscissa shows the $B$-band optical disk radius.}
\label{UGCA116}
\end{figure}

\begin{figure*}
\begin{center}
\includegraphics[angle=0,width=1\linewidth]{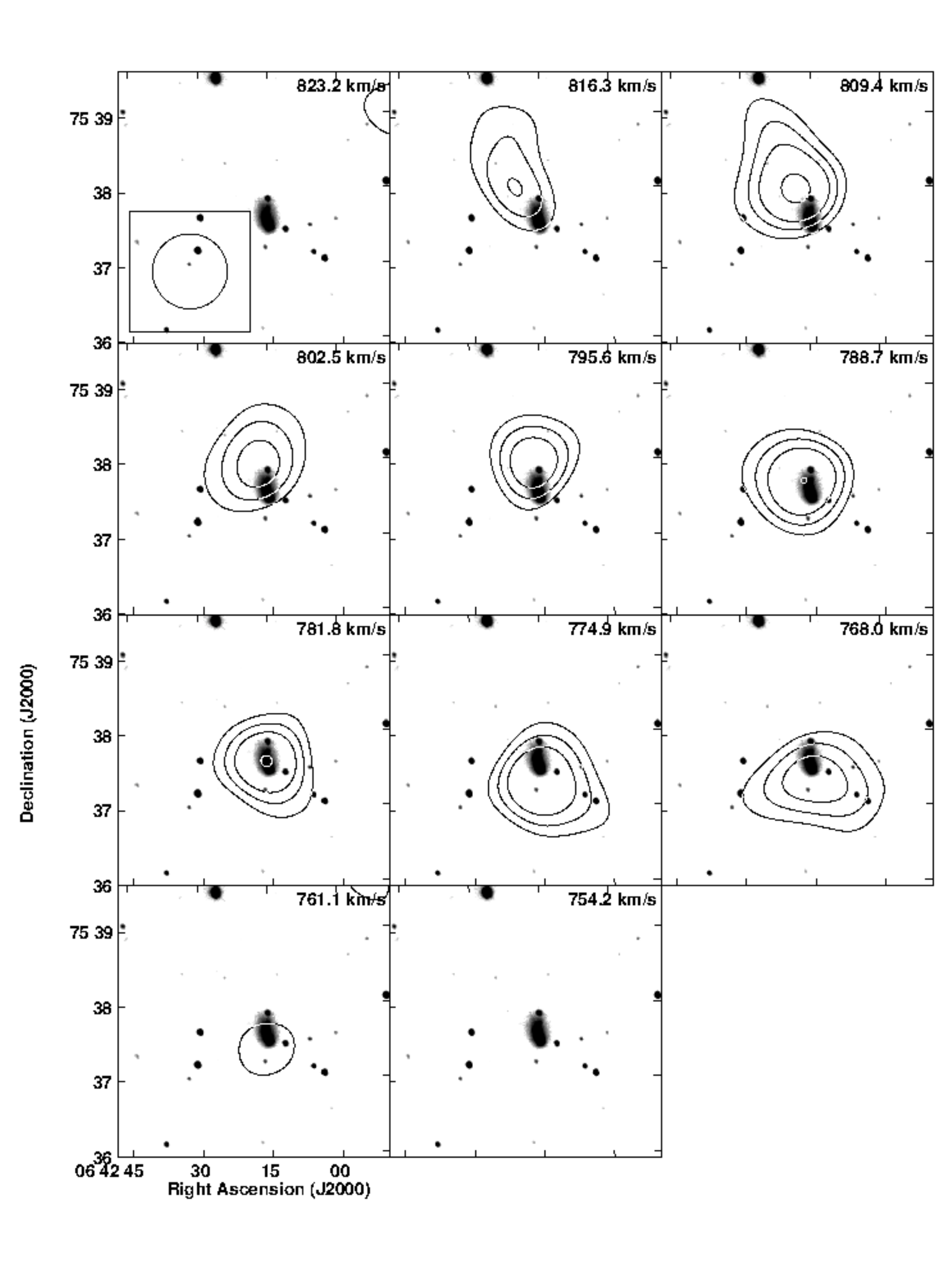}
\end{center}
\caption{The \HI contours from the low resolution channel images overlaid upon the grey scale optical $r$-band image of UGCA~130. The contours representing \HI emission flux are drawn at $2.5\sigma\times n$~mJy/Beam; n=1,1.5,2,3,4,6.}
\end{figure*}
\begin{figure}
\setlength{\unitlength}{1cm}
\begin{picture}(12,25) 
\put(0,17.9){\hbox{\includegraphics[scale=0.40]{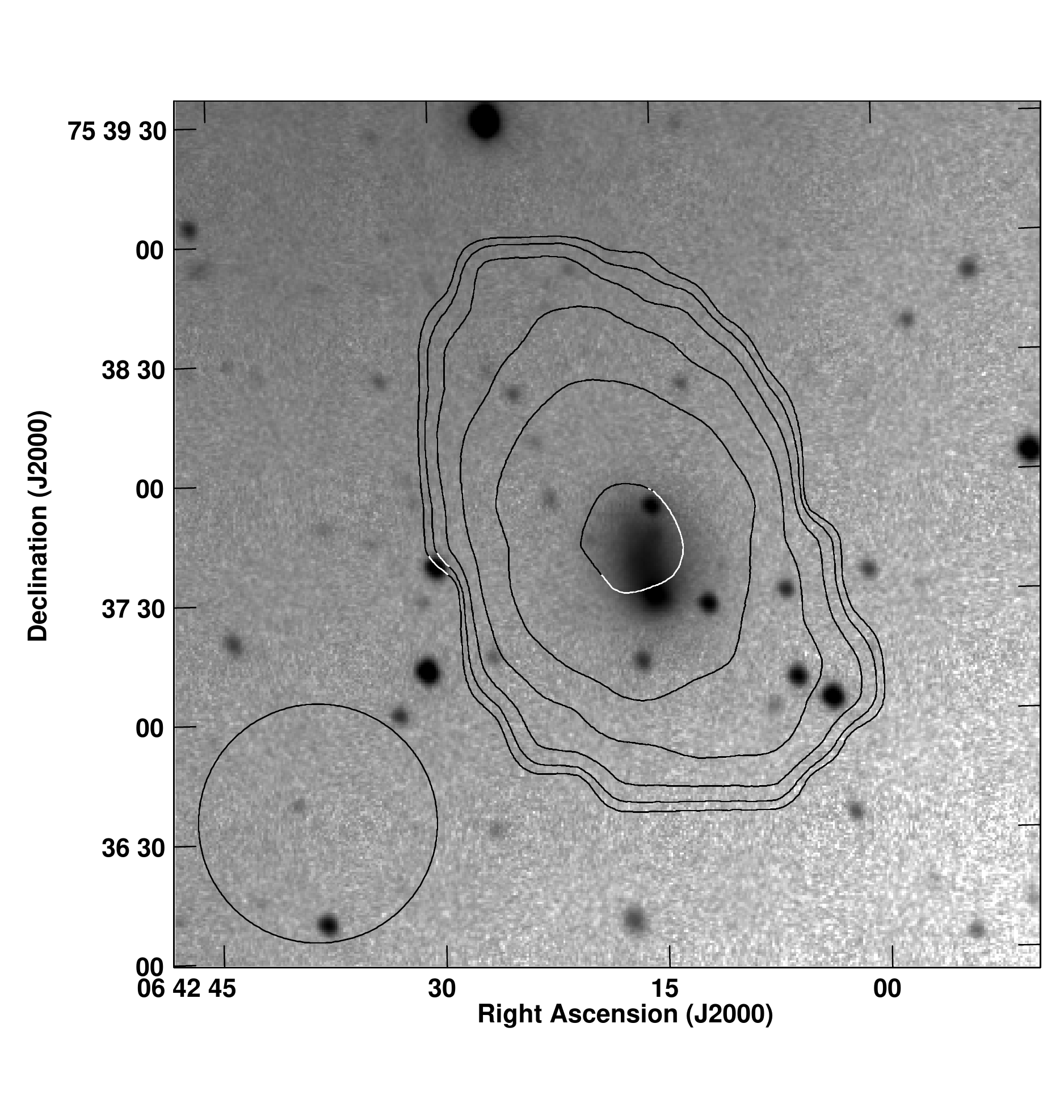}}} 
\put(1.5,24.8){(a)} 
\put(8.3,17.87){\hbox{\includegraphics[scale=0.40]{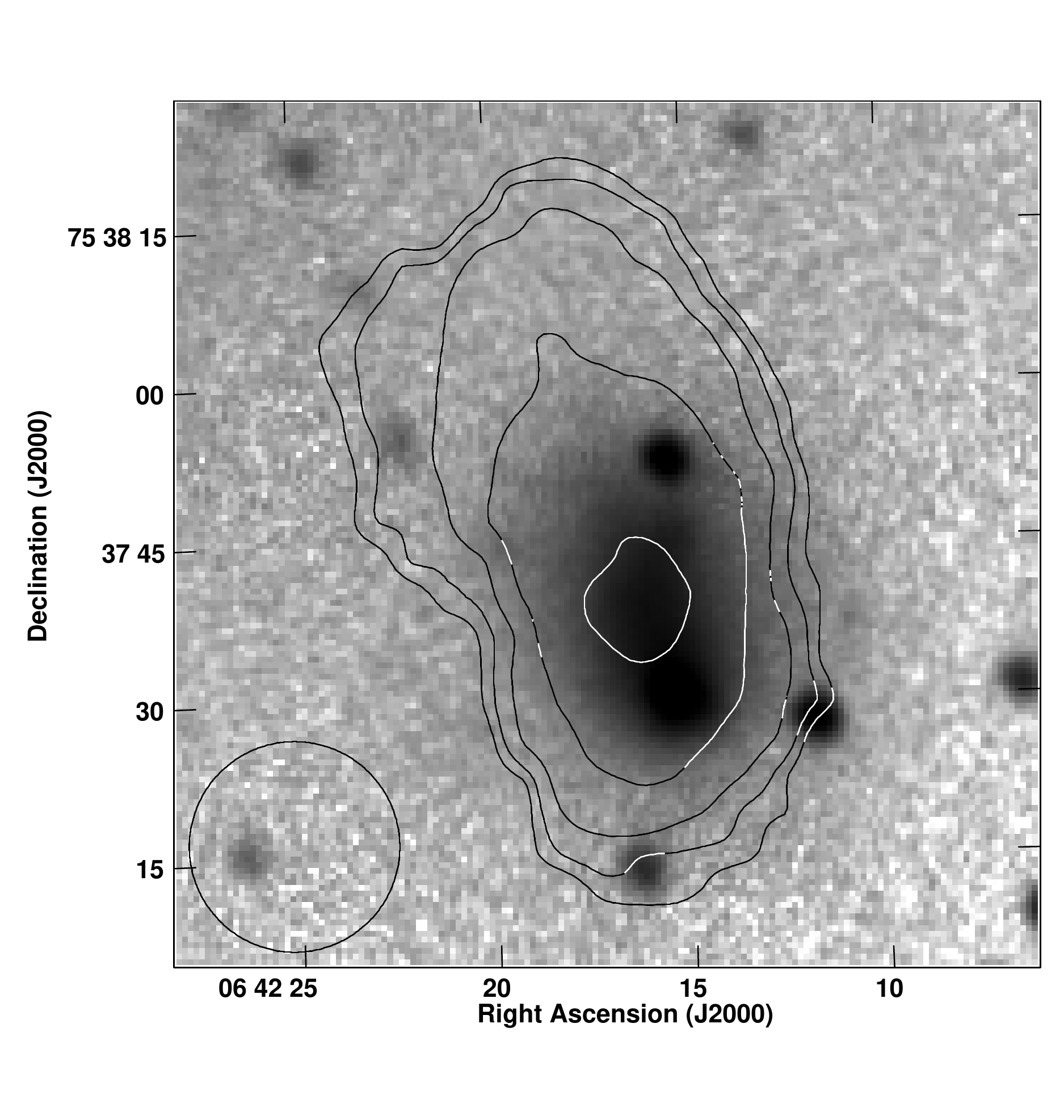}}}
\put(9.8,24.8){(b)} 
\put(0,10.8){\hbox{\includegraphics[scale=0.40]{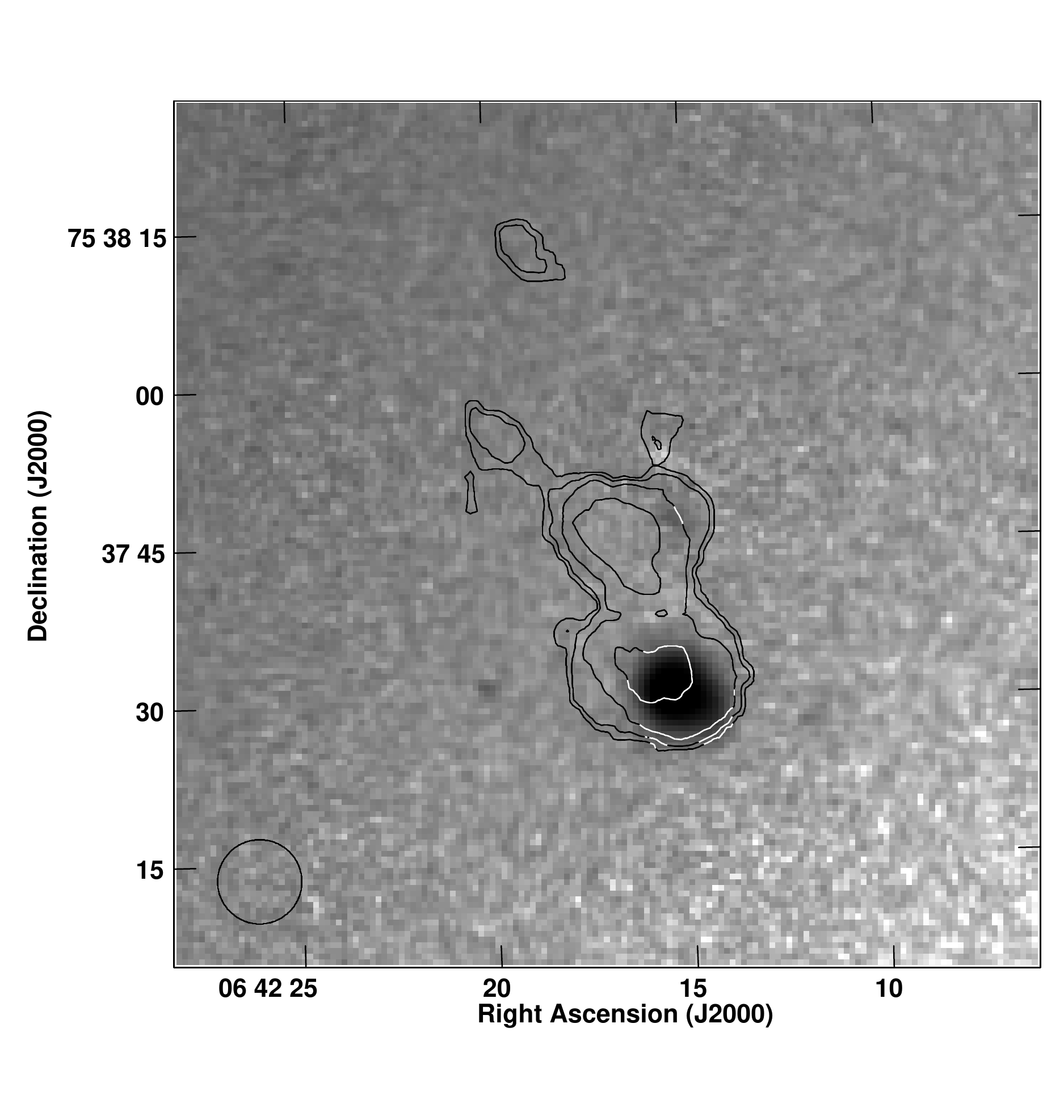}}}
\put(1.5,17.7){(c)} 
\put(7.95,11.0){\hbox{\includegraphics[scale=0.599]{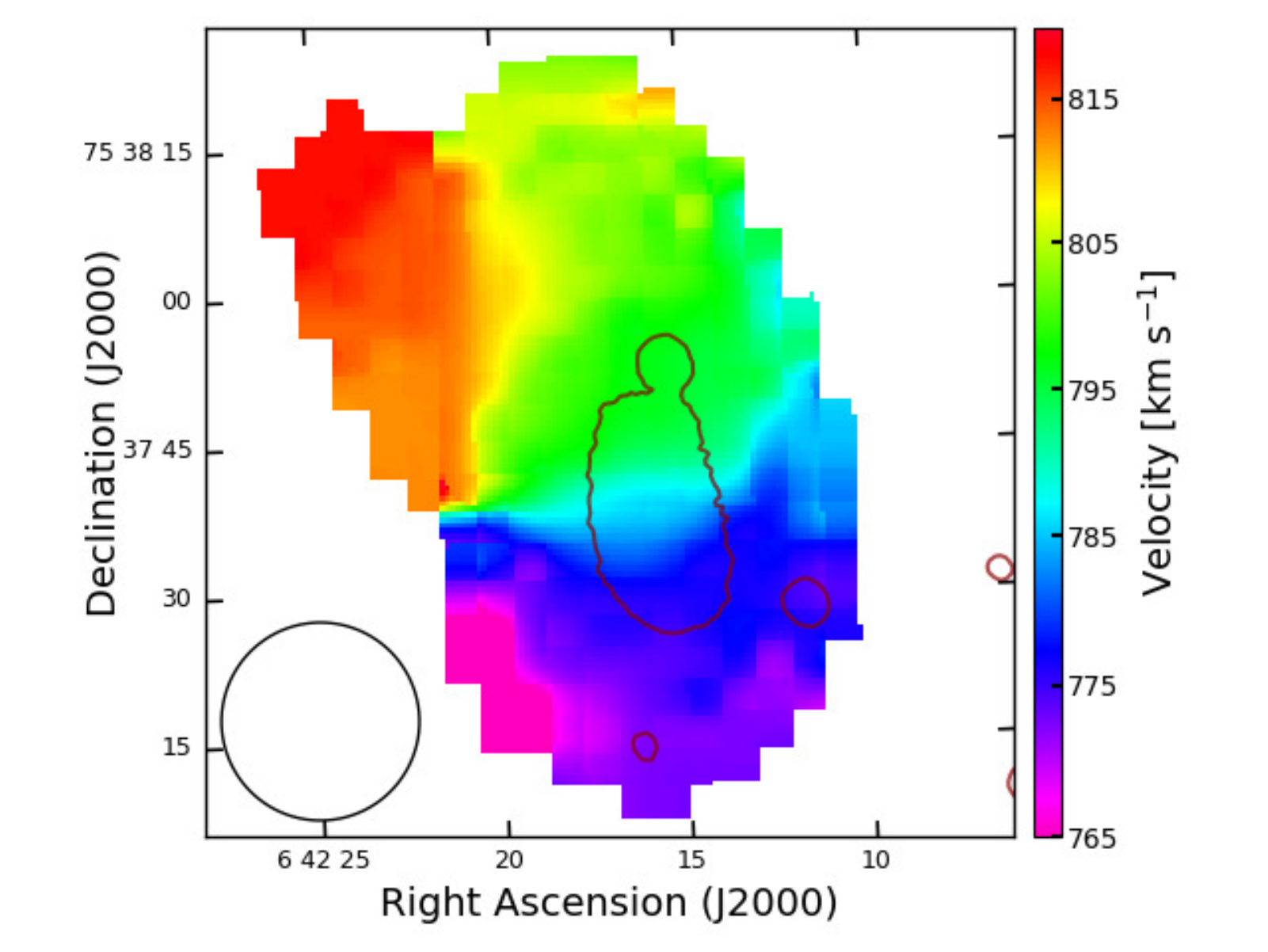}}}
\put(9.8,17.7){(d)} 
\put(0.55,3.9){\hbox{\includegraphics[scale=0.286]{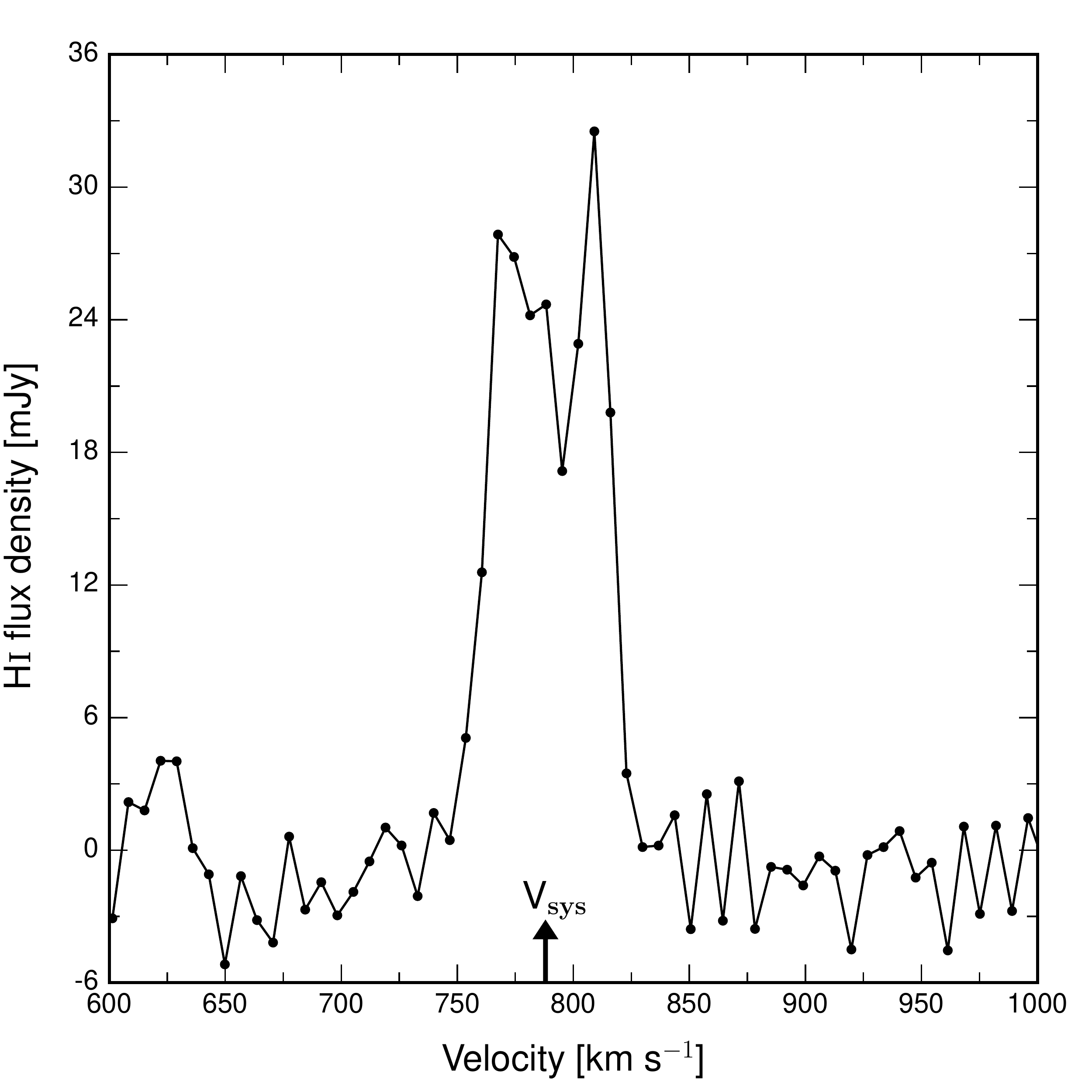}}} 
\put(1.5,10.4){(e)} 
\put(8.64,3.9){\hbox{\includegraphics[scale=0.286]{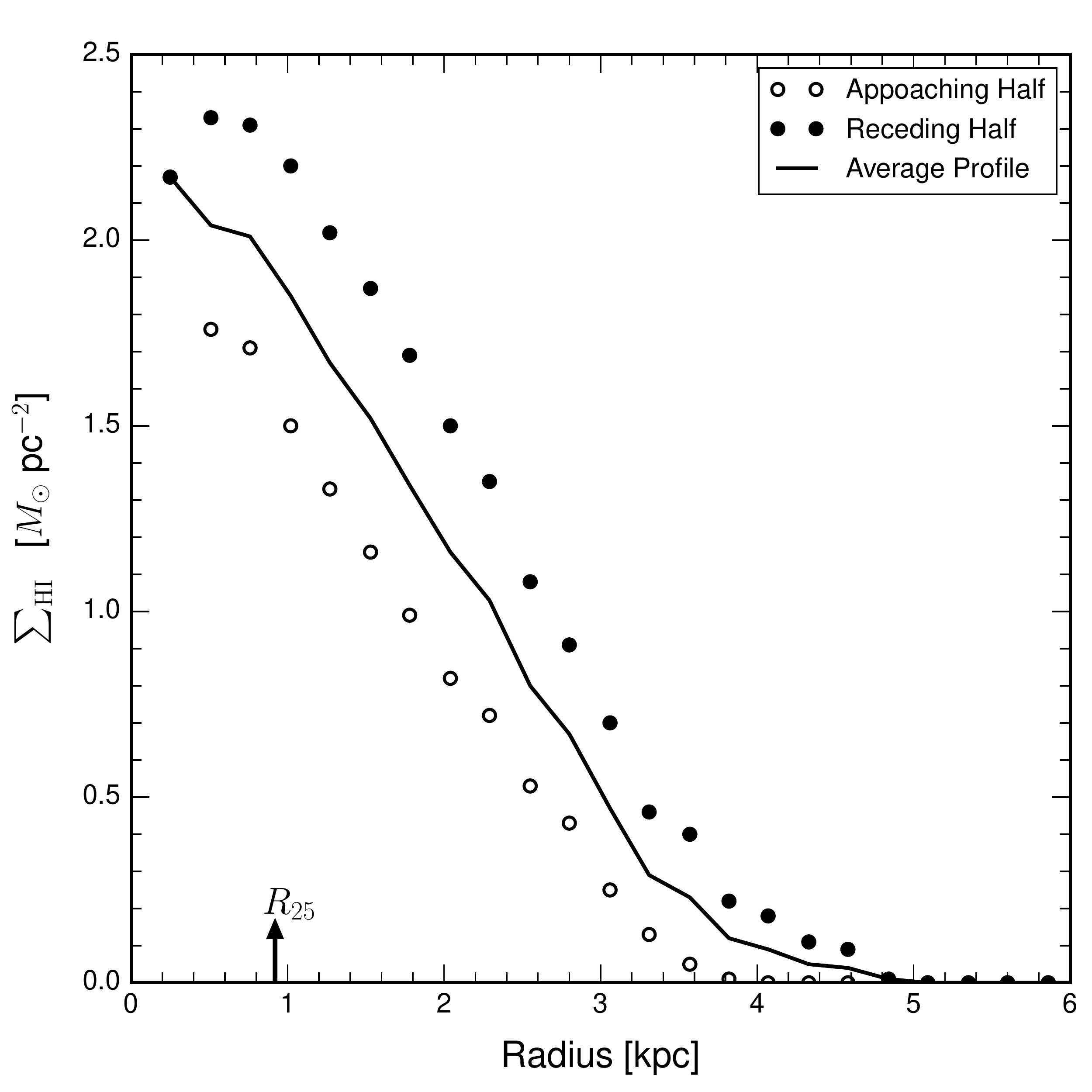}}} 
\put(9.8,10.4){(f)} 
\end{picture} 
\vspace{-4.2cm}
\caption{(a) The low resolution \HI column density contours of UGCA~130 overlaid upon its grey scale optical $r$-band image. The contour levels are $0.8 \times n$, where $n=1,2,4,8,16,32$ in units of $10^{19}$~cm$^{-2}$. (b) The intermediate resolution \HI column density contours overlaid upon the grey scale optical $r$-band image. The contour levels are $7.2 \times n$ in units of $10^{19}$~cm$^{-2}$. (c) The high resolution \HI column density contours overlaid upon the grey scale \Ha line image. The contour levels are $26.1 \times n$ in units of $10^{19}$~cm$^{-2}$. (d) The intermediate resolution moment-1 map, showing the velocity field, with an overlying optical $r$-band outer contour. The circle at the bottom of each image is showing the synthesized beam. The average FWHM seeing during the optical observation was $\sim 2''.6$. (e) The global \HI profile obtained using the low resolution \HI images. The arrow at the abscissa shows the systemic \HI velocity. (f) The \HI mass surface density profile obtained using the low resolution \HI map. The arrow at the abscissa shows the $B$-band optical disk radius.}
\label{UGCA130}
\end{figure}

\begin{figure*}
\begin{center}
\includegraphics[angle=0,width=1\linewidth]{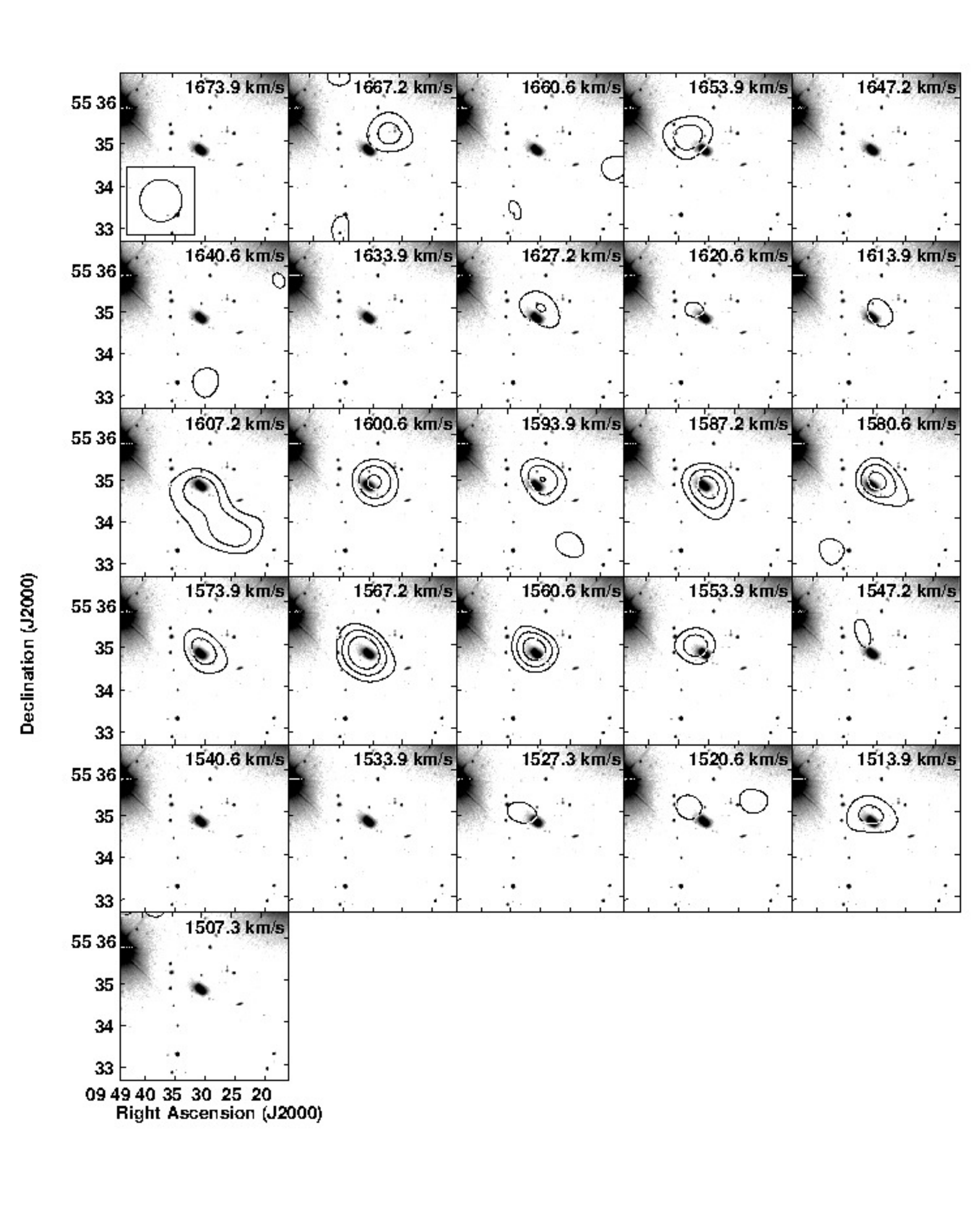}
\end{center}
\caption{The \HI contours from the low resolution channel images overlaid upon the grey scale optical $r$-band image of MRK~22. The contours representing \HI emission flux are drawn at $2.5\sigma\times n$~mJy/Beam; n=1,1.5,2,3,4,6.}
\end{figure*}
\begin{figure}
\setlength{\unitlength}{1cm}
\begin{picture}(12,25) 
\put(0,17.9){\hbox{\includegraphics[scale=0.39]{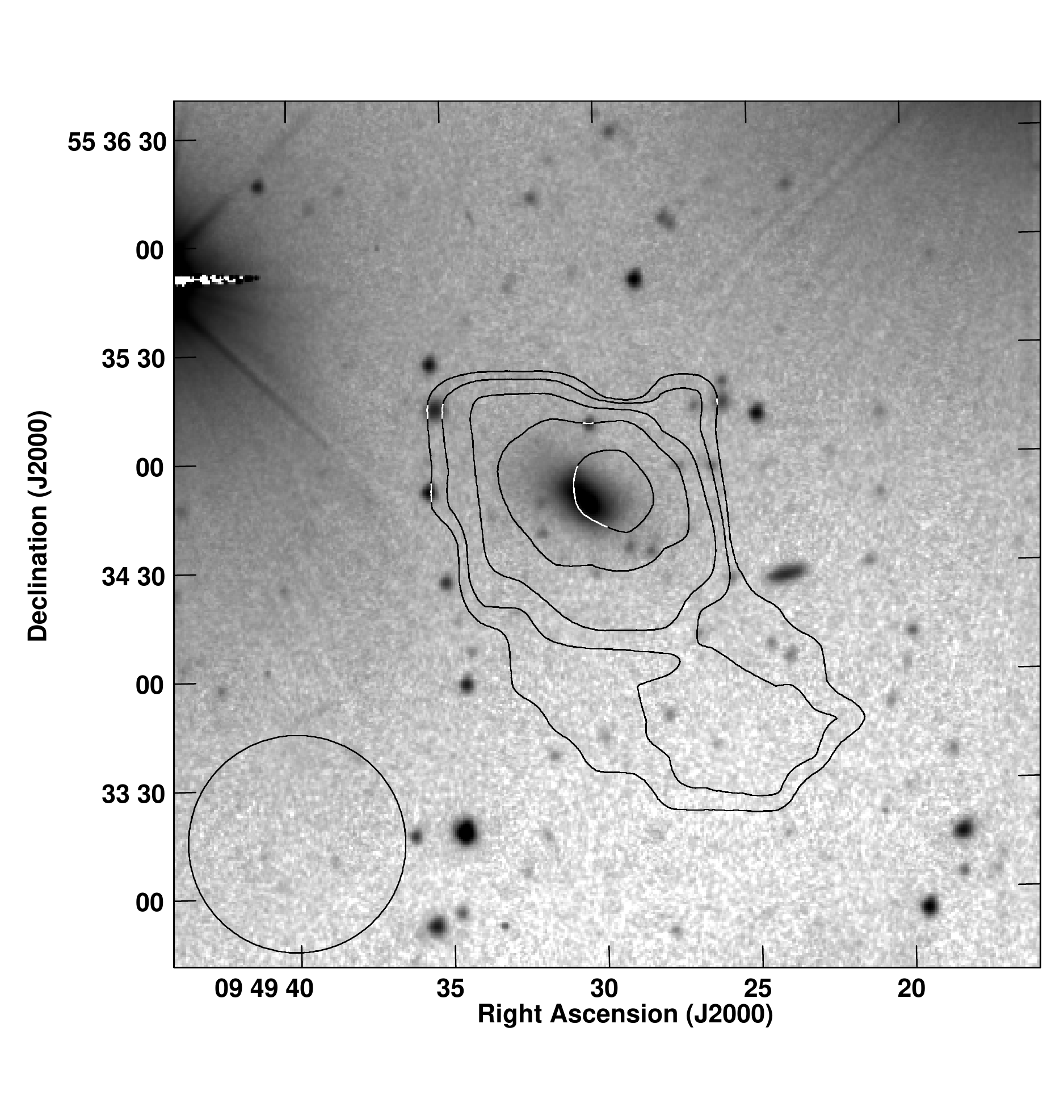}}} 
\put(1.5,24.6){(a)} 
\put(8.3,17.87){\hbox{\includegraphics[scale=0.391]{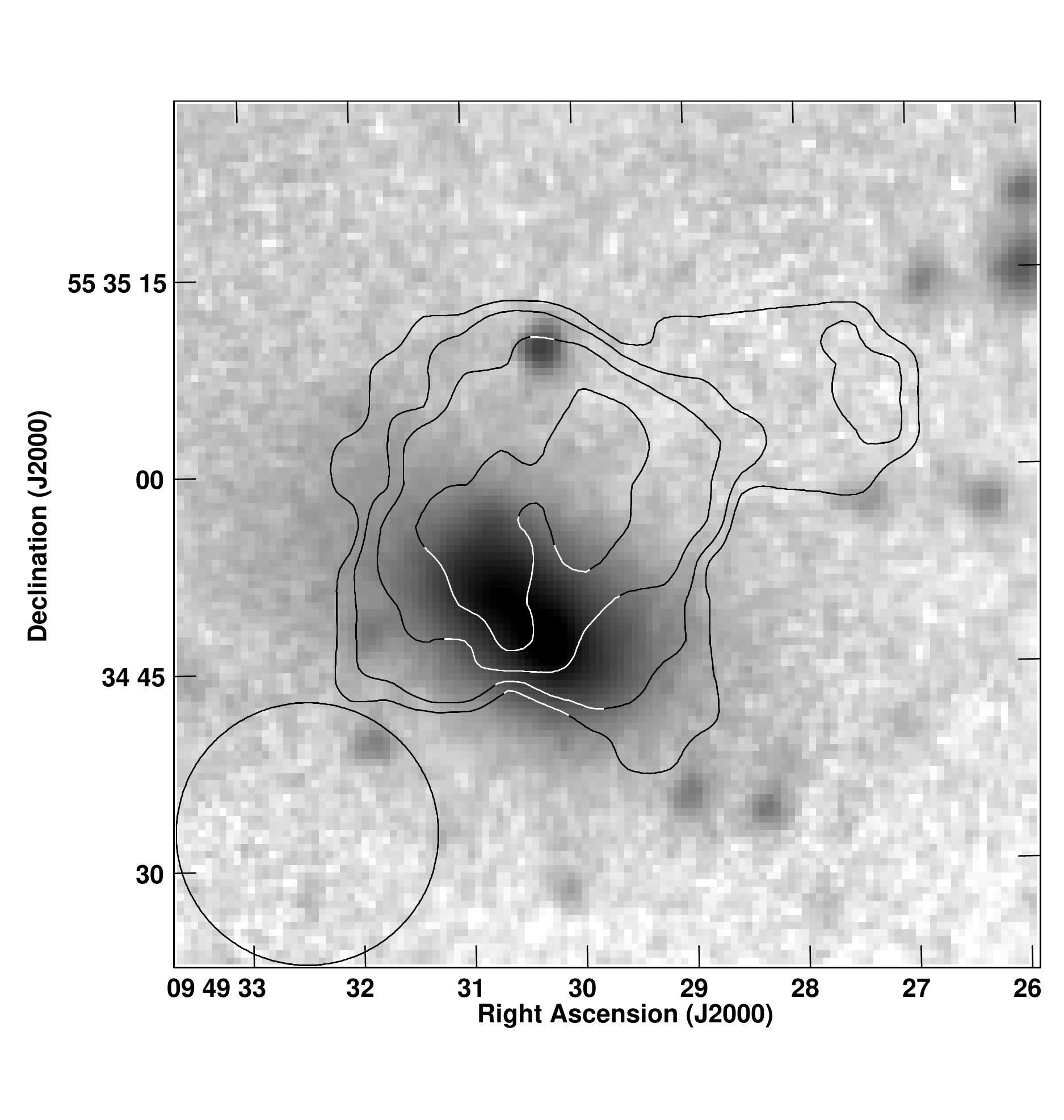}}}
\put(9.8,24.6){(b)} 
\put(0,10.8){\hbox{\includegraphics[scale=0.391]{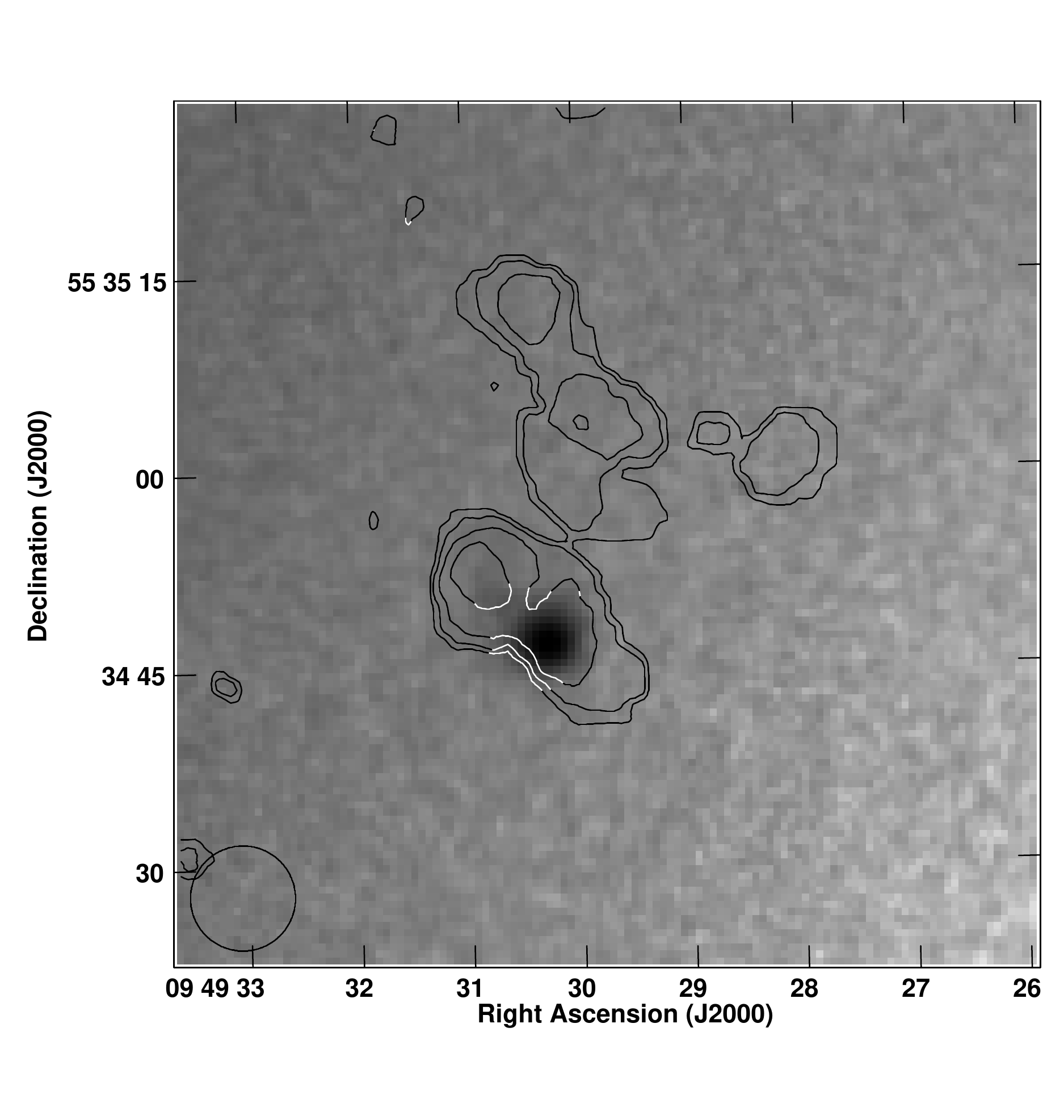}}}
\put(1.5,17.5){(c)} 
\put(8.02,11.0){\hbox{\includegraphics[scale=0.587]{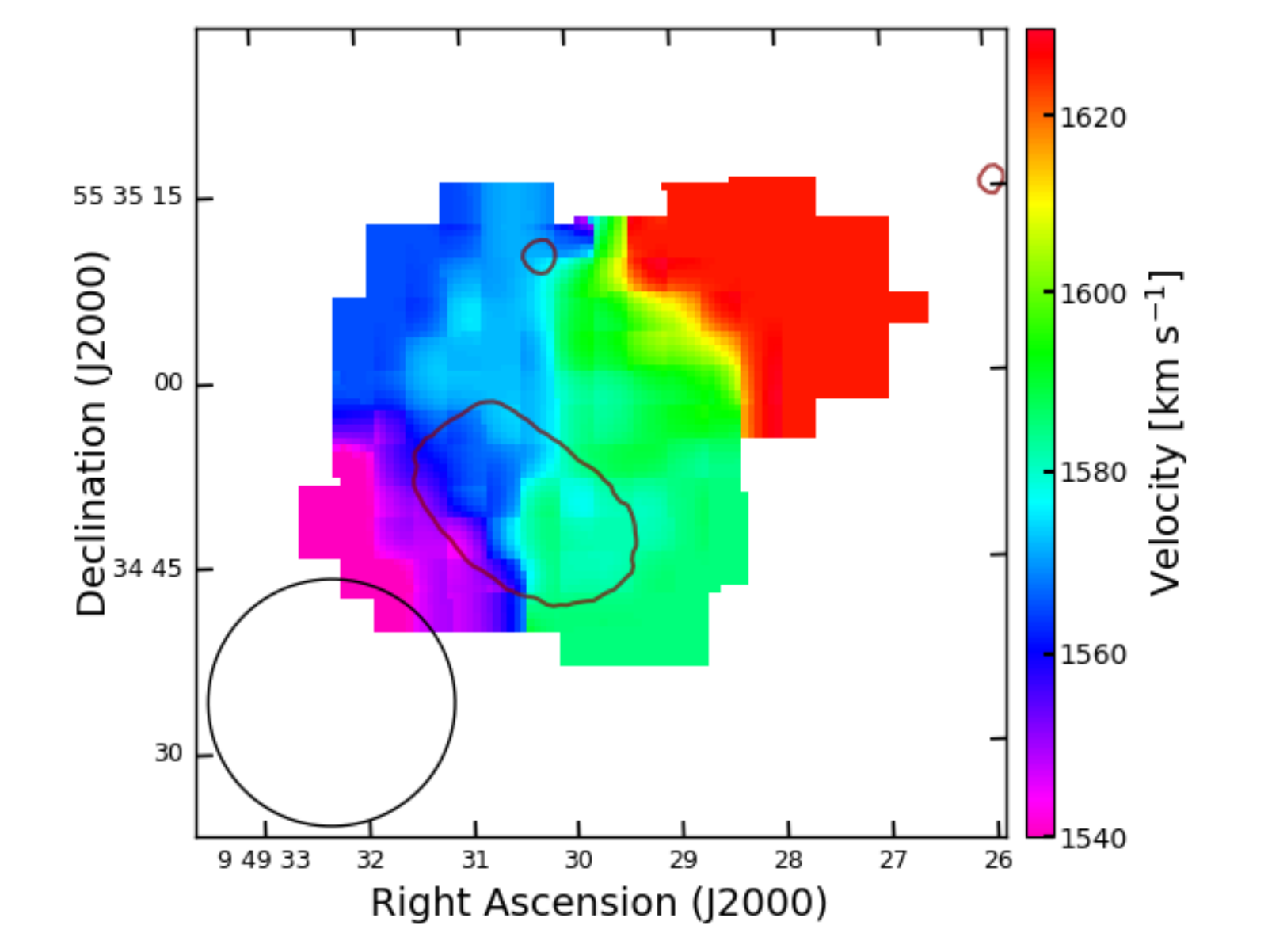}}}
\put(9.8,17.5){(d)} 
\put(0.51,3.9){\hbox{\includegraphics[scale=0.28]{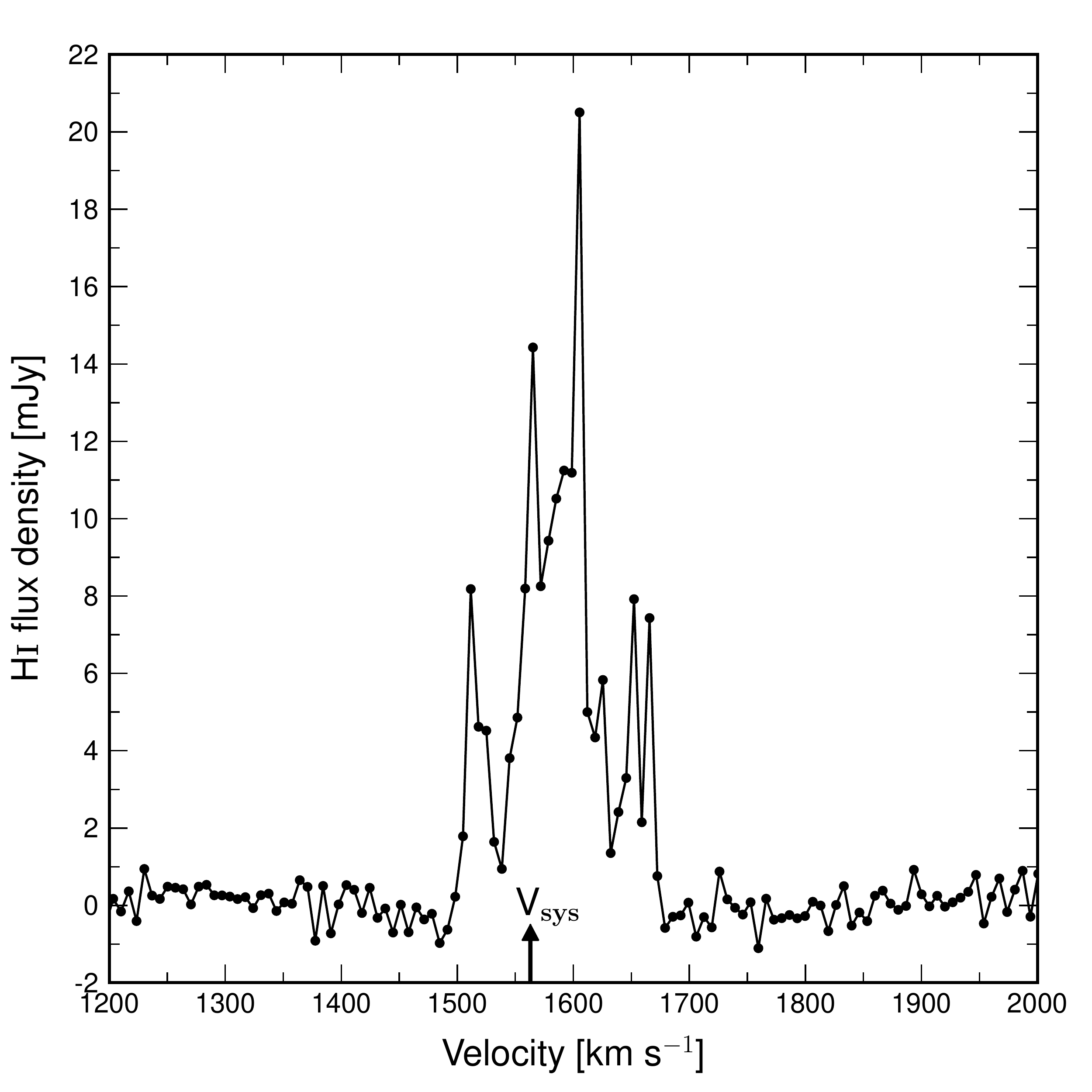}}} 
\put(1.5,10.2){(e)} 
\put(8.64,3.9){\hbox{\includegraphics[scale=0.28]{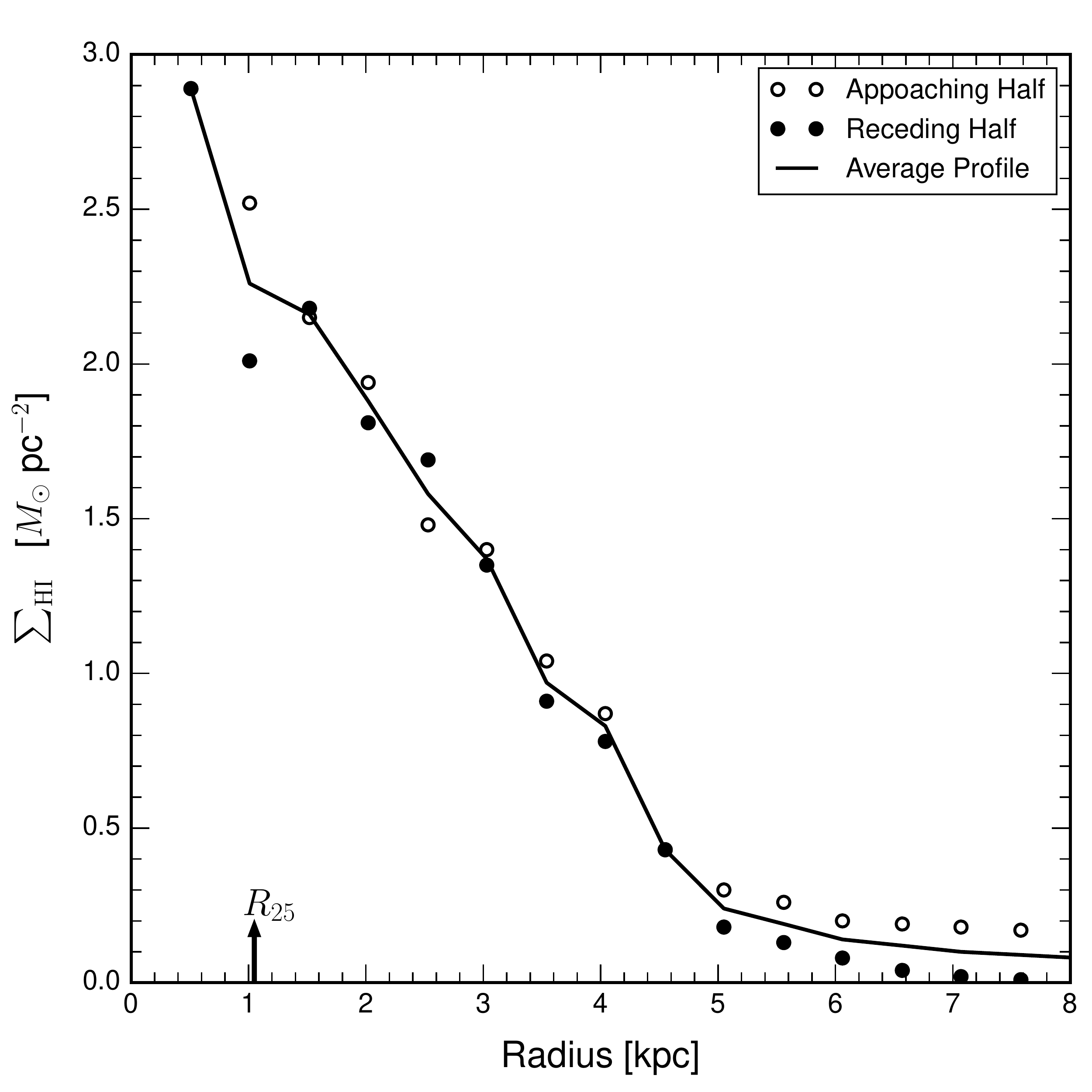}}} 
\put(9.7,10.2){(f)} 
\end{picture} 
\vspace{-4.2cm}
\caption{(a) The low resolution \HI column density contours of MRK~22 overlaid upon its grey scale optical $r$-band image. The contour levels are $0.6 \times n$, where $n=1,2,4,8,16,32$ in units of $10^{19}$~cm$^{-2}$. (b) The intermediate resolution \HI column density contours overlaid upon the grey scale optical $r$-band image. The contour levels are $4.5 \times n$ in units of $10^{19}$~cm$^{-2}$. (c) The high resolution \HI column density contours overlaid upon the grey scale \Ha line image. The contour levels are $20.9 \times n$ in units of $10^{19}$~cm$^{-2}$. (d) The intermediate resolution moment-1 map, showing the velocity field, with an overlying optical $r$-band outer contour. The circle at the bottom of each image is showing the synthesized beam. The average FWHM seeing during the optical observation was $\sim 2''.1$. (e) The global \HI profile obtained using the low resolution \HI images. The arrow at the abscissa shows the systemic \HI velocity. (f) The \HI mass surface density profile obtained using the low resolution \HI map. The arrow at the abscissa shows the $B$-band optical disk radius.}
\label{MRK22}
\end{figure}

\begin{figure*}
\begin{center}
\includegraphics[angle=0,width=1\linewidth]{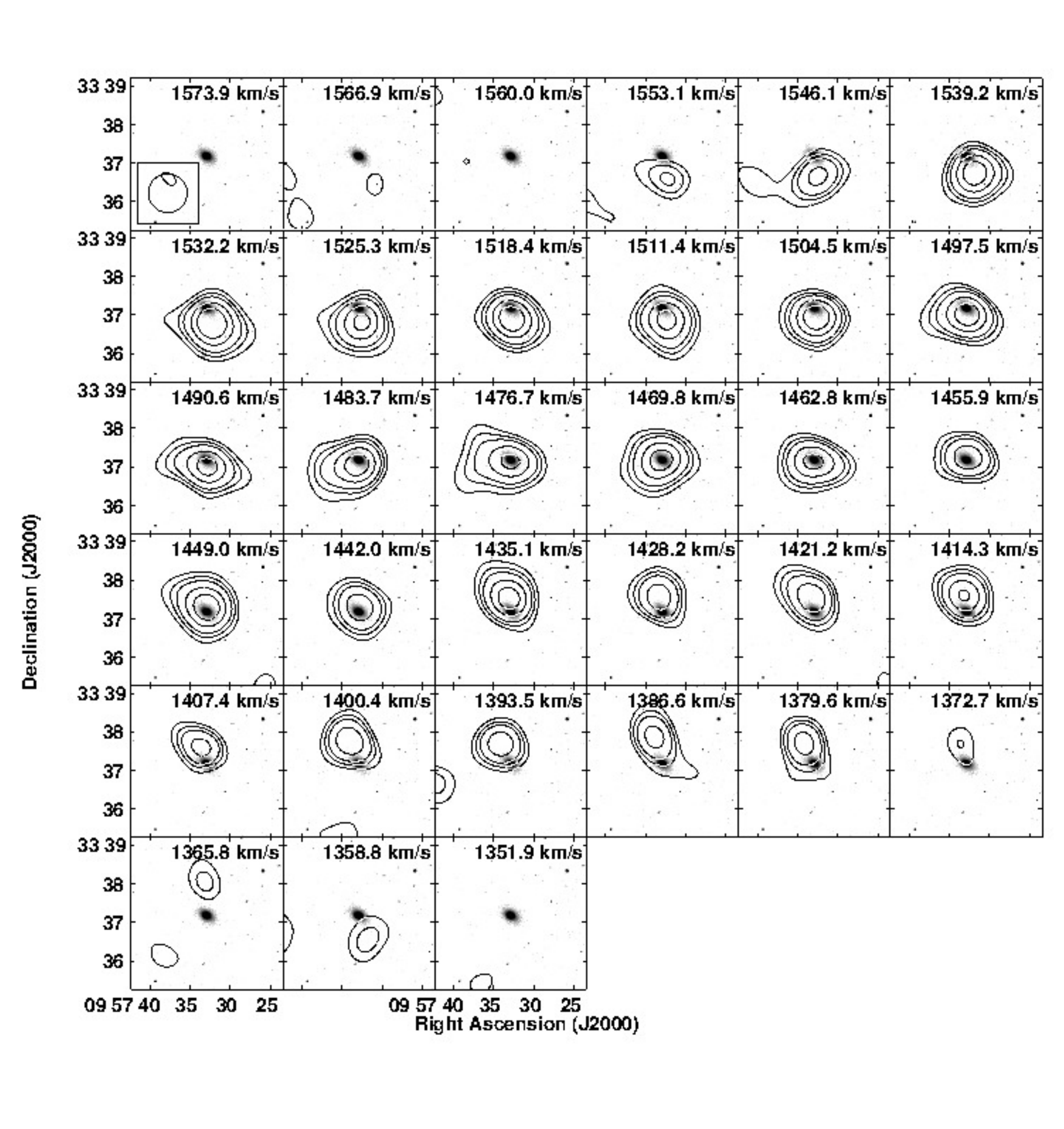}
\end{center}
\caption{The \HI contours from the low resolution channel images overlaid upon the grey scale optical $r$-band image of IC~2524. The contours representing \HI emission flux are drawn at $2.5\sigma\times n$~mJy/Beam; n=1,1.5,2,3,4,6.}
\end{figure*}
\begin{figure}
\setlength{\unitlength}{1cm}
\begin{picture}(12,25) 
\put(0,18.1){\hbox{\includegraphics[scale=0.39]{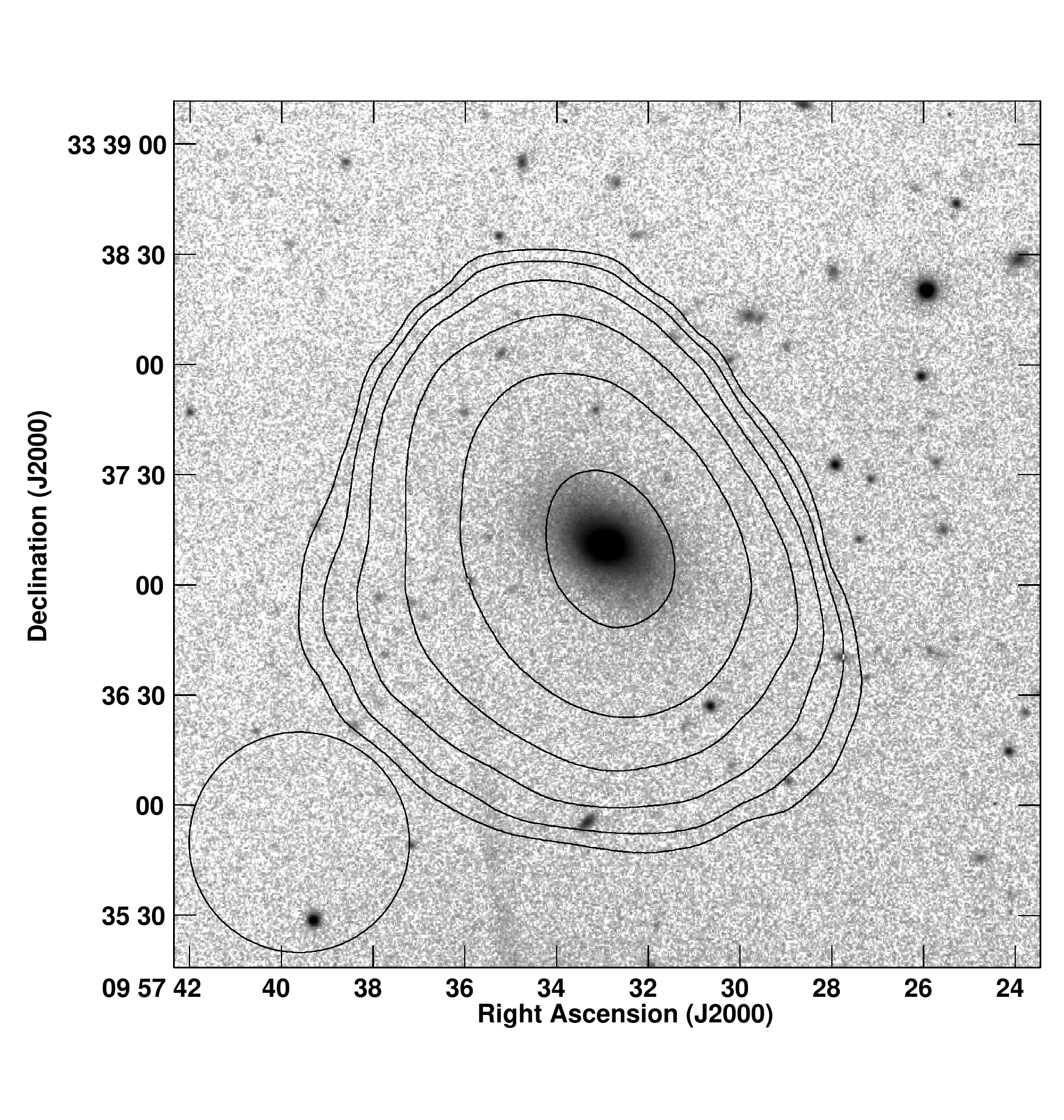}}} 
\put(1.4,24.8){(a)} 
\put(8.3,18.1){\hbox{\includegraphics[scale=0.39]{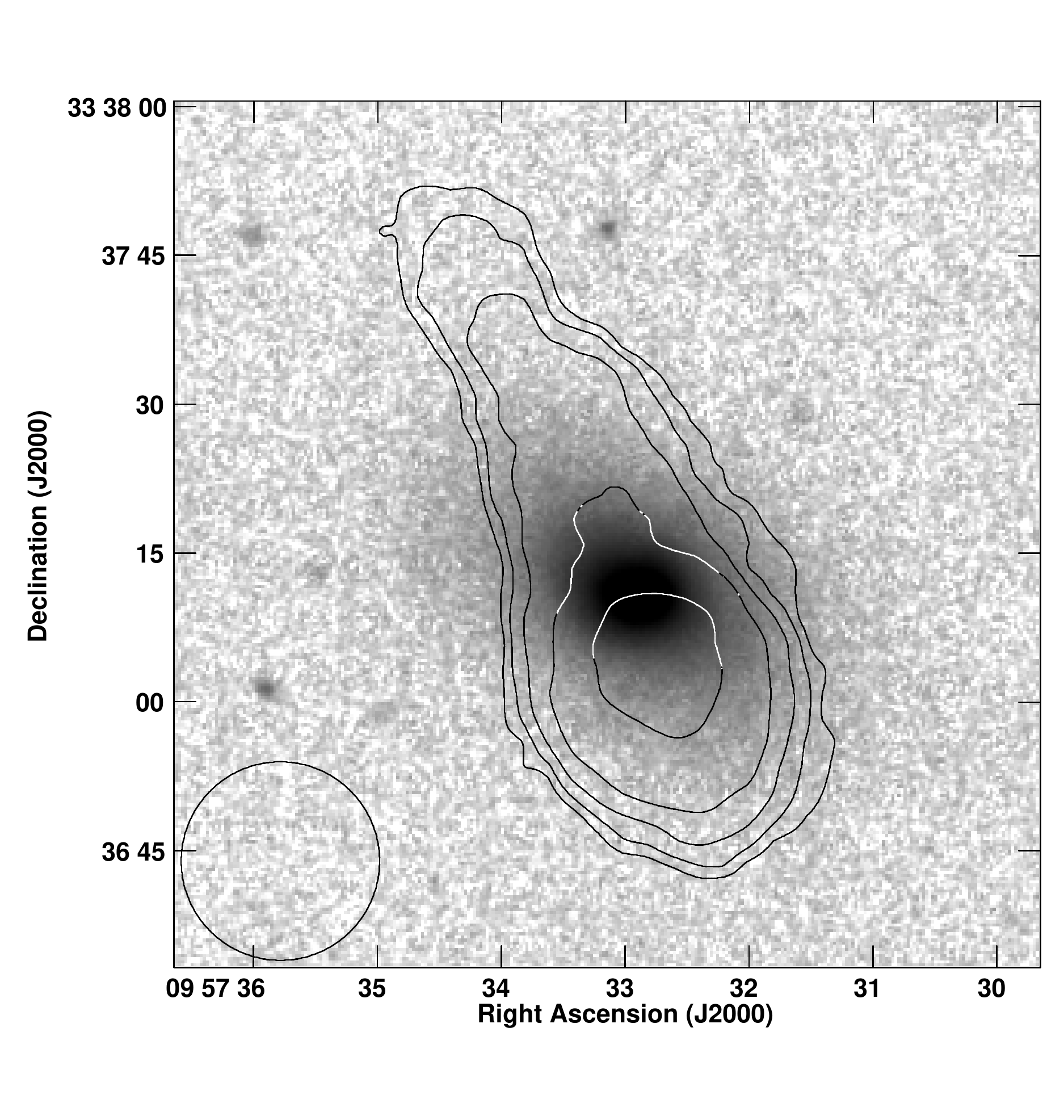}}}
\put(9.8,24.8){(b)} 
\put(0,10.9){\hbox{\includegraphics[scale=0.39]{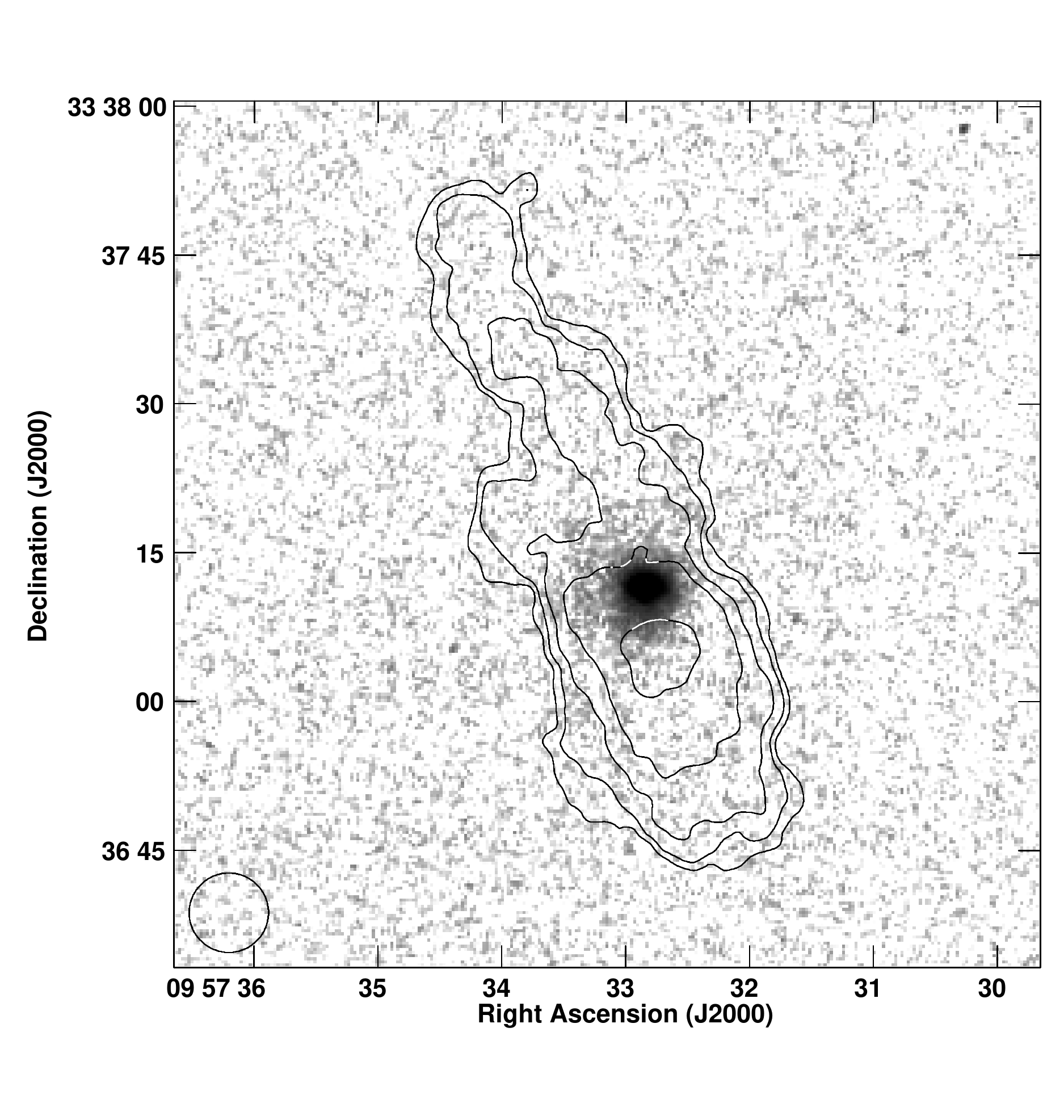}}}
\put(1.4,17.6){(c)} 
\put(8.0,11.1){\hbox{\includegraphics[scale=0.59]{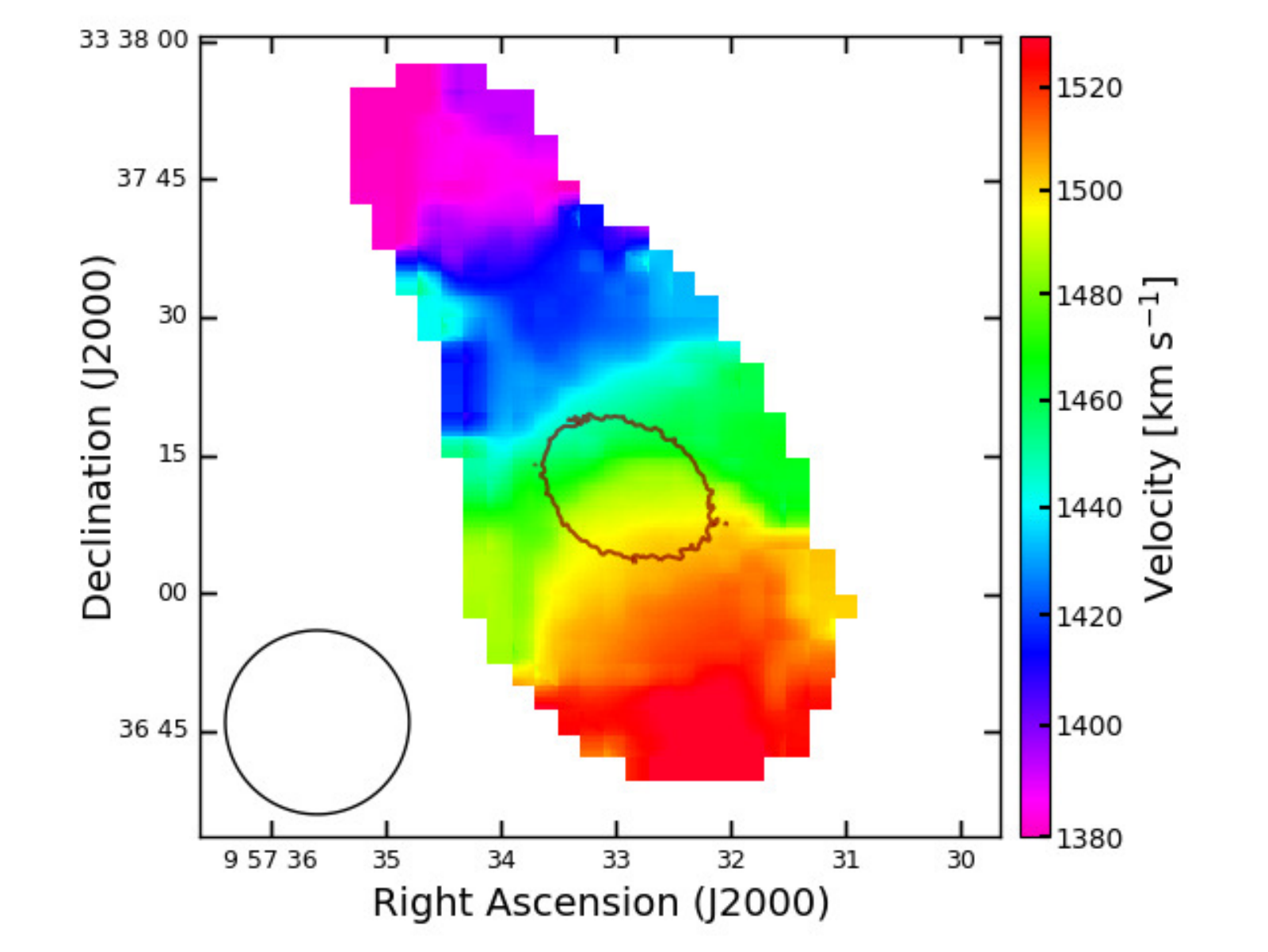}}}
\put(9.8,17.6){(d)} 
\put(0.5,3.9){\hbox{\includegraphics[scale=0.279]{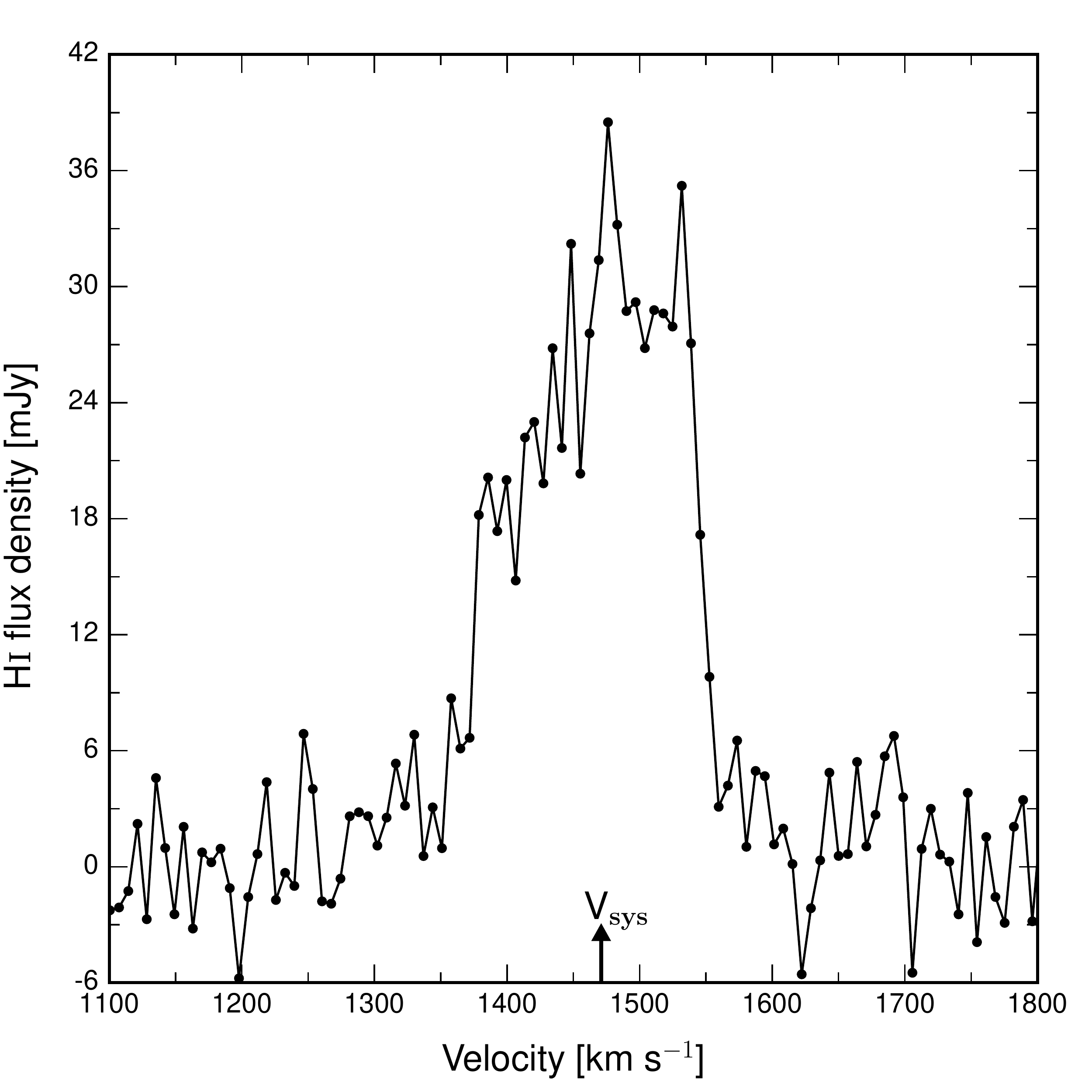}}} 
\put(1.4,10.2){(e)} 
\put(8.65,3.9){\hbox{\includegraphics[scale=0.279]{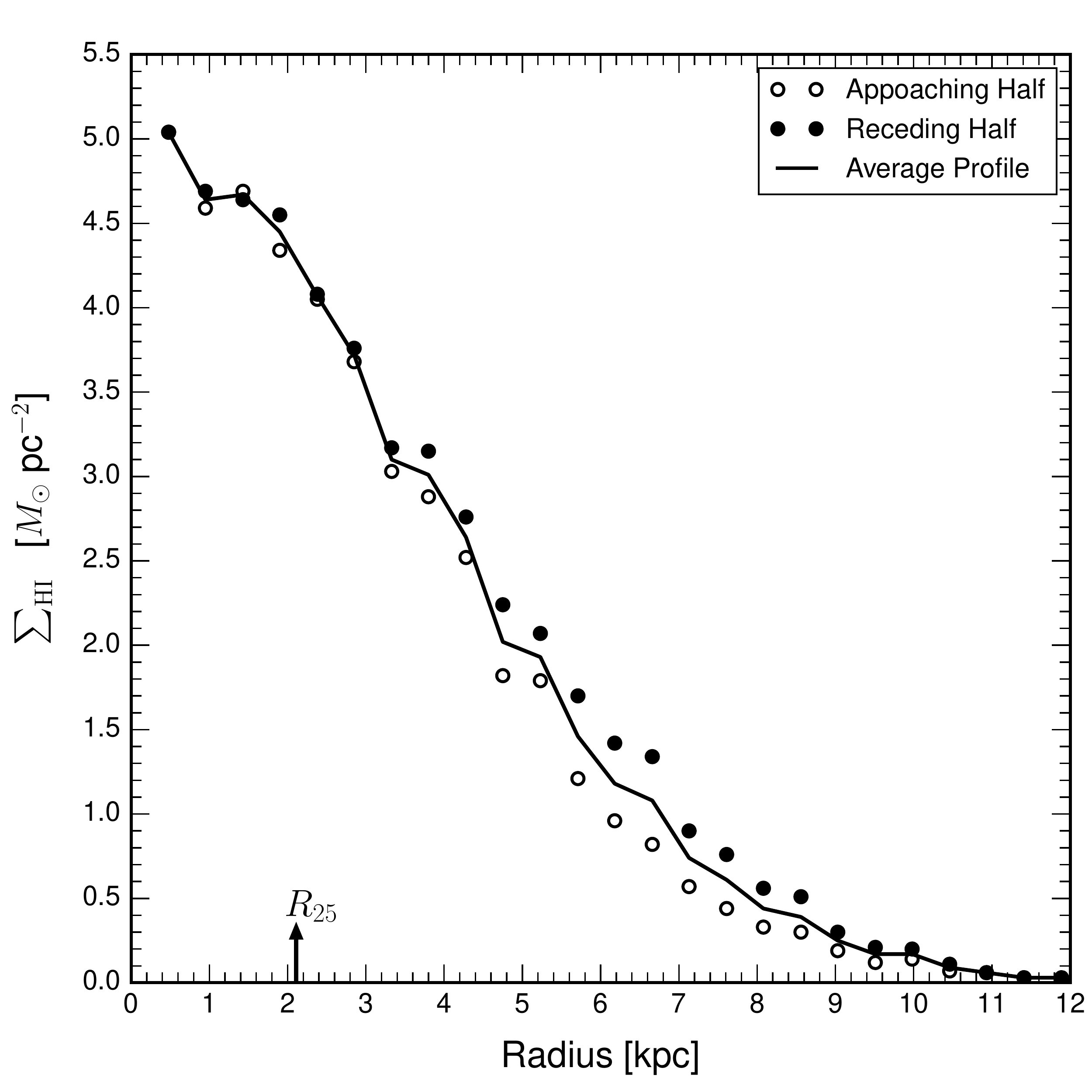}}} 
\put(9.8,10.2){(f)} 
\end{picture} 
\vspace{-4.2cm}
\caption{(a) The low resolution \HI column density contours of IC~2524 overlaid upon its grey scale optical $r$-band image. The contour levels are $2.3 \times n$, where $n=1,2,4,8,16,32$ in units of $10^{19}$~cm$^{-2}$. (b) The intermediate resolution \HI column density contours overlaid upon the grey scale optical $r$-band image. The contour levels are $13.9 \times n$ in units of $10^{19}$~cm$^{-2}$. (c) The high resolution \HI column density contours overlaid upon the grey scale \Ha line image. The contour levels are $34.8 \times n$ in units of $10^{19}$~cm$^{-2}$. (d) The intermediate resolution moment-1 map, showing the velocity field, with an overlying optical $r$-band outer contour. The circle at the bottom of each image is showing the synthesized beam. The average FWHM seeing during the optical observation was $\sim 1''.3$. (e) The global \HI profile obtained using the low resolution \HI images. The arrow at the abscissa shows the systemic \HI velocity. (f) The \HI mass surface density profile obtained using the low resolution \HI map. The arrow at the abscissa shows the $B$-band optical disk radius.}
\label{WR402}
\end{figure}

\begin{figure*}
\begin{center}
\includegraphics[angle=0,width=1\linewidth]{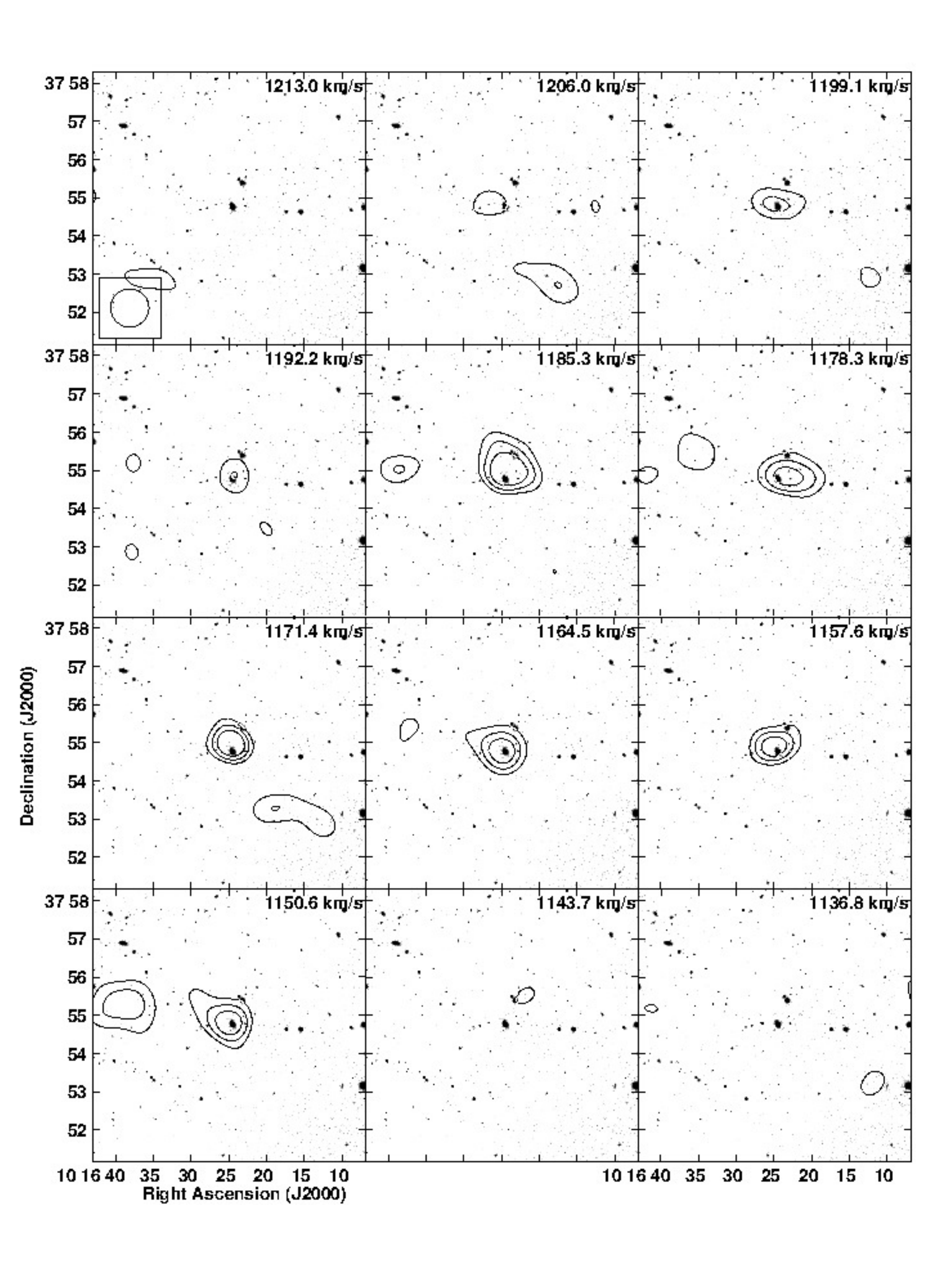}
\end{center}
\caption{The \HI contours from the low resolution channel images overlaid upon the grey scale optical $r$-band image of KUG~1013+381. The contours representing \HI emission flux are drawn at $2.5\sigma\times n$~mJy/Beam; n=1,1.5,2,3,4,6.}
\end{figure*}
\begin{figure}
\setlength{\unitlength}{1cm}
\begin{picture}(12,25) 
\put(0,18.1){\hbox{\includegraphics[scale=0.38]{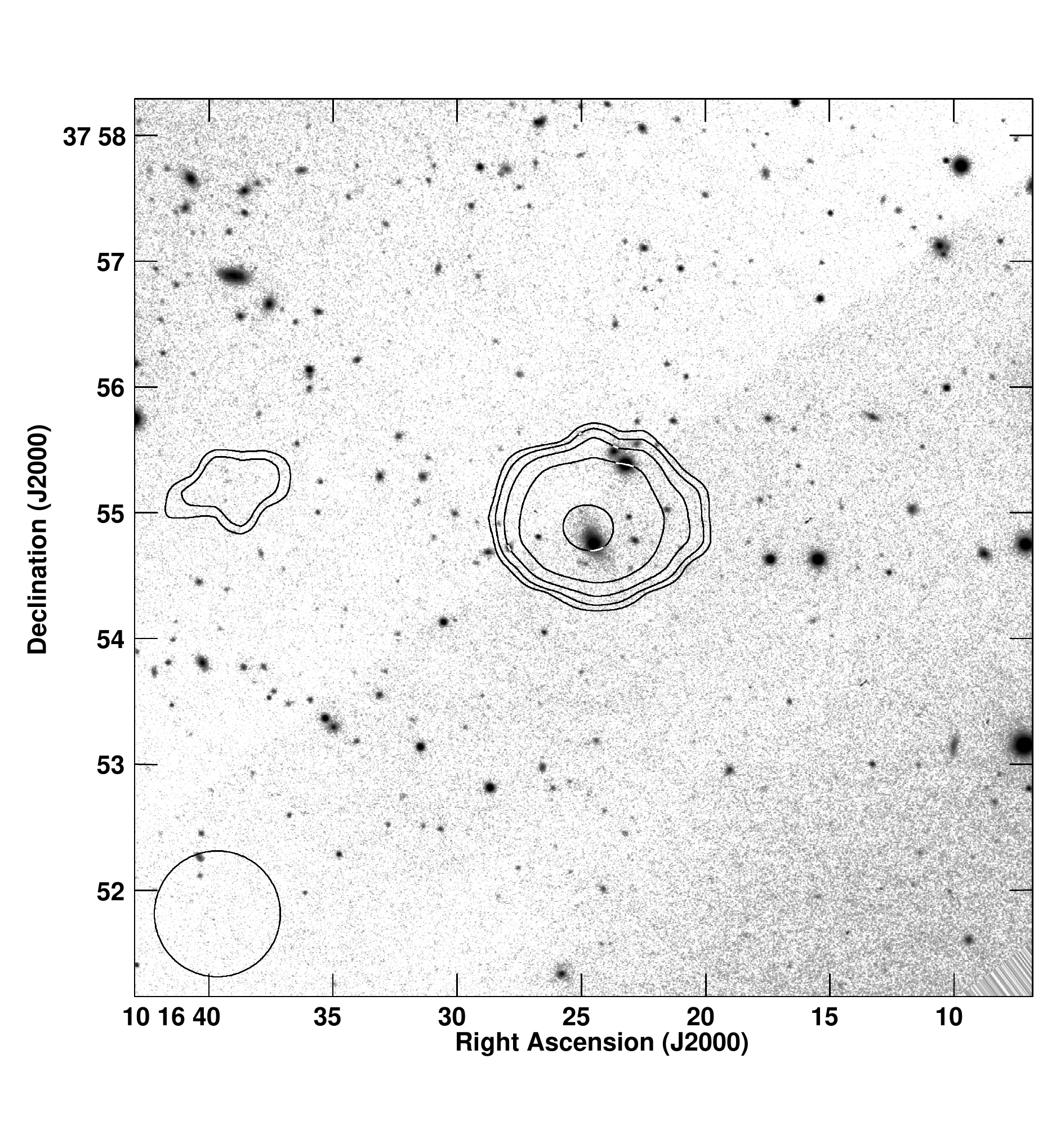}}} 
\put(1.2,24.9){(a)} 
\put(8.3,18.07){\hbox{\includegraphics[scale=0.396]{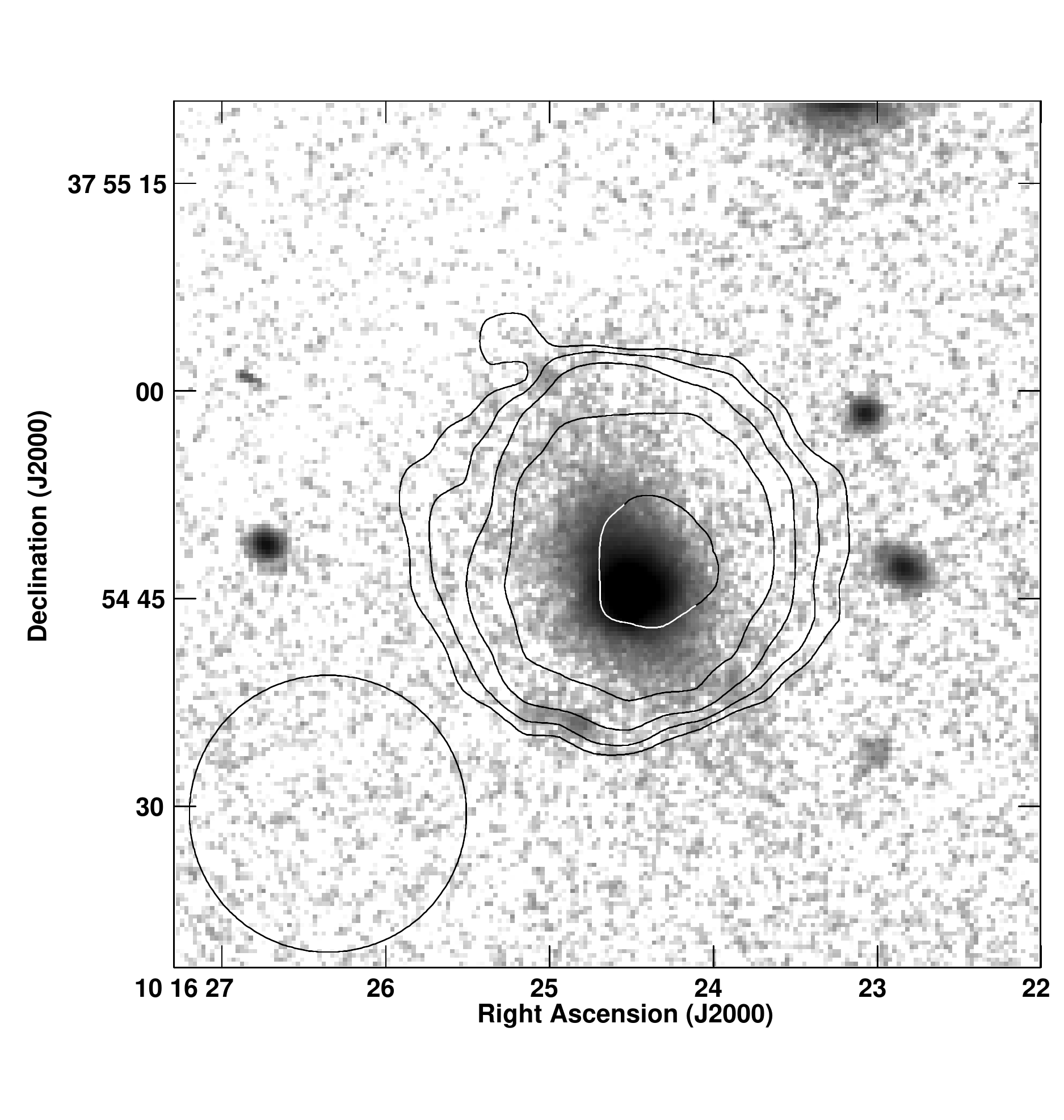}}}
\put(9.8,24.9){(b)} 
\put(-0.3,11.0){\hbox{\includegraphics[scale=0.396]{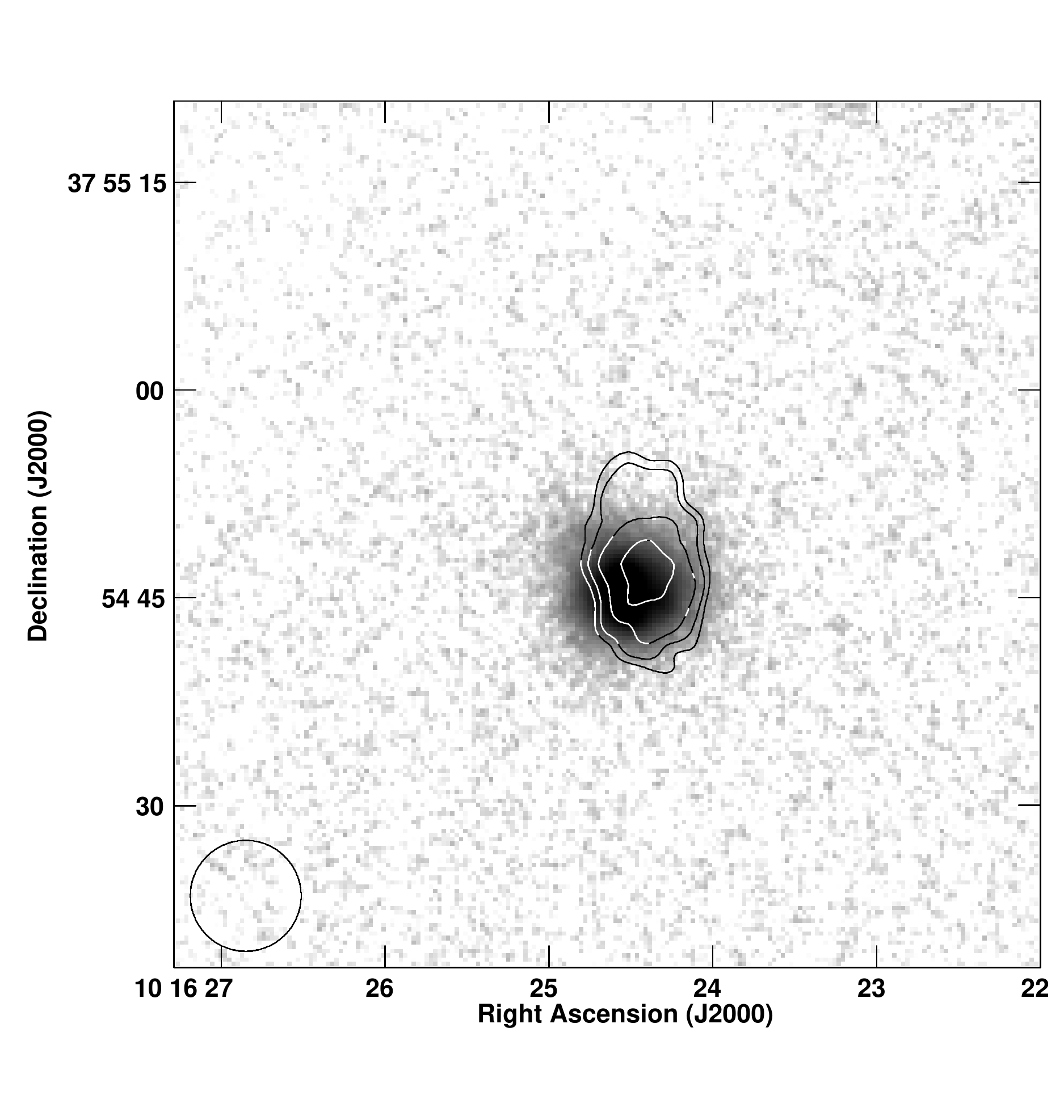}}}
\put(1.2,17.8){(c)} 
\put(8.05,11.22){\hbox{\includegraphics[scale=0.592]{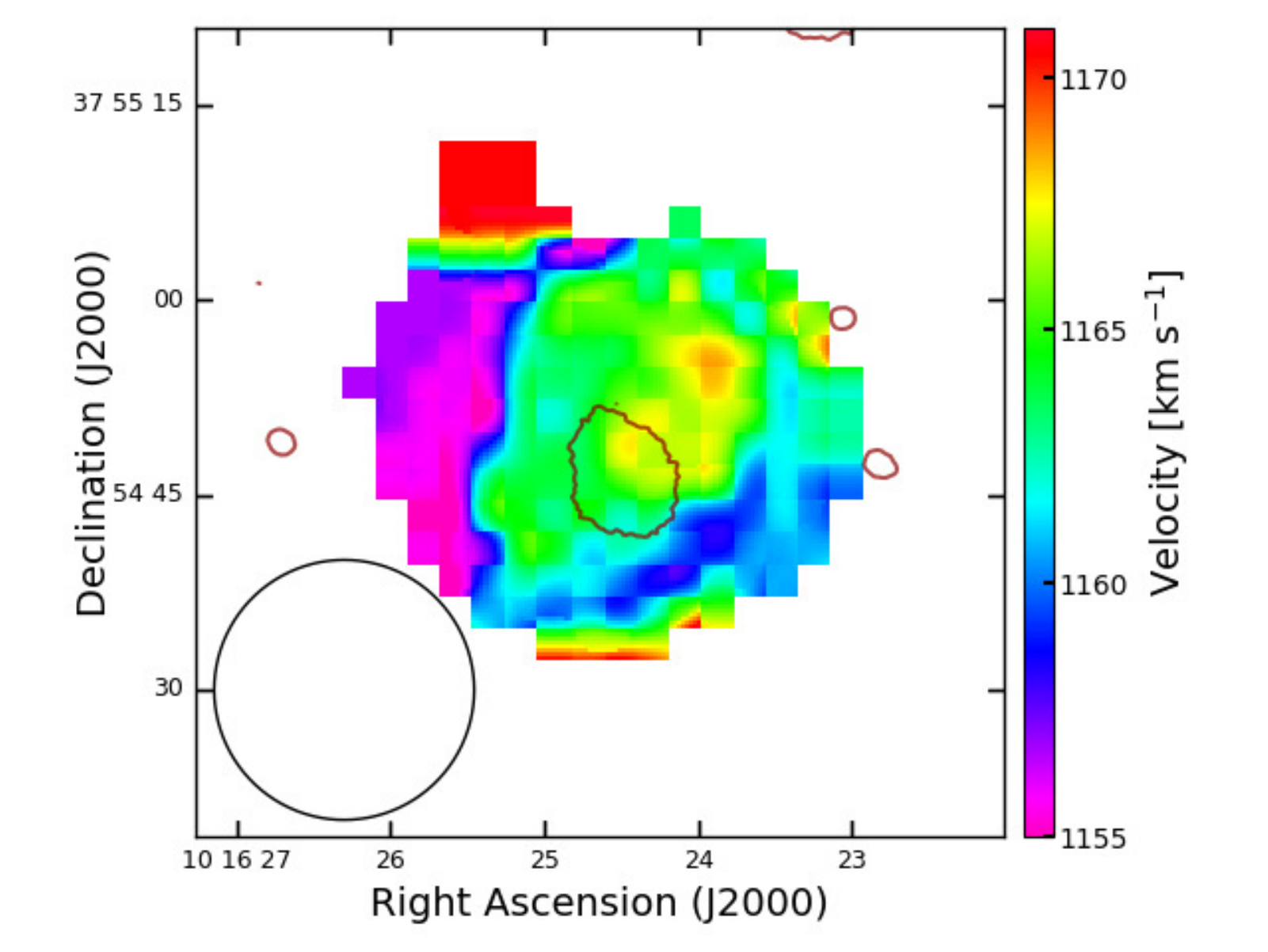}}}
\put(9.8,17.8){(d)} 
\put(0.21,4.2){\hbox{\includegraphics[scale=0.284]{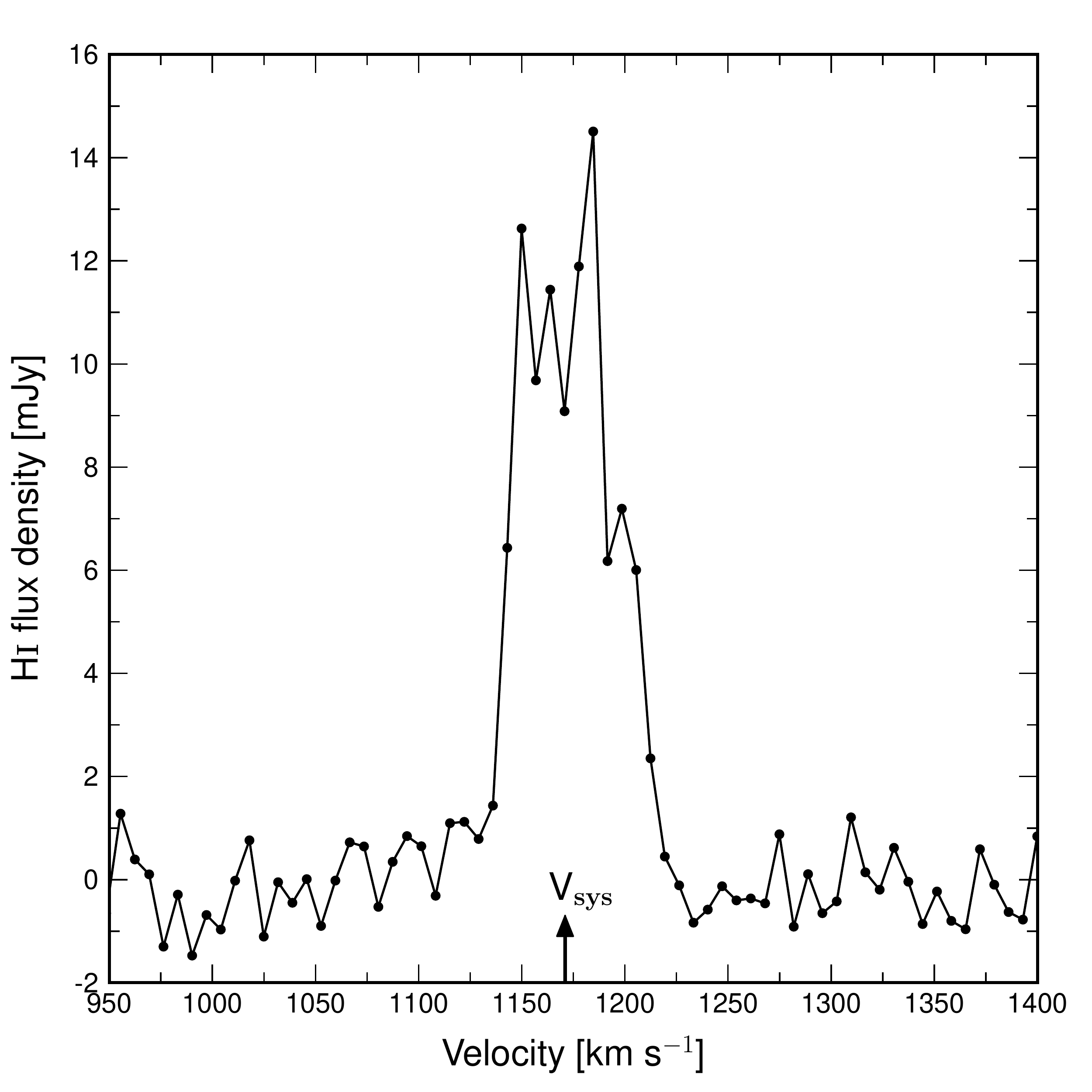}}} 
\put(1.2,10.6){(e)} 
\put(8.65,4.2){\hbox{\includegraphics[scale=0.284]{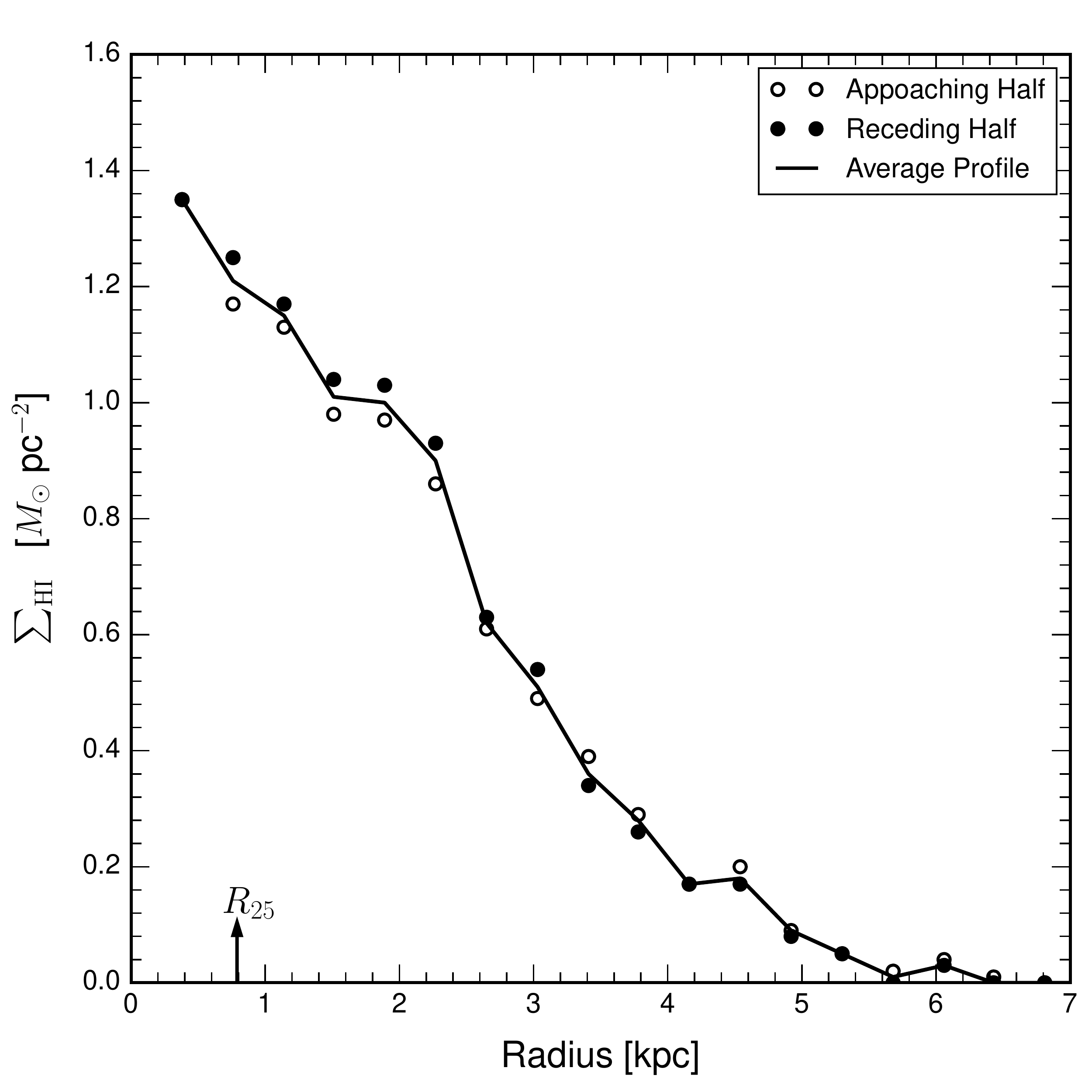}}} 
\put(9.8,10.6){(f)} 
\end{picture} 
\vspace{-4.4cm}
\caption{(a) The low resolution \HI column density contours of KUG~1013+381 overlaid upon its grey scale optical $r$-band image. The contour levels are $0.7 \times n$, where $n=1,2,4,8,16,32$ in units of $10^{19}$~cm$^{-2}$. (b) The intermediate resolution \HI column density contours overlaid upon the grey scale optical $r$-band image. The contour levels are $3.1 \times n$ in units of $10^{19}$~cm$^{-2}$. (c) The high resolution \HI column density contours overlaid upon the grey scale \Ha line image. The contour levels are $10.5 \times n$ in units of $10^{19}$~cm$^{-2}$. (d) The intermediate resolution moment-1 map, showing the velocity field, with an overlying optical $r$-band outer contour. The circle at the bottom of each image is showing the synthesized beam. The average FWHM seeing during the optical observation was $\sim 1''.4$. (e) The global \HI profile obtained using the low resolution \HI images. The arrow at the abscissa shows the systemic \HI velocity. (f) The \HI mass surface density profile obtained using the low resolution \HI map. The arrow at the abscissa shows the $B$-band optical disk radius.}
\label{WR295}
\end{figure}

\begin{figure*}
\begin{center}
\includegraphics[angle=0,width=1\linewidth]{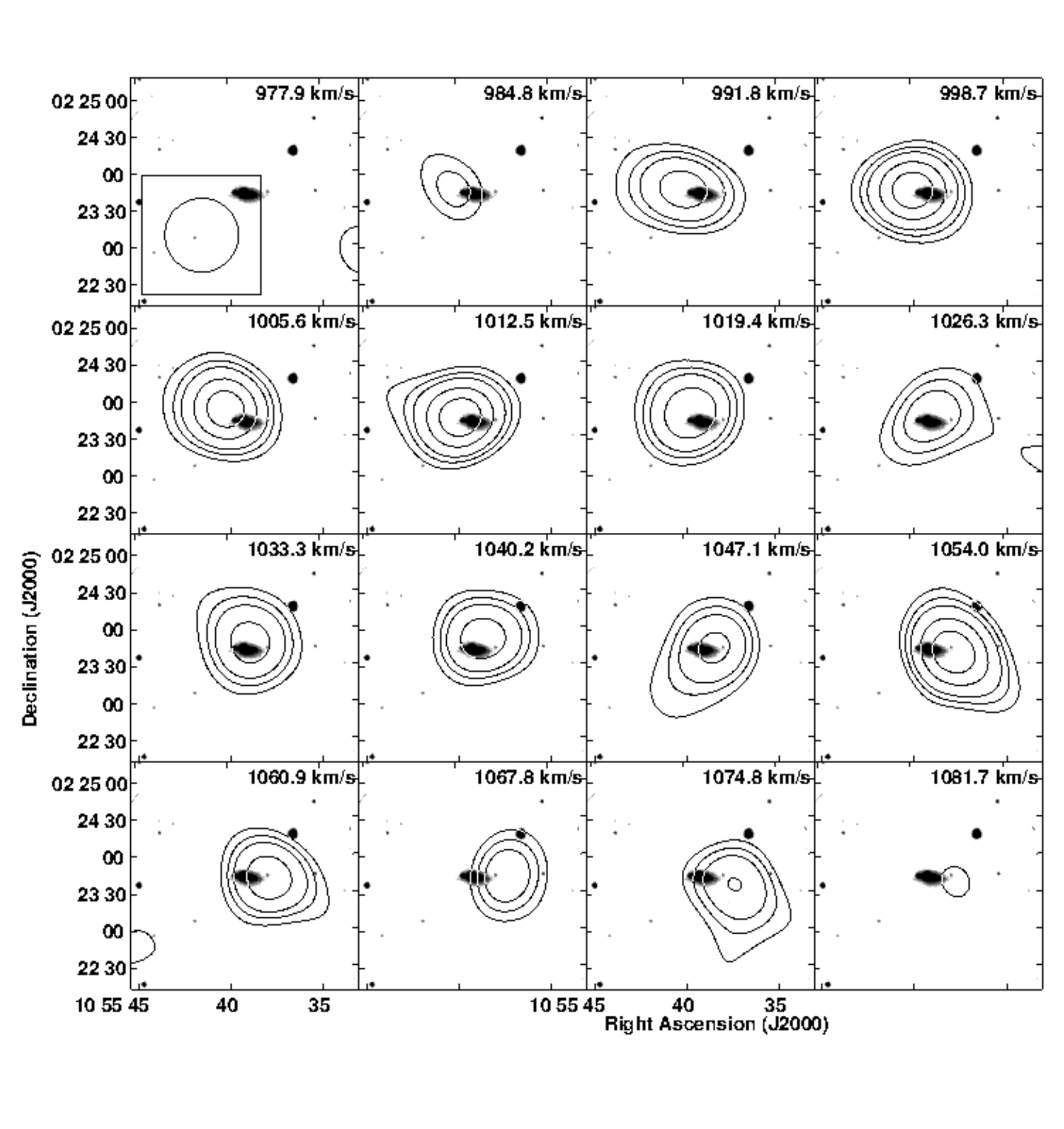}
\end{center}
\caption{The \HI contours from the low resolution channel images overlaid upon the grey scale optical $r$-band image of CGCG~038-051. The contours representing \HI emission flux are drawn at $2.5\sigma\times n$~mJy/Beam; n=1,1.5,2,3,4,6.}
\end{figure*}
\begin{figure}
\setlength{\unitlength}{1cm}
\begin{picture}(12,25) 
\put(0,18.1){\hbox{\includegraphics[scale=0.39]{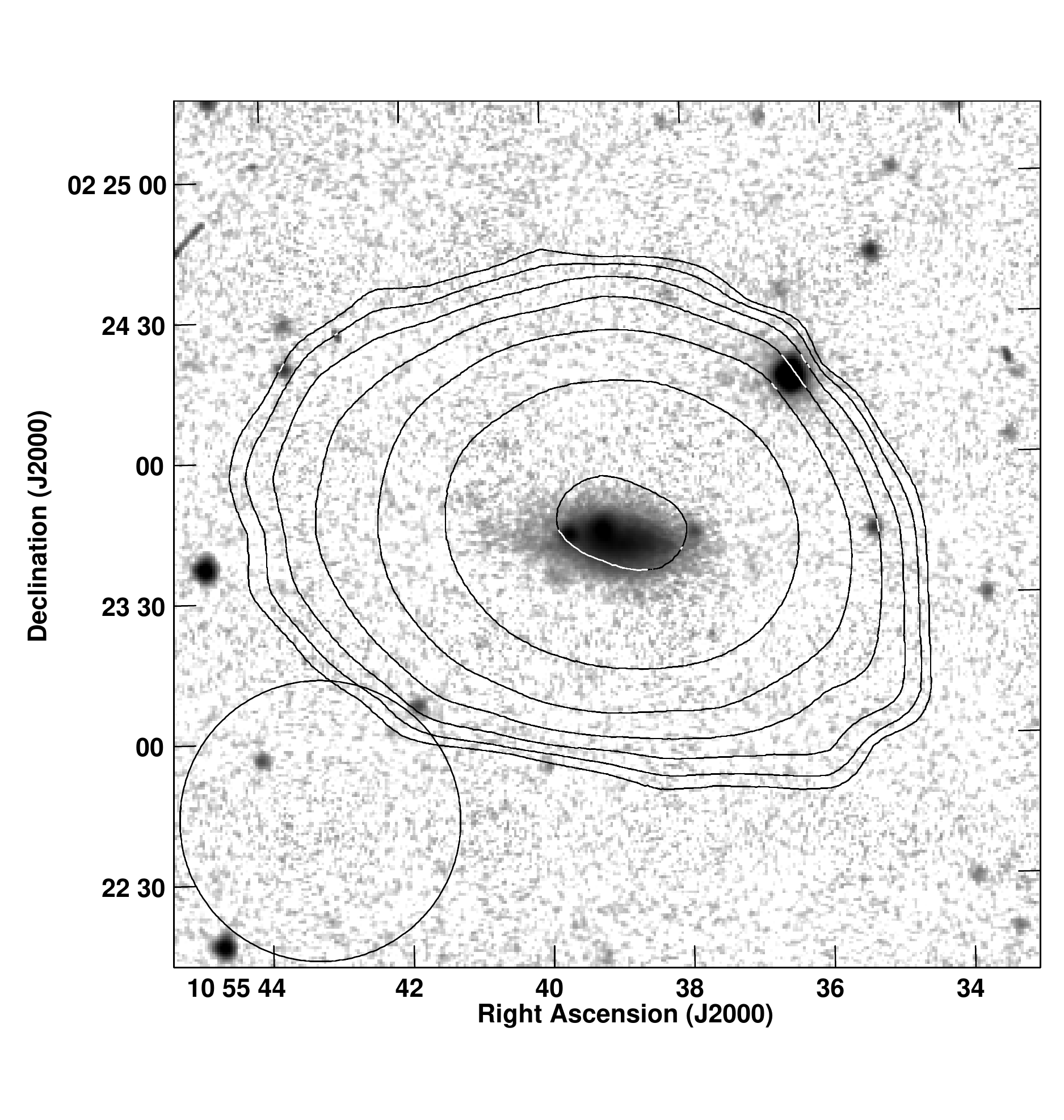}}} 
\put(1.4,24.8){(a)} 
\put(8.3,18.07){\hbox{\includegraphics[scale=0.39]{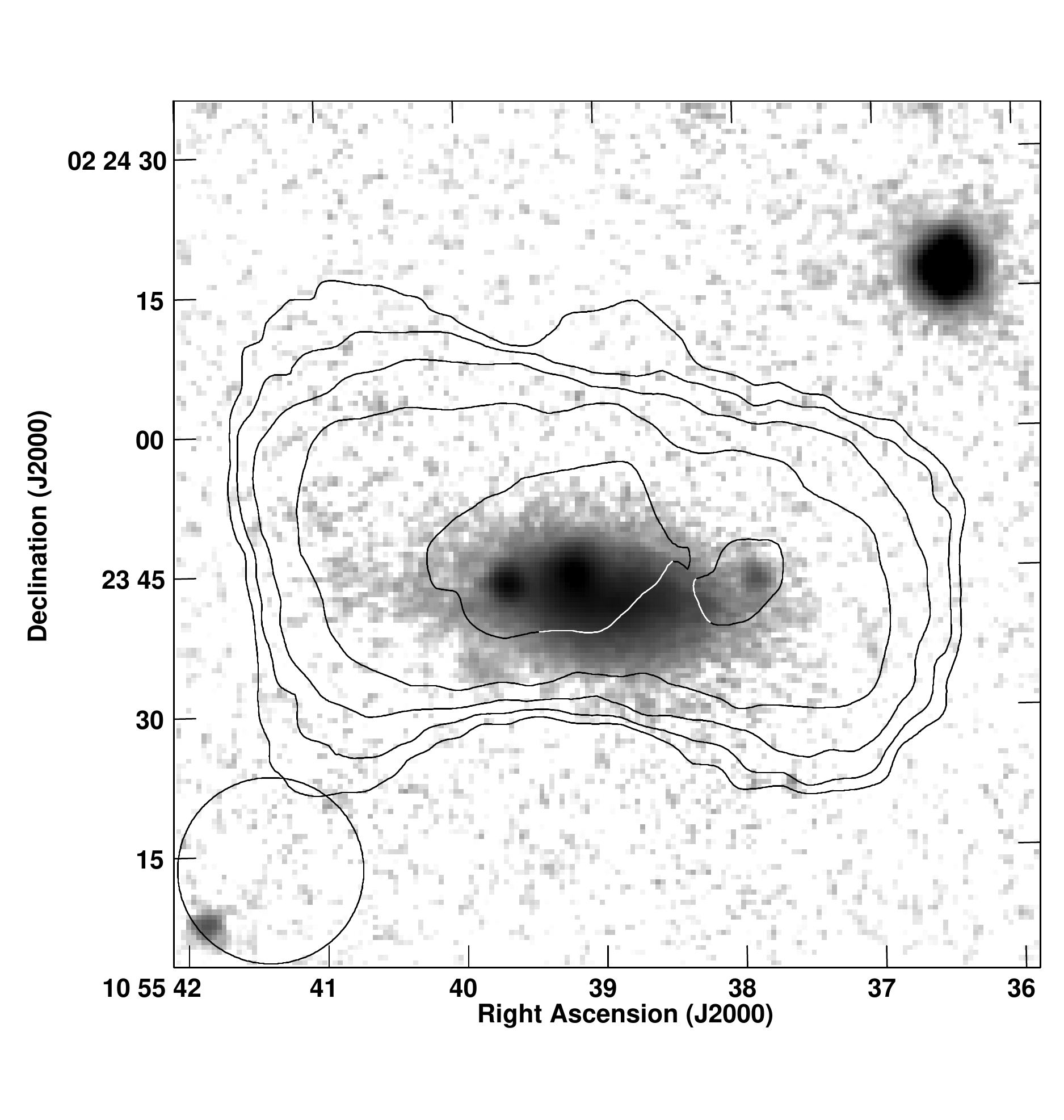}}}
\put(9.8,24.8){(b)} 
\put(0,11.1){\hbox{\includegraphics[scale=0.39]{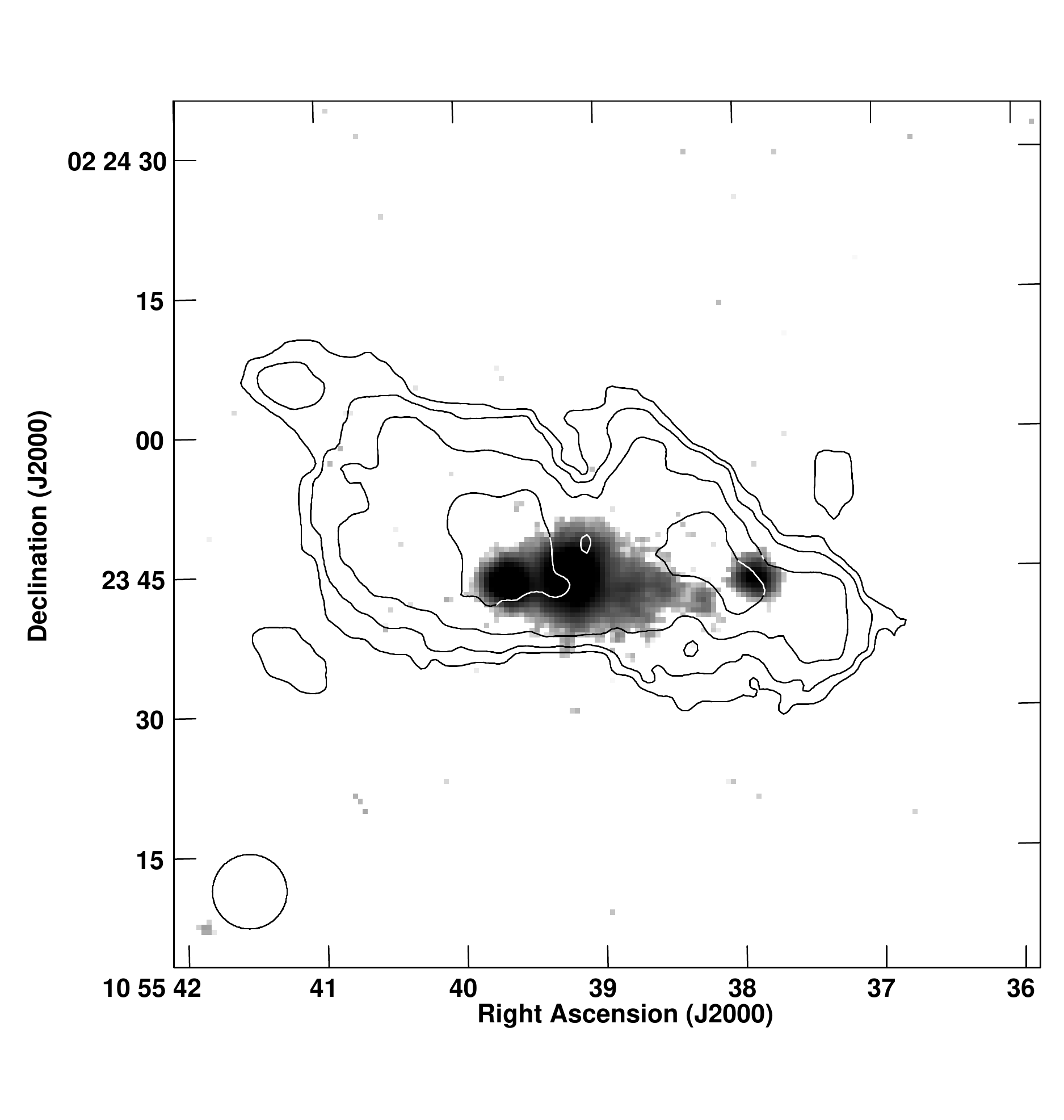}}}
\put(1.4,17.8){(c)} 
\put(7.95,11.28){\hbox{\includegraphics[scale=0.585]{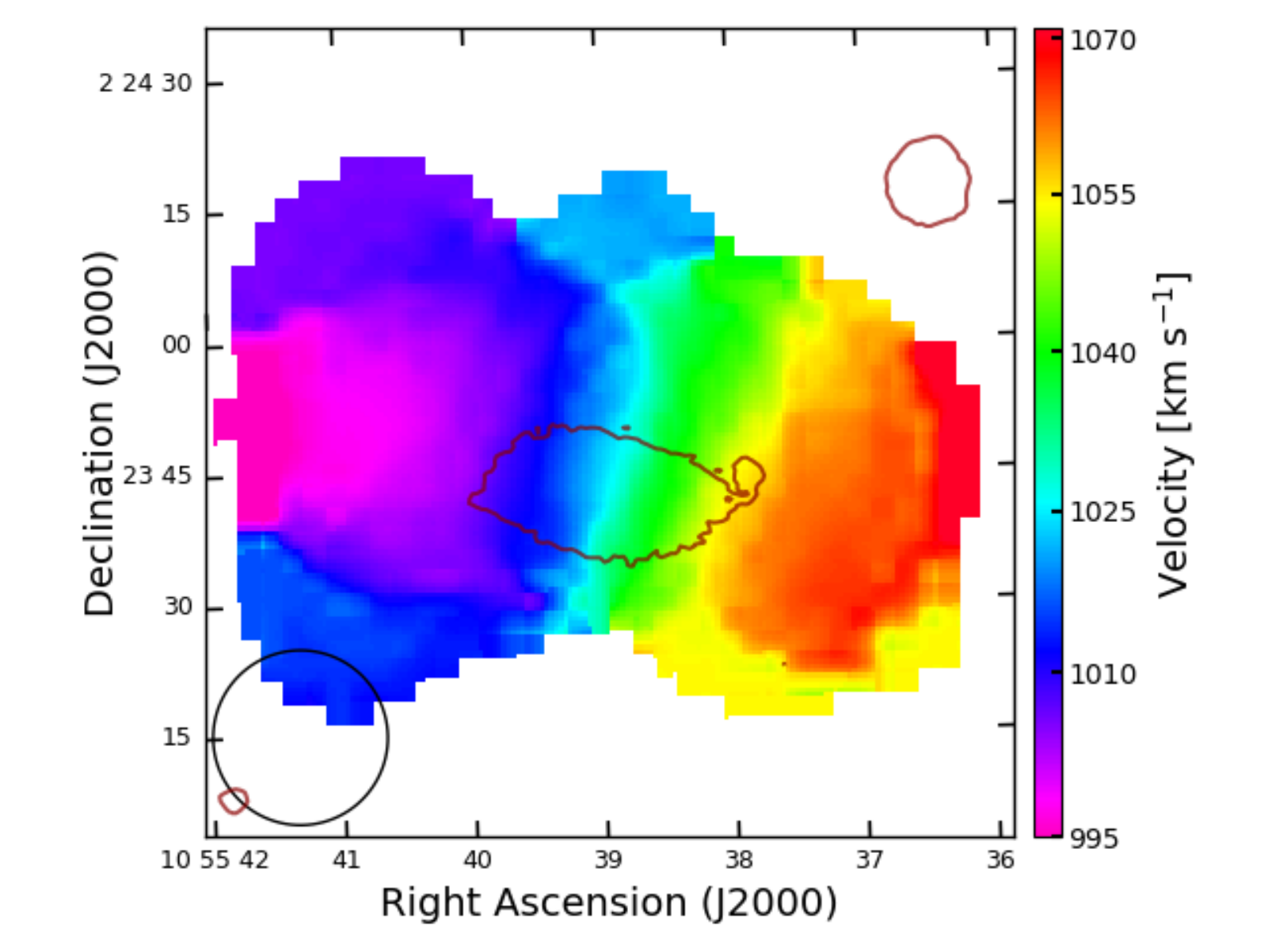}}}
\put(9.8,17.8){(d)} 
\put(0.53,4.3){\hbox{\includegraphics[scale=0.279]{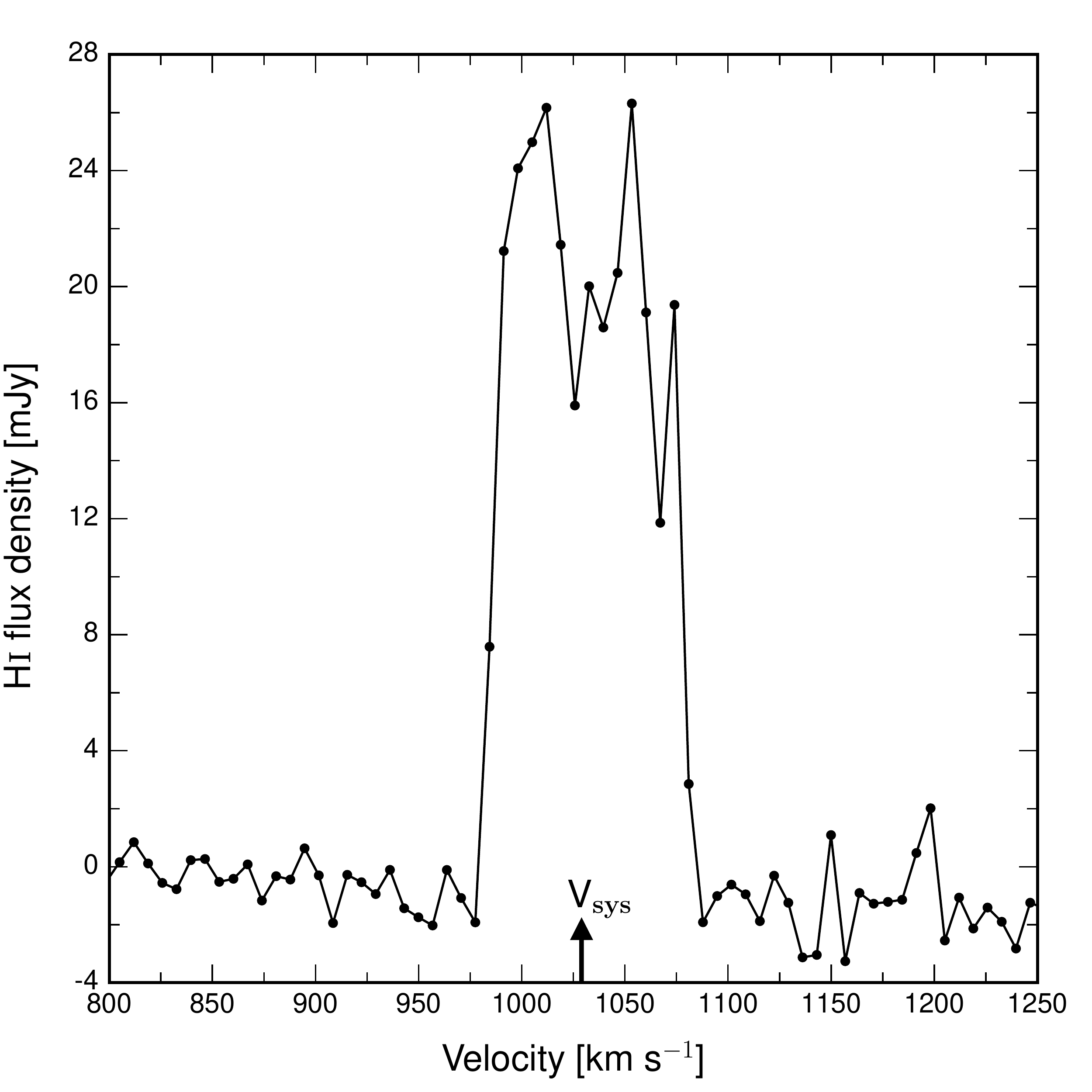}}} 
\put(1.4,10.6){(e)} 
\put(8.65,4.3){\hbox{\includegraphics[scale=0.279]{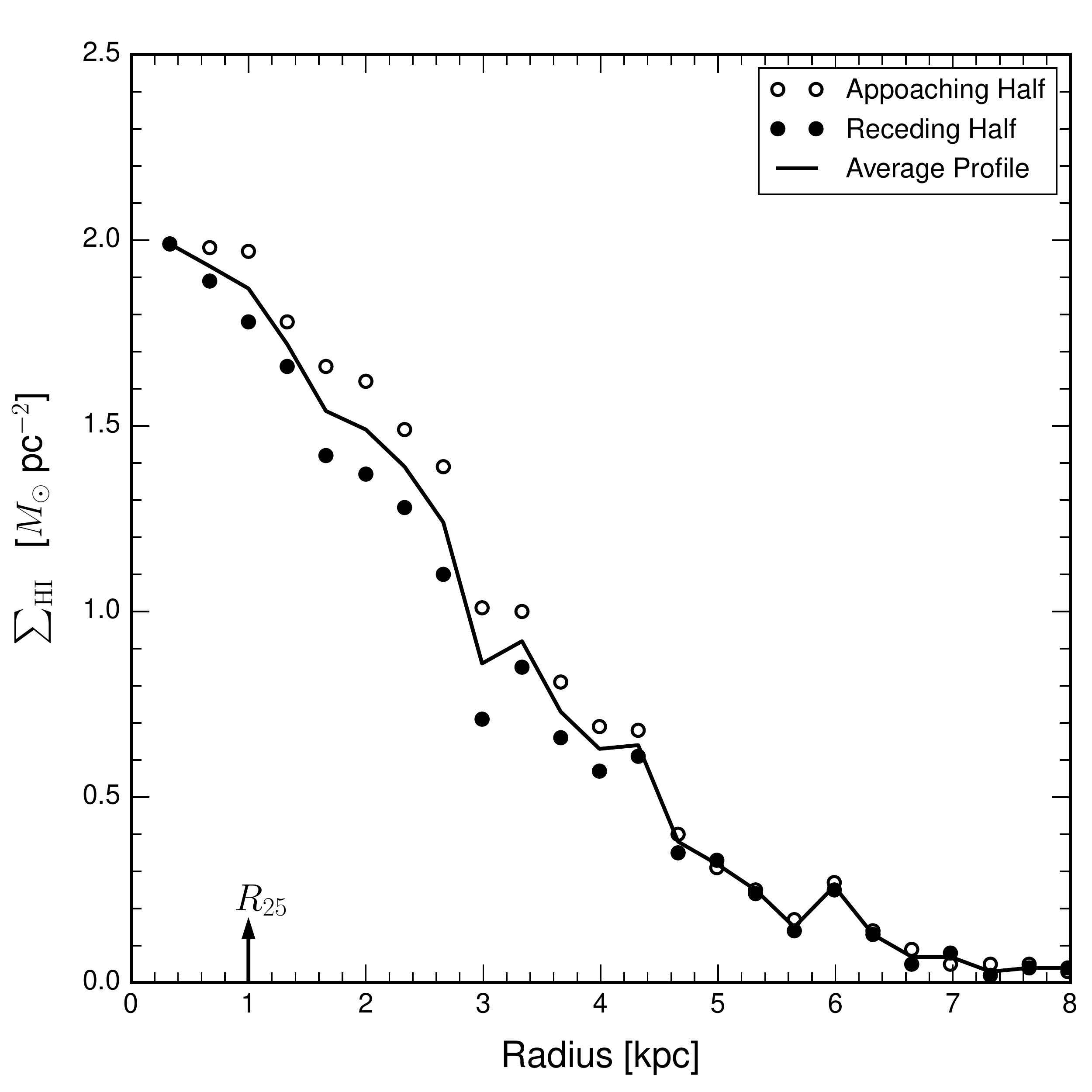}}} 
\put(9.8,10.6){(f)} 
\end{picture} 
\vspace{-4.5cm}
\caption{(a) The low resolution \HI column density contours of CGCG~038-051 overlaid upon its grey scale optical $r$-band image. The contour levels are $0.7 \times n$, where $n=1,2,4,8,16,32$ in units of $10^{19}$~cm$^{-2}$. (b) The intermediate resolution \HI column density contours overlaid upon the grey scale optical $r$-band image. The contour levels are $8.1 \times n$ in units of $10^{19}$~cm$^{-2}$. (c) The high resolution \HI column density contours overlaid upon the grey scale \Ha line image. The contour levels are $26.1 \times n$ in units of $10^{19}$~cm$^{-2}$. (d) The intermediate resolution moment-1 map, showing the velocity field, with an overlying optical $r$-band outer contour. The circle at the bottom of each image is showing the synthesized beam. The average FWHM seeing during the optical observation was $\sim 2''.2$. (e) The global \HI profile obtained using the low resolution \HI images. The arrow at the abscissa shows the systemic \HI velocity. (f) The \HI mass surface density profile obtained using the low resolution \HI map. The arrow at the abscissa shows the $B$-band optical disk radius.}
\label{WR083}
\end{figure}

\clearpage

\begin{figure*}
\begin{center}
\includegraphics[angle=0,width=1\linewidth]{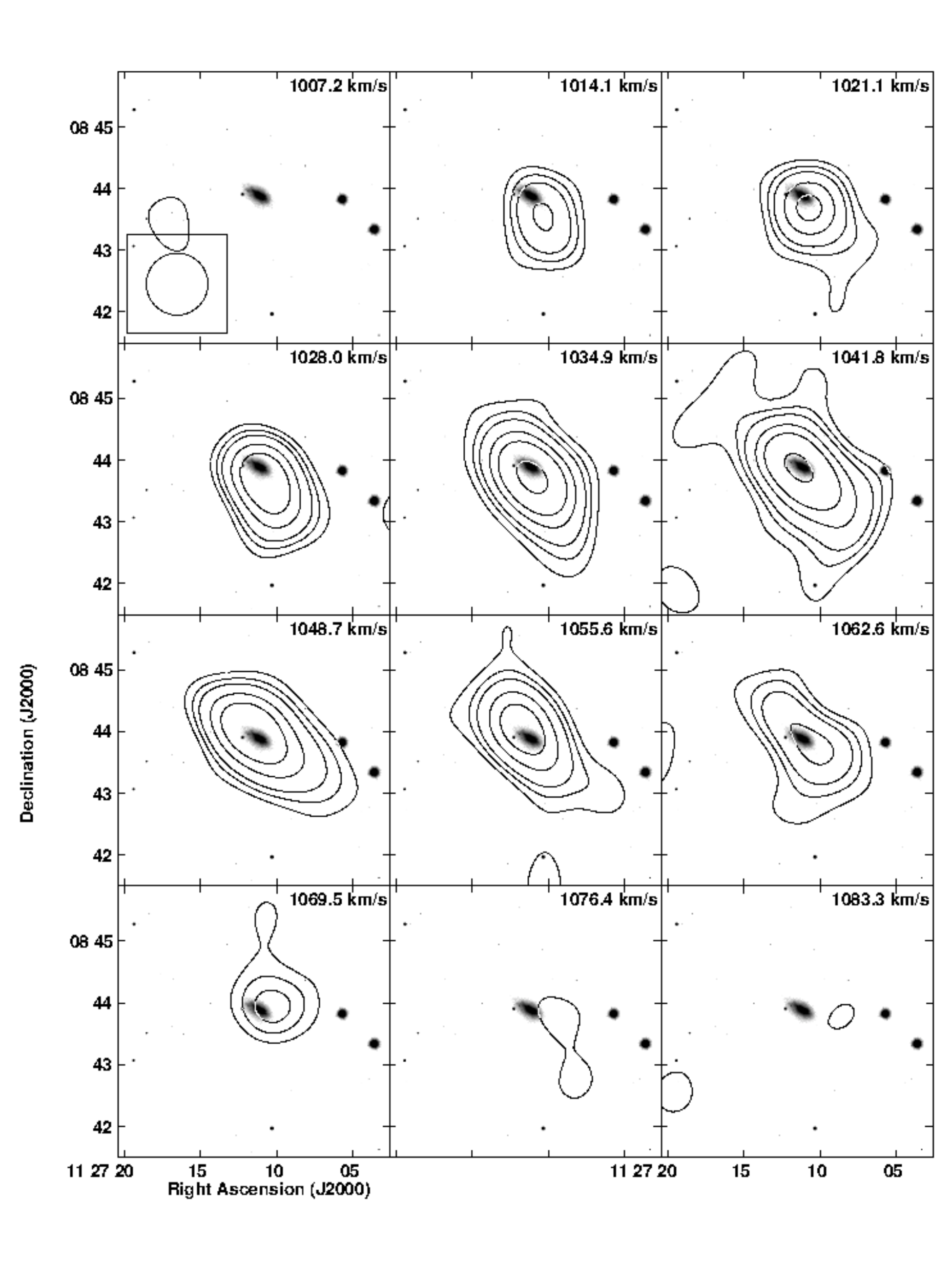}
\end{center}
\caption{The \HI contours from the low resolution channel images overlaid upon the grey scale optical $r$-band image of IC~2828. The contours representing \HI emission flux are drawn at $2.5\sigma\times n$~mJy/Beam; n=1,1.5,2,3,4,6.}
\end{figure*}
\begin{figure}
\setlength{\unitlength}{1cm}
\begin{picture}(12,25) 
\put(0,17.87){\hbox{\includegraphics[scale=0.40]{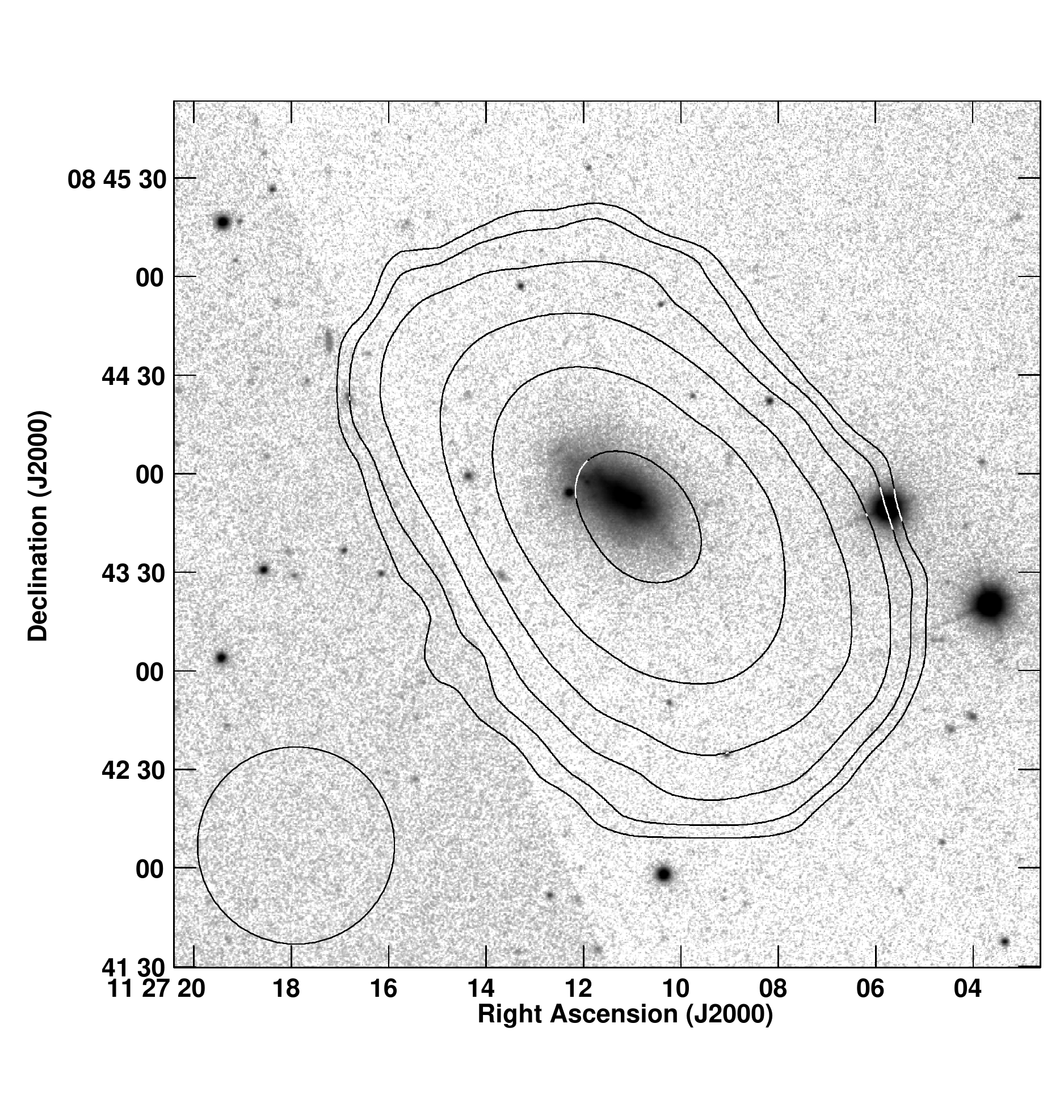}}} 
\put(1.5,24.8){(a)} 
\put(8.3,17.87){\hbox{\includegraphics[scale=0.40]{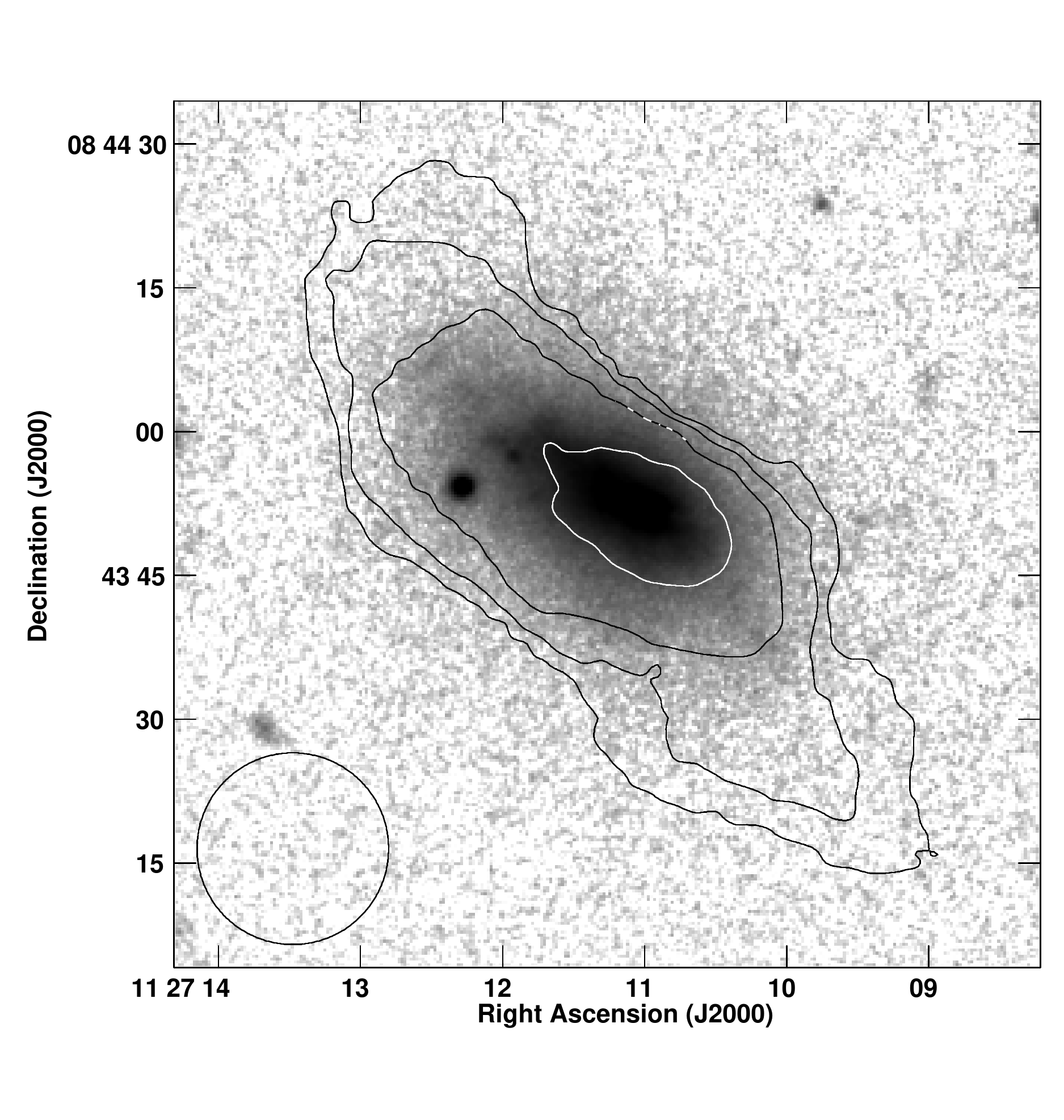}}}
\put(9.8,24.8){(b)} 
\put(0,10.85){\hbox{\includegraphics[scale=0.40]{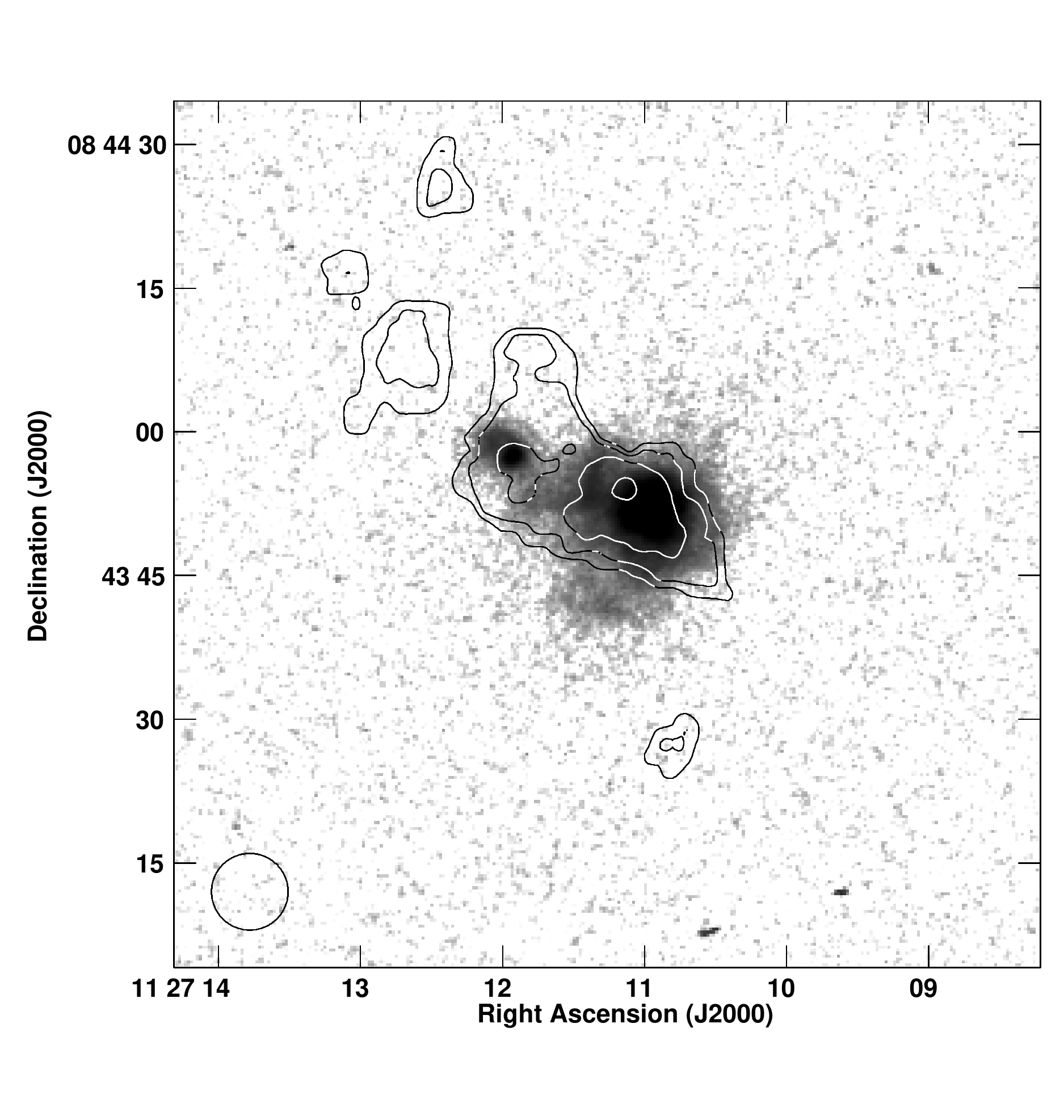}}}
\put(1.5,17.8){(c)} 
\put(8.05,11.05){\hbox{\includegraphics[scale=0.60]{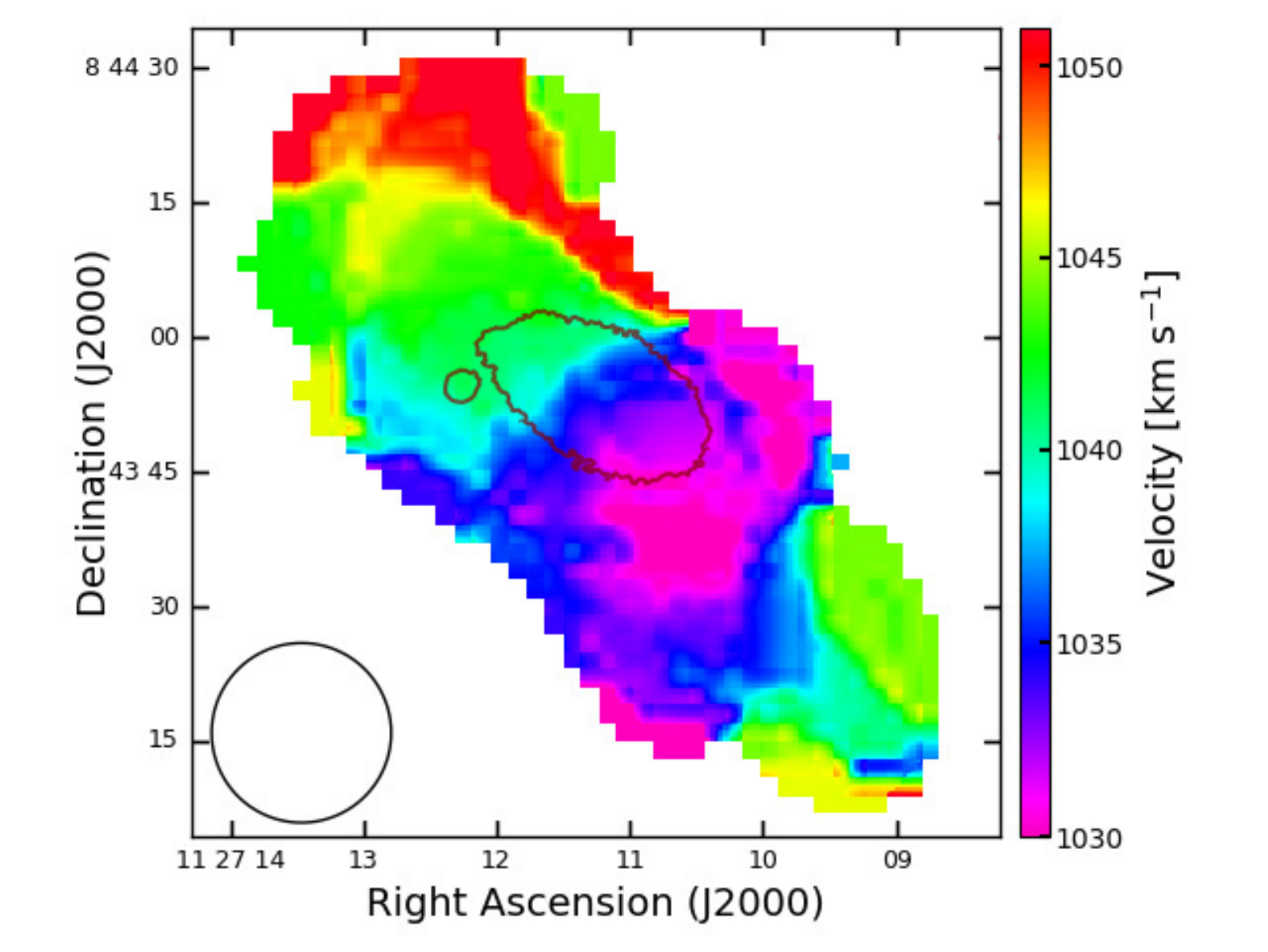}}}
\put(9.8,17.8){(d)} 
\put(0.47,3.9){\hbox{\includegraphics[scale=0.29]{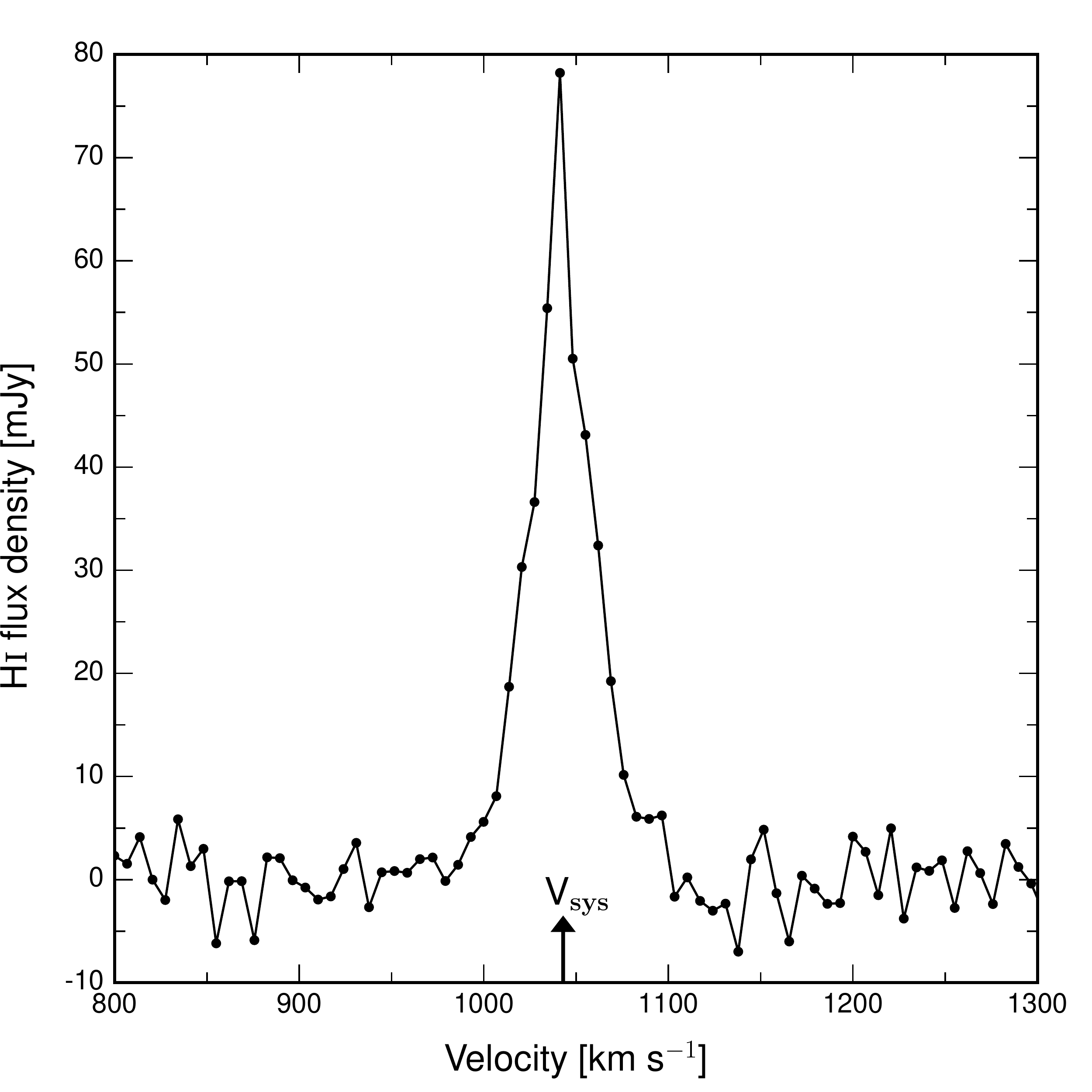}}} 
\put(1.5,10.5){(e)} 
\put(8.65,3.9){\hbox{\includegraphics[scale=0.29]{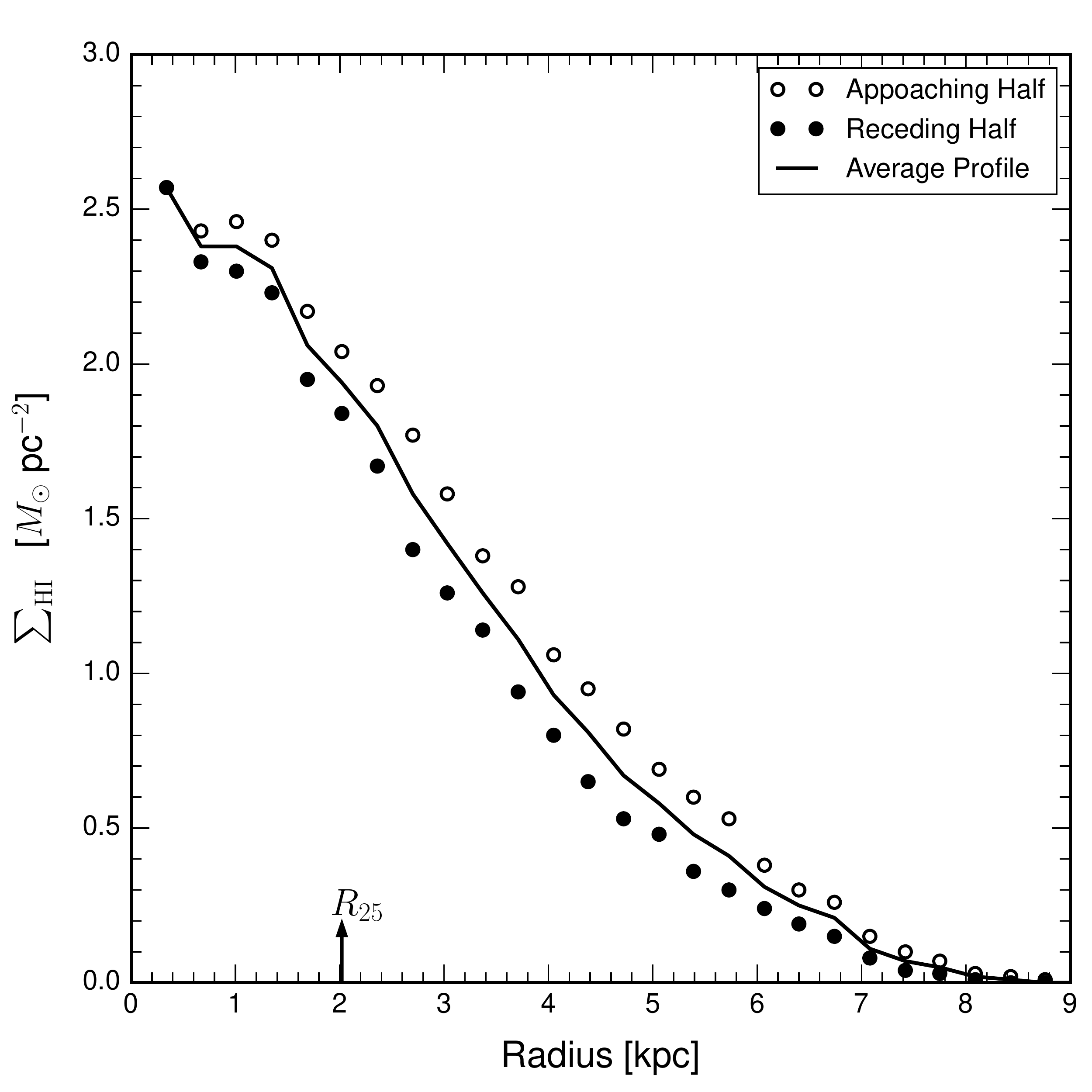}}} 
\put(9.8,10.5){(f)} 
\end{picture} 
\vspace{-4.2cm}
\caption{(a) The low resolution \HI column density contours of IC~2828 overlaid upon its grey scale optical $r$-band image. The contour levels are $1.1 \times n$, where $n=1,2,4,8,16,32$ in units of $10^{19}$~cm$^{-2}$. (b) The intermediate resolution \HI column density contours overlaid upon the grey scale optical $r$-band image. The contour levels are $11.2 \times n$ in units of $10^{19}$~cm$^{-2}$. (c) The high resolution \HI column density contours overlaid upon the grey scale \Ha line image. The contour levels are $22.6 \times n$ in units of $10^{19}$~cm$^{-2}$. (d) The intermediate resolution moment-1 map, showing the velocity field, with an overlying optical $r$-band outer contour. The circle at the bottom of each image is showing the synthesized beam. The average FWHM seeing during the optical observation was $\sim 1''.2$. (e) The global \HI profile obtained using the low resolution \HI images. The arrow at the abscissa shows the systemic \HI velocity. (f) The \HI mass surface density profile obtained using the low resolution \HI map. The arrow at the abscissa shows the $B$-band optical disk radius.}
\label{WR233}
\end{figure}

\begin{figure*}
\begin{center}
\includegraphics[angle=0,width=1\linewidth]{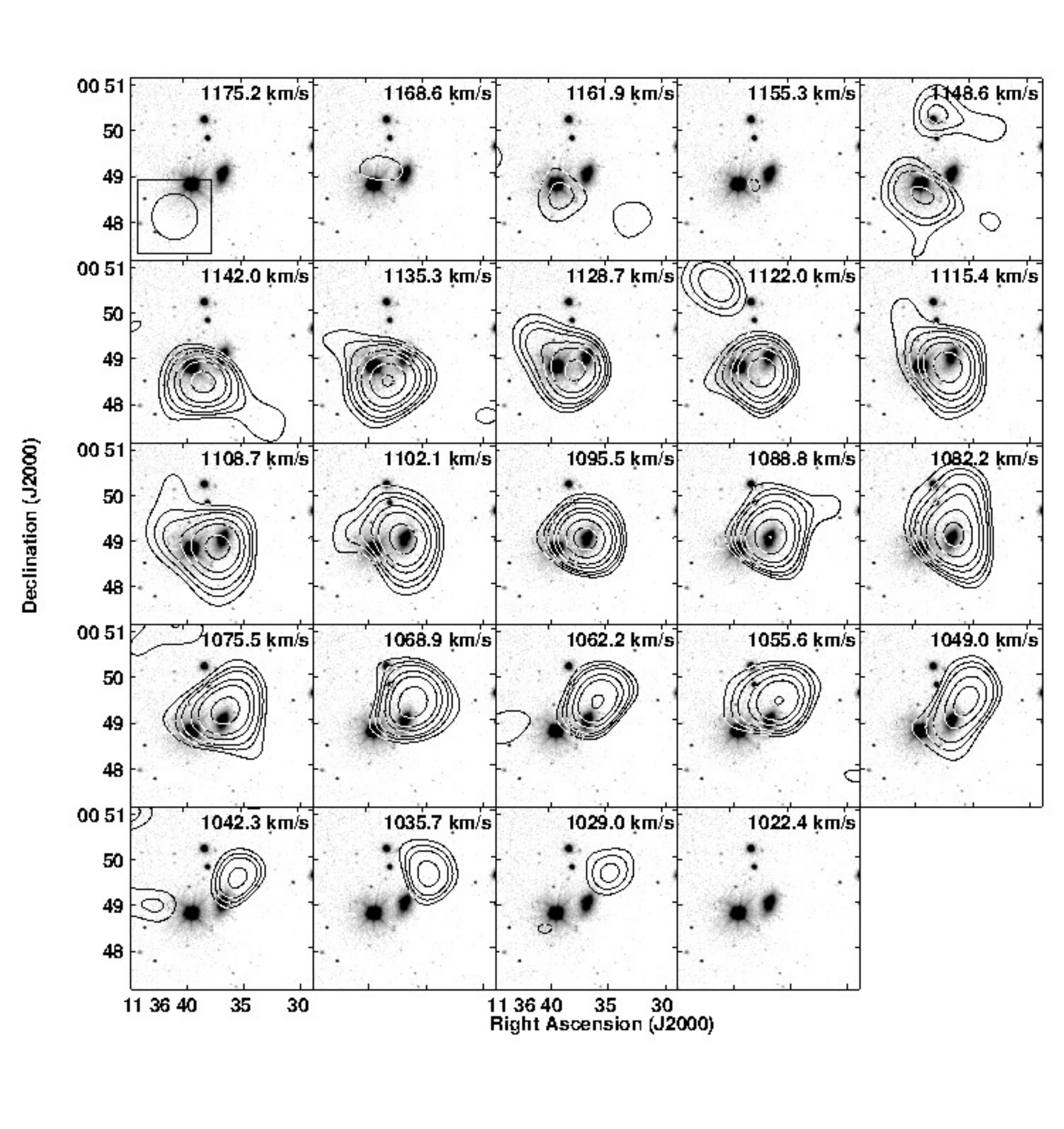}
\end{center}
\caption{The \HI contours from the low resolution channel images overlaid upon the grey scale optical $r$-band image of UM~439. The contours representing \HI emission flux are drawn at $2.4\sigma\times n$~mJy/Beam; n=1,1.5,2,3,4,6.}
\end{figure*}
\begin{figure}
\setlength{\unitlength}{1cm}
\begin{picture}(12,25) 
\put(0,18.0){\hbox{\includegraphics[scale=0.39]{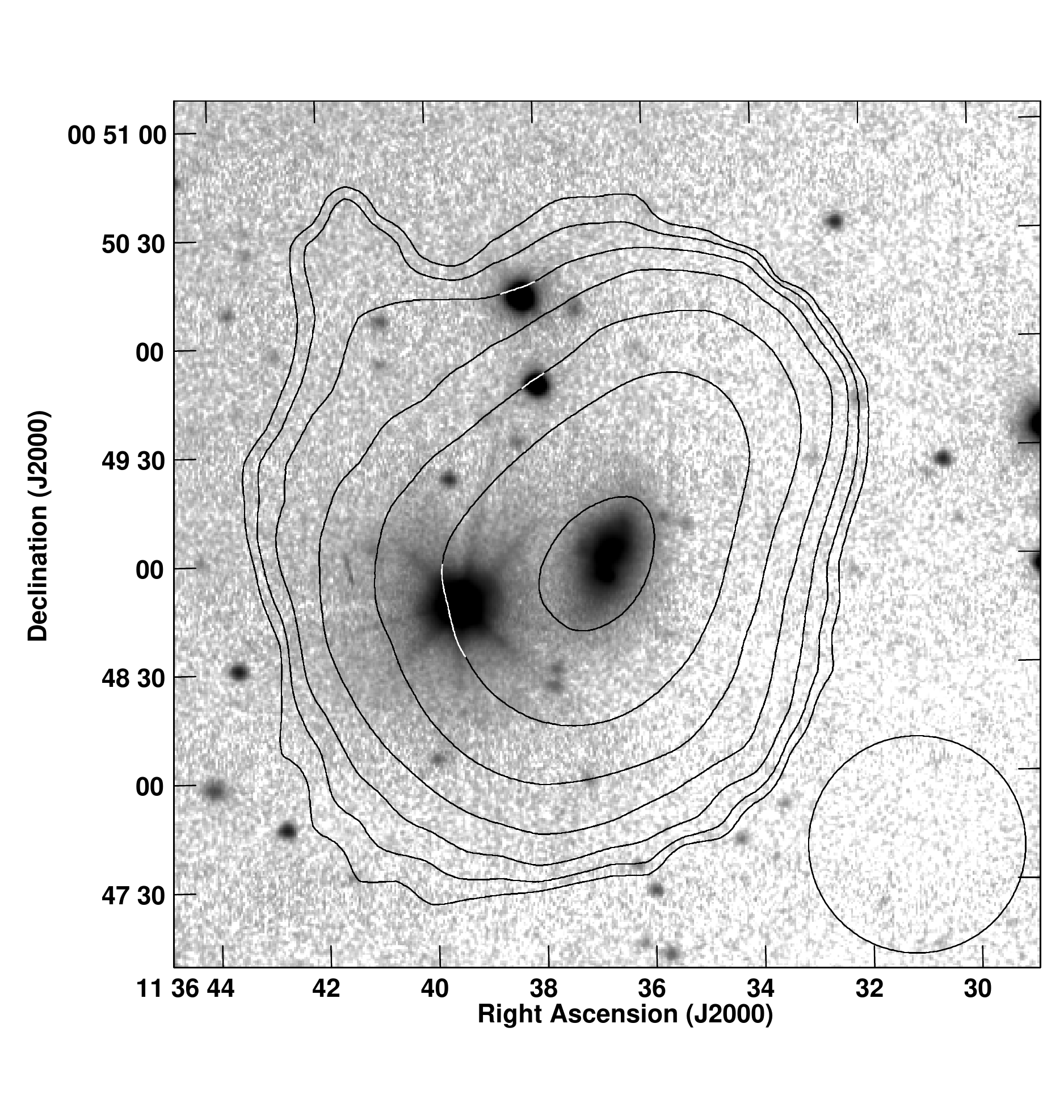}}} 
\put(1.5,24.7){(a)} 
\put(8.3,18.0){\hbox{\includegraphics[scale=0.39]{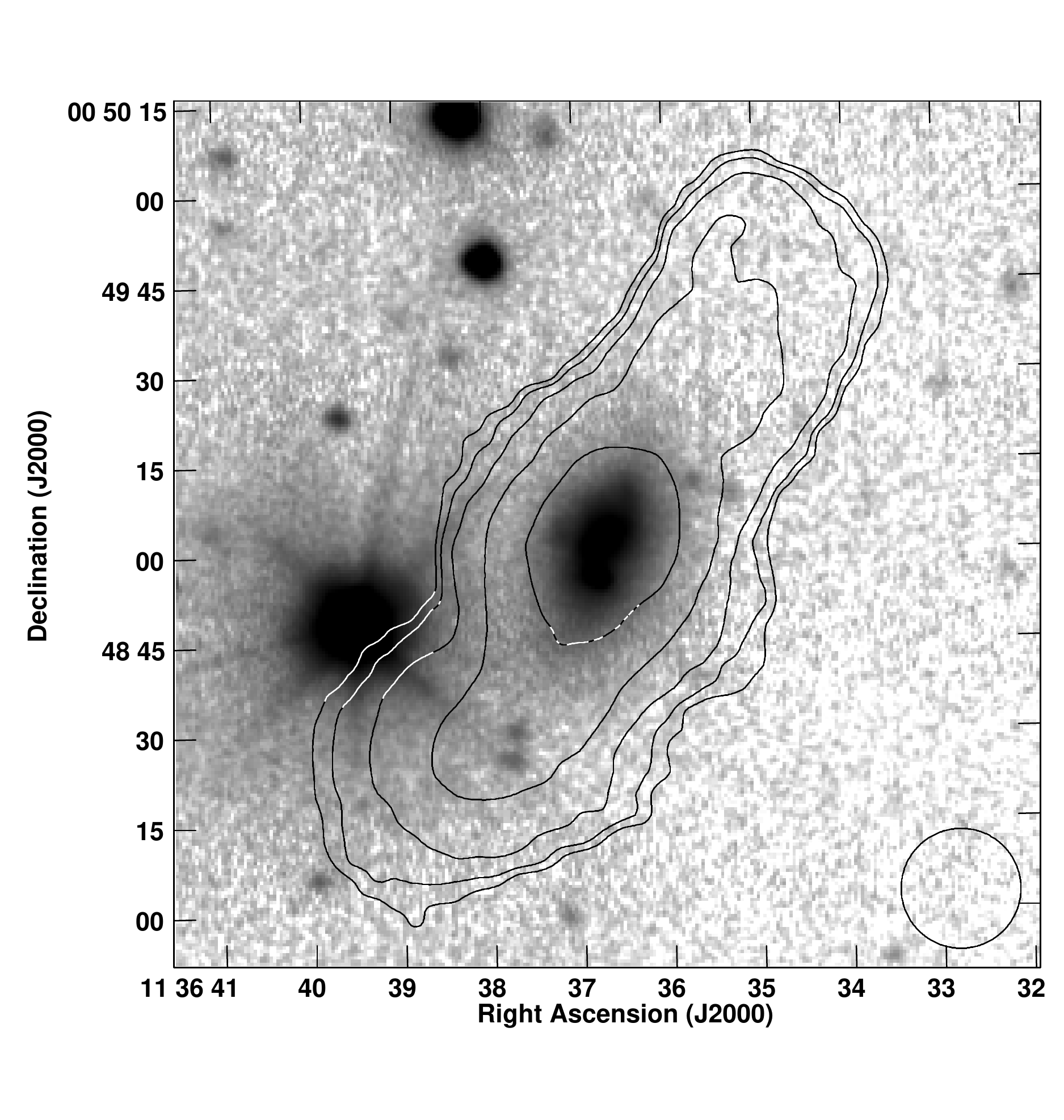}}}
\put(9.8,24.7){(b)} 
\put(0,11.0){\hbox{\includegraphics[scale=0.39]{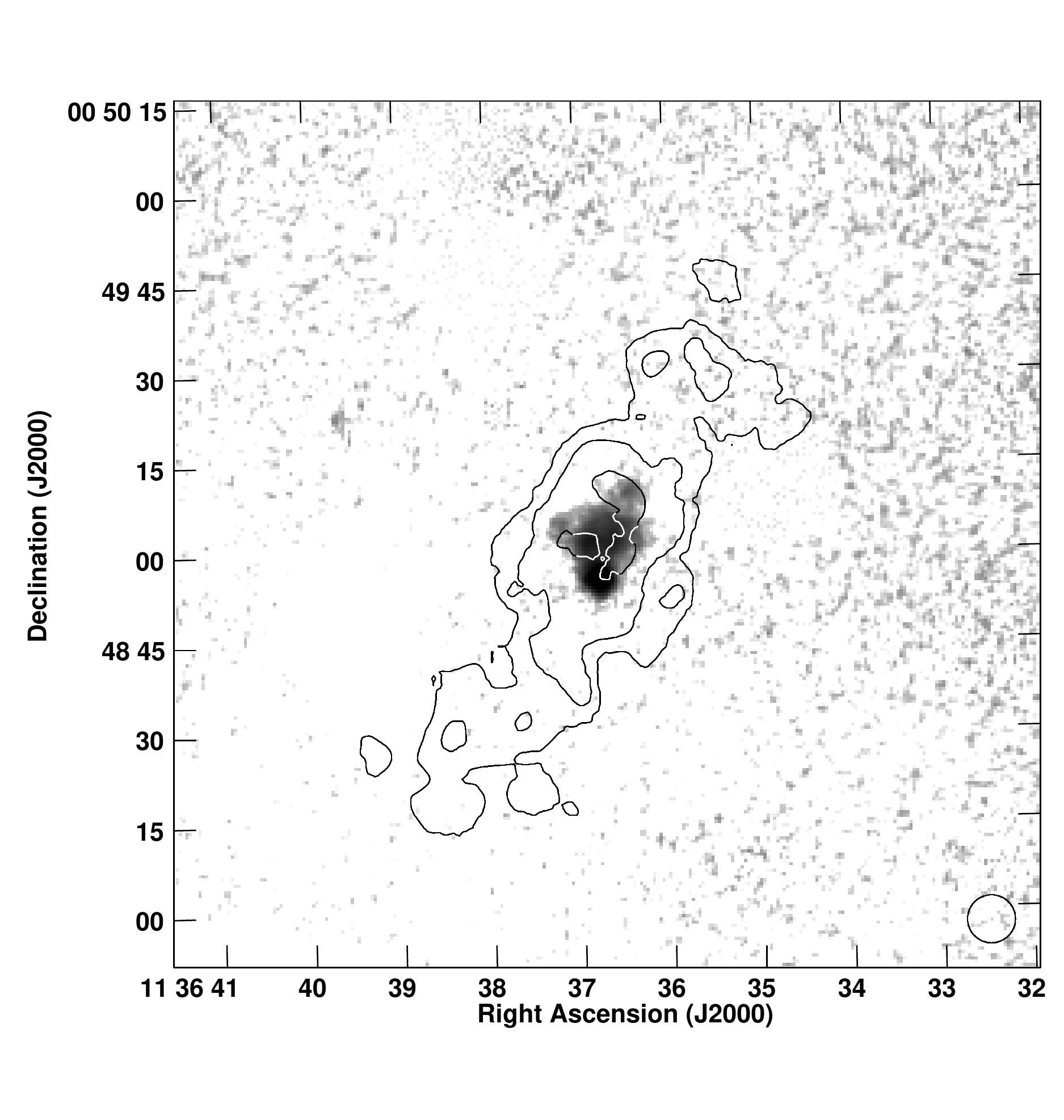}}}
\put(1.5,17.7){(c)} 
\put(7.95,11.2){\hbox{\includegraphics[scale=0.587]{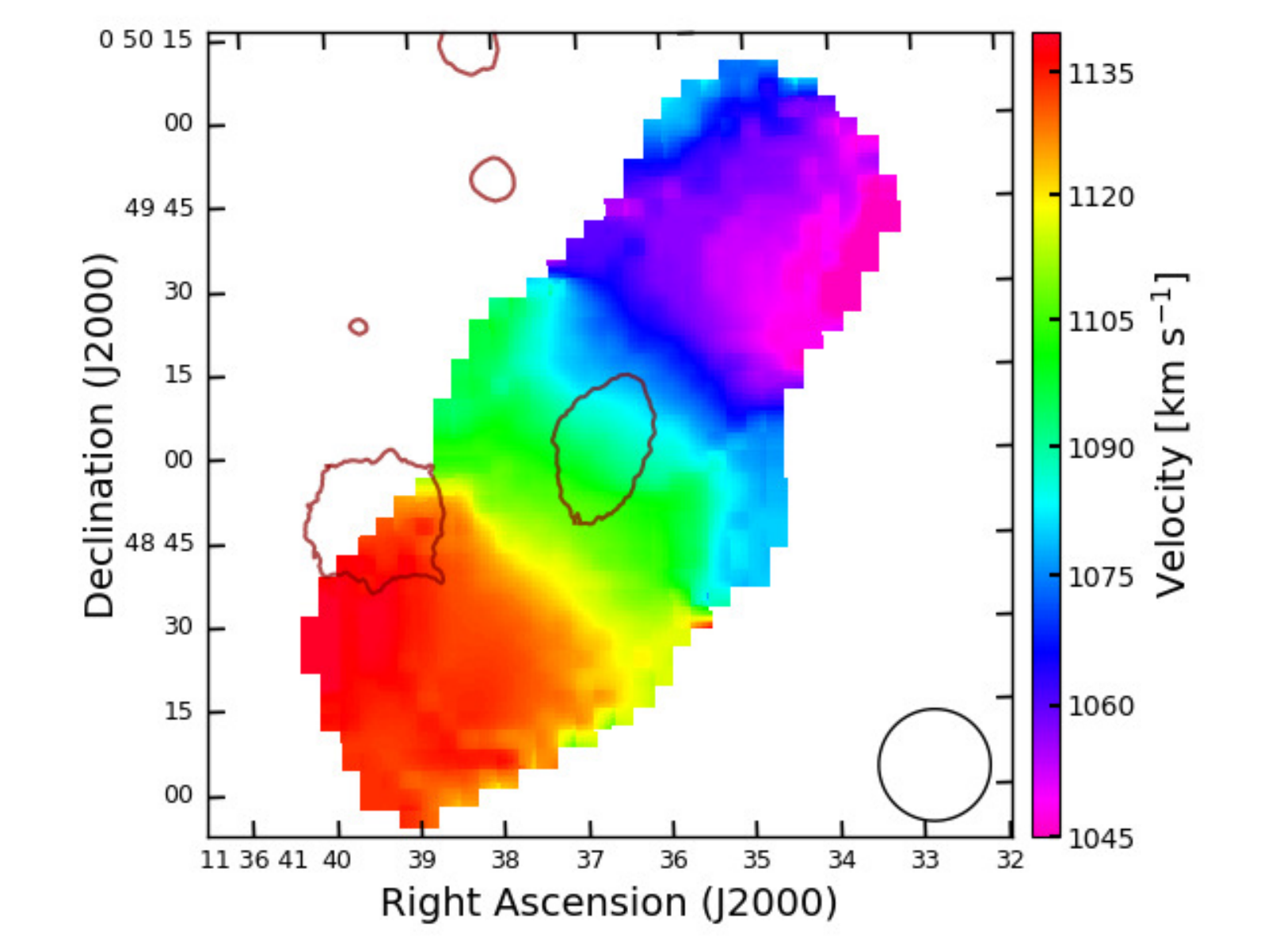}}}
\put(9.8,17.7){(d)} 
\put(0.5,4.2){\hbox{\includegraphics[scale=0.28]{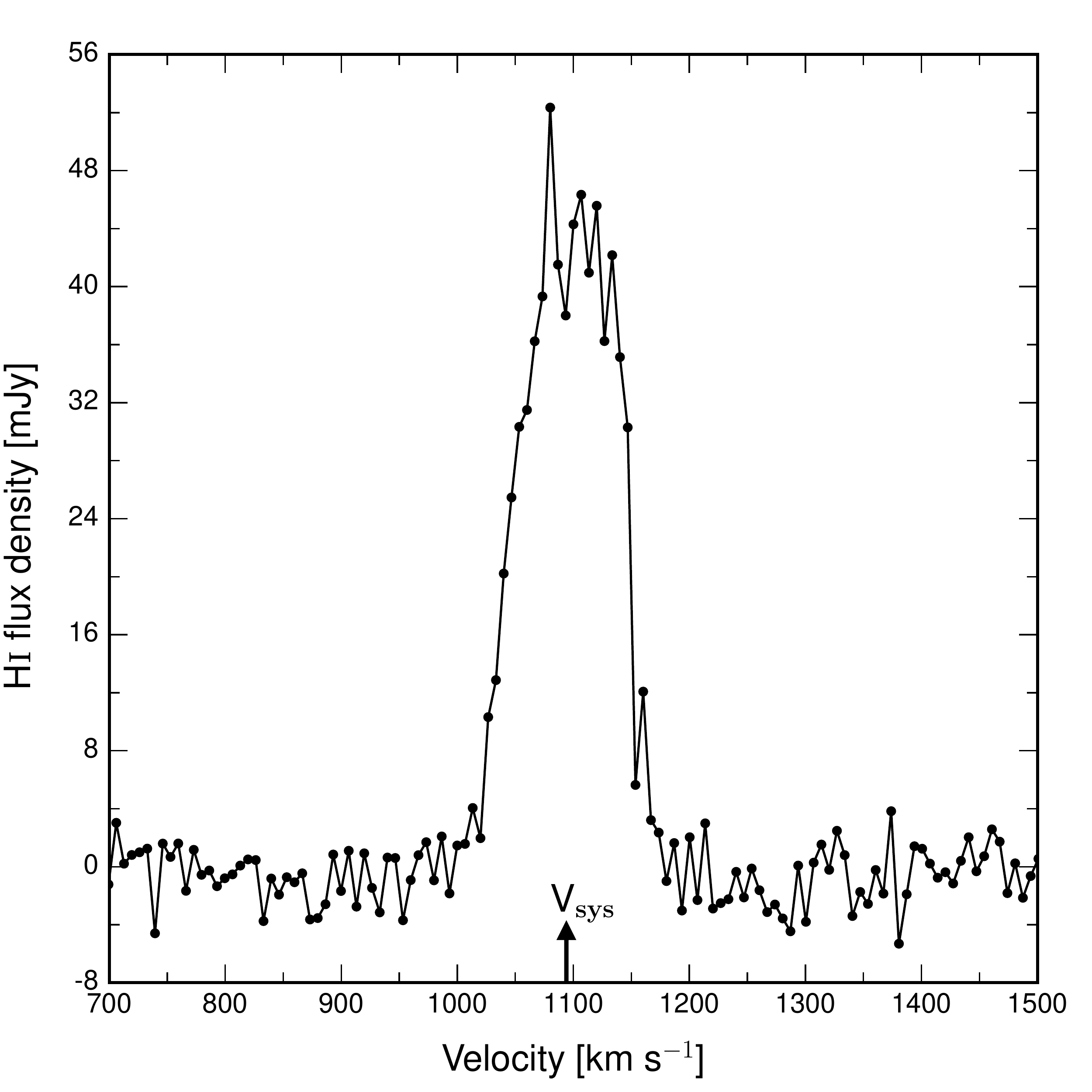}}} 
\put(1.5,10.6){(e)} 
\put(8.65,4.2){\hbox{\includegraphics[scale=0.28]{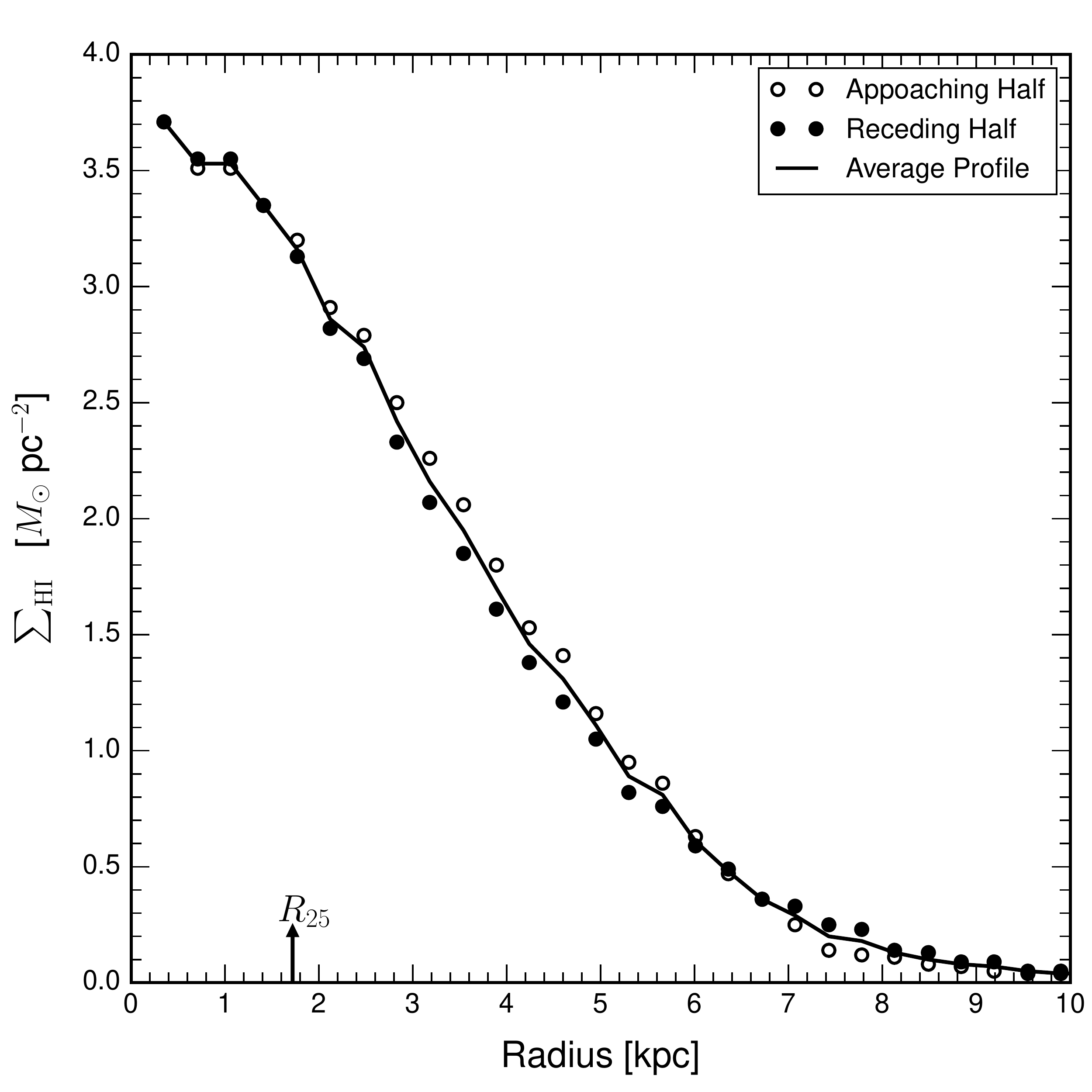}}} 
\put(9.8,10.6){(f)} 
\end{picture} 
\vspace{-4.3cm}
\caption{(a) The low resolution \HI column density contours of UM~439 overlaid upon its grey scale optical $r$-band image. The contour levels are $0.9 \times n$, where $n=1,2,4,8,16,32$ in units of $10^{19}$~cm$^{-2}$. (b) The intermediate resolution \HI column density contours overlaid upon the grey scale optical $r$-band image. The contour levels are $10.3 \times n$ in units of $10^{19}$~cm$^{-2}$. (c) The high resolution \HI column density contours overlaid upon the grey scale \Ha line image. The contour levels are $122.0 \times n$ in units of $10^{19}$~cm$^{-2}$. (d) The intermediate resolution moment-1 map, showing the velocity field, with an overlying optical $r$-band outer contour. The circle at the bottom of each image is showing the synthesized beam. The average FWHM seeing during the optical observation was $\sim 2''.1$. (e) The global \HI profile obtained using the low resolution \HI images. The arrow at the abscissa shows the systemic \HI velocity. (f) The \HI mass surface density profile obtained using the low resolution \HI map. The arrow at the abscissa shows the $B$-band optical disk radius.}
\label{UM439}
\end{figure}

\begin{figure*}
\begin{center}
\includegraphics[angle=0,width=1\linewidth]{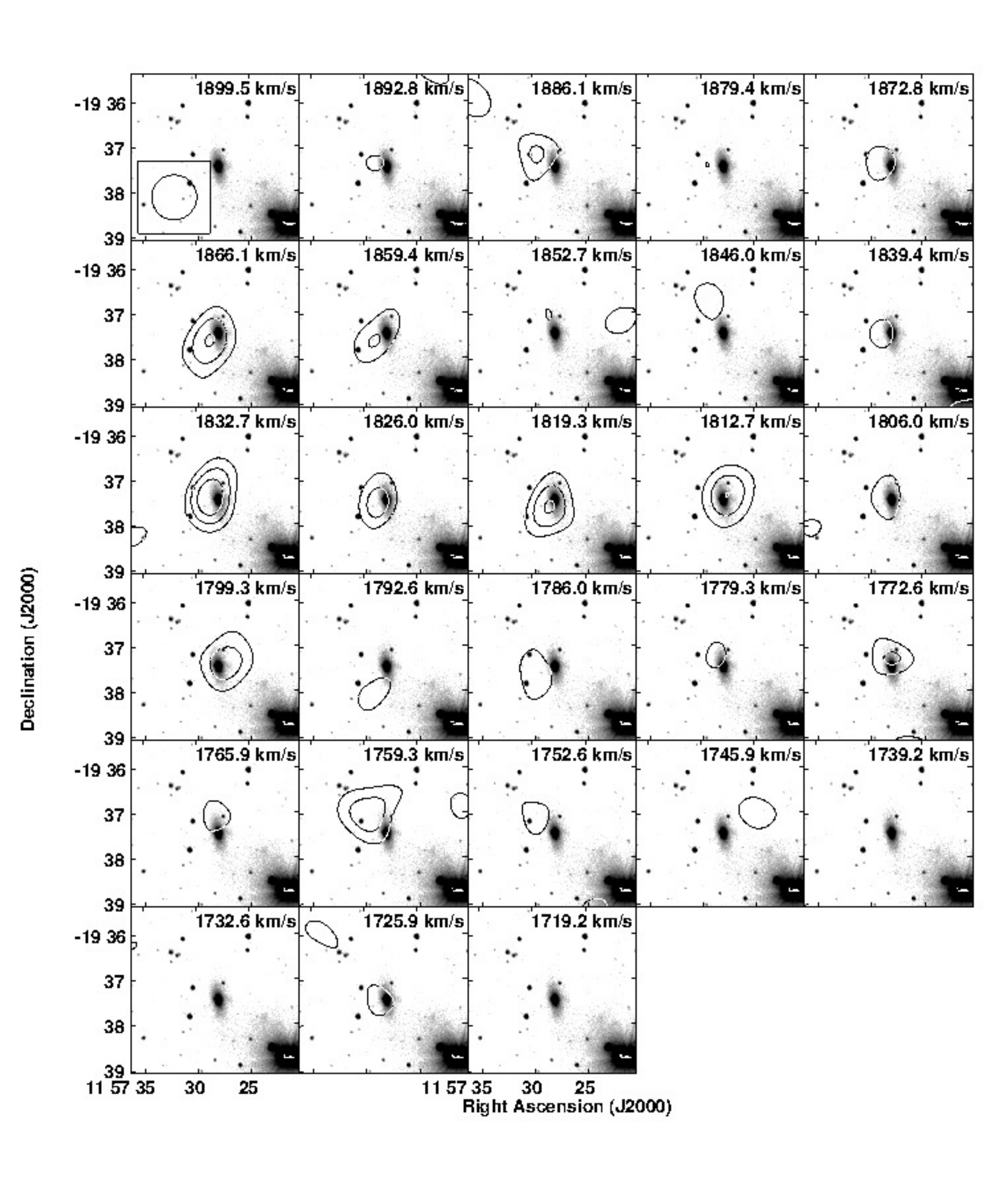}
\end{center}
\caption{The \HI contours from the low resolution channel images overlaid upon the grey scale optical $r$-band image of I~SZ~59. The contours representing \HI emission flux are drawn at $2.5\sigma\times n$~mJy/Beam; n=1,1.5,2,3,4,6.}
\end{figure*}
\begin{figure}
\setlength{\unitlength}{1cm}
\begin{picture}(12,25) 
\put(0,17.98){\hbox{\includegraphics[scale=0.40]{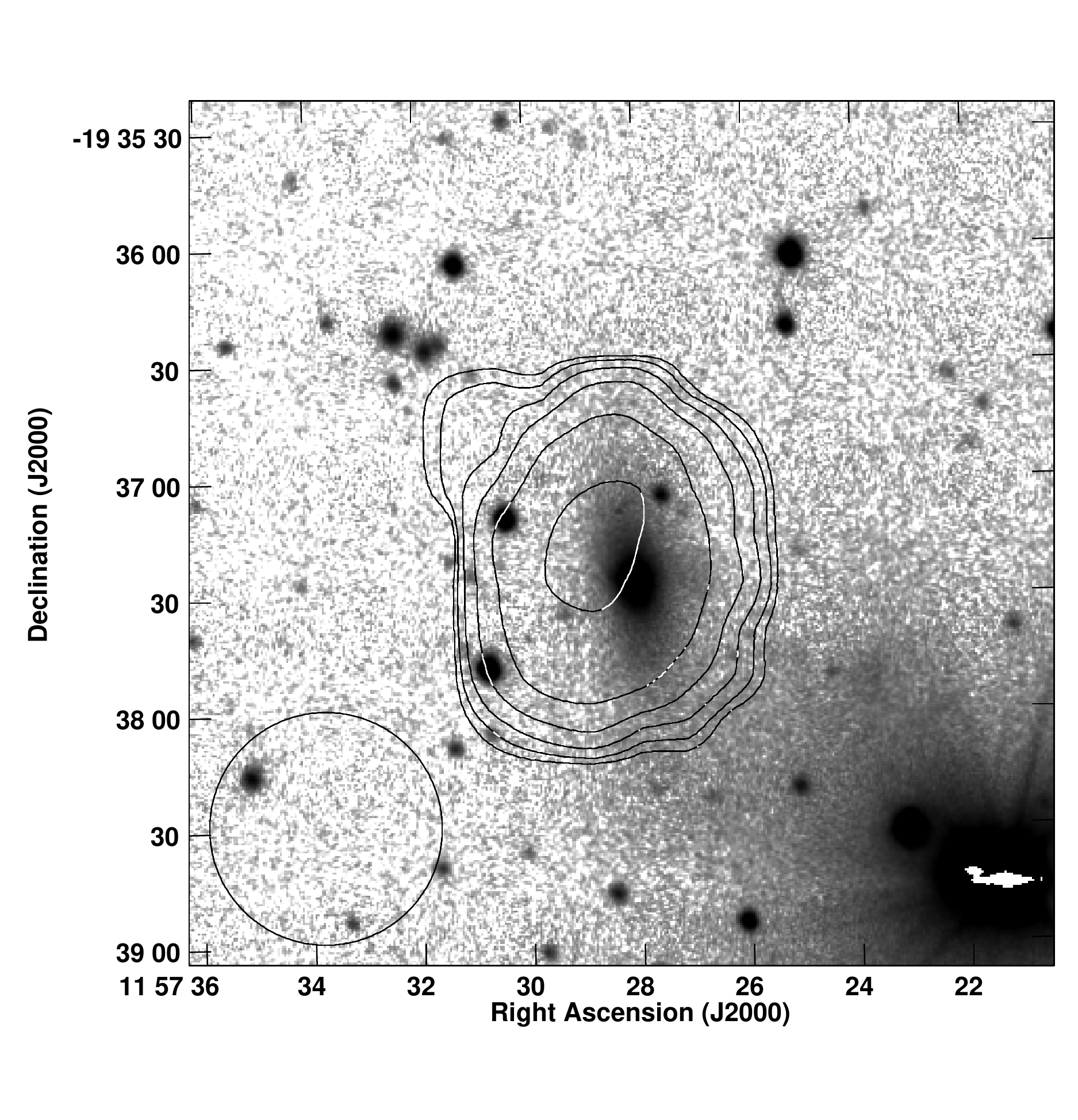}}} 
\put(1.5,24.75){(a)} 
\put(8.3,17.98){\hbox{\includegraphics[scale=0.40]{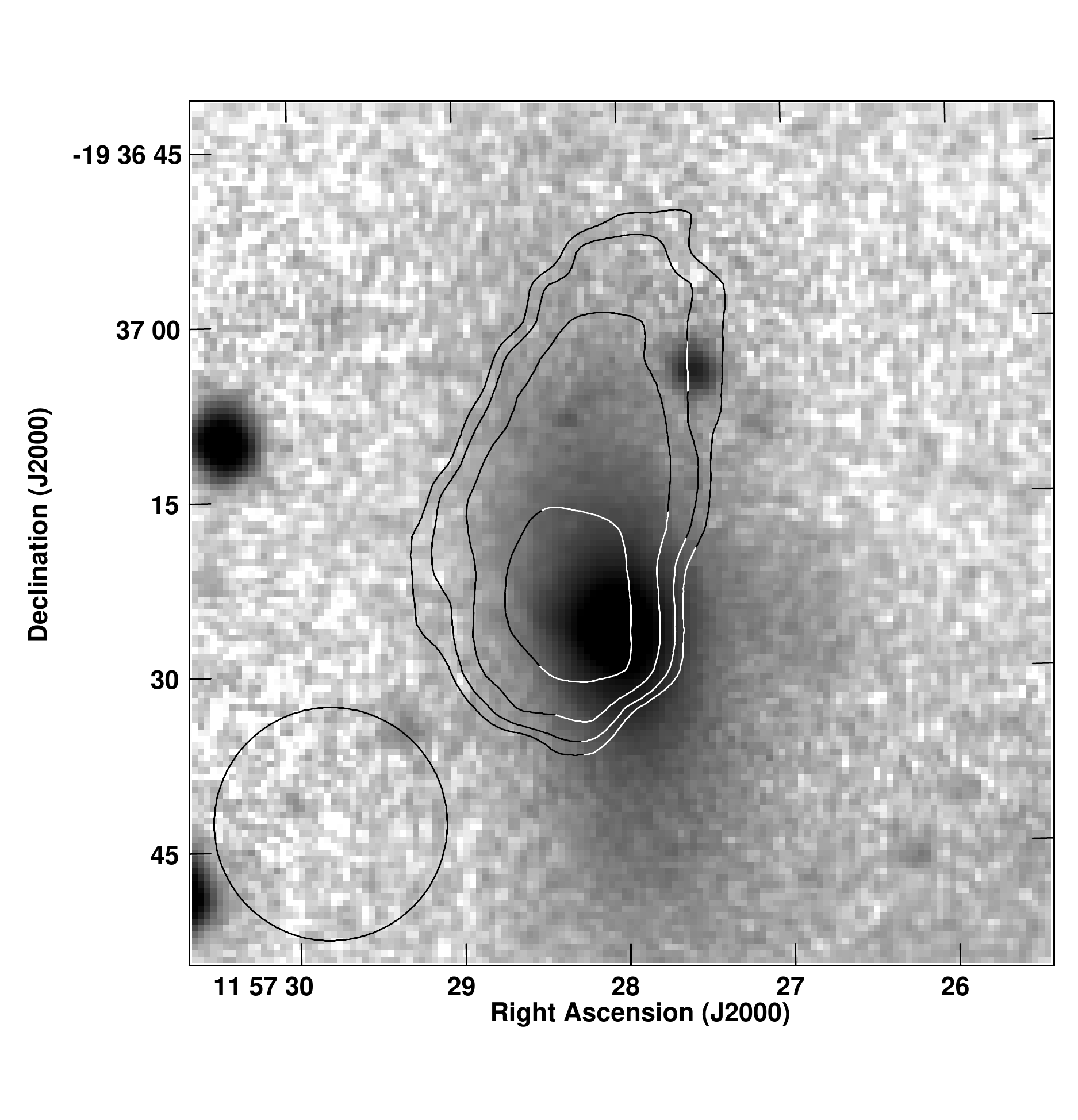}}}
\put(9.8,24.75){(b)} 
\put(0,10.95){\hbox{\includegraphics[scale=0.40]{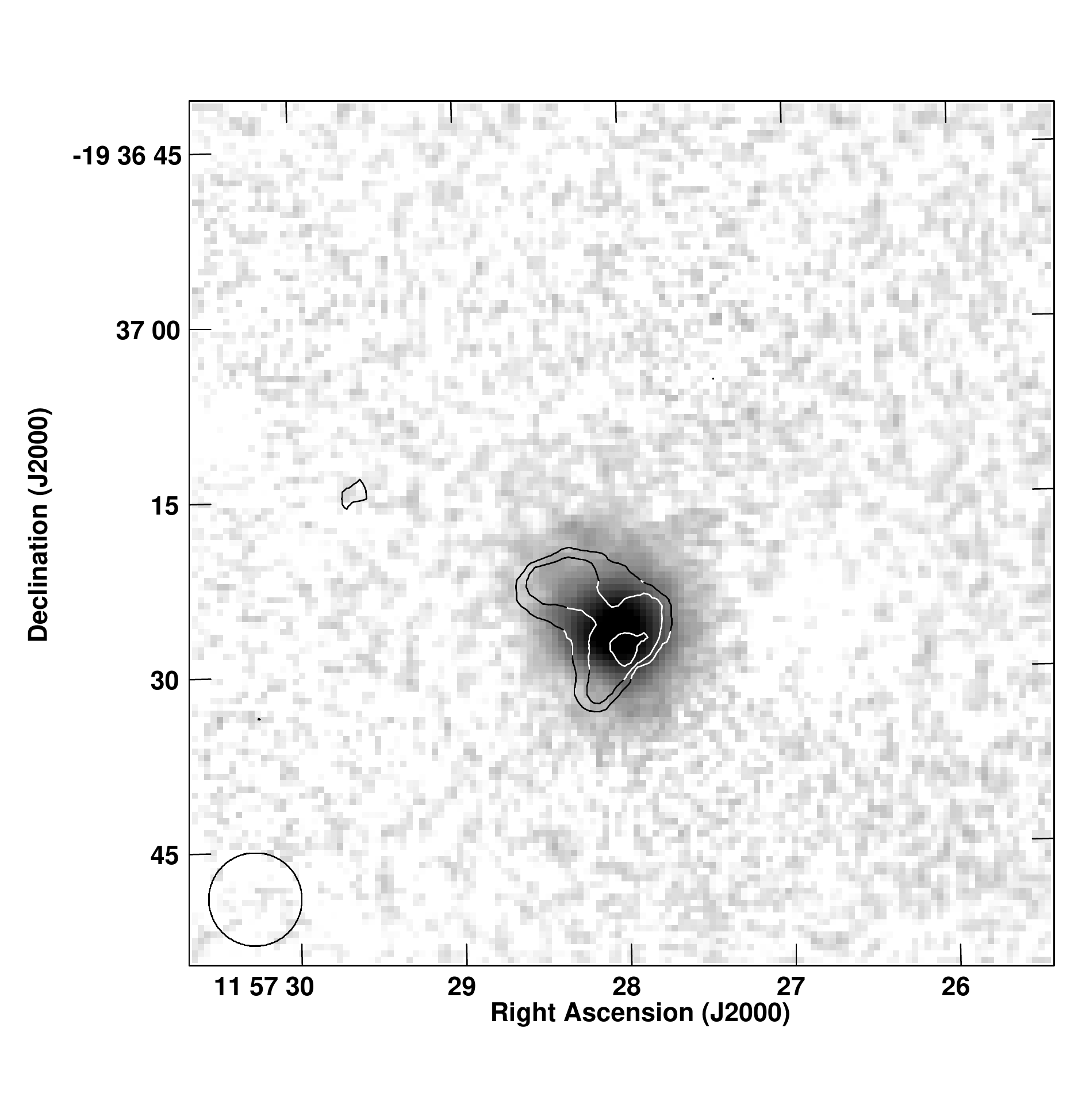}}}
\put(1.5,17.7){(c)} 
\put(8.1,11.18){\hbox{\includegraphics[scale=0.59]{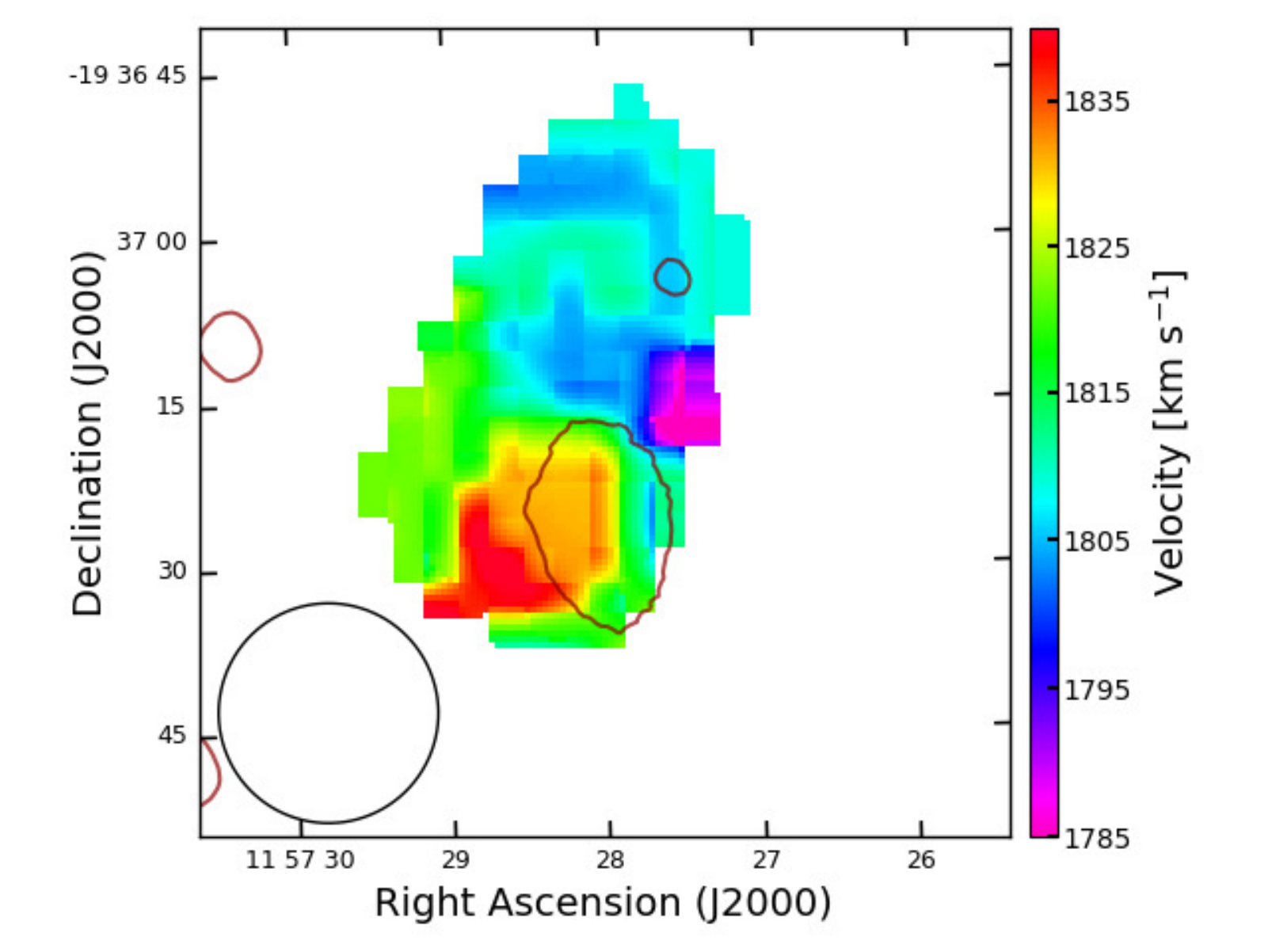}}}
\put(9.8,17.7){(d)} 
\put(0.6,4.0){\hbox{\includegraphics[scale=0.284]{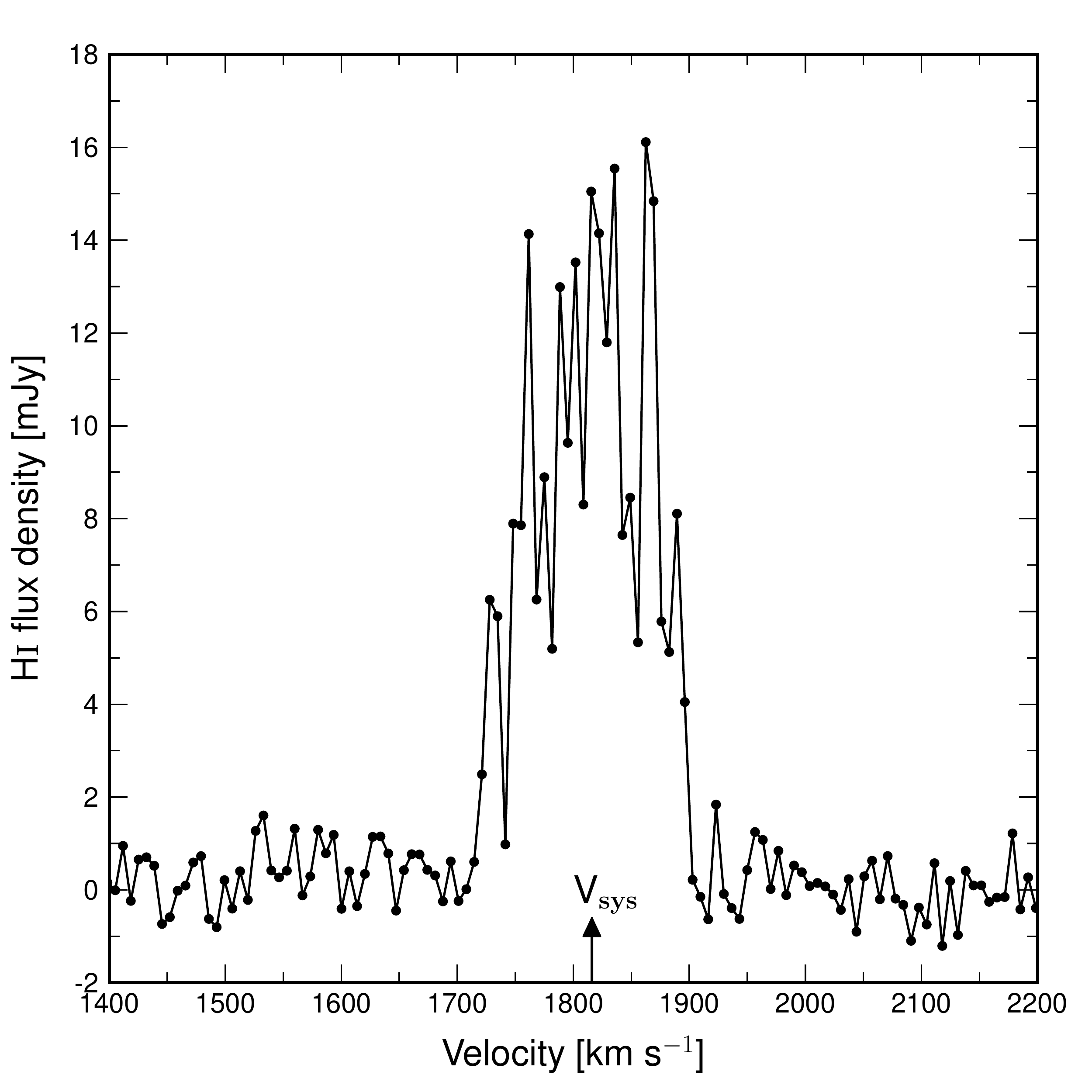}}} 
\put(1.5,10.4){(e)} 
\put(8.75,4.0){\hbox{\includegraphics[scale=0.284]{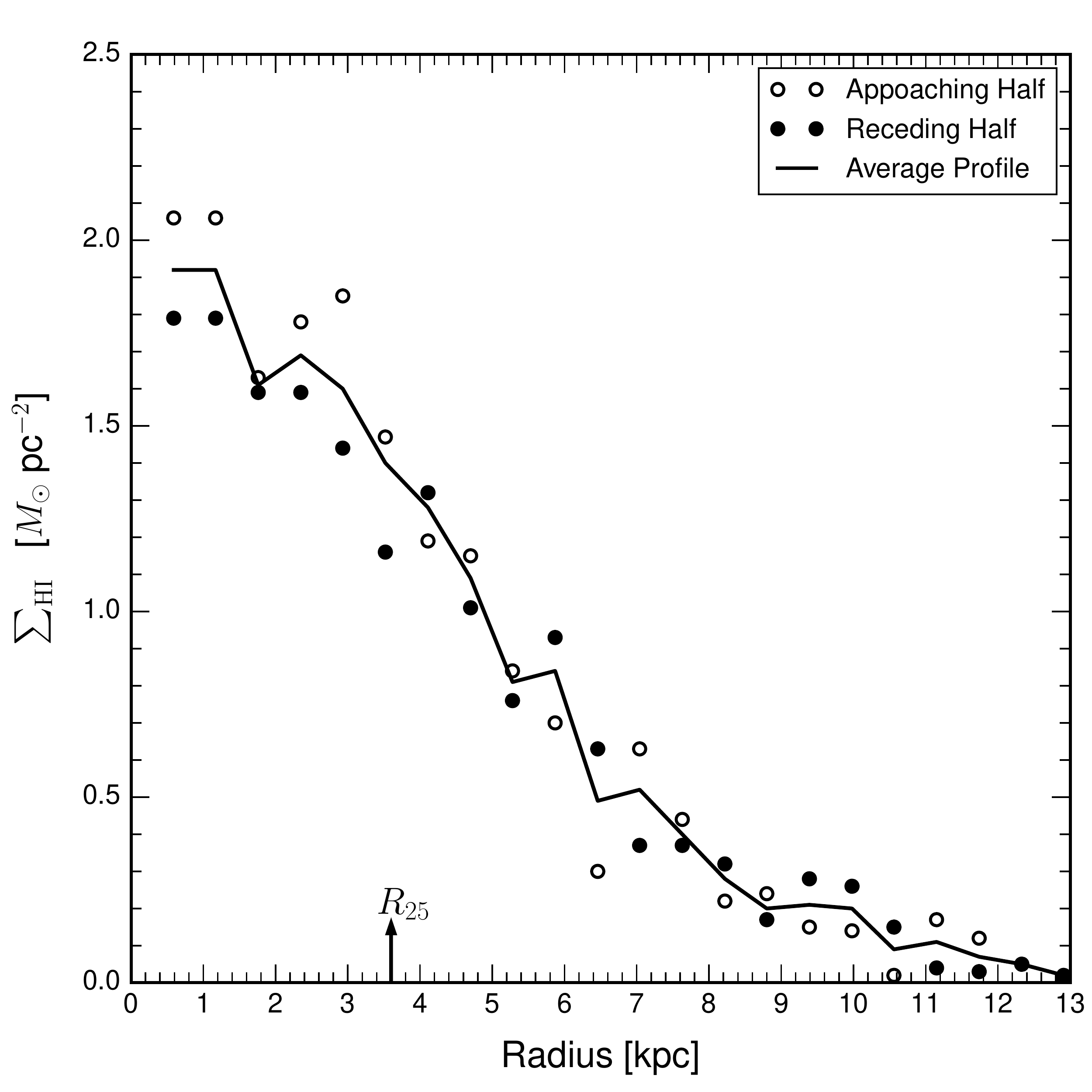}}} 
\put(9.8,10.4){(f)} 
\end{picture} 
\vspace{-4.3cm}
\caption{(a) The low resolution \HI column density contours of I~SZ~59 overlaid upon its grey scale optical $r$-band image. The contour levels are $1.1 \times n$, where $n=1,2,4,8,16,32$ in units of $10^{19}$~cm$^{-2}$. (b) The intermediate resolution \HI column density contours overlaid upon the grey scale optical $r$-band image. The contour levels are $7.0 \times n$ in units of $10^{19}$~cm$^{-2}$. (c) The high resolution \HI column density contours overlaid upon the grey scale \Ha line image. The contour levels are $50.5 \times n$ in units of $10^{19}$~cm$^{-2}$. (d) The intermediate resolution moment-1 map, showing the velocity field, with an overlying optical $r$-band outer contour. The circle at the bottom of each image is showing the synthesized beam. The average FWHM seeing during the optical observation was $\sim 2''.2$. (e) The global \HI profile obtained using the low resolution \HI images. The arrow at the abscissa shows the systemic \HI velocity. (f) The \HI mass surface density profile obtained using the low resolution \HI map. The arrow at the abscissa shows the $B$-band optical disk radius.}
\label{ISZ59}
\end{figure}

\begin{figure*}
\begin{center}
\includegraphics[angle=0,width=1\linewidth]{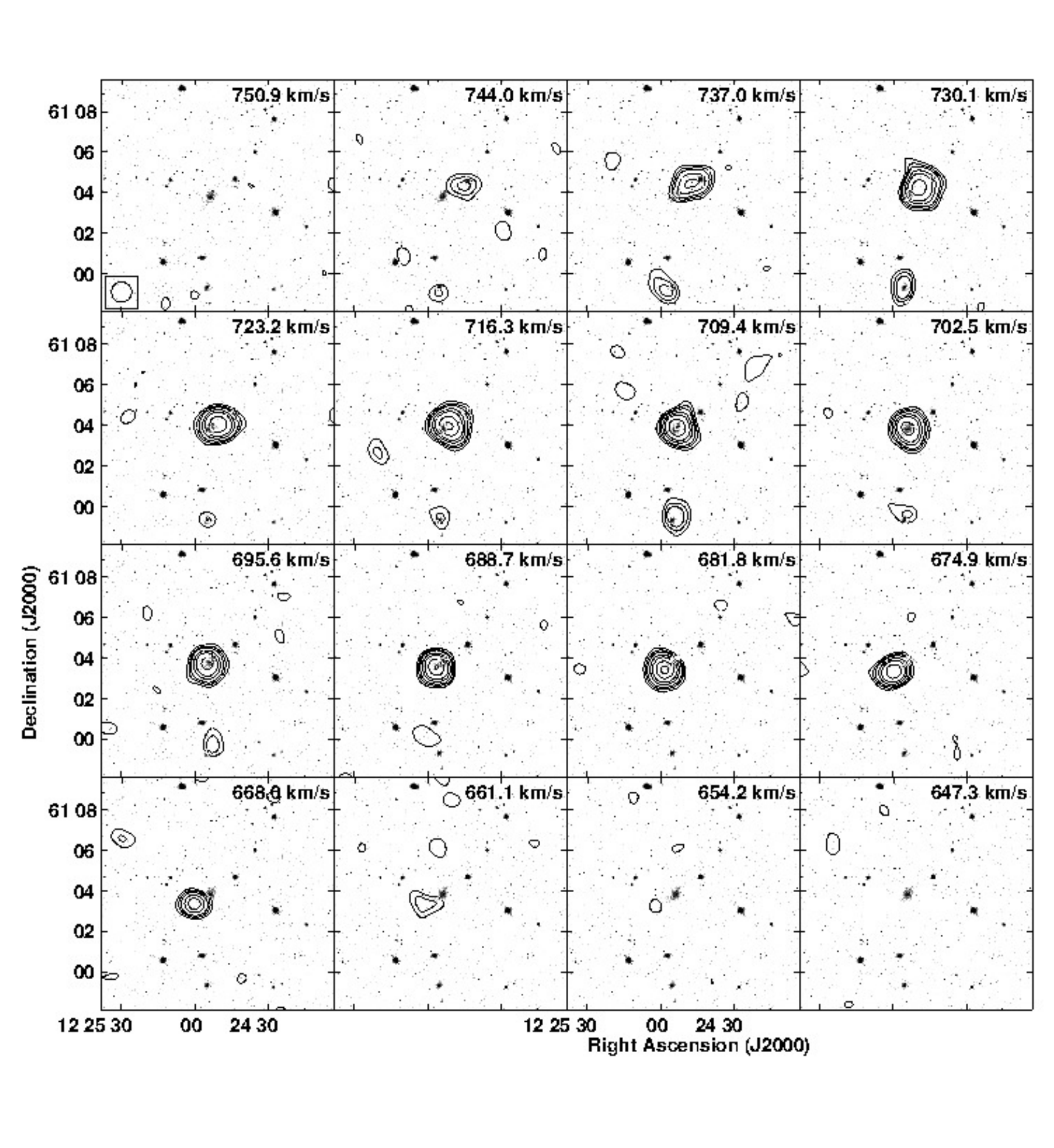}
\end{center}
\caption{The \HI contours from the low resolution channel images overlaid upon the grey scale optical $r$-band image of MCG~+10-18-044. The contours representing \HI emission flux are drawn at $2.5\sigma\times n$~mJy/Beam; n=1,1.5,2,3,4,6,8,12.}
\end{figure*}
\begin{figure}
\setlength{\unitlength}{1cm}
\begin{picture}(12,25) 
\put(0,17.95){\hbox{\includegraphics[scale=0.38]{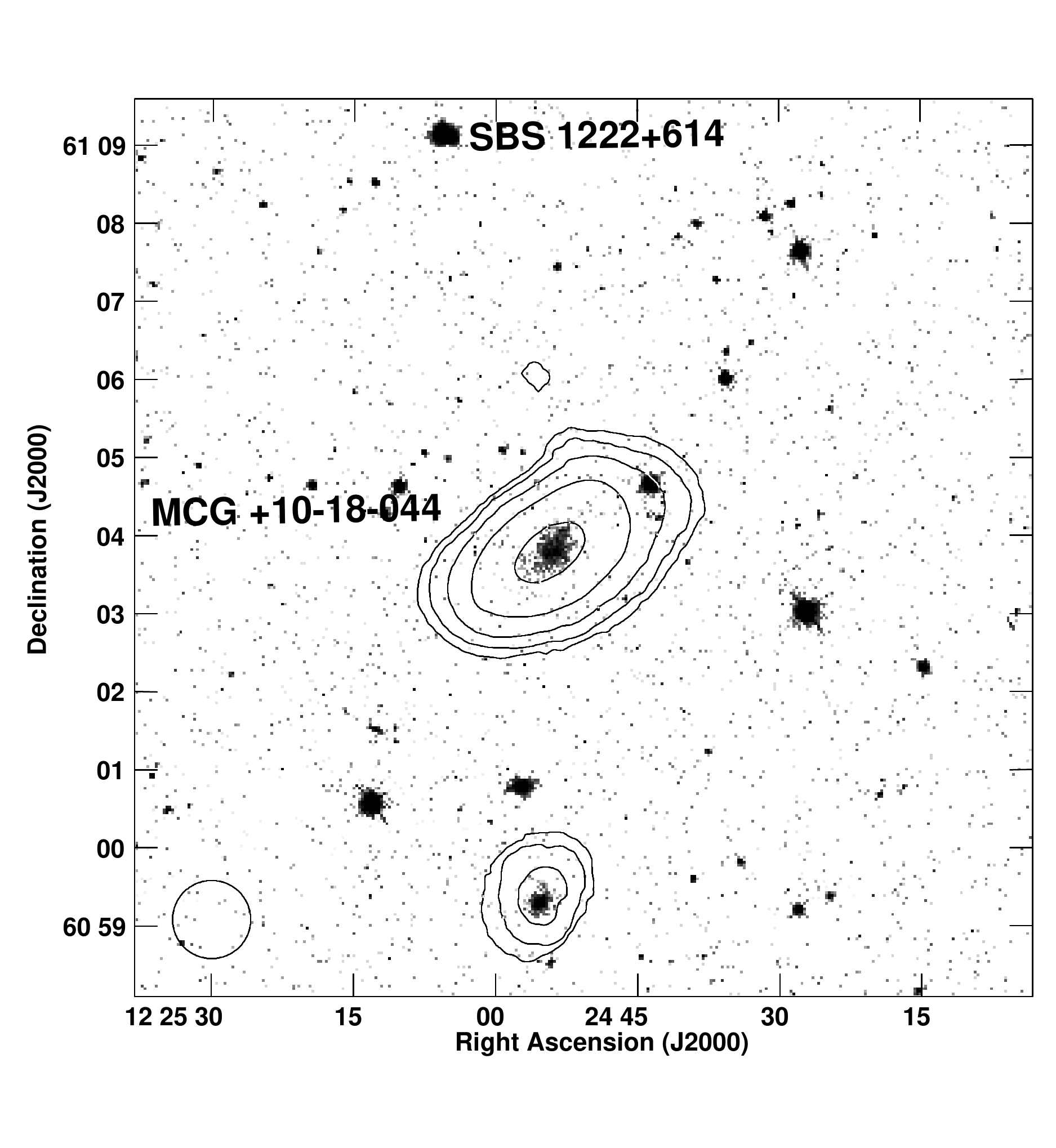}}} 
\put(1.2,24.8){(a)} 
\put(8.3,17.92){\hbox{\includegraphics[scale=0.397]{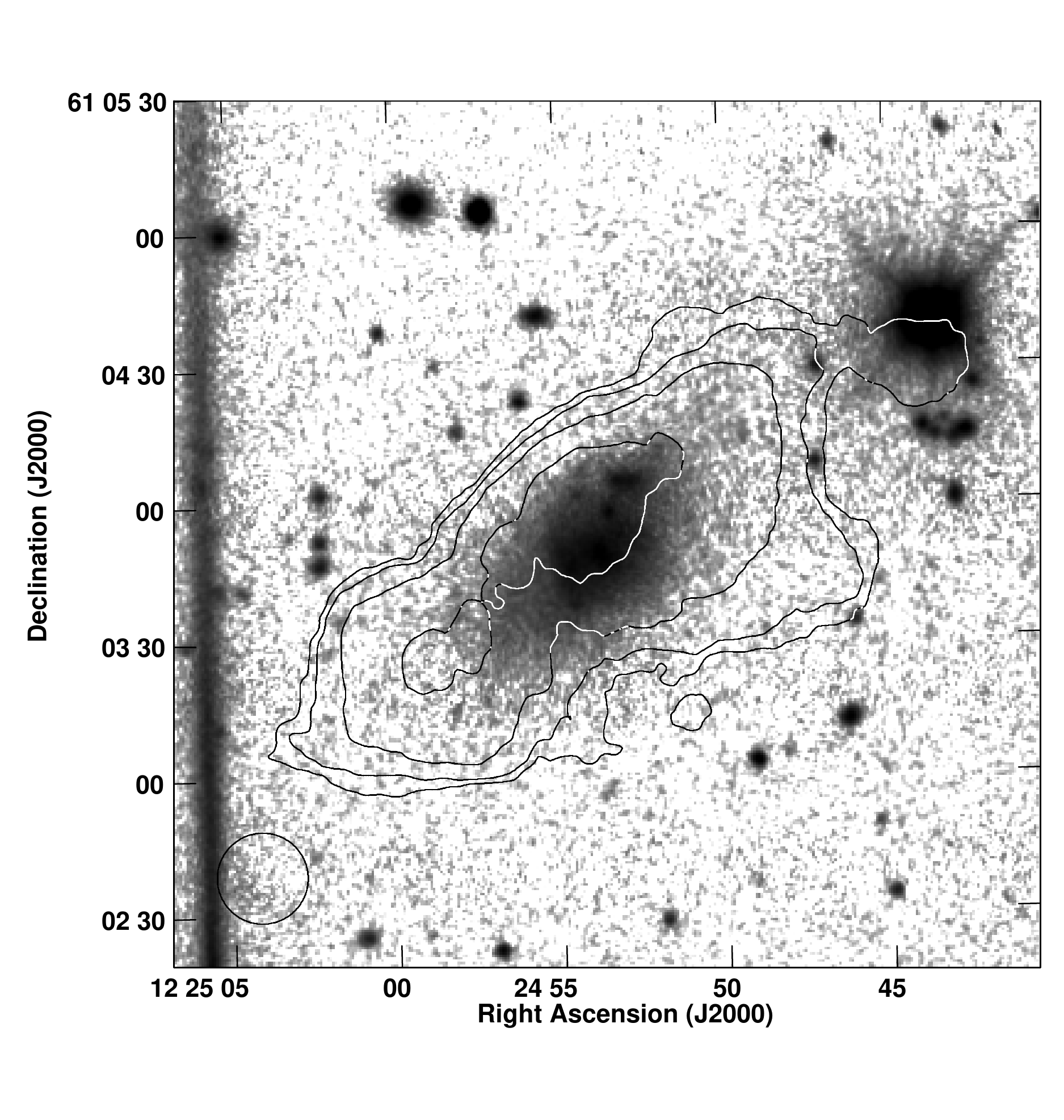}}}
\put(9.8,24.8){(b)} 
\put(-0.3,10.95){\hbox{\includegraphics[scale=0.397]{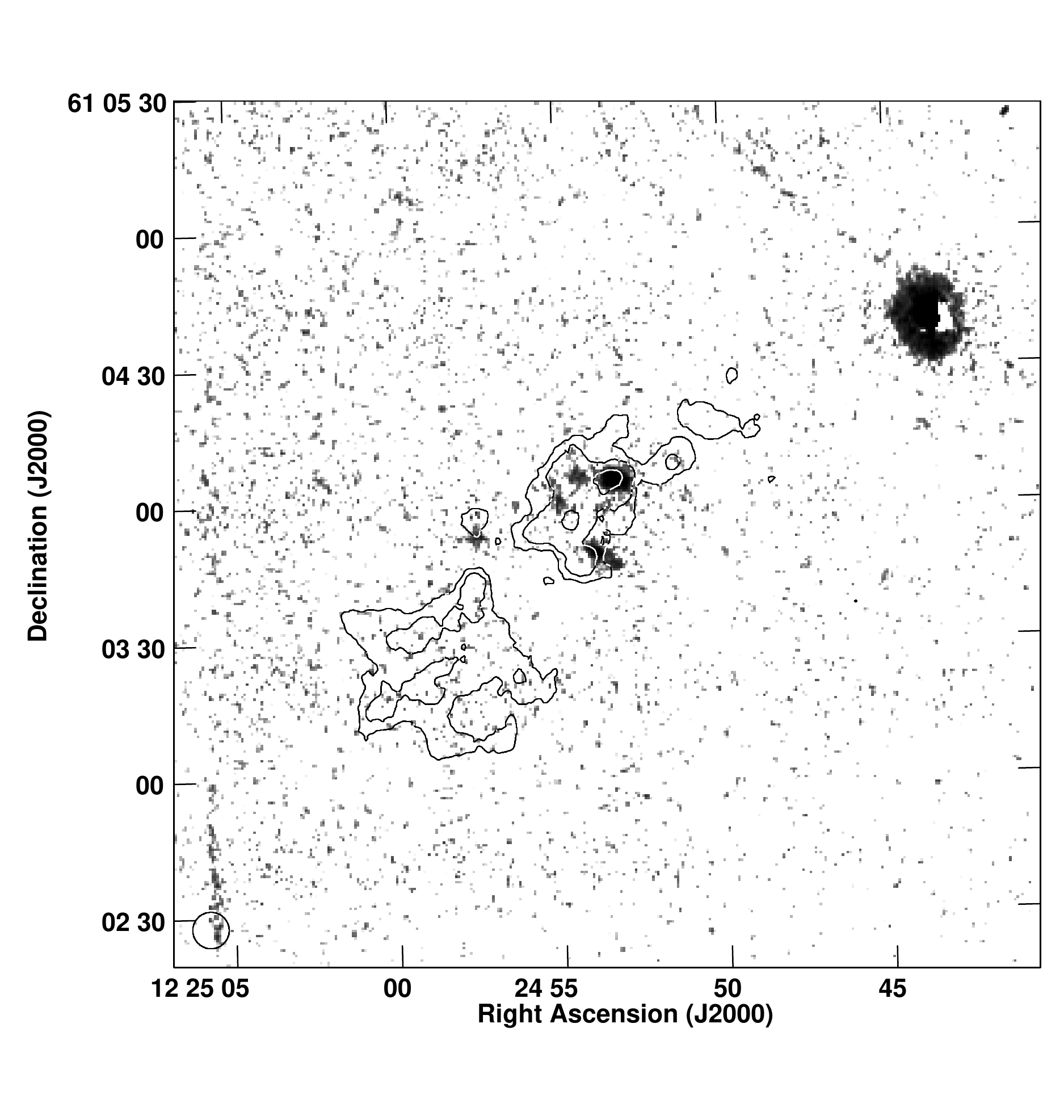}}}
\put(1.2,17.8){(c)} 
\put(7.9,11.15){\hbox{\includegraphics[scale=0.602]{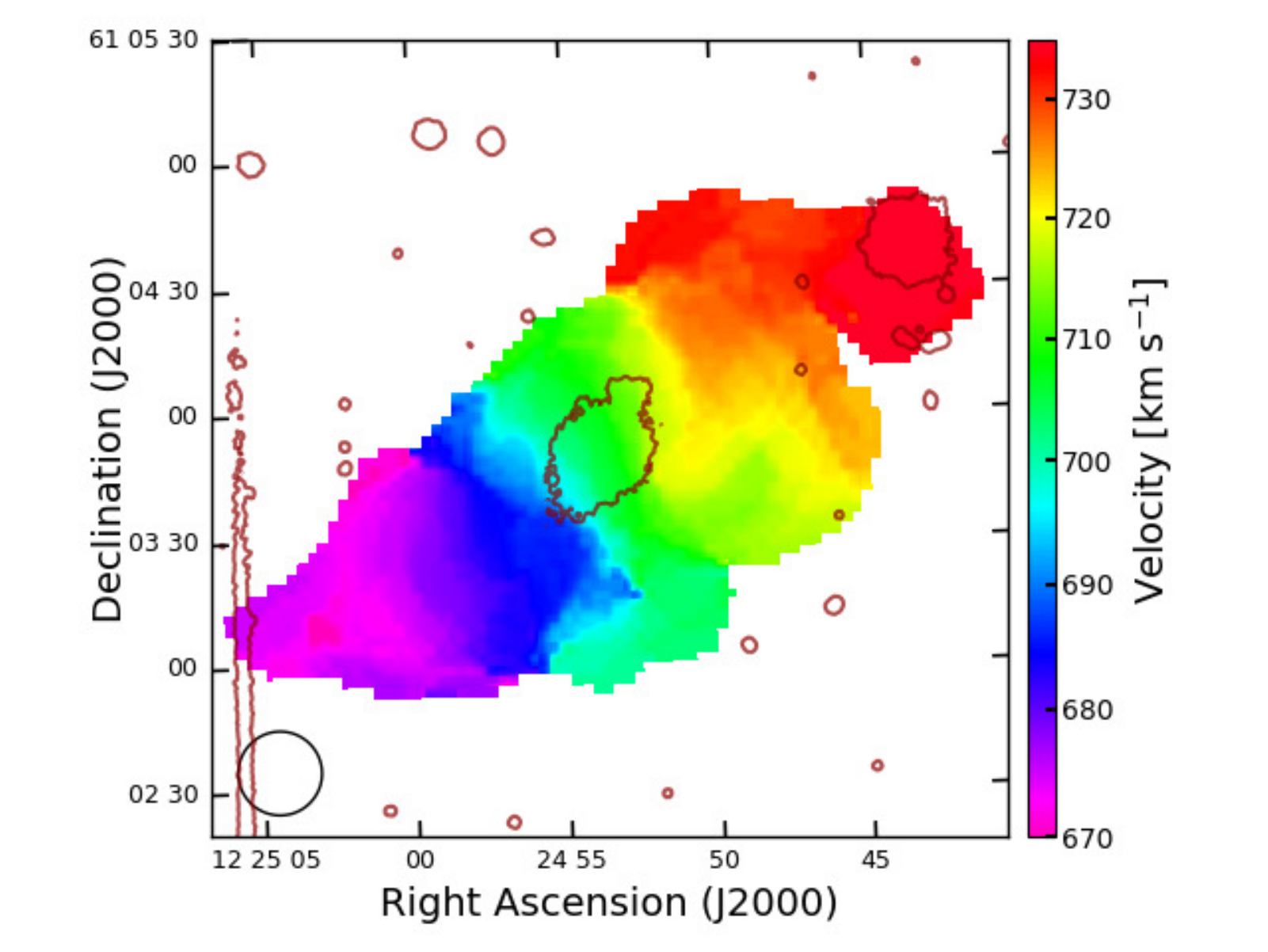}}}
\put(9.8,17.8){(d)} 
\put(0.17,4.08){\hbox{\includegraphics[scale=0.287]{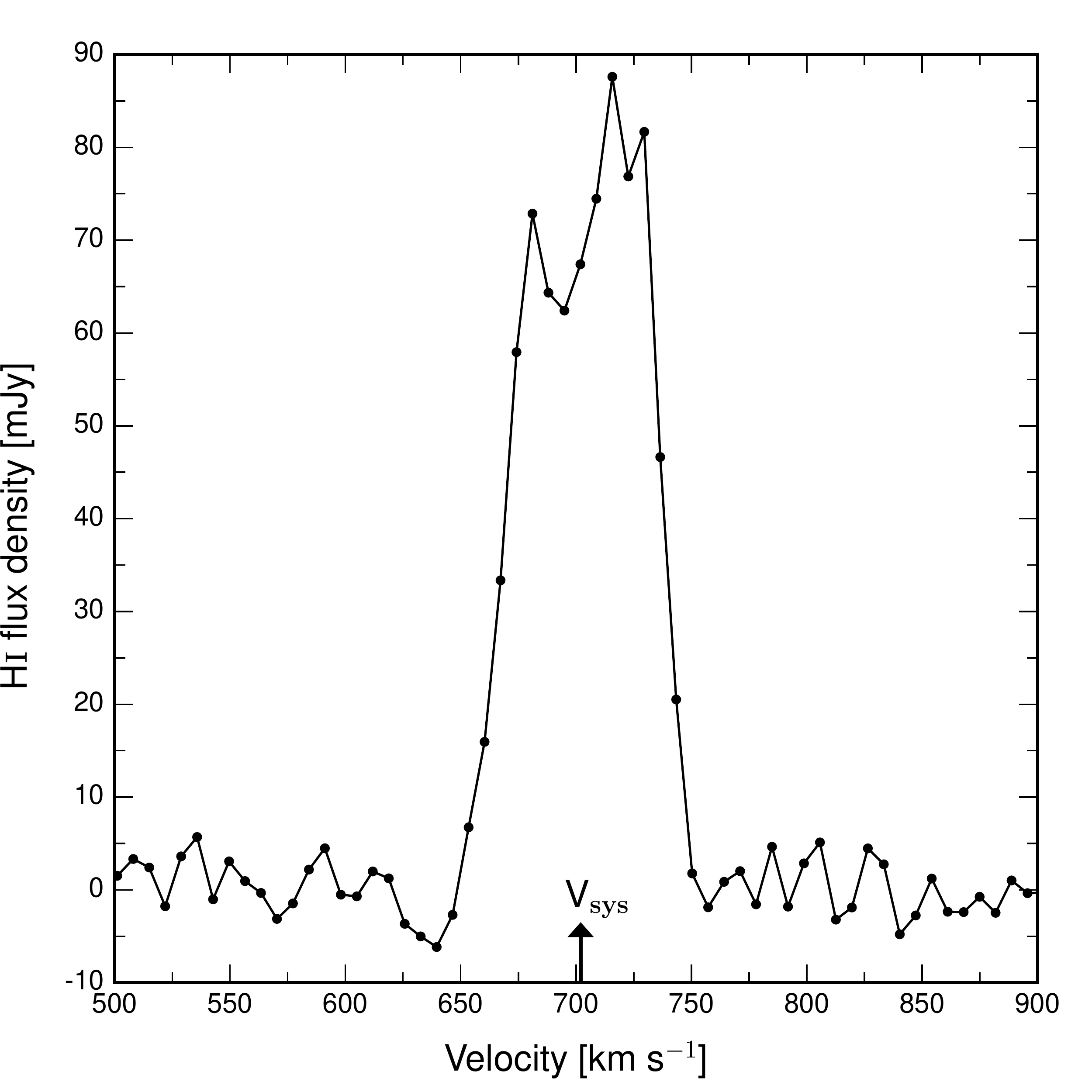}}} 
\put(1.2,10.6){(e)} 
\put(8.66,4.08){\hbox{\includegraphics[scale=0.287]{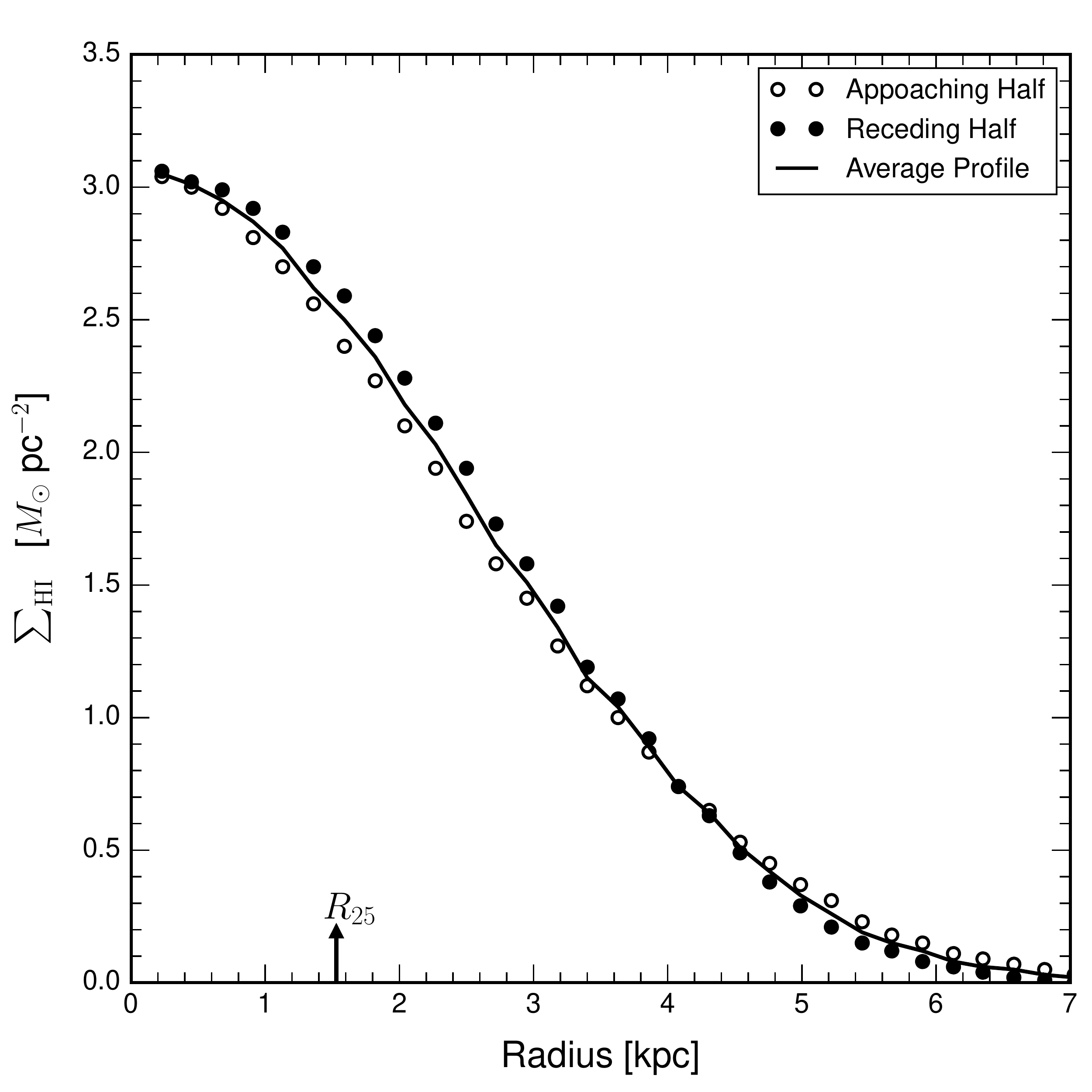}}} 
\put(9.8,10.6){(f)} 
\end{picture} 
\vspace{-4.4cm}
\caption{(a) The low resolution \HI column density contours of MCG~+10-18-044 overlaid upon its grey scale optical $r$-band image. The contour levels are $3.6 \times n$, where $n=1,2,4,8,16,32$ in units of $10^{19}$~cm$^{-2}$. (b) The intermediate resolution \HI column density contours overlaid upon the grey scale optical $r$-band image. The contour levels are $13.0 \times n$ in units of $10^{19}$~cm$^{-2}$. (c) The high resolution \HI column density contours overlaid upon the grey scale \Ha line image. The contour levels are $65.1 \times n$ in units of $10^{19}$~cm$^{-2}$. (d) The intermediate resolution moment-1 map, showing the velocity field, with an overlying optical $r$-band outer contour. The circle at the bottom of each image is showing the synthesized beam. The average FWHM seeing during the optical observation was $\sim 2''.1$. (e) The global \HI profile obtained using the low resolution \HI images. The arrow at the abscissa shows the systemic \HI velocity. (f) The \HI mass surface density profile obtained using the low resolution \HI map. The arrow at the abscissa shows the $B$-band optical disk radius.}
\label{WR195}
\end{figure}

\begin{figure*}
\begin{center}
\includegraphics[angle=0,width=1\linewidth]{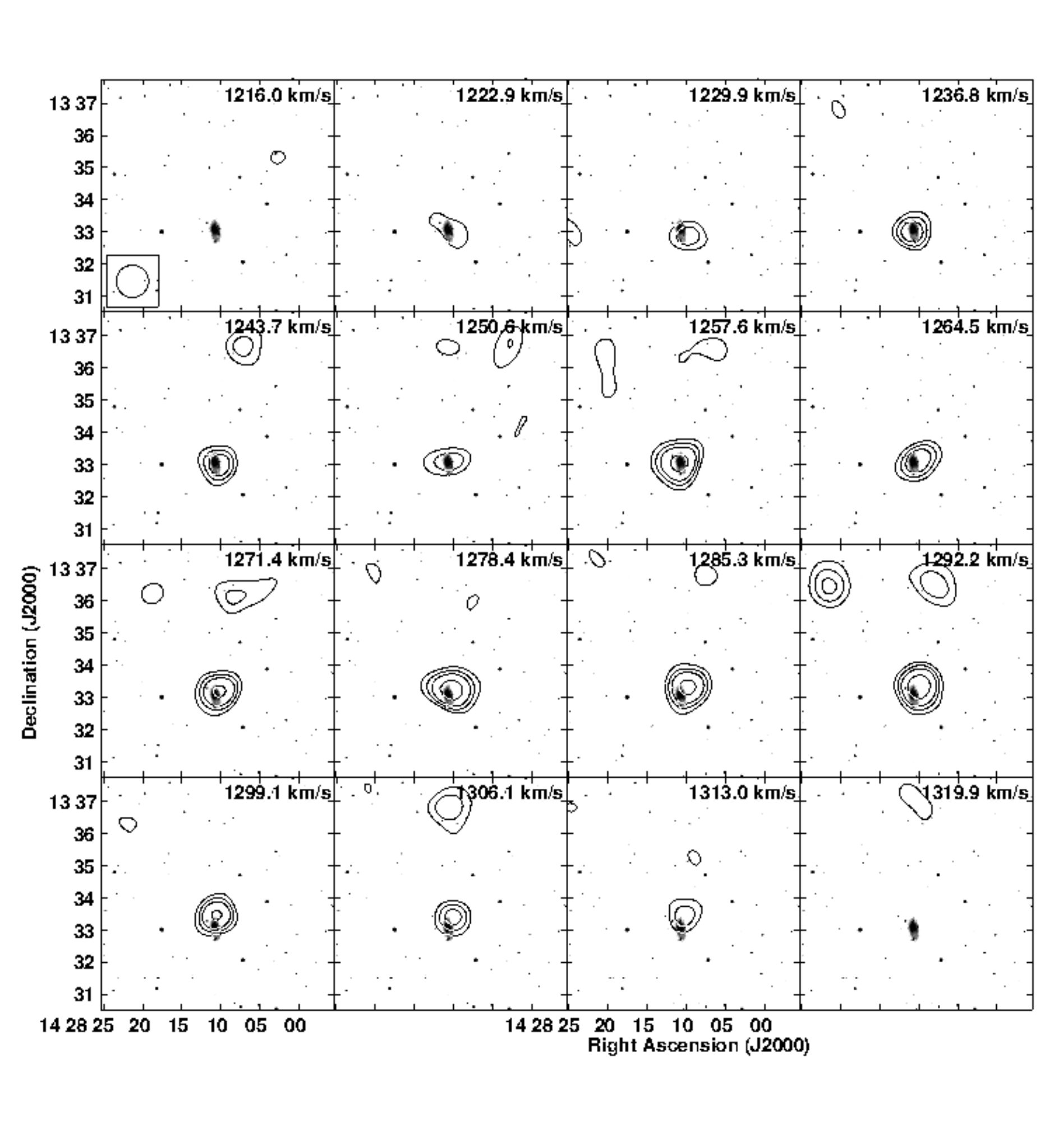}
\end{center}
\caption{The \HI contours from the low resolution channel images overlaid upon the grey scale optical $r$-band image of UGC~9273. The contours representing \HI emission flux are drawn at $2.6\sigma\times n$~mJy/Beam; n=1,1.5,2,3,4,6.}
\end{figure*}
\begin{figure}
\setlength{\unitlength}{1cm}
\begin{picture}(12,25) 
\put(0,17.8){\hbox{\includegraphics[scale=0.39]{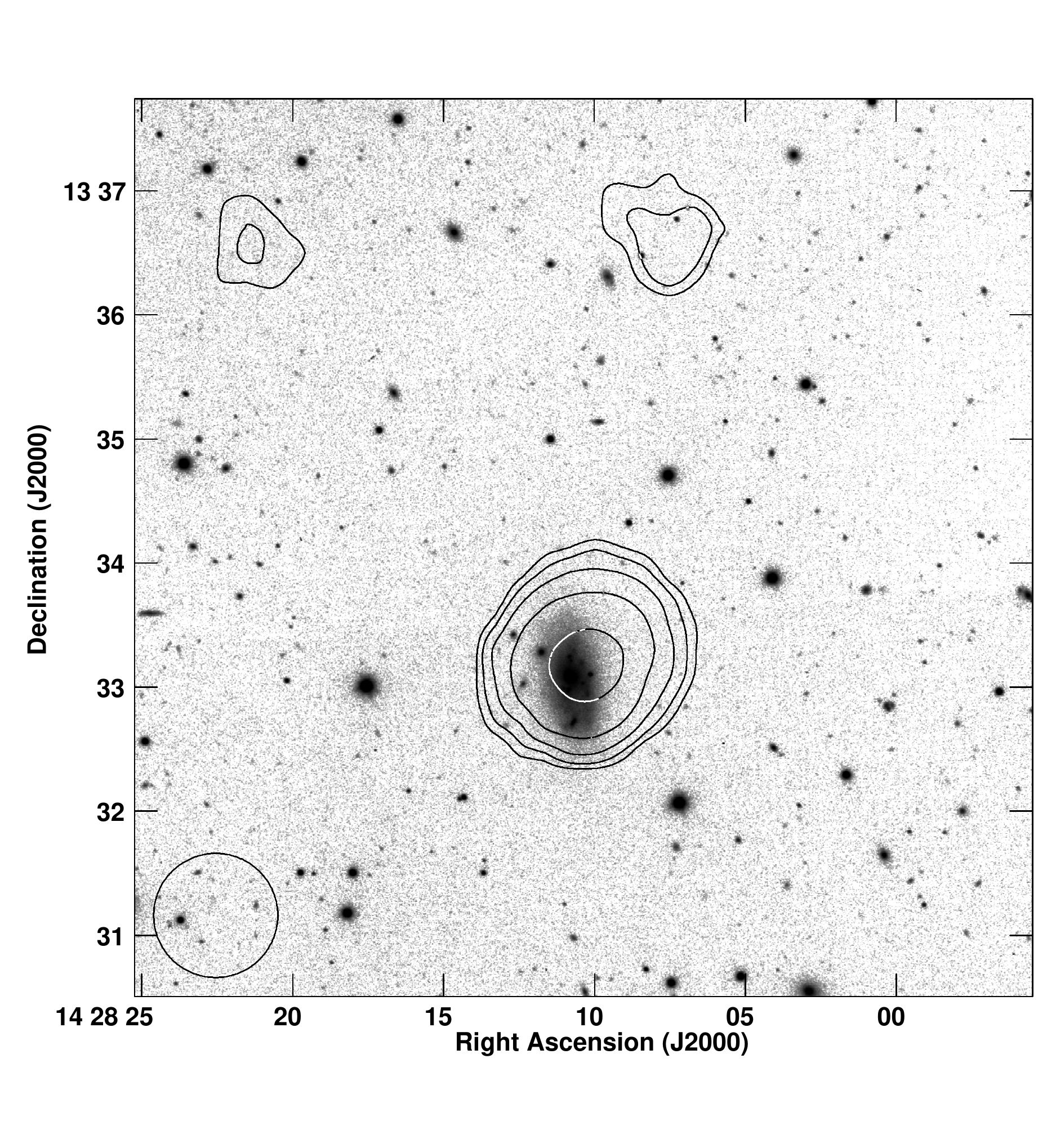}}} 
\put(1.2,24.8){(a)} 
\put(8.3,17.75){\hbox{\includegraphics[scale=0.408]{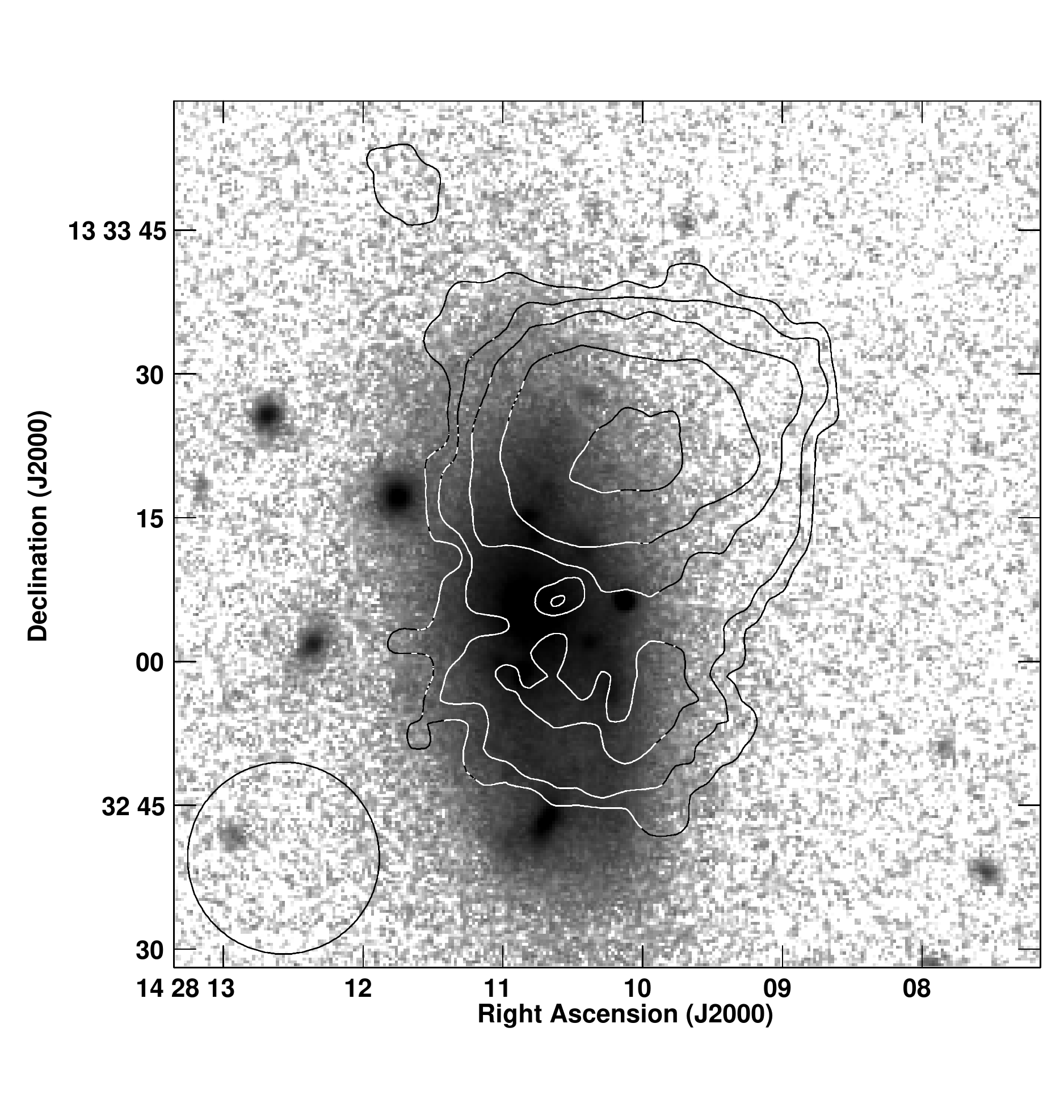}}}
\put(9.8,24.8){(b)} 
\put(-0.3,10.81){\hbox{\includegraphics[scale=0.405]{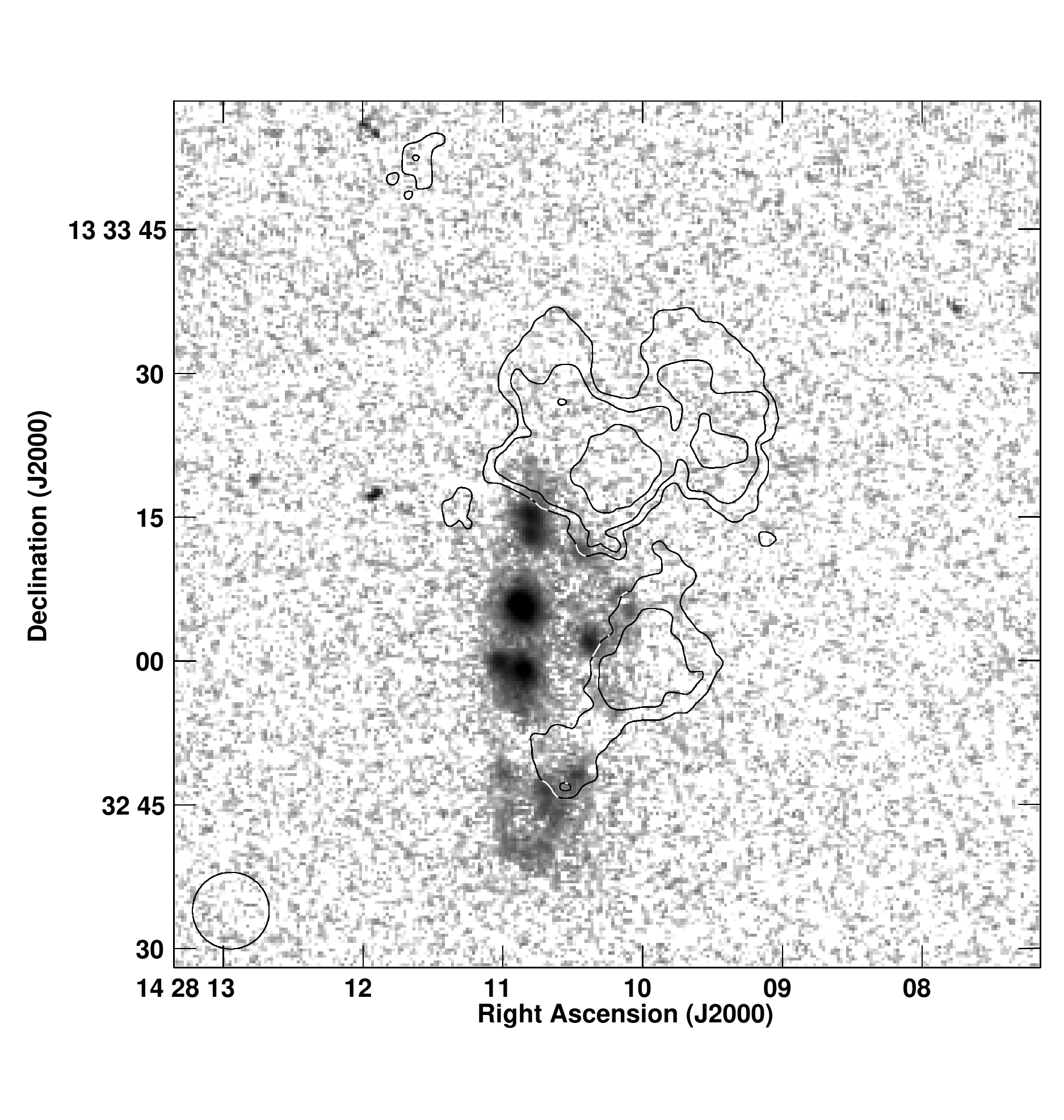}}}
\put(1.2,17.8){(c)} 
\put(8.03,10.98){\hbox{\includegraphics[scale=0.61]{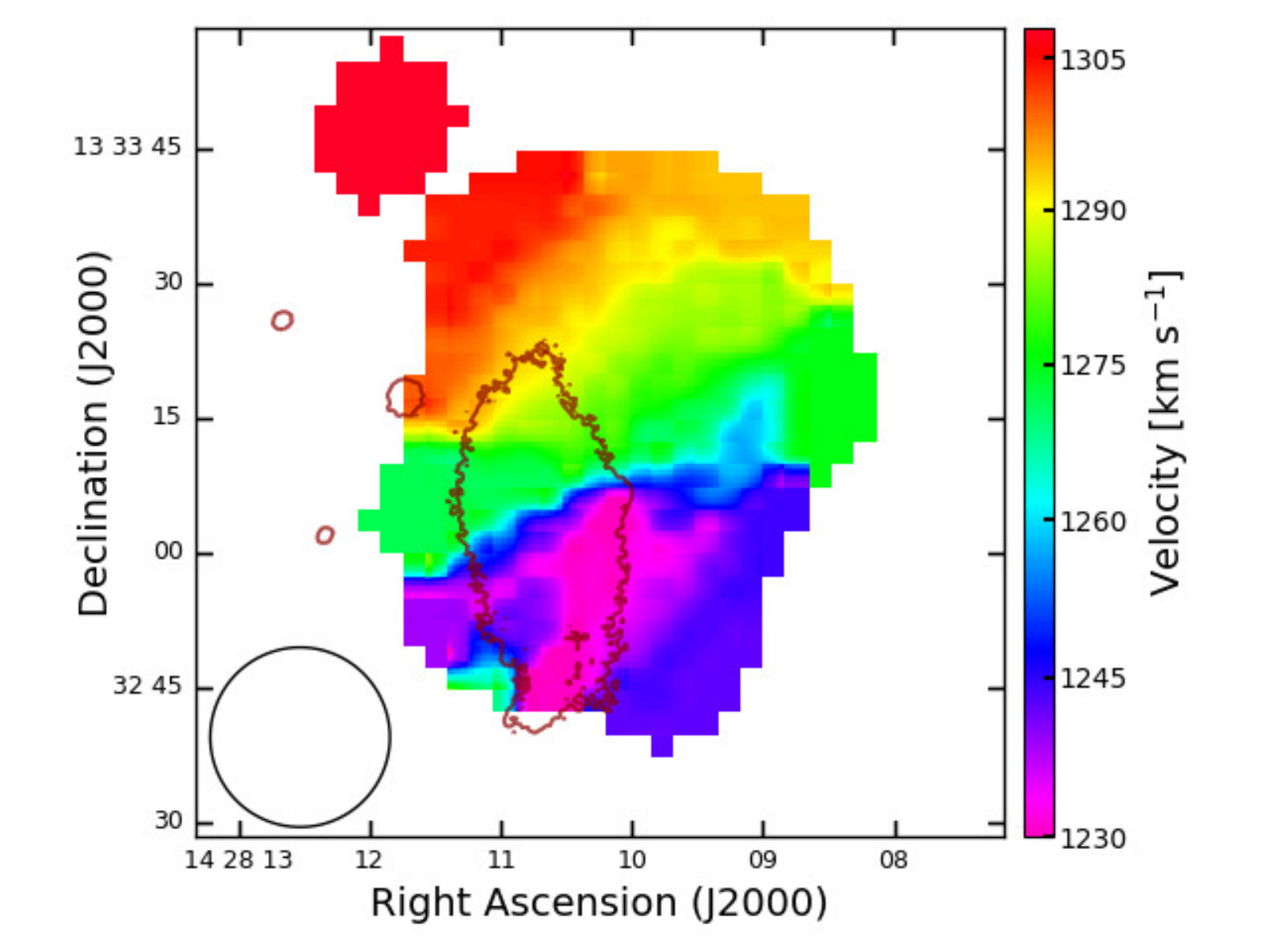}}}
\put(9.8,17.8){(d)} 
\put(0.21,3.8){\hbox{\includegraphics[scale=0.291]{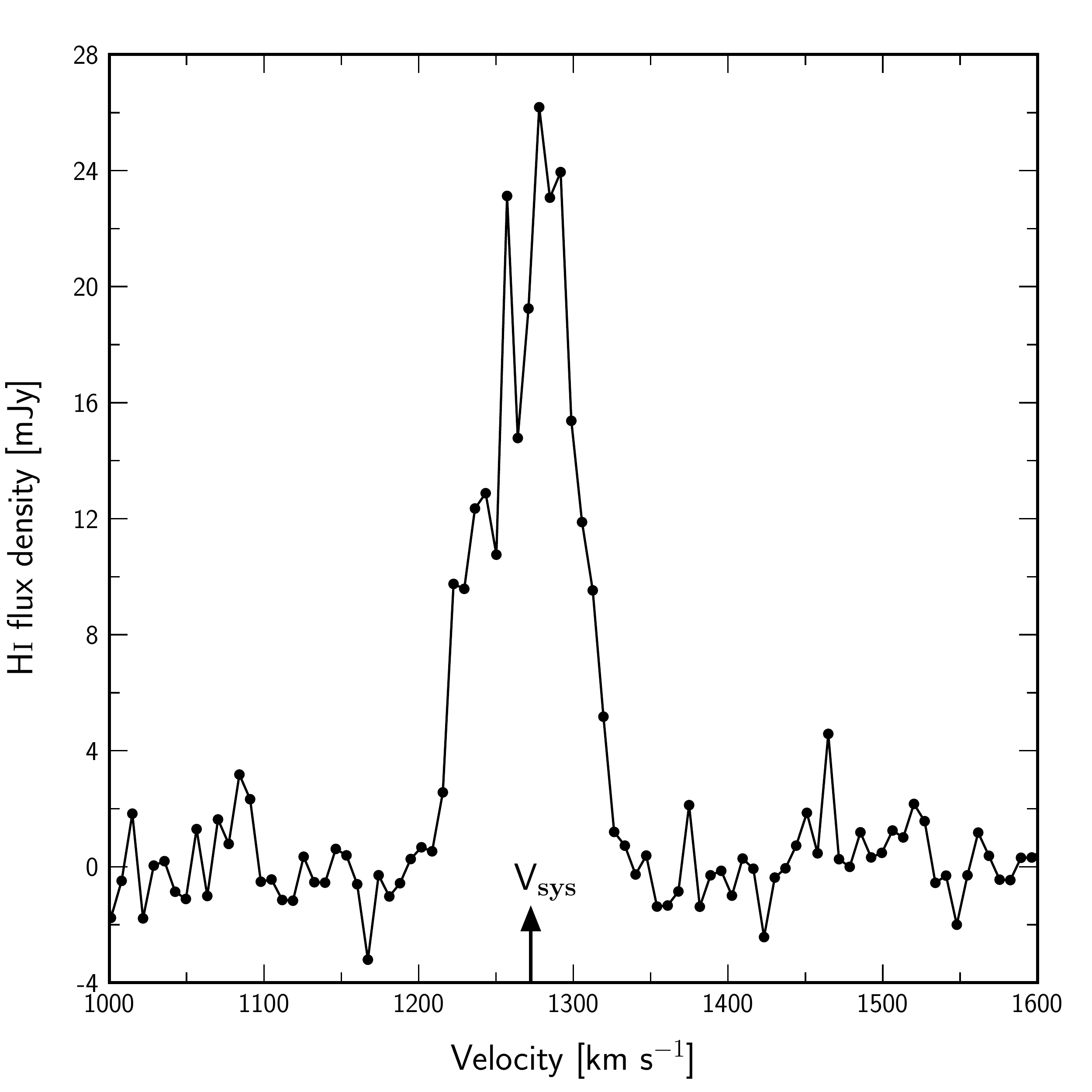}}} 
\put(1.2,10.6){(e)} 
\put(8.65,3.8){\hbox{\includegraphics[scale=0.291]{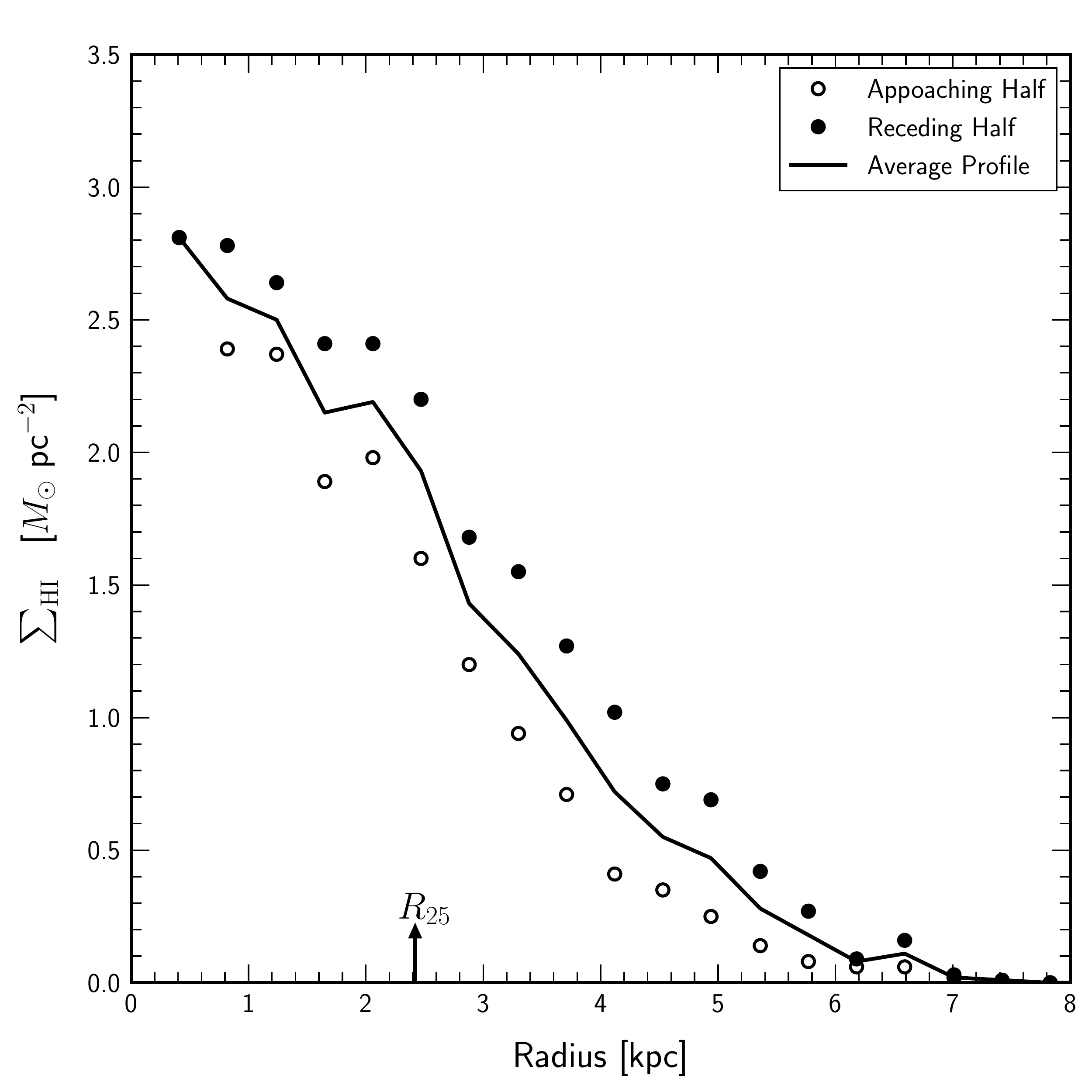}}} 
\put(9.8,10.6){(f)} 
\end{picture} 
\vspace{-4.2cm}
\caption{(a) The low resolution \HI column density contours of UGC~9273 overlaid upon its grey scale optical $r$-band image. The contour levels are $1.9 \times n$, where $n=1,2,4,8,16,32$ in units of $10^{19}$~cm$^{-2}$. (b) The intermediate resolution \HI column density contours overlaid upon the grey scale optical $r$-band image. The contour levels are $4.9 \times n$ in units of $10^{19}$~cm$^{-2}$. (c) The high resolution \HI column density contours overlaid upon the grey scale \Ha line image. The contour levels are $20.9 \times n$ in units of $10^{19}$~cm$^{-2}$. (d) The intermediate resolution moment-1 map, showing the velocity field, with an overlying optical $r$-band outer contour. The circle at the bottom of each image is showing the synthesized beam. The average FWHM seeing during the optical observation was $\sim 1''.2$. (e) The global \HI profile obtained using the low resolution \HI images. The arrow at the abscissa shows the systemic \HI velocity. (f) The \HI mass surface density profile obtained using the low resolution \HI map. The arrow at the abscissa denotes the $B$-band optical disk radius.}
\label{WR378}
\end{figure}

\begin{figure*}
\begin{center}
\includegraphics[angle=0,width=1\linewidth]{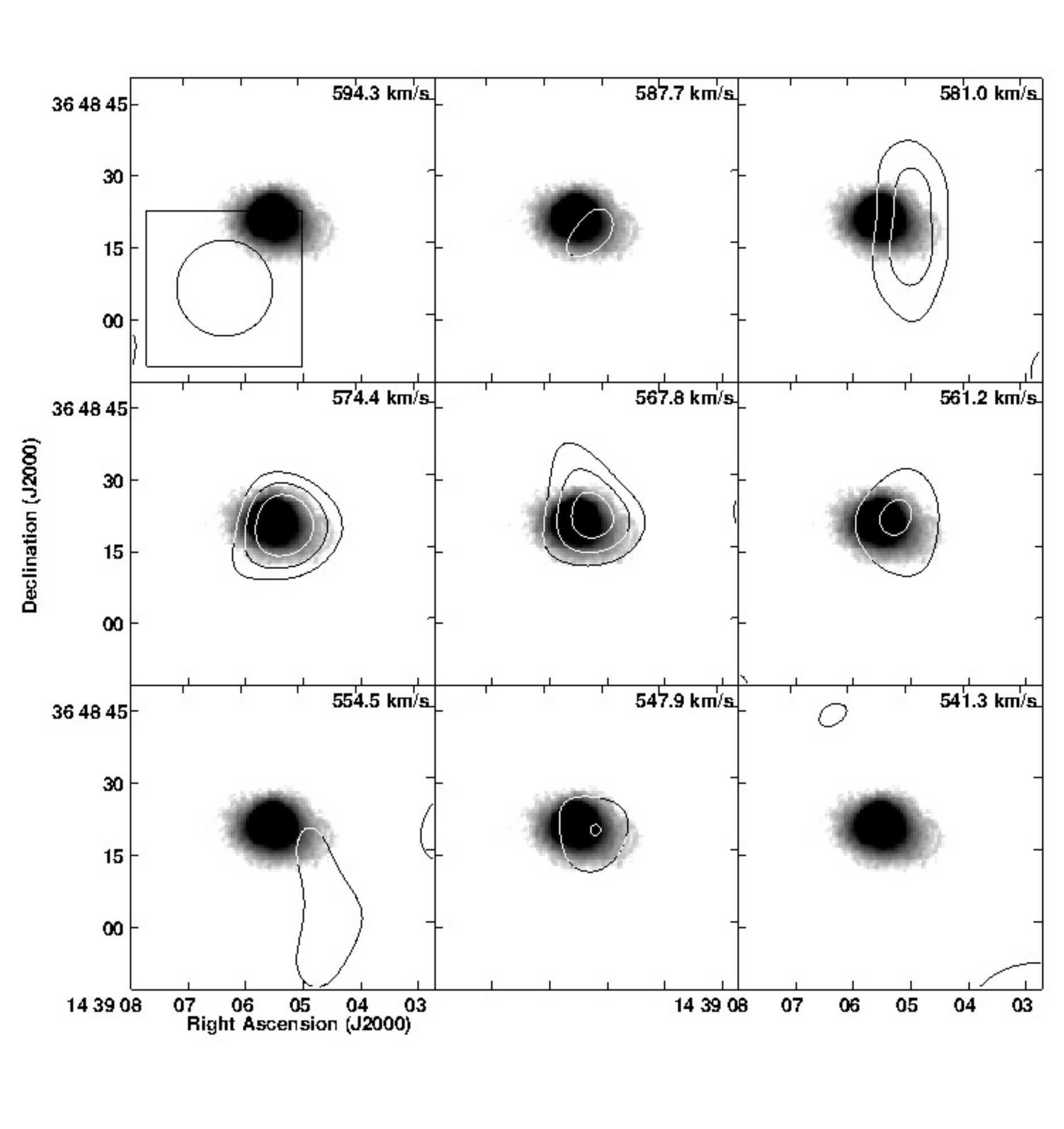}
\end{center}
\caption{The \HI contours from the intermediate resolution channel images overlaid upon the grey scale optical $r$-band image of MRK~475. The contours representing \HI emission flux are drawn at $2.7\sigma\times n$~mJy/Beam; n=1,1.5,2,3,4,6.}
\label{MRK475-1}
\end{figure*}
\begin{figure}
\setlength{\unitlength}{1cm}
\begin{picture}(12,25) 
\put(0,17.9){\hbox{\includegraphics[scale=0.40]{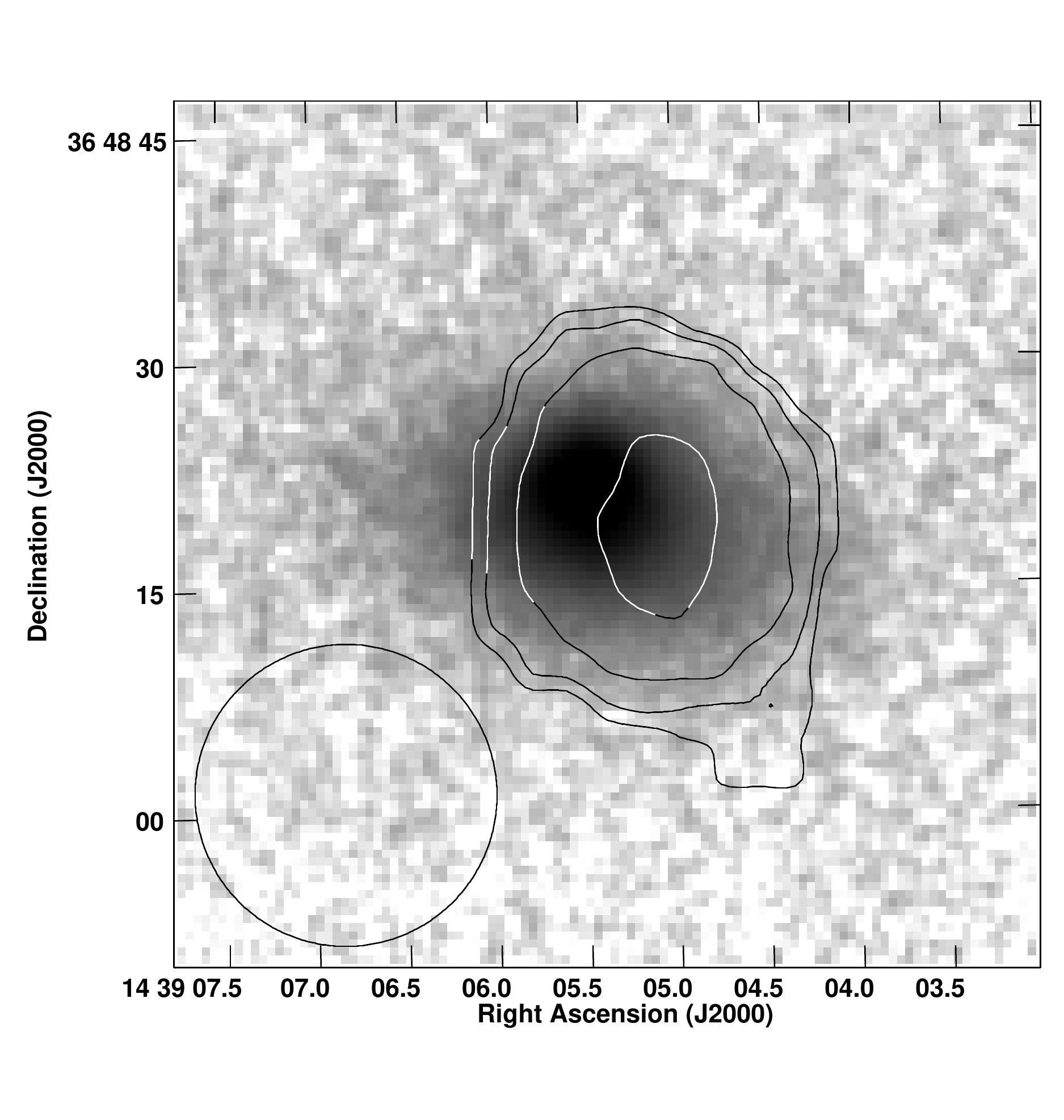}}} 
\put(1.5,24.7){(a)} 
\put(8.3,17.9){\hbox{\includegraphics[scale=0.40]{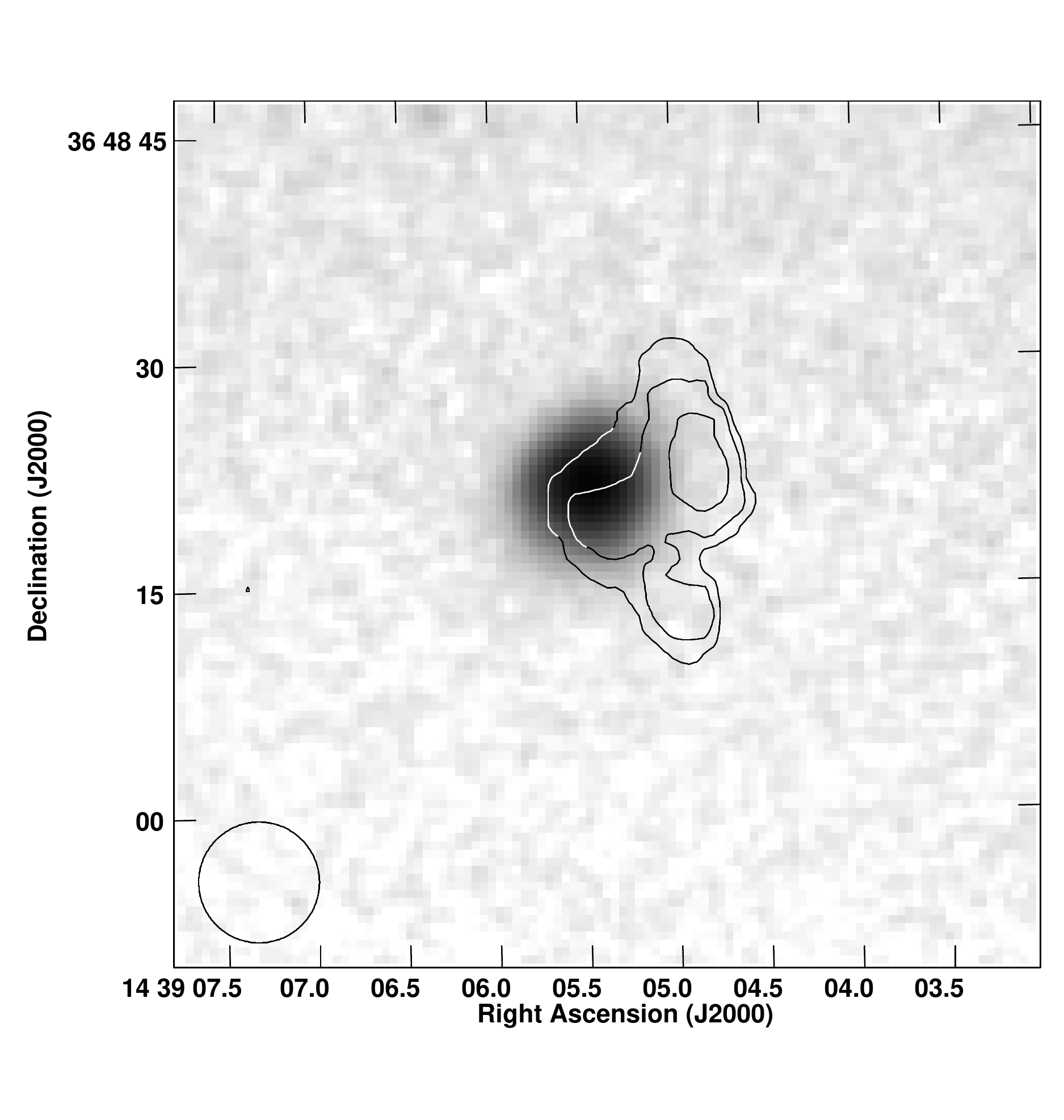}}}
\put(9.8,24.7){(b)} 
\put(-0.32,11.1){\hbox{\includegraphics[scale=0.596]{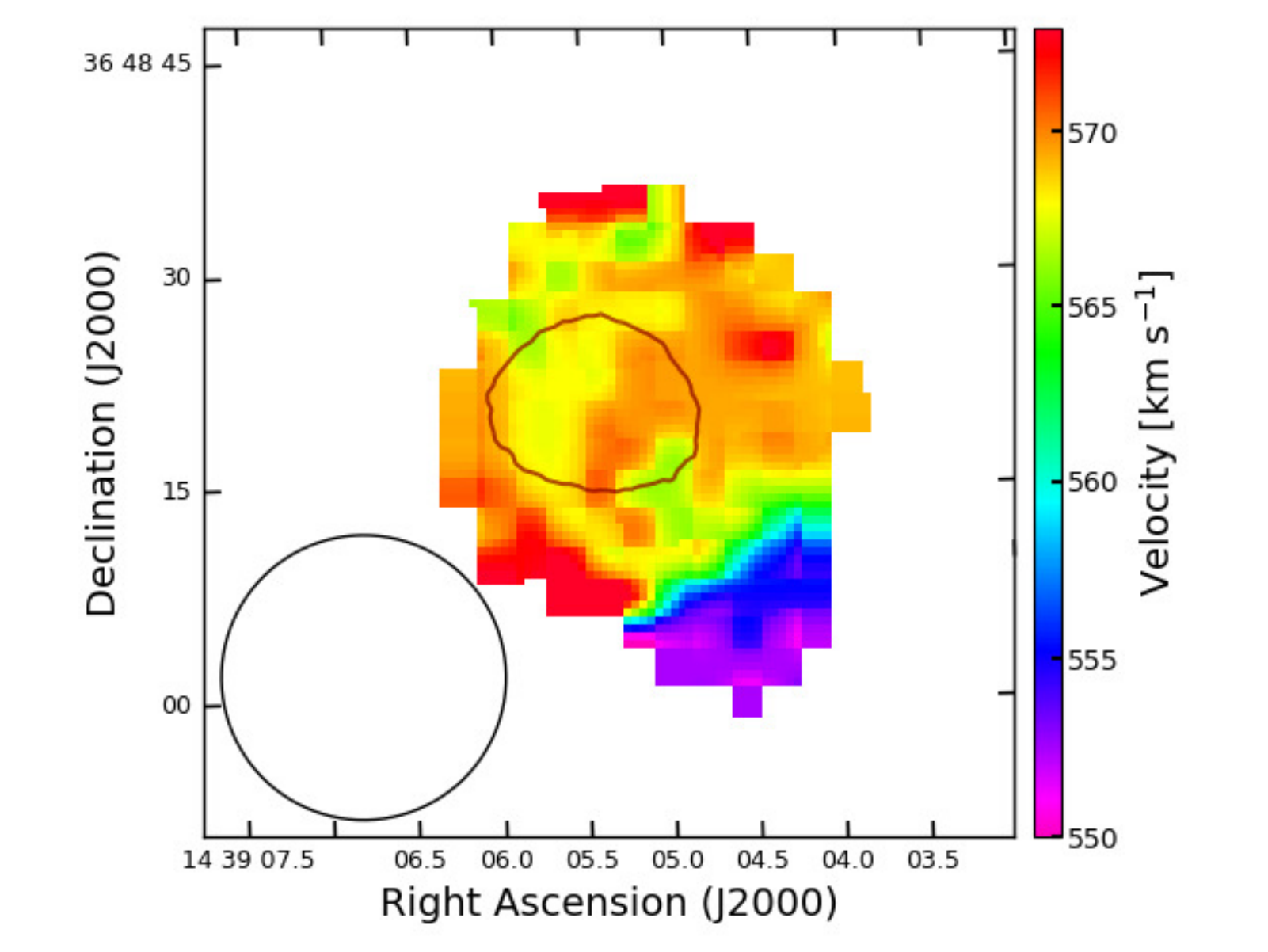}}}
\put(1.5,17.75){(c)} 
\put(8.8,11.2){\hbox{\includegraphics[scale=0.288]{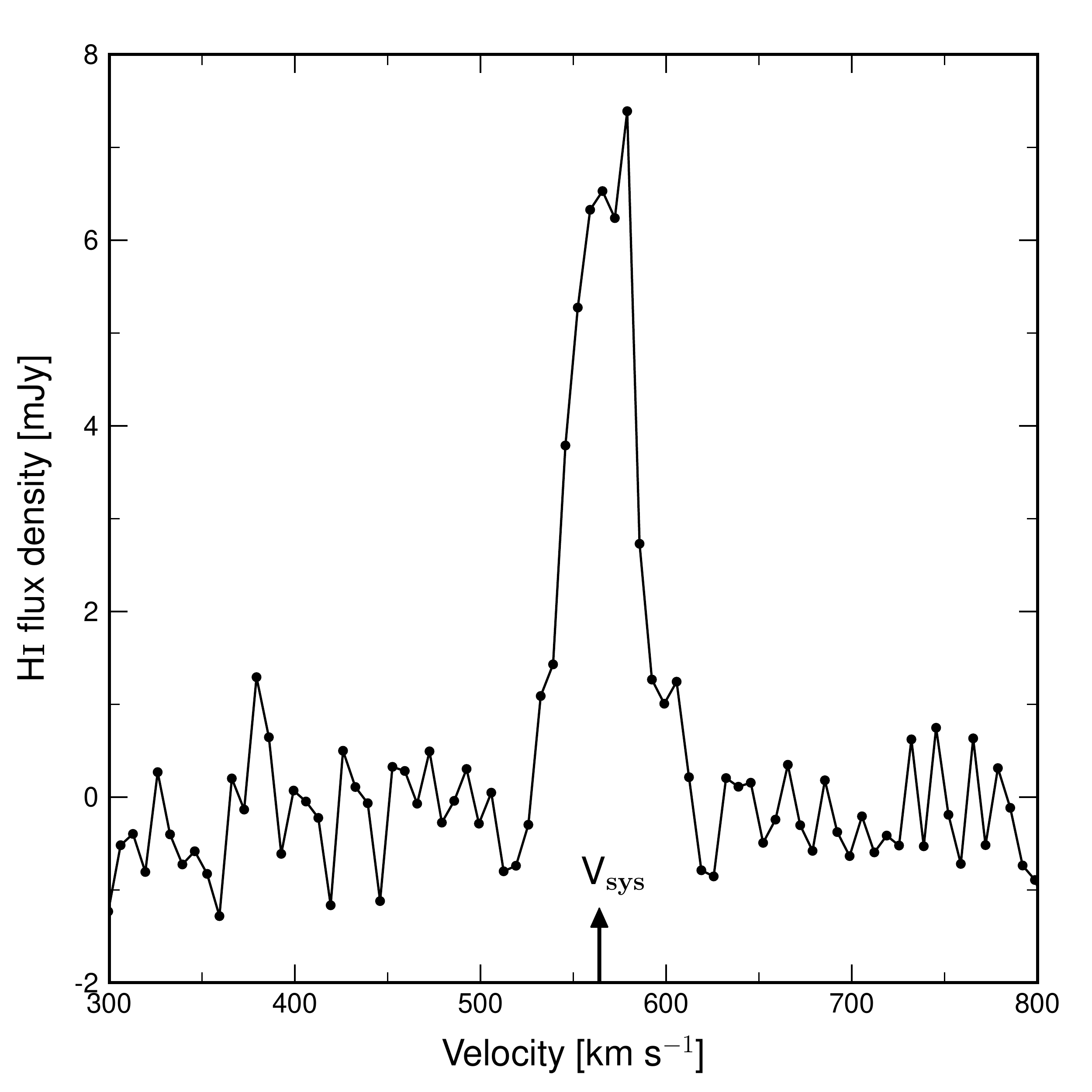}}}
\put(9.8,17.75){(d)} 
\put(0.4,4.0){\hbox{\includegraphics[scale=0.288]{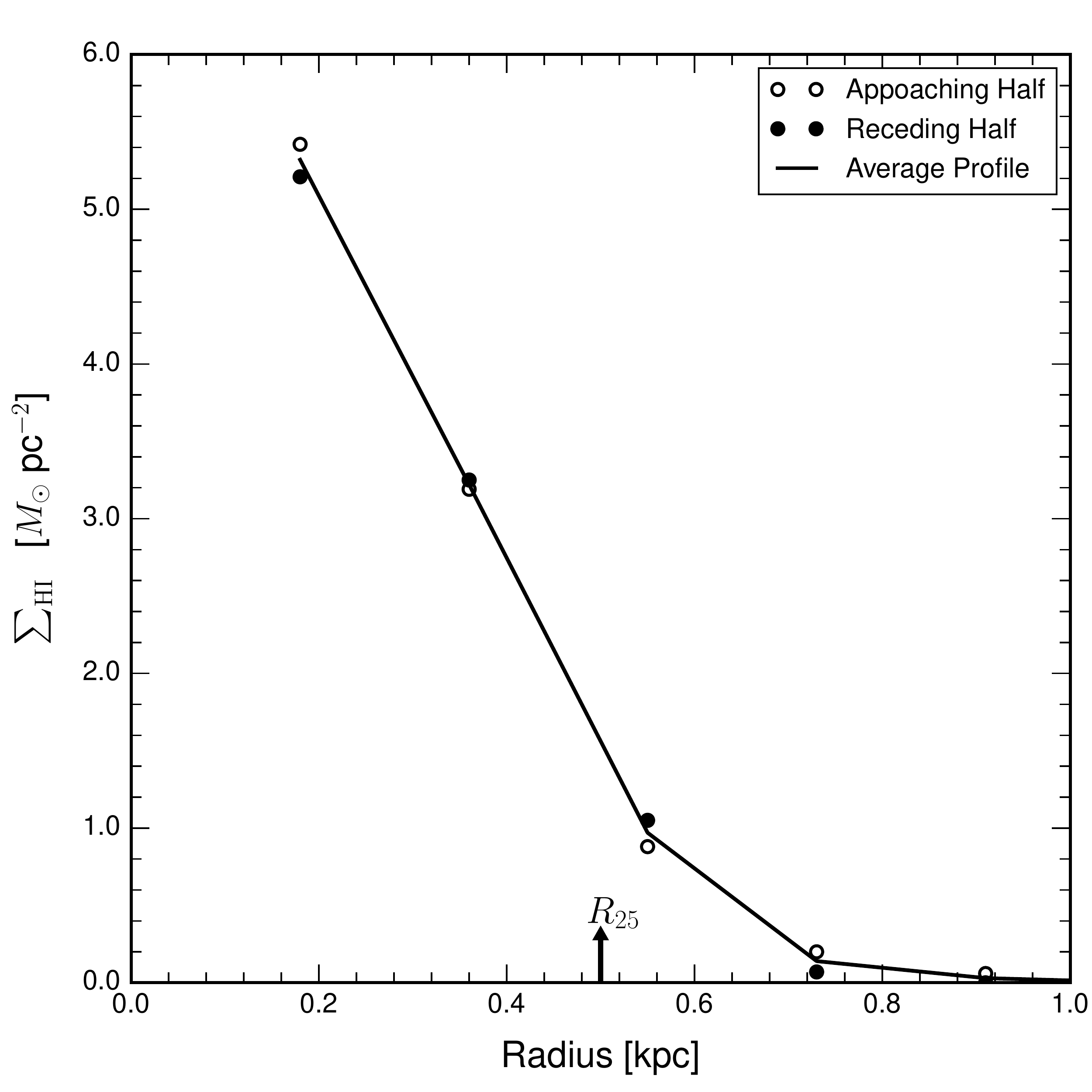}}} 
\put(1.5,10.55){(e)} 
\end{picture} 
\vspace{-4.1cm}
\caption{(a) The intermediate resolution \HI column density contours of MRK~475 overlaid upon its grey scale optical $r$-band image. The contour levels are $6.1 \times n$, where $n=1,2,4,8,16,32$ in units of $10^{19}$~cm$^{-2}$. (b) The high resolution \HI column density contours overlaid upon the grey scale \Ha line image. The contour levels are $42.7 \times n$ in units of $10^{19}$~cm$^{-2}$. (c) The intermediate resolution moment-1 map, showing the velocity field, with an overlying optical $r$-band outer contour. The circle at the bottom of each image is showing the synthesized beam. The average FWHM seeing during the optical observation was $\sim 2''.2$. (d) The global \HI profile obtained using the intermediate resolution \HI images. The arrow at the abscissa denotes the systemic \HI velocity. (e) The \HI mass surface density profile obtained using the intermediate resolution \HI map. The arrow at the abscissa denotes the $B$-band optical disk radius.}
\label{MRK475-2}
\end{figure}


\end{document}